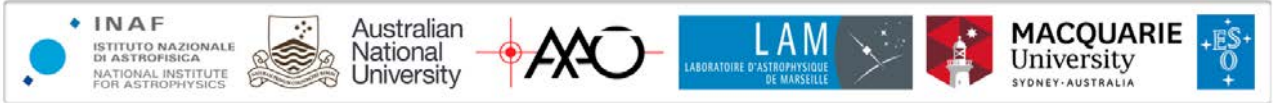

# MAVIS

# Phase A Science Case

Public Release v1.0, September 19th 2020

# List of Contributors

The MAVIS Phase A Science Case was prepared by the MAVIS Consortium Core Science Team, together with key co-authors from the MAVIS community. The case comprises a distilled set of thematically linked science cases drawn from the ***MAVIS White Papers*** submitted by the astronomy community in 2018 (published via the MAVIS blog: www.mavis-ao.org/whitepapers), and subsequently refined during the Phase A study via community science workshops (see http://mavis-ao.org/Workshops/). Specific papers used are highlighted below, selected to illustrate the driving requirements of the instrument. However, *all* the original White Papers present important and innovative science enabled by MAVIS, and provide a crucial benchmark for demonstrating the compliance of the instrument design with the community's science goals. We refer the reader to the links above for further information.

The MAVIS Project gratefully acknowledges the contributions of the many researchers from across the international community who have engaged with the MAVIS Project to date, and whose names and institutions are listed on the following page.

**MAVIS Science Case Prepared by:** Richard McDermid, Giovanni Cresci, Jean-Claude Bouret, Gayandhi De Silva, Marco Gullieuszik, Laura Magrini, Trevor Mendel, Francois Rigaut, Simone Antoniucci, Giuseppe Bono, Devika Kamath, Stephanie Monty, Holger Baumgardt, Luca Cortese, Deanne Fisher, Filippo Mannucci, Alessandra Migliorini, Sarah Sweet, Eros Vanzella, Stefano Zibetti

**White Papers featured:**

"*Mass accretion in low-metallicity environments of the outer Galaxy*", S. Antoniucci et al.

"*Frequency of accreting sub-stellar companions of low-mass stars*", S. Antoniucci et al.

"*Jets from young stars at high angular resolution: the launching mechanism as a solution to the angular momentum problem*", F. Bacciotti et al.

"*Resolving the physics of ram-pressure stripping*", L. Cortese et al.

"*Dynamical measurements of supermassive black holes and nuclear star clusters*", E. Dalla Bonta et al.

"*Probing the cores of globular clusters: Abundance anomalies on the unevolved main sequence stars and impact of binarity*", G. De Silva et al.

"*Probing stellar clumps in distant star-forming galaxies around z~1*", M. Dessauges-Zavadsky et al.

"*Resolving Super-Massive Black Holes in Compact, Low-Mass Galaxies*", A. Ferre-Mateu

"*Extreme Star Forming Regions in Galaxies at Cosmic Noon and the Nearby Universe*", D. Fisher

"*Star-forming clumps*", M. Gullieuszik

"*Spectroscopy of distant Cepheids with MAVIS: A detailed chemical mapping of spiral galaxies in the Local Universe*", L. Inno et al.

"*Tracing the chemical evolution of nearby galaxies with star clusters*", L. Magrini et al.

"*A MAVIS deep field*", F. Mannucci et al.

"*Resolved stellar populations in distant galaxies*", R. McDermid

"*Evolution of the Interstellar Medium with MAVIS*", T. Mendel et al.

"*Exploiting strong lensing clusters with MAVIS*", P. Rosati et al.

"*Star and planet formation at low metallicities*", G. Sacco et al.

"*Young Massive Clusters in Metal-Poor Starburst Dwarf Galaxies*", R. Sanchez-Janssen et al.

"*Direct detection of UV-upturn hot, old stars in nearby ETGs and bulges*", S. di Serego Alighieri et al.

"*Resolved Angular Momentum with MAVIS: towards a physical morphological classification*". S. Sweet et al.

"*Resolved stellar populations and star formation histories in the Local Universe: Star-forming dwarf galaxies in the field and in clusters*", M. Tosi et al.

"*Physical properties and origin of classical bulges & pseudo-bulges in the local Universe*", S. Zibetti et al.

**Additional Contributors:** Angela Adamo, Alberto Adriani, Guido Agapito, Ricardo Amorin, Francesca Annibali, Maryam Arabsalmani, Carmelo Arcidiacono, Francesca Bacciotti, Gabriel Bartosch Caminha, Stephanie Basa, Rob Bassett, Giuseppina Battaglia, Luigi Bedin, Michele Bellazzini, Andrea Bellini, Daniela Bettoni, Katia Biazzo, Samuel Boissier, Dylan Bollen, Alessandro Boselli, Veronique Buat, Lucio Buson, Nell Byler, Francesco Calura, Fabrizio Capaccioni, Luca Casagrande, Giada Casali, Barbara Catinella, Laure Ciesla, Michele Cignoni, Fraser Clarke, Claudio Codella, Edvige Corbelli, David Corre, Enrico Corsini, Luca Costantin, Maria Cristina De Sanctis, Virginia Cuomo, Valentina D'Orazi, Elena Dalla Bonta, Ric Davies, Richard de Grijs, Orsola De Marco, Mirka Dessauges-Zavadsky, Sperello di Serego Alighieri, Simon Driver, Simon Ellis, Benoit Epinat, Steve Ertel, Renato Falomo, Davide Fedele, Andrea Ferrara, Anna Ferre-Mateu, Gianrico Filacchione, Giuliana Fiorentino, Duncan Forbes, Elena Franciosini, Ken Freeman, Johan Fynbo, Carme Gallart, Anna Gallazzi, Teresa Giannini, Karl Glazebrook, Davide Grassi, Andrea Grazian, Laura Greggio, Claudio Grillo, Marco Grossi, Brent Groves, Leslie Hunt, Laura Inno, Michael Ireland, Colin Jacobs, Helmut Jerjen, Erkki Kankare, Lisa Kewley, Jacques Kluska, Rubina Kotak, Bertrande Lemasle, Kieran Leschinski, Chengyuan Li, Mattia Libralato, Juan Manuel Alcalá, Dan Maoz, Antonio Marasco, Alessandro Marconi, Elena Masciadri, Davide Massari, Fabrizio Massi, Seppo Mattila, Lucio Mayer, Nicola Menci, Massimo Meneghetti, Amata Mercurio, Matteo Monelli, Lorenzo Morelli, Alessia Moretti, Jeremy Mould, Brunella Nisini, Matthew O'Dowd, Tino Oliva, Ronaldo Oliveira da Silva, Ilari Pagotto, Elena Pancino, Vera Patricio, Michele Perna, Celine Peroux, Giuseppe Piccioni, Alessandro Pizzella, Linda Podio, Bianca Poggianti, Elisa Portaluri, Sofia Randich, Veronica Roccatagliata, Donatella Romano, Piero Rosati, Michela Rubino, Stuart Ryder, Germano Sacco, Stefania Salvadori, Jorge Sanchez Almeida, Ruben Sanchez-Janssen, Nicoletta Sanna, Daniel Schaerer, Nick Seymour, Mojtaba Taheri, Valentina Tamburello, Maria Teresa Capria, Eline Tolstoy, Crescenzo Tortora, Monica Tosi, Andrea Tozzi, Paolo Tozzi, Alessio Turchi, Jesse van de Sande, Mathieu van der Swaelmen, Schuyler van Dyk, Hans Van Winckel, Benedetta Vulcani, Frederic Vogt, Rachel Webster, Emily Wisnioski, Tayyaba Zafar, Anita Zanella

**Contributing Institutions**: Australian Astronomical Optics (AAO) - Macquarie University; Australian National University; Astronomisches Rechen-Institut - Heidelberg University; California Institute of Technology; CEA Saclay; City University of New York (CUNY); Curtin University; ETH Zurich; Federal University Rio de Janeiro; INAF Arcetri Astrophysical Observatory; INAF Astronomical Observatory of Padova; INAF Astronomical Observatory of Rome; INAF Astrophysics and Space Science Observatory of Bologna; INAF Capodimonte Astronomical Observatory; INAF Catania Astrophysical Observatory; INAF Istituto di Astrofisica e Planetologia Spaziali (IAPS); Institut de Ciencies del Cosmos Universitat de Barcelona; Instituto de Astrofisica de Canarias (IAC); International Centre for Radio Astronomy Research (ICRAR); Kapteyn Institute University of Groningen; Katholieke Universiteit Leuven; Kavli Institute for Cosmology University of Cambridge; Laboratoire d'Astrophysique de Marseille (LAM); Macquarie University Research Centre for Astronomy, Astrophysics and Astrophotonics (MQAAAstro); Max Planck Institute for Astrophysics; Max Planck Institute for Extraterrestrial Physics; MeteoSwiss; Niels Bohr Institute University of Copenhagen; NRC-Hertzberg; Oskar Klein Centre University of Stockholm; Oxford University; Queen's University Belfast; Space Telescope Science Institute (STScI); Swinburne University of Technology (SUT); Tel Aviv University; UK Astronomy Technology Centre; University of Arizona; University of Ferrara; University of Geneva; University of Melbourne; University of Milan; University of Padova; University of Pisa; University of Sydney; University of Turku; University of Vienna; University of Western Australia; University of Zurich.

# TABLE OF CONTENTS

# 1 Introduction

## 1.1 MAVIS Instrument Overview

The Multi-conjugate Adaptive-optics Visible Imager-Spectrograph (MAVIS) operating on the Adaptive Optics Facility (AOF) of the Very Large Telescope (VLT) is a general-purpose instrument for exploiting the highest possible angular resolution of any single optical telescope available in the next decade, either on Earth or in space, and with sensitivity comparable to (or better than) larger aperture facilities. MAVIS comprises three principal modules (see Rigaut et al. *in prep.* for a technical description of MAVIS):

- An **Adaptive Optics Module** (AOM), responsible for wavefront sensing and correction. This features an innovative transmissive design, which includes two post focal deformable mirrors totalling about 4,250 actuators (5,420 including the deformable secondary mirror of the AOF), eight 40x40 Shack-Hartmann LGS wavefront sensors, and three near-infrared natural guide star wavefront sensors, which also provide low order truth sensing and slow focus corrections.

- An **imager module**, Nyquist sampling the near-diffraction limited optical beam over a 30"x30" field of view. The imager is equipped with a range of broad- and narrow-band filters, with throughput maximised via minimal re-imaging optics following the AOM.

- An image-slicing **integral field spectrograph (IFU) module**, with flexible spatial and spectral configurations. Exchangeable fore-optics and dispersing elements provide two spatial modes (0.025" and 0.050" square spaxels, covering respectively 3.6"x2.5" and 7.2"x5" fields of view) each of which have four spectral modes, covering 370nm-1μm at resolutions from 5,000-15,000.

The instrument specifications are summarised in Table 1.1, with the top-level requirements and science driver traceability presented in Section 6. Sensitivity estimates are documented in Appendix A.

The MAVIS technical field, from which Natural Guide Stars (NGS) may be selected, covers 120" (see Fig. 1.1), and benefits from some off-axis correction from the post-focal deformable mirrors, enabling at least 50% sky coverage at the Galactic pole. MAVIS will nominally deliver a mean V-band Strehl ratio of >10% (goal >15%), varying less than 10% across the 30"x30" science field. This equates to FWHM-like resolution of <20mas at 550nm. Even in the limiting sky coverage case, MAVIS will ensure >15% of the transmitted V-band light falls within a 50mas square aperture.

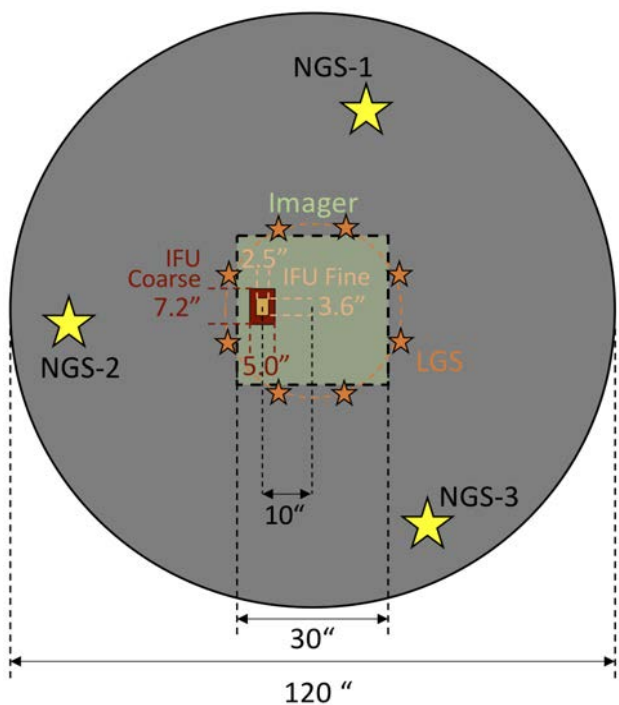

**Figure 1.1:** *MAVIS fields of view, showing the technical (grey), imaging (green), and offset spectrograph/IFU fields (red/peach). Orange stars and the dashed circle indicate LGS positions and their track with field rotation, respectively. As the NGS are sensed in the near infrared, NGS selection does not obstruct the science field and does not impact on transmission in the visible science band. The Laser Guide Star (LGS) asterism is located at a fixed radius of 17.5", and comprises 8 individual beacons. These are generated by splitting the beams from the 4 existing AOF LGS units, and whose light is blocked from the science beam via a notch filter. The off-centred IFU provides additional flexibility in guide star selection by rotating the science field position angle.*

**Table 1.1:** *An overview of MAVIS characteristics. Performance numbers are for the Paranal standard conditions, seeing=0.8"@500nm, zenith angle=30 degrees (effective seeing=0.87").*

| MAVIS General Properties | |
|---|---|
| **Focus** | Nasmyth A VLT-AOF (UT4) |
| **NGS Field of View** | 120" diameter circle |
| **Number of NGS** | Up to 3 |
| **Limiting magnitude** | Hmag ≥ 18.5 |
| **LGS beacons** | 8 |
| **Sky coverage** | ≥ 50% at Galactic pole |
| **Ensquared Energy** | ≥ 15% within 50mas at 550nm |
| **Strehl** | ≥ 10% (15% goal) in V-band, <10% variation (RMS) |

| MAVIS Imager | |
|---|---|
| **Field of View** | 30" x 30" |
| **Pixel Scale** | 7.4 mas/pix |
| **Filters** | UBVRI, ugriz, various narrow bands |
| **Sensitivity** | V ≥ 29mag (5σ) in 1hr |

| MAVIS Integral Field Spectrograph | | | |
|---|---|---|---|
| **IFU Spaxel Size and Field of View** | **Fine Sampling:** 20-25mas spaxels, 2.5" x 3.6" FoV | | |
| | **Coarse Sampling**: 40--50mas spaxels, 5.0" x 7.2" FoV | | |
| **Spectral Configurations** | **LR-BLUE** | **LR-RED** | **HR-BLUE** | **HR-RED** |
| **Median Resolution** ($\lambda/\lambda\Delta$) | 5,900 | 5,900 | 14,700 | 11,500 |
| **Wavelength** | 370-720nm | 510-1000nm | 425-550nm | 630-880nm |
| **Sensitivity** (ABmag, 10σ in 1hr, Fine Sampling) | 21@550nm | 21.5@750nm | 19.6@475nm | 20.7@725nm |

## 1.2 MAVIS Science Overview

By probing the frontier of angular resolution and sensitivity across a large portion of the observable sky, MAVIS will enable progress on an array of scientific topics, from our own planetary system to those around other stars, and from the physics of star formation in the Milky Way to the first star clusters forming in the early Universe (Fig. 1.2).

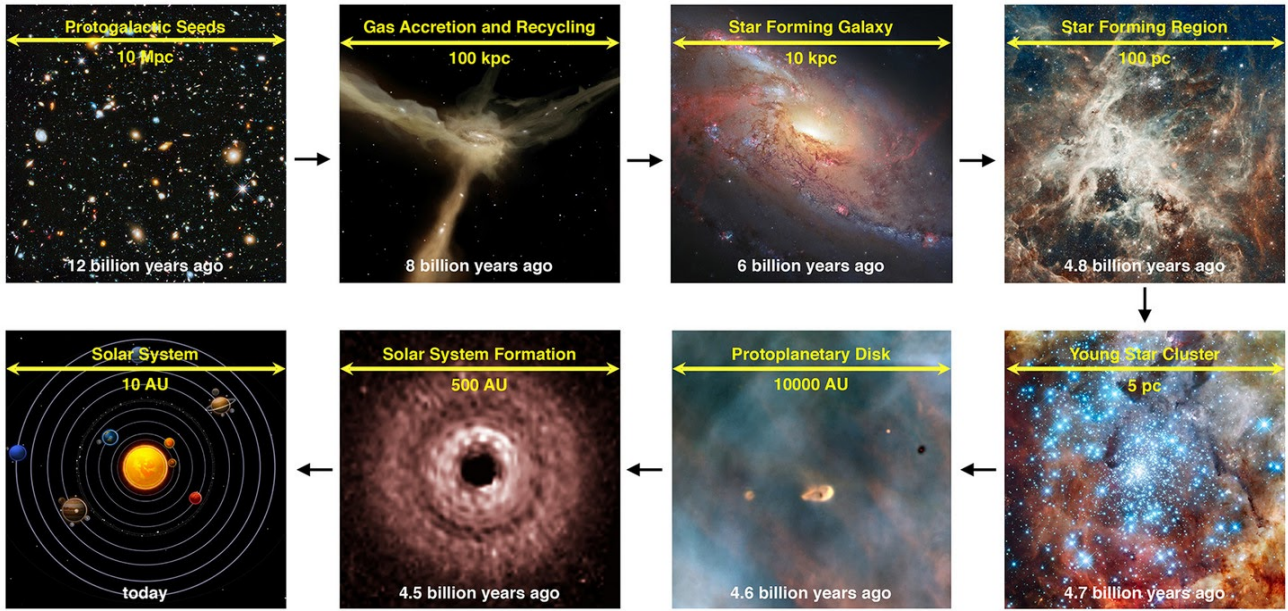

**Figure 1.2:** *Overview of the cosmic history of our Universe in relation to the relevant physical scales at different epochs[1]. By providing high angular resolution, sensitivity, and spectral resolving power across most of the sky, MAVIS will enable new discoveries across each of these regimes.*

Combined imaging and spectroscopic capabilities are crucial to the overall scientific utility and impact of MAVIS, making full use of the unique access to very high angular resolutions across most of the sky. *The point-source imaging sensitivity of MAVIS will exceed that of the Hubble Space Telescope, giving an order of magnitude higher depth, and with higher angular resolution*. The comparatively large field of view of MAVIS and high sky coverage furthermore open up the possibility for exploration and discovery surveys of these regimes, for example via "deep field" blind surveys, as well as targeted follow-up campaigns. Integral-field spectroscopy will make full use of the exquisite MAVIS image quality, giving the full range of optical diagnostics tracing chemical and dynamical properties across spatially complex and crowded fields, and with spectral resolutions matched to the characteristic mass and physical scale regimes that MAVIS will uniquely probe.

MAVIS will enable access to the new regimes of faint sources and crowded fields that will become commonplace at longer wavelengths in the era of ELTs, but with the diagnostic power of optical wavelengths. This will be particularly important in the low-redshift Universe (z<0.5), where the most well-calibrated and understood physical diagnostics are found at wavelengths below 1μm. In this way, *MAVIS will be a crucial complement to ELT capabilities*, delivering comparable angular resolution in the optical to that delivered by ELTs in the infrared. A key example of this synergy is in studying resolved stellar populations beyond the Local Group. While ELTs will be able to detect individual stars in galaxies beyond several Mpc with modest integrations times, infrared wavelengths probe mainly the Rayleigh-Jeans tail of the stellar black body spectrum, making infrared colours largely degenerate to key stellar population parameters such as age and metallicity. By contrast, the shallower depths accessible on an 8m telescope are fully compensated by the increased diagnostic power of optical colours, making MAVIS a critical tool for capitalising on ELT science.

---

[1] From the 2015 AURA report "From Cosmic Birth to Living Earths", www.hdstvision.org

The astrometric precision and magnitude limits of MAVIS will exceed that of HST, and extend the ultra-precise local framework from GAIA out to the furthest reaches of the Milky Way and into the Local Group. MAVIS will provide proper motion accuracies of 5-10km/s out to distances of ~100kpc over 5 year timescales, approaching the radial velocity accuracy from MAVIS spectroscopy. The combined spectral and imaging capabilities will provide full 6D phase-space information of individual stars, giving a new precision to studies of intermediate mass black holes in globular clusters, and unique dynamical information to constrain the dark matter properties of local group dwarf galaxies.

The unprecedented combination of sensitivity and angular resolution afforded by MAVIS and the AOF will allow, for the first time, a complete picture of the life of stars across cosmic time, and galaxy morphology, from the earliest phases of the formation of individual stars and planetary systems in our galaxy, to the detailed properties of ancient and chemically distinct stars in massive galaxies, and the fundamental nature and fate of the first star clusters in galaxies at early times. Exploration of these topics aligns well with the science interests of the broader ESO community, and with the robust and flexible operational model envisioned for MAVIS, can be addressed through a variety of observing campaigns, both as dedicated surveys and Large Programs, and through shorter allocations to small teams, as expected for a general purpose facility instrument.

## 1.3 MAVIS Operational Context

The next 5-10 years will see the fruition of decades of technological development, planning, and investment in astronomy, with the commissioning of the next generation of major optical-infrared astronomical observatories, including the 6m James Webb Space Telescope (JWST, due to launch in 2021), the 25m Giant Magellan Telescope (GMT, first light 2029), the Thirty Metre Telescope (TMT, first light 2027), and the 39m European Extremely Large Telescope (ELT, first light ~2025). The timeline of these and other facilities at different wavelengths is summarised in Fig. 1.3.

Thanks to their increased aperture size, these much-anticipated facilities will open new regimes of sensitivity and angular resolution, allowing astronomers to probe new parameter space in faintness, distance, mass scales, and ultimately in our understanding of the Universe. However, each of these facilities is optimised primarily for infrared wavelengths, due to both a scientific motivation to push to larger cosmological distances (and therefore red-shifted light), but also largely through the technical demands of engineering tolerances at optical wavelengths on large facilities.

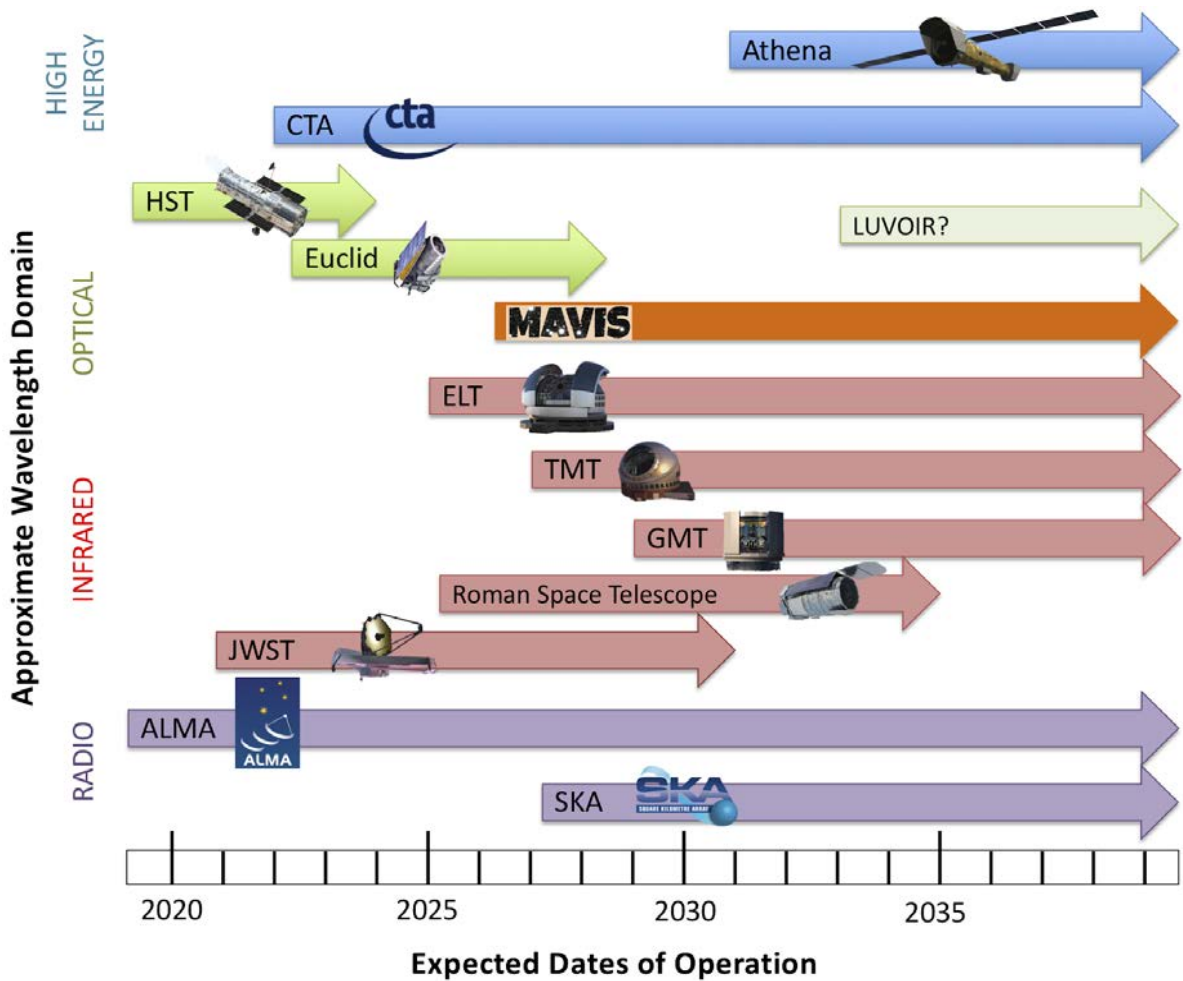

***Figure 1.3:*** *Timeline of current and future large observing facilities. All dates are based on published information from the corresponding observatories at the time of writing.*

Fig. 1.4 summarises the capabilities of existing and planned facilities in the parameters of sensitivity and angular resolution at different wavelengths. While the increased aperture sizes of next-generation facilities like JWST and ELT give a consummate increase in sensitivity across the optical/infrared regime, Fig. 1.4 demonstrates that this is not achieved uniformly at all angular scales. Diffraction-limited performance from these large aperture observatories is limited to wavelengths longer than 1 micron. Shortward of this, the best angular resolution available from a general-purpose observatory is that of the Hubble Space Telescope (HST) – a diffraction-limited telescope, but with a comparatively modest 2.4m aperture, providing angular resolution around 0.1".

The scientific impact of HST has been nothing short of revolutionary over its three decades of operation, thanks to its exquisite image quality which is, until now, unparalleled from the ground. At the time of writing, however, the future of this great observatory is very uncertain. In 2016, NASA extended the science operations support contract for HST to 2021, at the time to provide overlap with the James Webb Space Telescope (JWST). In March 2018, however, the JWST launch was postponed from 2018 to 2021, effectively eliminating this overlap. Operations can be extended, but the hardware is aging. Since October 2018, HST is down to three operational gyroscopes[2] - the minimum needed for stable pointing and optimal image quality. While mitigation plans are in place to maintain reduced-performance operations with a smaller number of gyroscopes, no service missions are currently planned to replace HST hardware.

---

[2] STScI Newsletter, 2019, Volume 36, Issue 03

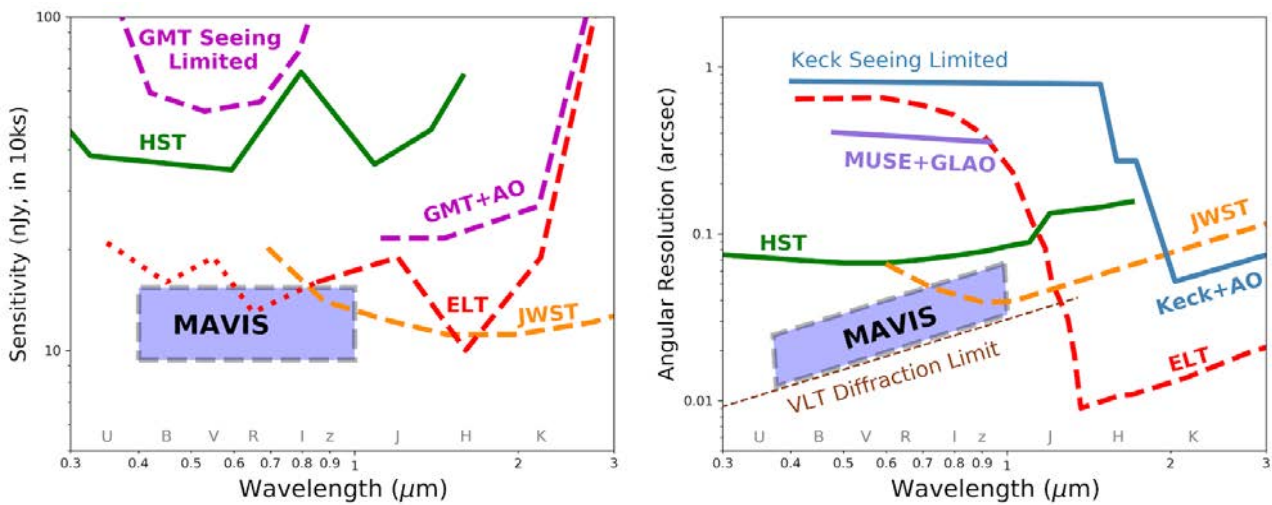

***Figure 1.4:*** *Comparison of MAVIS with existing (solid line) and future (dashed line, or dotted where performance is less certain) general-purpose observing facilities. Data were extracted from public sources (instrument handbooks, performance simulators, and exposure time calculators). Left: Limiting point-source sensitivity as a function of wavelength for a 10σ measurement with 10ksec of integration. The approximate region for MAVIS is indicated, based on the MAVIS ETC (see Appendix A). Right: Angular resolution as a function of wavelength, with MAVIS operating at or near the VLT diffraction limit. MAVIS compares favourably with ELT-class facilities for sensitivity at optical wavelengths, and uniquely combines this with high angular resolution.*

Uncertainties around the future of HST notwithstanding, the limitations of a 2.4m diffraction-limited aperture will be starkly apparent in the era of JWST and ELTs, as illustrated by the significant difference in depth between HST and these future facilities (Fig. 1.4). And while space, with its drastically reduced atmospheric background, will remain the most competitive and sensitive location for infrared observations via JWST, ground-based ELTs will play a crucial role in clarifying the detailed spatial distribution of sources through exquisite angular resolution on 10mas scales. By achieving the diffraction limit at optical wavelengths, MAVIS naturally allows this role to be complemented in the blue with the existing AOF facility.

## 1.4 MAVIS as a General Purpose Instrument

In preparation for this Phase A design study, the MAVIS Consortium initiated an open call for scientific White Papers, launched in April 2018. The MAVIS White Paper process resulted in 57 complete individual science applications being proposed, involving 150 researchers from across the international astronomy community (see Fig. 1.5). The original papers are publicly available via the MAVIS website (www.mavis-ao.org/whitepapers).

Section 6.12 summarises how the adopted baseline MAVIS design aligns with the White Paper proposals, showing that it is fully compliant with the vast majority of science cases. Scientifically, the White Papers fall into roughly physical scale-specific regimes, evenly spanning the full range of categories considered by ESO's Observing Programmes Committee (OPC, see Fig. 1.5). This indicates that MAVIS, with its flexible combination of imaging and spectroscopic capabilities, provides a general purpose capability that will satisfy a broad range of science interests in the ESO community. To put this in context, Fig. 1.6 shows the distribution of MAVIS White Paper topics across OPC categories compared with the proposal requests for existing ESO instruments during a recent observing period. In this figure, all instruments have been ordered by the variance in their distribution across OPC categories - a coarse figure of merit for how 'general purpose' the instrument is. This clearly shows that MAVIS sits squarely with other flexible facility instruments like XShooter, HAWKI, FORS2, and MUSE.

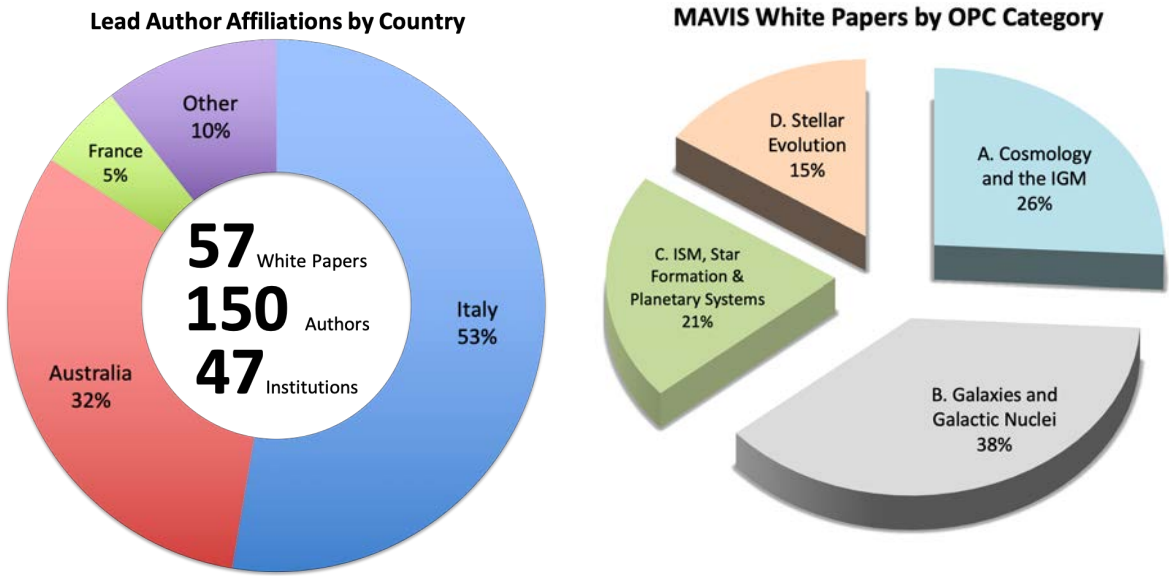

**Figure 1.5:** *Statistics derived from the MAVIS White Papers. Left: Break-down of lead author affiliation countries, and additional statistics. Right: Distribution of MAVIS White Papers across the ESO OPC sub-panel categories, showing a fairly even spread.*

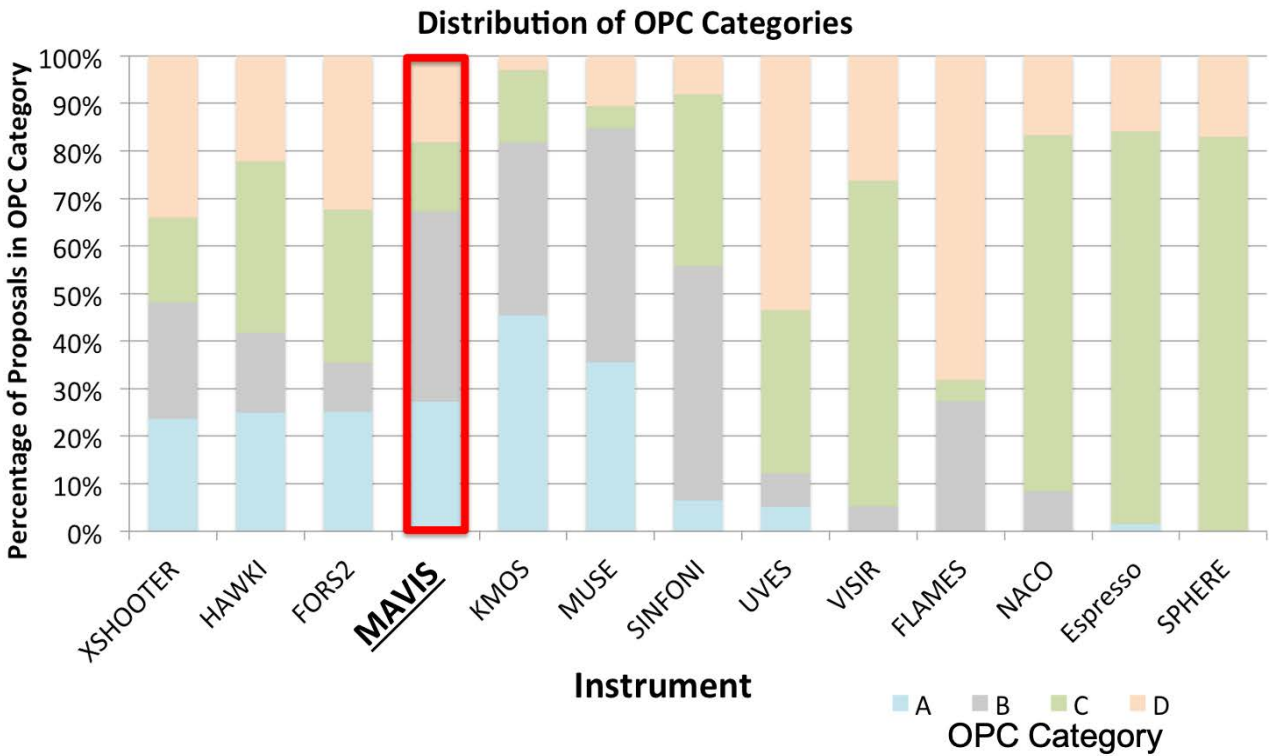

**Figure 1.6:** *Example distribution of proposals from ESO observing Period 102 across OPC categories (category names are as in Figure 1.5). The instruments are listed from left to right in order of increasing variance in the percentages across the categories, with left-most instruments being most evenly spread across the four OPC categories. MAVIS is included here using the distribution of White Paper topics. MAVIS sits squarely in the group of general purpose instruments towards the left of the figure.*

## 1.5 The MAVIS Science Case - Structure of this document

This document presents an overview of priority science cases that have been identified by the Consortium through open consultation with the ESO community (MAVIS White Papers - see Section 1.4), and further refined by the MAVIS Science Team through several interactive workshops with the consortium and external stakeholders (See www.mavis-ao.org/Workshops for details, including talk slides). These priority cases are presented in the context of four thematic chapters, namely:

- ***Emergence of the Hubble Sequence***
- ***Resolving the Contents of Galaxies Beyond the Local Group***
- ***Star Clusters Over Cosmic Time***
- ***The Birth, Life and Death of Stars and their Planets***

Each chapter comprises related subsections, each with their own key open questions and distinct applications of MAVIS. Taken together, these combine to illustrate how MAVIS will advance our understanding of the overall chapter topics. In addition, a number of 'Case Studies' are provided, to indicate what actual MAVIS observing programmes will look like, in terms of targets with suitable guide stars, sample sizes, and exposure times.

These main science chapters are followed by a description of the top level science requirements, and how these have been derived from the science cases. This includes a summary of requirements requested from the White Paper process, and of how compliant the baseline instrument is against those same White Paper science cases. Finally, an assessment of how competitive and complementary MAVIS is with existing and planned facilities is presented.

# 2 Emergence of the Hubble Sequence

## 2.1 Introduction

The physical underpinnings of morphological differences observed among galaxies in the local Universe and the origin of the so-called "Hubble sequence" is one of the most outstanding yet unsolved problems in astrophysics. Despite the existence of a relatively well understood theoretical framework that describes the assembly of gravitationally bound structures following the dynamics of dark matter halos, the complexity of baryonic physics, with its dissipative and radiative processes and the many flavours of feedback mechanisms, still hampers the quest for a complete and coherent picture.

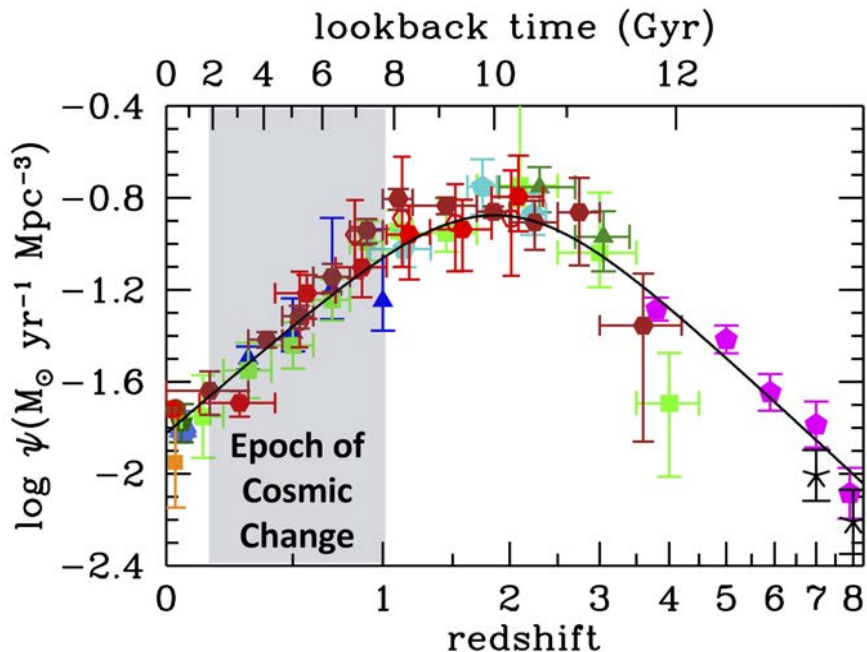

***Figure 2.1: Epoch of Cosmic Change.*** *The figure shows the star formation rate density of the Universe as a function of cosmic time (from Madau & Dickinson 2014). From 8 billion years ago until just recently, the population of massive galaxies underwent a dramatic change in fundamental properties. The diagram shows that galaxies decrease their star formation rate by roughly an order of magnitude over this period.*

Cutting edge instruments of the last 15 years have shown that the ancient Universe was turbulent and messy. Galaxies at z~1−3 experienced 5-10 times greater star formation rates (Fig. 2.1), had amorphous morphology (e.g. Abraham et al. 1996, see Fig. 2.2), and were fueled with an order-of-magnitude larger gas reservoirs than local Universe galaxies (e.g. Tacconi et al. 2013). Perhaps the most transformative observations came from the technical innovation of adaptive optics IFU instruments, like SINFONI. Systematic surveys (e.g Förster Schreiber et al. 2009, Cresci et al. 2009) found that these clumpy systems are not necessarily merging galaxies. Many have ordered, rotationally-dominated kinematics, yet with very high gas velocity dispersions. These kinematics suggest thick, puffy disks unlike that of the Milky Way. KMOS enabled very large surveys at z > 1 (Fig. 2.3) showing that ~1/2 of galaxies at this epoch are consistent with this picture of "turbulent, clumpy disk galaxy". Adaptive optics imaging indicates that the size of star forming regions was 10 times larger than in the Milky Way. The aggregate of these observations motivated a paradigm shift in galaxy evolution models (e.g. Dekel et al. 2009), moving towards one in which smoother yet efficient modes of gas accretion, as well as feedback effects due to outflowing gas, play an important if not dominant role at high-z.

Then 8 billion years ago, at z=1, everything began to change. HST observations reveal a rapid decline in clumpy galaxies from z~1 to z~0.5 (Guo et al. 2015). There is a simultaneous rapid decline in both star formation (Fig. 2.1) and gas content. At z ≈ 0 these turbulent, clumpy disk galaxies are all but absent. Galaxies in the local Universe are well described by spiral morphologies, reminiscent of the well known Hubble sequence (review by van der Kruit & Freeman 2011). Local galaxies have very low gas velocity dispersions, suggesting that thin, high angular momentum disk galaxies are now dominant. It stands to reason that the underlying drivers of galaxy evolution likely experience a change in nature or abundance that begins roughly at z~1 and is complete before z~0.

The epoch of cosmic time from 8 to 3 billion years ago thus marks a dramatic change in the properties of galaxies, and is one of the most critical times for galaxy evolution. These changes in observational properties likely represent a fundamental shift in the mechanisms driving galaxy evolution. However, due to a lack of appropriate instrumentation, the properties of galaxies during this period of change, from z ≈ 0.5 to z ≈ 1, are very poorly constrained. ***MAVIS will be the only instrument on any telescope that is capable of measuring the morphology, kinematics and the spatially resolved ISM properties of galaxies at the required spatial and spectral resolution during this epoch in which the main modes of galaxy evolution are changing.***

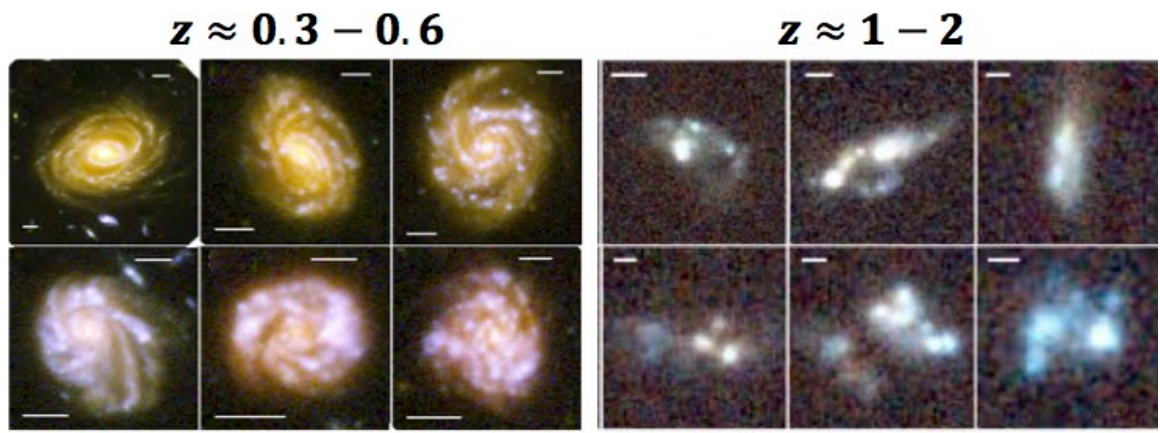

***Figure 2.2: The Morphological Change in Galaxies***: *Thousands of hours of observations with HST make it very clear that galaxies undergo a drastic morphological change between z~1 and z~0.5. Because morphology is not an easily quantified concept, we are not yet able to understand the drivers of this change. Moreover, the available resolution with HST is not sufficient for detailed measurement of galaxy sub-structures. Resolved morphology, kinematics and chemistry are key. MAVIS will be uniquely transformative to our picture of galaxy evolution in this epoch of cosmic change.*

Fundamental clues to the buildup of the Hubble sequence we observe in the present Universe can also be found by resolving morphological details of the first galaxies beyond z~3, during the earliest stages of galaxy assembly. As discussed above, the fraction of irregular galaxies increases going from low (z ∼ 0) to higher redshifts (z ∼ 3−5; e.g. Brinchmann et al. 1998, Ribeiro et al. 2017), possibly due to merging events or galaxy interactions. Galaxies at z~1 are also much smaller in size than present-day galaxies (from < 1 kpc to ∼ 5 kpc; e.g., van der Well et al. 2014), making their observation challenging. In the rest-frame UV, the largest high-z galaxies display multiple clumps of enhanced star formation (e.g. Elmegreen et al. 2005, Genzel et al. 2011, Forster-Schreiber et al. 2009, 2017). However, tracing galaxy properties out to the highest redshifts is nearly impossible using current instruments, with $R_e$~500 pc (~80 mas) for the z~6 i-band dropouts: ***only diffraction limited images at optical wavelengths using an 8m telescope, like the ones provided by MAVIS, will be able to observe such high-z galaxies with sufficient resolution to reveal significant details of their morphology, and to trace back their evolutionary path.***

In addition to the direct study of high-z galaxies, the bulges of local disk/spiral galaxies and their diskless counterparts, the elliptical galaxies, may indeed retain some clues to understand the origin of the Hubble sequence by providing the evidences for the dynamical processes that create such density-enhanced regions (see "Galactic Bulges", 2015, Laurikainen, Peletier & Gadotti eds. and the contribution by J. Kormendy in particular, for extensive reviews and discussions). Indeed, a detailed analysis of the physical properties of bulges, of their dynamical state and of their multiphase and stellar content, per se and in relation to the hosted SMBH and to the environment in which a galaxy has been living, is crucial to disentangle the physical mechanisms that lead to the bulge formation/growth and the morphological shaping of galaxies. However, given that the bulges of spirals range between a few tens of pc up to 1-2 kpc half light radius, we are currently reaching the limits of available instrumentation, as angular resolution <100 mas would be needed to sample the physical properties of these objects even at distances of ~20 Mpc. *MAVIS is uniquely suited for this challenge, as it is the only instrument that will simultaneously provide exquisite angular resolution, competing with HST; coverage of the visible wavelength range up to ~1 µm; integral field spectroscopy in the visible with a relatively extended FoV; and enormous sensitivity thanks to the VLT*.

## 2.2 Kinematics and structural properties of the first disk and compact galaxies in the epoch of Cosmic Change

> **Science goal:** what physical drivers lead to the emergence of thin, high angular momentum disk galaxies in the epoch from z ≈ 0.4 to z = 1.0? Which is the origin of the most compact massive galaxies?
>
> **Program details:** High resolution imaging and IFU spectroscopy of samples of intermediate redshift galaxies in different mass ranges with the LR-Red and HR-Red gratings, to follow Hα or [OIII].
>
> **Key observation requirements:** spectral resolution R>5000; wavelength coverage 4500-10000Å; spatial resolution ~ 50 mas.
>
> **Uniqueness of MAVIS:** large sky coverage in the main extragalactic fields required, in combination with high spectral resolution to resolve the intrinsic velocity dispersion combined with high spatial resolution.

As discussed above, the redshift epoch just shy of z~1.0 may provide a critical linchpin for testing models of galaxy evolution, given the fundamental changes in the properties of the galaxy population: nevertheless, this same period remains very poorly constrained. Wisnioski et al. (2015) present a comprehensive analysis of the redshift evolution of the kinematics of galaxies from IFU surveys (see Fig. 2.3). There is an 8 billion year gap from local Universe surveys to those of the distant past. This gap starts at exactly the redshift interval where morphological and molecular gas surveys tell us that galaxies are changing. Just as it was critical to our picture of z > 1 galaxy evolution, we need kinematic information to understand why galaxies undergo this transition from clumpy, turbulent disks to Hubble sequence galaxies.

Moreover, as thin disks become dominant the substructure of galaxies changes as well. We must be able to characterise the smaller scale structures that are forming at this epoch (e.g. pseudobulges, bars, rings). Current resolution from HST is insufficient for this goal. We need to link the kinematic measurements to the structure to make a full test of models of galaxy evolution.

*This huge gap in our picture of galaxy evolution is because there is no instrument, on any telescope, sufficient to measure the kinematics of galaxies in this epoch.* For this reason, a fundamental question remains unanswered: what physical drivers lead to the emergence of thin, high angular momentum disk galaxies during the epoch from z ≈ 0.4 to z = 1.0? To answer this question we must measure the kinematics of disk galaxies at sufficient spatial resolution (of order ~1 kpc) in this redshift window for a sufficient sample of galaxies. Moreover, we must simultaneously be able to measure their structure/morphology and characterize their local environment. At z ≈ 1 we must be able to resolve at least

0.15 arcsecond to separate out bulges. Higher resolution is preferable as it allows for characterization of the bulge kinematics themselves (e.g. Fabricius et al. 2012).

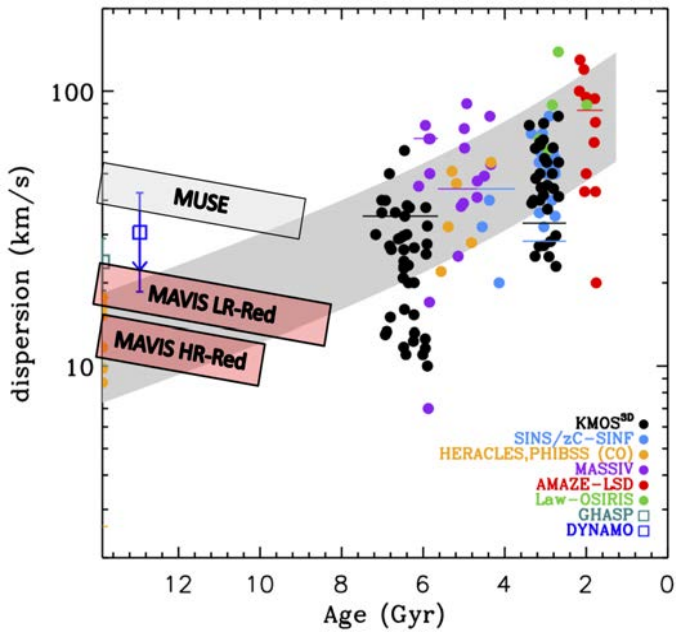

***Figure 2.3: A huge knowledge gap in the kinematic transformation of galaxies**: Observed velocity dispersion of star-forming gas in galaxies over cosmic time, overlaid with the instrumental resolutions of MUSE and MAVIS for redshifted Hα. Prior to z=1, the gas dispersion is much higher than observed today, indicating enhanced dynamical motion and/or feedback. At z=0, galaxy disks are dynamically colder, and morphologically more regular. There is, however, currently no instrument that is capable of tracking the kinematics of galaxies during this 8 billion year period of cosmic change. It is a critical epoch of galaxy evolution, and one where MAVIS will make ground breaking observations. Adapted from Wisnioski et al. 2015.*

The strongest emission lines from which kinematics can be measured in galaxies are Hα, Hβ and [OIII]. Current adaptive optics instruments with sky coverage large enough to observe faint galaxies are only able to achieve sufficient FWHM and sensitivity at J band or redder wavelengths. The reddest of these lines, Hα, is not observable with traditional adaptive optics techniques for redshifts less than z ≈ 0.9. Even with currently planned improvements, at these distances using fainter rest-frame near-IR lines, such as Paα and Brγ, is impractical for a significant number of galaxies due to sensitivity limits. Therefore, to study the kinematics of galaxies during this epoch of cosmic transformation requires an optical, AO enabled integral field spectrograph. Because there are multiple bright ionized gas emission lines available, this research can be achieved with spectral coverage of 6300~10000Å.

Moreover, such an instrument requires a spectroscopic resolution that can adequately identify thin disks with lower velocity dispersion ($\sigma_{gas} \approx 10-25$ km s$^{-1}$). Studies of ionized gas in the local Universe (z ≈ 0) find that disk galaxies can have velocity dispersion as low as $\sigma_{gas} \approx 10$ km s$^{-1}$, which is a fundamental lower limit set by the thermal broadening of ionized gas in HII regions. This requires a spectral resolution of R~10000. JWST and MUSE may have sufficient sensitivity, but they will not have sufficient spectral resolution. Other than MAVIS there is no present or planned instrument that can study the kinematics of disks at this critical epoch. A high spectral resolution IFU on MAVIS adds a unique, fundamental capability that will open up a critical redshift window for galaxy evolution.

Typical disk galaxies have half-light radii ranging from ~3-15 kpc, at z = 0.5 this corresponds to diameters of order 5 arcsec, making the MAVIS IFU perfectly suited for the task. The critical addition of high spatial resolution imaging makes MAVIS the ideal instrument for studies of the first thin disk galaxies.

A large body of observations and theory suggest that this same epoch, z~0.3−1.0, is a critical time in which disk galaxies secularly evolve, often producing pseudobulges (reviews in Kormendy & Kennicutt 2004, Athanassoula 2005, Fisher & Drory 2016). This theory of secularly evolving disks makes direct predictions for the change in bulge properties, and how they relate to structure and kinematics of the disk. Pseudobulges, which exist within roughly 80% of disk galaxies in the local Universe (Fisher & Drory 2012), are thought to be a natural part of disk galaxy evolution. Based on present-day star-formation rates and stellar masses Fisher et al. (2009) argue that, if secular evolution is responsible for pseudobulges, then pseudobulges must have formed much of their mass during the same epoch of cosmic change.

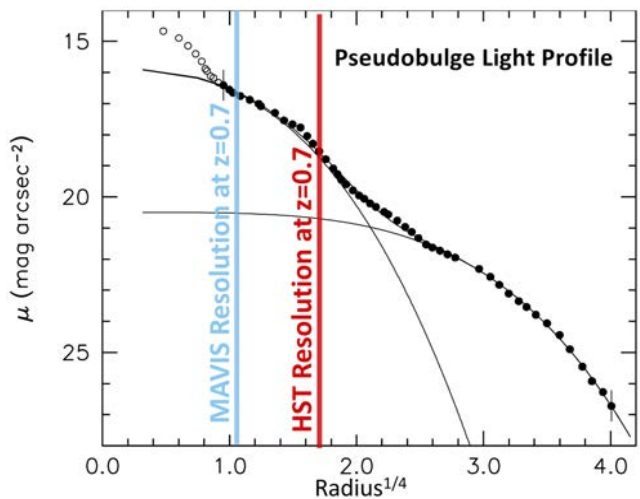

**Figure. 2.4: The initial phase of disk galaxies secular evolution.** The figure shows the V band stellar light surface brightness profile of a typical nearby pseudobulge galaxy (NGC 3169). The red and blue lines indicate what the MAVIS (blue) and HST (red) would be when this disk galaxy is observed at z~0.7. Using the current state of the art instrumentation (HST), the bulge is unresolved. MAVIS opens a new window to this critical aspect of disk galaxies evolution

Due to the need for spatial resolution, MAVIS will be the only instrument capable of measuring the optical light of the pseudobulges at the point in time when they form. It is this combined need to measure both the kinematics and imaging at fine-spatial scales that makes MAVIS a one-stop-shop for this critical epoch of galaxy evolution. Currently the highest spatial resolution images can reliably achieve of order 0.1 arcsec FWHM. As we show in Fig. 2.4, this is insufficient to properly decompose stellar light profiles at z > 0.5. At z ~ 0.5 − 1.0 this resolution is only good enough to identify the presence of a bulge, not to characterise any of its properties. Moreover, high spatial resolution is needed to break the known degeneracies between the parameters of Sersic function. The problem becomes even more complicated by the presence of extra components that are widespread in disks, such as bars and rings. Fisher & Drory (2016) argue that a practical resolution limit of 200-300 pc is needed to measure bulges and disks. MAVIS will deliver this with 20-30 mas imaging, in reasonable exposure times (of order 30 min to 1 hour per target).

Similarly to the disk galaxy population, also the ***massive quiescent galaxies*** have undergone a significant evolution in their structure and size since high redshift (z>2) to the present-day (see e.g. Toft et al. 2007, van Dokkum et al. 2008, van der Wel et al. 2014). In particular, massive quiescent galaxies (with masses of $>10^{11} M_{Sun}$) were much more compact at z~2, with typical effective radii of 1-1.5 kpc, a factor of ~5 smaller than equally massive quiescent galaxies at z=0. The most plausible scenario to explain the significant increase in size, along with a modest mass growth, of quiescent galaxies is evolution through a series of minor gas-poor merging (e.g. Bezanson et al. 2009, Oser 2012, Cimatti et al. 2012). Such a scenario will have to be confronted with the kinematic properties of high and intermediate redshift galaxies as well, of which little is currently known because of observational limitations due to their compact sizes. Indeed, recent observations from deep, high-resolution imaging surveys based on HST H-band observations have revealed the existence at z>~2 of a population of compact star-forming galaxies that could represent the progenitors of compact quiescent galaxies (Barro et al. 2013, 2017). Compact galaxies (first blue, mainly at high redshift, then red) are thus thought to mark an evolutionary phase that follows the quick accretion of gas in the inner regions of galaxies, possibly as a consequence of a gas-rich merger, and precedes growth through minor gas-poor merging.

A key quantity to constrain this proposed evolutionary sequence is the central stellar mass density (within 1 kpc, corresponding to the typical effective radius of compact quiescent galaxies and the core of their star forming progenitors and their quiescent descendants; see e.g. Barro et al. 2017). Thanks to its exquisite spatial resolution and deep imaging, MAVIS will finally allow to resolve the inner structure and color gradients of compact massive galaxies (quiescent and star-forming) from intermediate redshift (z~0.4) up to high redshift (~4) in order to understand i) if the compact star-forming and quiescent galaxies are consistent with being the initial and final phase of a merger remnant and of which kind, ii) how the relation between age and structure varies with redshift, and iii) where do the high redshift compact quiescent galaxies end up in the local Universe (e.g. cores of classical bulges, see next section). These data will be a crucial complement to existing HST NIR observations of compact high- redshift galaxies in order to characterise their structure as traced by stellar populations of different ages. The typical effective radii of intermediate- and high-z compact galaxies are in fact 1-1.5 kpc (0.16"-0.24" at z=0.5; 0.12"-0.18" at z=2): such small sizes require exquisite spatial resolution in order to resolve galaxies on scales of fraction of their effective radii and of their central ~1 kpc core. MAVIS will thus be a unique instrument to characterize the structure of massive compact galaxies in their rest-frame UV-optical on sub-kpc scales, as a resolution of 20-25 mas (FWHM) would allow to resolve ~5-10 spatial bins within 1 Re (and/or within the compact core). In order to sample galaxies with stellar mass >~$10^{10.5}$ $M_{Sun}$ up to z~2 we need to go as deep as R~26 mag. This is achievable in ~5-10 hr of integration time. Suitable targets of compact quiescent and star-forming galaxies can be extracted from already existing catalogs at intermediate redshift (e.g. BOSS, Damjanov et al. 2014, Tortora et al. 2018) and at high redshift (e.g. GOODS, UKIDSS-UDS, CANDELS, Barro et al. 2017).

Moreover, for the brightest/most-massive (R<23) intermediate-z (z<0.8) compact galaxies MAVIS will allow to obtain spatially resolved, moderate spectral resolution spectroscopy. For the typical magnitude and size of our targets, the average surface brightness within Re is expected to be $\mu_r$ ~ 21 mag/arcsec$^2$, with a central peak possibly going down to <~18 mag/arcsec$^2$. Therefore, with ~10 hrs of integration we should be able to reach a SNR=10 per resolution element in the central regions, while the same goal can be reached in the outer parts applying spatial binning. These observations would therefore allow to measure the velocity and velocity dispersion profiles (and hence the galaxy rotational support) as well as stellar population gradients from stellar continuum and absorption line fitting. This kind of observations will finally allow to measure the central mass density and stellar populations gradients within the typical effective radius of this kind of sources, tracing their inner structure through cosmic time to understand whether the proposed evolutionary sequence can hold also at intermediate redshifts.

The area of science we describe in this section is fundamental to studies of galaxies in general. We have outlined some key studies that on their own would fill a critical gap in our model of galaxy evolution. However, at its base these observations would have endless applicability. Currently SAMI, MaNGA and CALIFA represent many thousands of hours of observations with hundreds of team members and a principal focus of characterising the properties of main-sequence disks at z = 0. It is almost certain that MAVIS will have similarly broad impacts. MAVIS makes it possible to study thin disk galaxies as pristine objects in formation. This is a fundamental shift in how we would address the studies of disk galaxies.

Since Edwin Hubble first published Extragalactic Nebulae in 1926, astronomers have been trying to understand the formation of thin disk galaxies, and their relationship to the Hubble Sequence. Indeed, almost every major instrument in astronomy has attempted to explain the Hubble Sequence. The missing piece of this critical question has been the ability to make resolved observations of the first thin disk galaxies. Studies of galaxies in the local Universe with large surveys like SAMI and CALIFA have made drastic leaps forward. These surveys, however, must unravel billions of years of evolution, we still do not know what the conditions at their formation was. Studies of galaxies at larger redshift with instruments like SINFONI and KMOS have likewise been critical, as they tell us what disks were like before they settled into the thin systems like our Milky Way. ***MAVIS will fill in the missing link to galaxy evolution, allowing us to finally connect these two epochs.***

## 2.3 Resolving ISM variations across cosmic times

**Science goal:** How do the properties of the ISM in galaxies evolve between z=1 and z=0?

**Program details:** IFU spectroscopy across of samples of intermediate redshift galaxies in different mass ranges, with the LR-Red and LR-Blue gratings

**Key observation requirements:** spectral resolution R~5000 and wavelength coverage 4000-10000 Å; spatial resolution ~ 50 mas.

**Uniqueness of MAVIS:** large sky coverage in the main extragalactic fields required, in combination with high spectral resolution to resolve the intrinsic velocity dispersion combined with high spatial resolution.

Scaling relations among global properties of galaxies (mass, SFR, chemical abundance, dynamics, dark matter ecc.) have provided a wealth of information about the processes leading to galaxy formation and driving their evolution. Nevertheless, the global galaxy properties are not enough to tell the full story because galaxies are also shaped by a number of internal processes, i.e., affecting different parts of the galaxies in different ways and at different times. Galaxies are regulated by a complex interplay between cosmic gas accretion, merger events, star formation, gas redistribution and gas outflows due to the stellar- and AGN-driven winds (e.g. Somerville & Davé 2015) . All these processes affect different parts of the galaxies in different ways, creating gradients that evolve with time.

Specifically, these events leave imprints on the chemical and dynamical properties of the ISM and on its distribution inside the galaxy and in its circum-galactic medium (CGM). Obtaining the spatially-resolved properties of ISM and CGM across cosmic time is a crucial step-forward for a complete understanding of many processes involved.

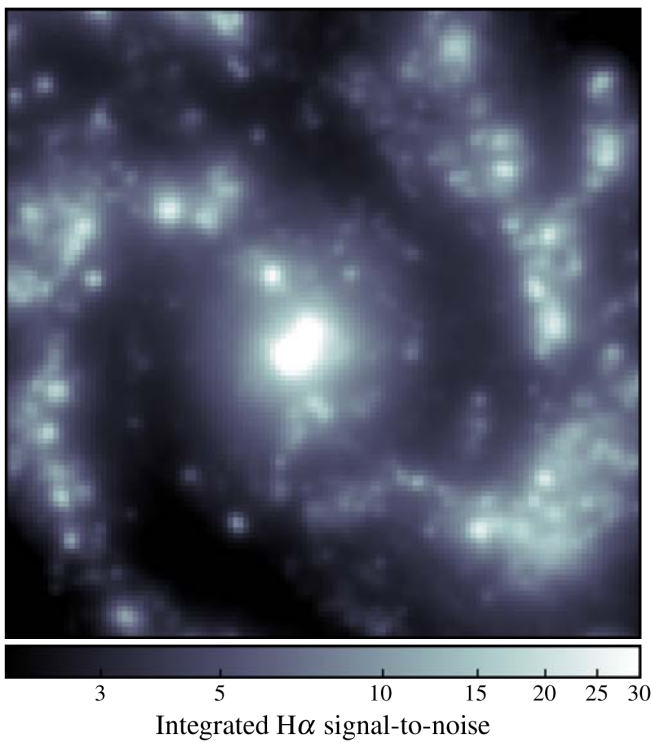

*Integrated Hα signal-to-noise*

**Figure 2.5: Simulation of the MAVIS view of a z~0.1 galaxy:** *Map of the expected S/N ratio on the Hα line for a main-sequence galaxy at z~0.1. Assumes 25mas spatial sampling and 3 hours on source integration. Based on high-resolution data for M83, observed as part of the TYPHOON survey.*

Gas-phase Metallicity, i.e., the content of heavy elements in the ISM, is linked to all the processes above and, therefore, is one of the most revealing quantities to be observed (see, e.g., Maiolino & Mannucci 2019). Many nearby galaxies were observed with 3D, integral-field spectroscopy with few arcsec of resolution (e.g., with MANGA or VLT/MUSE) and their large-scale distribution of metals is well known (e.g., Belfiore et al 2017). MAVIS can obtain detailed metallicity maps of a large number of nearby (z<0.5) galaxies with ~10 times better resolution, probing the scales of about 100pc where the single star-forming regions and their interaction with the surrounding ISM can be probed. This will allow us to finally probe the inner cycle of gas and metal flows in the galaxies.

As an example, MAVIS will be able to finally probe the extent and shape of metallicity gradient at intermediate redshift (z~0.5). These are still almost unexplored, despite the fact that this epoch represents the still missing link between the negative metallicity gradients observed in nearby disc galaxies and the positive or flat gradient observed beyond z~1 (see Fig. 2.6). This fundamental change in the gas enrichment is key to understanding the chemical evolution process in galaxies, and the exquisite spatial resolution, wavelength coverage and sensitivity of MAVIS will finally allow to probe the gas phase abundances in this cosmic epoch using a variety of diagnostic line ratios.

A spectral resolution of R~5000 is required to avoid sky lines in the red part of the spectrum. Simultaneous coverage of the optical lines from [OII] to [SII] is preferable to increase the observing efficiency, while preserving the coverage of the multiple lines needed for ISM investigation. A wavelength range between 4000-10000 Å would be therefore ideal to cover the galaxies at z~0.5.

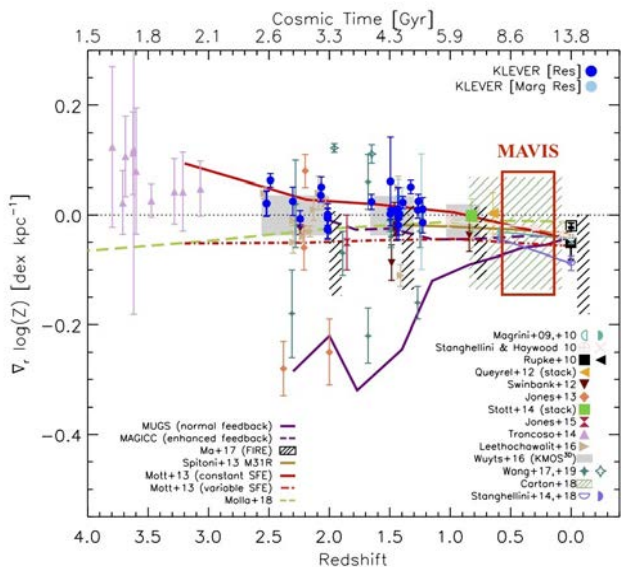

***Figure 2.6****: Compilation of metallicity gradients from different surveys, plotted as a function of redshift (or equivalently the age of the Universe). MAVIS will cover the crucial epoch between the negative gradients observed in local galaxies and the flat or positive gradients observed at z>1 (adapted from Curti et al. 2019).*

Equally important is the dynamical state of the gas in different parts of the galaxy. This is particularly interesting to understand the origin and the properties of the feedback processes controlling the efficiency of star formation, and the complex interactions between stellar and AGN winds and star formation activity. These winds can produce both negative and positive feedback (e.g., Cresci et al 2015a,b) and even star formation inside the outflowing gas (Maiolino et al 2017). In this case, spectral resolution of R=10000 would be needed to study stellar features, e.g. to search for blue shifted young stars inside outflows.

To really understand how feedback works and measure relevant quantities such as the mass-loading factor and the star-formation efficiency it is necessary to spatially resolve the outflows and the star-forming regions, study the dynamics and ionization parameters of the ISM, investigate the presence of fast-moving stars, and correlate these properties with the nearby AGNs and sites of star formation.

**Case Study: Resolving ISM evolution over the past 8 billion years**

**Key questions:** What drives turbulence in the interstellar medium of star-forming galaxies? What physical processes govern the emergence of thin disks at z < 1?

**Why MAVIS?** The redshift range from 0.2 < z < 0.8 represents a crucial time in the evolution of star-forming galaxies during which they transform from the clumpy, turbulent disks observed during cosmic noon into the thin disks characteristic of local late-type galaxies. With its combination of high spatial and spectral resolutions, MAVIS provides a unique opportunity to trace evolution of the ISM during this critical time period and uncover the mechanisms driving the evolution of disk galaxies over the last 8 billion years.

**How will MAVIS be used:** The IFS mode will be used with the 50mas spaxel scale and multiple spectral resolutions (LR/HR-Red) to study properties of the ISM as they evolve from z~0.2 out to z > 0.5. MAVIS will trace key strong emission lines (e.g. [OII], Hbeta, [OIII], Halpha, and [NII]) on the scale of hundreds of parsecs. At high spectral resolution (R > 5000) these data will provide a detailed view of velocity structure in the ISM, including first and second moments for multiple kinematic components (e.g. star-formation, outflows, shocks, and AGN), and will be used to derive spatially-resolved maps of star-formation rate surface density, dust attenuation, and chemical abundance. These measurements are critical constraints for theoretical models of evolution in galactic disks, which predict that both radial transport and star formation feedback are required to reproduce the evolution ionized gas kinematics over cosmic time as well as small-scale correlations between star-formation rate and gas velocity dispersion.

**Target selection:** The extensive multi-wavelength imaging and spectroscopy available in the extragalactic deep fields (COSMOS, UDS, and GOODS-S) make them a natural choice for follow-up with MAVIS. Based on the CANDELS/3D-HST catalogs (Skelton et al. 2014; Momcheva et al. 2016), there are more than 500 galaxies at 0.2 < z < 0.8 spanning a range of stellar mass, star-formation rate, and environments, all with suitable tip-tilt/NGS asterisms.

**Exposure time:** For a typical main-sequence galaxy observed at z=0.4, a 3 hour on-source exposure with MAVIS is sufficient to detect Halpha emission at SN > 3 per spaxel out to the disk scale radius using the low-resolution spectral mode.

## 2.4 Resolving the physical properties of classical bulges and pseudo bulges

**Science goal:** disentangle the physical mechanisms that lead to the bulge formation/growth and the morphological shaping of galaxies

**Program details:** imaging in BVRIz to map the inner regions of large nearby galaxies.; spectroscopy between 4000-8500 Å to map velocity and velocity dispersion fields, and stellar population analysis.

**Key observation requirements:** imaging with large (30"x30") field of view and spatial resolution <100 mas; IFU spectroscopy with R~5000 and high sensitivity.

**Uniqueness of MAVIS:** higher sensitivity than HST, combination of imaging and spectroscopy at diffraction limit resolution.

In the past decades bulges have been shown to span a wide range in structural (i.e. light profile), morphological (presence of rings, spiral arms, bars etc) and kinematic (rotation vs dispersion) properties. The (simplified) picture emerging from this complex observational puzzle classifies bulges in two distinct classes: classical bulges and pseudo-bulges (e.g. Kormendy & Kennicutt 2004).

Classical bulges are like small elliptical galaxies embedded in the host disk. Contrary to the disk, their stars are old and are dynamically supported by velocity dispersion, rather than rotation. Pseudo-bulges, as opposed, display typical disk properties, such as large rotational-over-dispersion support, and structure like spiral arms and bars which derive from dynamical instabilities in disk-like systems (see Fig. 2.7). They appear as the result of secular dynamical processes that redistribute the matter in the disk. For the creation of classical bulges (and ellipticals), on the other hand, different channels are viable according to theoretical studies and simulations, essentially related to the primordial collapse of gas or merger of proto-galactic clumps, the major merger of two progenitor galaxies (wet, at high z, or dry, at low z) and AGN feedback (as a mechanism to inhibit gas collapse and take angular momentum away).

The link between bulge formation and AGN feedback is indeed suggested by the observational relations between the mass of the central supermassive black holes (SMBH) and the physical properties of the bulge (stellar mass, velocity dispersion). Evidence is rising, however, that such relations may not hold for all bulges, but for classical bulges only, i.e. for those bulges that were possibly formed in primordial phases and have experienced violent relaxation (e.g. mergers), but not for the bulges that result from secular dynamical evolution of disks.

In this context, a detailed analysis of the physical properties of bulges, of their dynamical state and of their multiphase and stellar content, per se and in relation to the hosted SMBH and to the environment in which a galaxy has been living, is crucial to disentangle the physical mechanisms that lead to the bulge formation/growth and the morphological shaping of galaxies. Understanding bulge formation is a key to understanding when and how different physical mechanisms that form and assemble the baryonic/stellar mass of galaxies come into place, and how these mechanisms are hindered by feedback processes.

### The observational challenge

The half-light radius of the (classical or pseudo) bulges of intermediate- to low-luminosity spirals ranges between a few tens of pc up to 1-2 kpc. The sphere of influence of the SMBH in such galaxies (approximated as $GM_{BH}/\sigma^2$) is typically a few tens of pc. If we consider that 10 pc correspond to 20 mas (100 mas) at 100 Mpc (20 Mpc), it appears obvious that even for a local sample of galaxies the only way to properly characterize structure, content and kinematics is to overcome the limitation of the atmospheric seeing and attain angular resolution <<100 mas, which is only doable from the space or using AO from the ground. Despite the relatively high surface brightness of these central galaxy regions ($_V$<~14-15 mag arcsec$^{-2}$ in the innermost pc's, down to ~19 at ~1-2 kpc), the high angular resolution results in very faint fluxes per resolution element (e.g. 19 mag arcsec$^{-2}$ corresponds to V~27 mag per 25mas-resolution element), which therefore requires the large collecting area of a 8m-class telescope to succeed.

Ideally one would like to assemble a set of multiband images in the optical/NIR range in order to characterize the structure of the bulge (size, shape), the rough distribution of stellar mass and dust. Even more importantly, obtaining IFS with such a high angular resolution and spectral resolution R~5000 ($\sigma$~25 km/s) over the optical/NIR range enables a full 2-D mapping of the kinematics and physical properties of the stars and (where present) ionized gas. The kinematic measurements would allow not only to characterize the dynamics of the bulges in terms of orbital decomposition and identify possible stream motions of the gas (dynamically induced or driven by the AGN), but also to obtain a dynamical estimate of the SMBH mass. From stellar population analysis it is also possible to make an "archaeological" dating and chemical tagging of the bulges and of their innermost regions in particular.

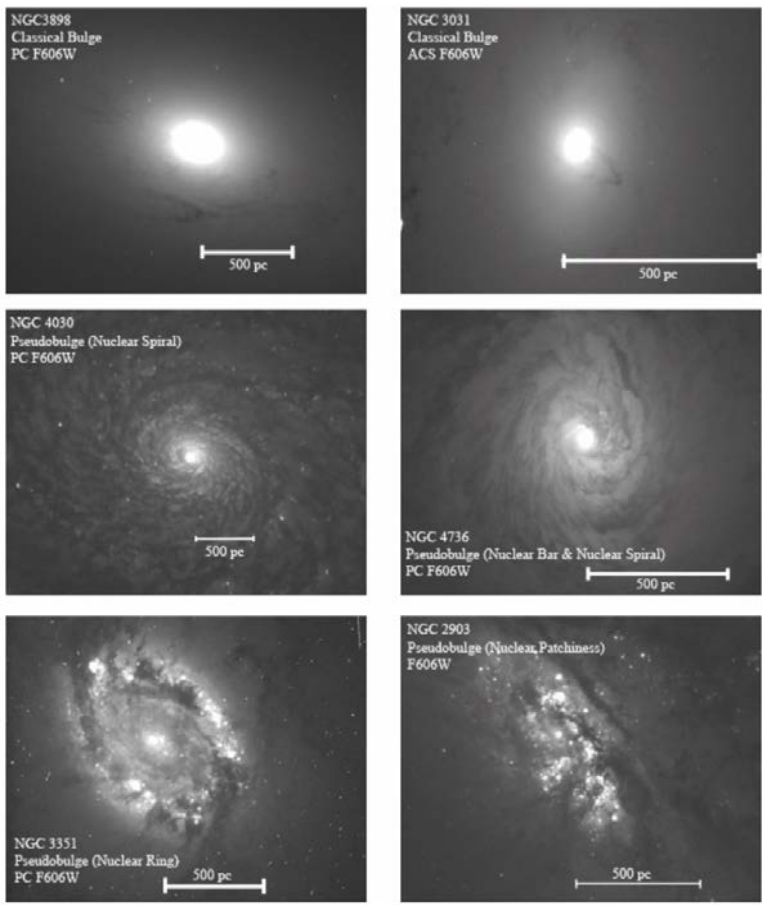

*Figure 2.7. Examples of morphologies of classical bulges (top two panels) and pseudobulges (bottom four panels).* All images are taken by HST in the F606W filter. On each panel, we draw a line representing 500 pc. Figure from Fisher & Drory (2010).

MAVIS is uniquely suited for this challenge because it is the only instrument that will provide simultaneously: 1) exquisite angular resolution, competing with HST; 2) coverage of the visible wavelength range up to ~900 nm; 3) integral field spectroscopy in the visible with a relatively extended FoV; 4) enormous sensitivity thanks to the VLT. The unprecedentedly large FoV in imaging will deliver multiband maps of galaxies within 100 Mpc (the distance to the Coma-Supercluster) out to >5 kpc, with a physical resolution better than 10 pc, thus providing an extremely detailed view of a galaxy from the global scale to the scale relevant to star formation. With the MAVIS IFU it will be possible to conduct the investigations outlined above with exposure times of a few hours using the low spectral resolution mode R~5000 and adjusting the spatial resolution configuration according to the distance of the galaxy. R=5000 is the minimum required for optimal sky line subtraction in the red and accurate stellar kinematic estimates. Specifically, we shall be able to map: 1) the inner pc's at high spatial resolution (spaxel 25 mas) and SNR~10 in about 1hr with LR Red; 2) a more extended region of 1-2 kpc around the nucleus with the low spatial resolution mode (spaxel 50 mas) and/or adopting spatial binning/smoothing in both LR Blue and Red with ~5 hr exposure.

MAVIS will therefore enable a thorough investigation of the nature and physical origin of different kinds of bulges for a sizable sample of galaxy, spanning a whole range of intrinsic (mass, morphology, SFR, SMBH…) and environmental properties. Such a survey is made possible by the high acuity and sensitivity of MAVIS, which basically expands the survey volume from a maximum distance of a few Mpc to ~100 Mpc.

## The connection between morphological structure and angular momentum

Angular momentum $j_*$ is a fundamental parameter in the evolution of galaxies, because it traces the tidal torques experienced during their lifetimes (Mo, van den Bosch & White, 2010). The total specific angular momentum $j_* = J/M$ of a galaxy disk is empirically similar to that of its dark matter halo (Fall 1983), but the distributions of $j_*$ are different, in that there is a lack of both low- and high-$j_*$ baryons with respect to the corresponding distribution for the halo (Catelan & Theuns 1996a,b; Barnes & Efstahiou 1987; van den Bosch et al. 2001). This discrepancy presumably arises during disk assembly, when the baryons experience additional physical processes. For example, angular momentum can be ejected by feedback or tidal stripping, rearranged due to mergers or migration of high-dispersion components, or lost due to biased collapse of the disk (van den Bosch et al. 2001). The physical processes that eject and rearrange angular momentum are also linked to merger-driven and secular evolution, respectively contributing to building a classical or pseudo- bulge. The distribution of angular momentum is also therefore intimately linked to morphological structure. One can calculate the probability density function of normalised specific angular momentum $s = j_*/j_*,mean$, PDF(s). Classical bulges have little or no ordered rotation, Gaussian-smeared by random motions, leading to a normal PDF(s) centred near s = 0; on the other hand, galaxy disks are dominated by rapidly rotating material which gives rise to a PDF(s) that is skewed towards high angular s (Sharma & Steinmetz 2005).

As discussed above, for nearby galaxies, it is common to perform photometric bulge-disk decompositions to measure the fraction of light that comprises the two main galaxy components (e.g. Kormendy 1977). However, most galaxies at high and intermediate redshifts have clumpy, compact and/or disturbed morphologies which cannot be described by familiar Hubble types (e.g. Glazebrook 2013), so traditional decomposition methods are not useful. The way forward is to develop a method of classifying galaxy morphology that is grounded in fundamental properties and which can be applied at all redshifts. The distribution of specific angular momentum PDF(s) is derived from both photometry and kinematics, so it contains more physical information than traditional empirical photometric bulge-disk decompositions. The different photo-kinematic signatures of the bulge and disk PDF(s) should allow to perform a decomposition based on fundamental properties of mass and angular momentum (Sweet et al., 2020; in prep.).

Recently PDF(s) has been calculated for a high-quality subset of the CALIFA survey of local galaxies (Sweet et al. 2020; in prep). The shape of the PDF(s) is quantified with the statistics of the distribution. Fig. 2.8 illustrates the correlation between skewness $b_1$ and both Hubble type T and bulge-to-total light ratio β. Earlier-type galaxies with larger bulges tend to have positively-skewed PDF(s), while the later types with smaller bulges have symmetric PDF(s). There are clear trends of the shape of the PDF(s) with morphology, but the spatial resolution is not sufficient to reliably recover a two component decomposition. What is needed is a significant advancement in our understanding of PDF(s) by calculating high-precision PDF(s) for a statistically significant spanning set of local and intermediate-redshift galaxies.

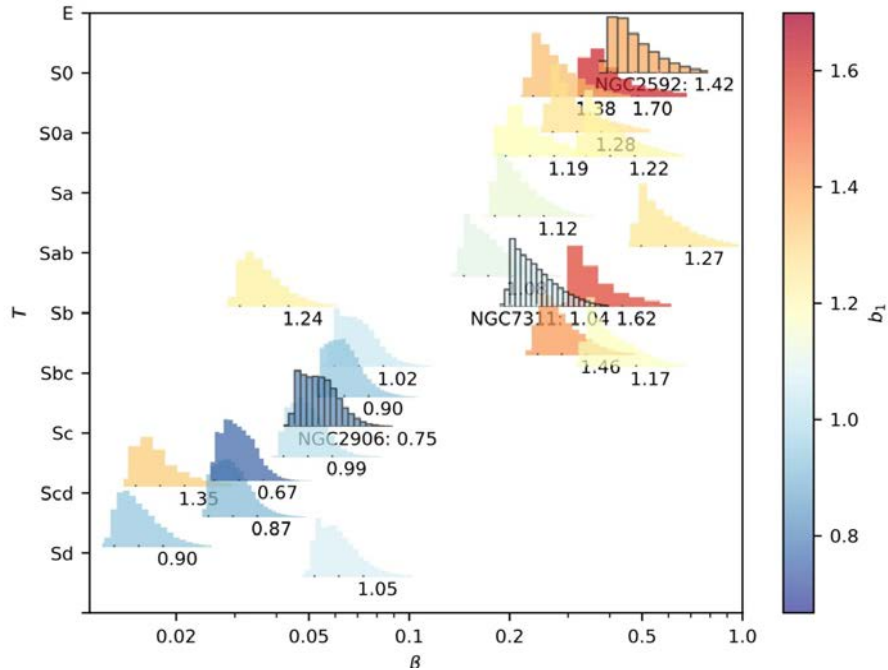

**Fig. 2.8: The skewness of PDF(s) correlates with bulge-to-total ratio β and Hubble type T**. *Probability density functions of normalised specific angular momentum (PDF(s)) for CALIFA galaxies, plotted in bulge-to-total light ratio β - Hubble type T space. An offset of up to ±0.5ΔT is added in the y-direction for clarity. Each PDF(s) has ticks at s=0, 1, and 2, and is labelled with the $b_1$ measurement for that galaxy. Highlighted with black outlines are three examples: 1) late-type spiral NGC 2906 with low β = 0.06 has a broad, symmetric PDF(s) centred on s = 1. 2) early type NGC 2592 with high β = 0.54 has a strongly-skewed PDF(s) which peaks nearer s = 0. 3) NGC 7311 with moderate β = 0.27 has a PDF(s) which is intermediate between the two extremes. The shape of PDF(s) is quantified by skewness $b_1 = \mu_3/\sigma_3$. More normal, symmetric distributions ($b_1 \approx 0.8$) are coloured blue, while strongly-skewed ($b_1 > 1.2$) distributions are coloured red. Galaxies with earlier types and larger β tend to have skewed PDF(s), while the later types with smaller β have symmetric PDF(s).*

The critical parameter for this work is the spatial resolution. To calculate and adequately characterise the structure of PDF(s), one requires high spatial resolution observations of stellar kinematics out to high multiples of effective radius (Sweet et al., 2019). High multiples of re are important to reach the flat part of the rotation curve where most $j_*$ is contained (e.g. 0.99$j_*$ at 3re), and so to provide a secure estimate for the total $j_*$. High spatial resolution is especially important in the central ~ 0.5re of a galaxy, where beam-smearing has the greatest effect on accurate determination of the velocity field. Improving the spatial resolution is also required to confirm that the scatter in the PDF(s) – morphology relation is due to the relative difficulties in performing photometric classifications such as β and T compared with the more physically-motivated kinematic classification b1. MAVIS will be the only instrument that can provide the wide-field / high spatial resolution integral field spectroscopy that is required to perform accurate PDF(s)-based decompositions and calibrate the PDF(s) – morphology relation.

To simulate the effects of the improved resolution provided by MAVIS we construct model example data cubes to approximate the three example galaxies presented in Fig 2.8, with photometric B/T = [0, 0.3, 0.5]. We convolve the recovered maps with nominal MAVIS AO PSF (Moffat profile with β = 1.8) at FWHM = 20 mas, and sample with 25-mas spaxels to construct the PDF(s). In Fig. 2.9 we compare CALIFA galaxy NGC 7311 to the analogous B/T = 0.3 model galaxy. The left-hand panel is the PDF(s) for NGC 7311, a Sab galaxy with intermediate B/T = 0.27. The two-component normal + log-normal fit does not well describe the data, particularly the 'bulge' at s 0. In the CALIFA dataset it is not possible to separate the PDF(s) of the bulge from PDF(s) of the disk due to the natural seeing and 0.1" spaxel sampling. In contrast, the simulated MAVIS dataset shown in the right-hand panel can clearly resolve the rapidly-rotating thin disk (PDF(s) mode > 1) and the dispersion- dominated bulge (PDF(s) mode ~ 0). The fine sampling and FWHM of MAVIS will allow such a physically-motivated kinematic decomposition. A 25 mas PSF FWHM is required to map the inner regions of the smallest bulges with <<100pc resolution.

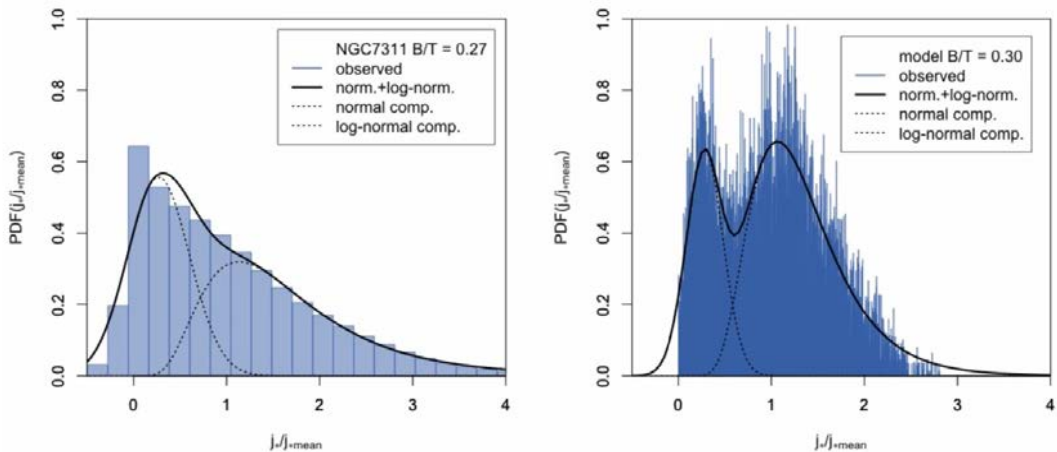

***Figure 2.9****: Left: PDF(s) for Sb galaxy NGC 7311, with intermediate bulge-to-total ratio 0.27. At this spatial resolution it is not possible to resolve separate profiles for the thin disk and the bulge. Right: PDF(s) for model galaxy with B/T = 0.3 at FWHM = 20 mas sampled by 25-mas spaxels. The disk and bulge components are clearly separable. Only MAVIS will be able to achieve this.*

## 2.5 Morphology of the first galaxies

**Science goal:** What is the faint end of the UV luminosity func on at the highest redshi s? Which are the structural propertes of the first galaxies?

**Program details:** deep imaging at the best resoluon obtainable in the opcal

**Key observa on requirements:** diffracon limited resoluon of a wide field of view and high sensivity for compact sources is required

**Uniqueness of MAVIS:** MAVIS will be the only opcal instrument capable of obtaining beer image quality and deeper exposures than HST

Astronomy changed when the directors of STScI decided to invest a significant amount of HST DDT to produce the Hubble Deep Field (HDF) and the Hubble Ultra Deep Field (UDF). For the first time high spatial resolution and high sensitivity images were used to obtain a statistically sound sample of normal, faint, high-z galaxies to study their characteristics (Williams et al. 1996, Beckwith et al. 2006). Initially, only optical cameras were used, then infrared bands were added after the WFC3 camera was installed. The amount of science obtained with these fields is impressive: the most distant objects are discovered in these fields, their morphology and sizes are studied, their stellar populations are investigated by their colours in multiple bands. Most of our knowledge of the universe at z>3 is based on these and other deep fields.

Among many other results, the faint-end slope α of the UV LF was measured for the first time in high-z galaxies, discovering that it increases with redshift (e.g., Bowens et al. 2015). This slope is a critical parameter of high-z galaxies, as it can constrain models of galaxy formation (e.g., Torrey et al. 2014) and dominates the number of ionizing photons produced that could produce the re-ionization of the universe (e.g., Robertson et al. 2013). The role of faint galaxies in the epoch of reionization is uncertain not only for the unknown escape fraction of ionizing photons (e.g., Driver et al. 2016), but also for the uncertain extrapolations of the LF toward faint galaxies (Robertson et al. 2013).

The most distant galaxies are selected with the drop-out technique, i.e., by comparing the colours in three adjacent filters. At high redshift, intergalactic hydrogen absorbs virtually all the continuum radiation below Lyα, producing a sharp edge with peculiar colours on three filters bridging the edge. Another technique to estimate redshifts and isolate distant galaxies is SED fitting, which uses all the available filters instead of just three. The results are very similar because these "photometric" redshifts are dominated by the colours on the same few filters across the Lyα break. The sensitivity in detecting high-redshift galaxies is limited by the depth of the images in the filters below the Lyα line because this limits the sharpness of the edge that can be detected and distinguished from the more common, smooth spectrum of lower-redshift, red galaxies.

The UDF, the deepest field, reaches 10σ limiting AB magnitudes of 29.45 at 606nm and 29.35 at 775nm. These limits are computed for point sources in square apertures of 200mas on a side, optimized for the HST PSF of about 80mas FWHM (Beckwith et al. 2006). Studies of high-z galaxies (e.g. Bowens et al. 2015) quote a slightly different limiting magnitude of 28.8AB 5σ in 350 diameter circular apertures at 775nm. Given that very red colors are needed for detection (e.g., i775-z850>1 for galaxies at z~6), only relatively "bright" galaxies (z850<27.8) can be safely selected as high-z objects, even if the z850 detects galaxies 1 mag fainter. This poses a serious limitation on the number of low-mass galaxies selected at these redshifts and, as a consequence, makes uncertainties of the low-mass end slope of the LF very large.

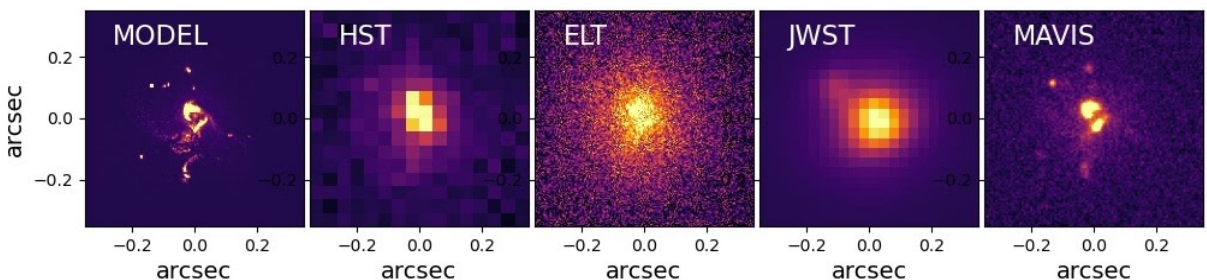

**Figure 2.10:** *Example of a morphological study of an high-z galaxy as observed with HST, ELT-MICADO, JWST-NIRCam and MAVIS in I band. Based on the z=5 high resolution simulated "Althaea" galaxy (Pallottini et al. 2017), shown in the first panel. The other panels show instead how the same target of 25 AB magnitude would appear in I-band if observed with different facilities for a fixed exposure time (1h). In the VLT/MAVIS image, clumpy regions as faint as 29th AB mag are detected with a signal-to-noise-ratio 5.*

Typical high-z galaxies are very compact. For example, the z~6 i-dropouts typically have effective radii Re~500pc, corresponding to about 80mas, at the limit of HST spatial resolution. ***MAVIS will be able to image these compact galaxies, studying their UV morphology better than any other optical instrument***. In particular the presence of star forming clumps will be investigated and their properties compared with models, and those observed at lower redshifts (see Section 4.6). Faint, high-redshift QSOs are also interesting targets that can be studied together with their host galaxies.

The optical imaging with MAVIS in different filters will be complemented with future high-resolution near-IR cameras such as ELT/MICADO, with similar resolution at longer wavelengths. MAVIS will allow us to select galaxies up to z~7, while only drop-out galaxies above z~11 can be selected by using MICADO images alone. Moreover, at lower redshift VLT/MAVIS observations are required to map the UV-optical emission of z~1-3 galaxies, sampling their young stellar populations. Therefore, the combination of near-IR imaging and multi-band MAVIS surveys will ultimately allow us to better understand the phenomena driving the morphological assembly across the whole galaxy assembly history, from the first galaxies to the local Universe, providing unique constraints to models of galaxy formation and evolution.

# Case study: A MAVIS Deep Field

**Key Questions:** What are the sizes and UV morphologies of the first galaxies? What are the galaxy's number counts at the faint end of the luminosity function up to z~7?

**Why MAVIS?** MAVIS will allow the deepest optical images ever obtained for point sources and compact galaxies. The expected 5σ limiting mag for a point source is 30.4 AB in the I band in 10 hours in a square aperture of 49mas$^2$ (7 pixels). ***This is about 1 mag deeper than the UDF in i775 even if the exposure time of the latter is about 10 times longer (96h)***. Extrapolating the counts in the UDF from Driver et al. (2016) and taking into account the smaller FoV, we expect to detect about 600 galaxies in a single 30"x30" field. The majority of these galaxies will be well resolved thanks to MAVIS exquisite resolution, and therefore we will be able to study their morphology and the distribution of colours and stellar populations. MAVIS capabilities are illustrated by Fig. 2.11. This shows a simulated deep field using 30h of exposure time in HST and MAVIS in the I814 filter. It is evident that MAVIS can provide better detection of faint, compact galaxies, and allows for a much better study of their morphologies.

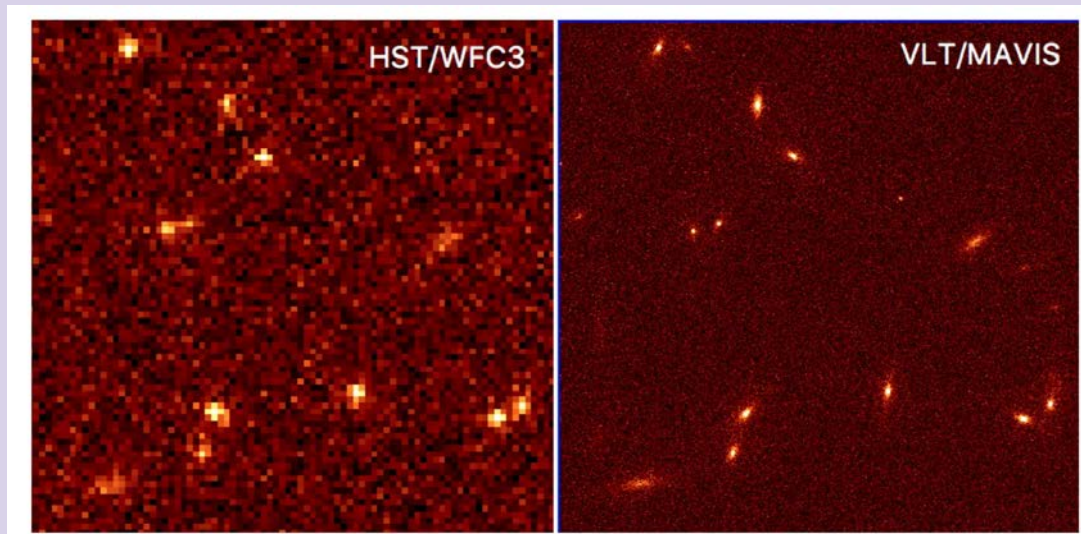

*Figure 2.11: A comparison between a simulated deep field with HST (left) and MAVIS (right). The field is 3.3"x3.3", and it contains a number of compact galaxies modeled with Sersic profiles observed for 30h in the same F814W filter to allow for an easier comparison.*

**How will MAVIS be used:** MAVIS in imaging mode will be used to obtain these ultradeep exposures, with the best angular resolution obtainable at optical wavelengths.

**Field selection:** There is ample freedom to choose for the best asterisms inside or outside existing extragalactic deep fields, to provide the best PSF conditions and maximize the performances of the AO system.

**Exposure time:** Standard photometric filters are suitable for source detection and characterisation, allowing the selection of dropout galaxies and to study their morphologies. The integration time required to obtain deeper images than HST UDF is 10 hrs per filter.

# 2.6 Spectroscopy of the first galaxies in the epoch of reionization

**Science goal:** Which are the sources that reionized the Universe?

**Program details:** IFU spectroscopy at high spatial resolution to detect and resolve Lyα in field and lensed galaxies at z>5

**Key observation requirements:** High sky coverage to target lensed galaxies; high sensitivity for compact sources; R~5000 to resolve the Lyα line profile; wavelength coverage up to 10000Å.

**Uniqueness of MAVIS:** MAVIS is the only optical IFU capable of observing with high sky coverage, high spatial and spectral resolution, and to access the 9300-10000Å region.

The epoch of reionization marked a major phase transition of the Universe, during which the intergalactic medium (IGM) became transparent to UV photons. Determining the timeline and topology of the reionization process, and the physical processes involved, represent the latest frontier in observational cosmology.

A substantial step in our knowledge of the reionization timeline has been recently given by the identification of a significant decrease of the fraction of Lyα emitters between z~6 and z~7 (e.g., Stark et al. 2010, Fontana et al. 2010, Pentericci et al. 2014, 2018, Ono et al. 2012, Schenker et al. 2012). The analysis of independent lines-of-sight also suggests a patchy reionization topology (Treu et al. 2012, Pentericci et al. 2014). However, our knowledge of the topology of reionization is very limited and only very few investigations on the relation between galaxy density, proto-clusters and Lyα visibility have been carried out so far (Castellano et al. 2016, 2018; Trebitsch et al. 2020). In fact, understanding reionization topology is likely to become a key topic in the future thanks to 21cm tomography (e.g., Elbers & van de Weygaert 2019) and the possibility to cross-correlate 21cm signal and galaxy density (e.g., Hutter et al. 2017). Constraints on the relation between galaxy density and neutral hydrogen fraction will allow us to discriminate between "inside-out" and "outside-in" reionization scenarios (e.g., Choudhury et al. 2009), and will enable a more in-depth understanding of the sources responsible for the reionization process.

The possibility to access redshift 7 sources is therefore a key feature for an instrument like MAVIS, while ELT will complement and extend to even higher redshift (z>>7). It implies the 1 micron wavelength should be approached in the red side of the MAVIS wavelength interval. Each Angstrom gained at lambda>9300A corresponds in fact to (2-3) $10^5$ years of the cosmic time, eventually reaching 720 Myr after the Big Bang at z=7.22 (Lya at 1 micron). The wavelength range 9300-10000Å will be therefore crucial to probe the tail-end of the reionization epoch (z ~ 6.5 - 7.2).

On one hand, diffraction limited AO imaging at lambda>9300Å allow us to finally unveil the ultraviolet morphology or put stringent limits on the sizes of galaxies at 6<z<7, for which diffraction limit optical imaging is required (Re~500pc, corresponding to about Re = 80mas).

On the other hand, IFU spectroscopy up to lambda~10000A allow us to probe UV high-ionization lines like CIV1550 up to redshift 5.4, and Lyα to z=7.22; IFU spectroscopy with FoV 7"x5" will enable for the first time 1) a two-dimensional mapping of such emission lines, especially for strongly lensed galaxies or arcs requiring flexibility when following their shape (see e.g. Fig 2.12, right); and 2) the possibility to blindly search for the presence of clustered line emitters at z~6.5-7, within a 25 x 35 physical kpc field of view (Fig. 2.12, left);

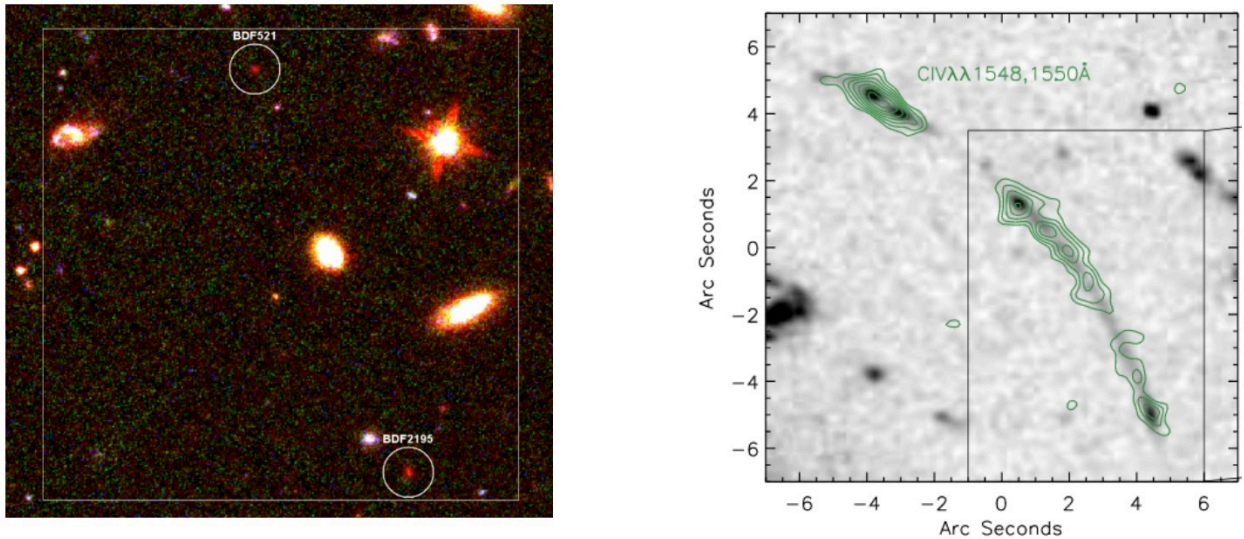

***Figure 2.12:*** *Left panel: from Castellano et al 2018, a pair of galaxies confirmed at z=7.1 with FORS instrument (Vanzella et al. 11) separated by 90 kpc (~17 arcsec); two IFU pointings with MAVIS can blindly search for clustered LAEs belonging to the same structure. Right panel: a z=4.88 strongly lensed galaxy in which the CIV 1550 nebular emission (observed at 9114Å) is spatially resolved and detected with MUSE-WFM (green contours, Smit et al. 2017). MAVIS can extend such studies to higher redshift (z~5.3), higher spectral resolution (R>5000) and higher angular resolution (a few tens of mas).*

A spectral resolution R>5000 is key to probe signatures of emission lines of down a few dozens km/s FWHM. In particular Lyα multi-peaked profiles (e.g. Vanzella et al. 2020) require R>5000 in order to allow proper radiation transfer modeling and recognize narrow features as signatures of transparent ionized channels for the escaping of ionizing photons (see Fig. 2.12, right; see also Fig. 2.13), a key feature to identify dense young star clusters that can contribute to the ionization of the IGM through holes created by stellar feedback.

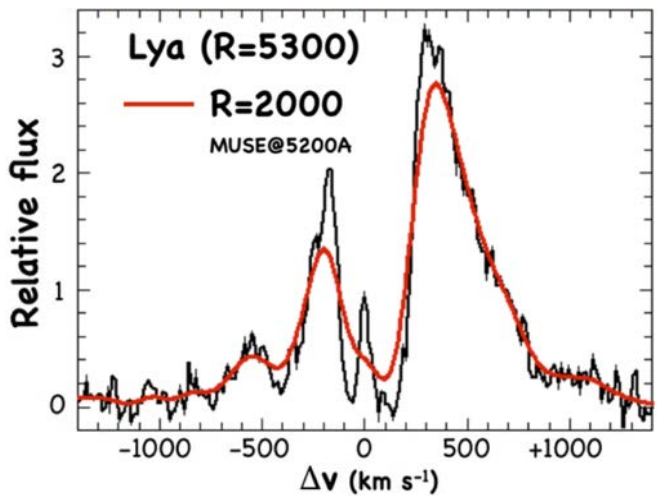

***Figure 2.13:*** *Lyα emission from 'Ion2' - a well-known LyC emitter at z = 3.2121 lying in the CDFS. The black line shows the XShooter spectrum with R=5300 at the observed wavelength (~5200Å), with multiple peaks clearly visible, indicating a 'leaky' IGM. At the lower resolution of MUSE (red line), this structure, and in particular the central peak, is not recovered. Higher redshift sources likely exhibit a similar structure. Adapted from Vanzella et al. 2020.*

# 3 Resolving the Contents of Nearby Galaxies

## 3.1 Introduction

By the mid 2020s when MAVIS is expected to become available, our view of our own Milky Way galaxy will have fundamentally and irreversibly changed. Through wide-field stellar spectroscopic surveys like GALAH, APOGEE, 4MOST, and the planned SDSS-V Galactic Genesis Survey, combined with exquisite photometry and astrometry from the then-completed GAIA mission, we will have mapped the entire hierarchy of structure within our Galaxy, gaining a view that is both global, yet densely sampled throughout the disk- and bulge-dominated regions. This will include high-quality spectra of individual stars, rich with information about their basic physical parameters, including their ages, chemistry, and kinematics; and from interstellar absorption, the detailed composition and dynamics of the Milky Way's gas and dust, from which new stars are forming.

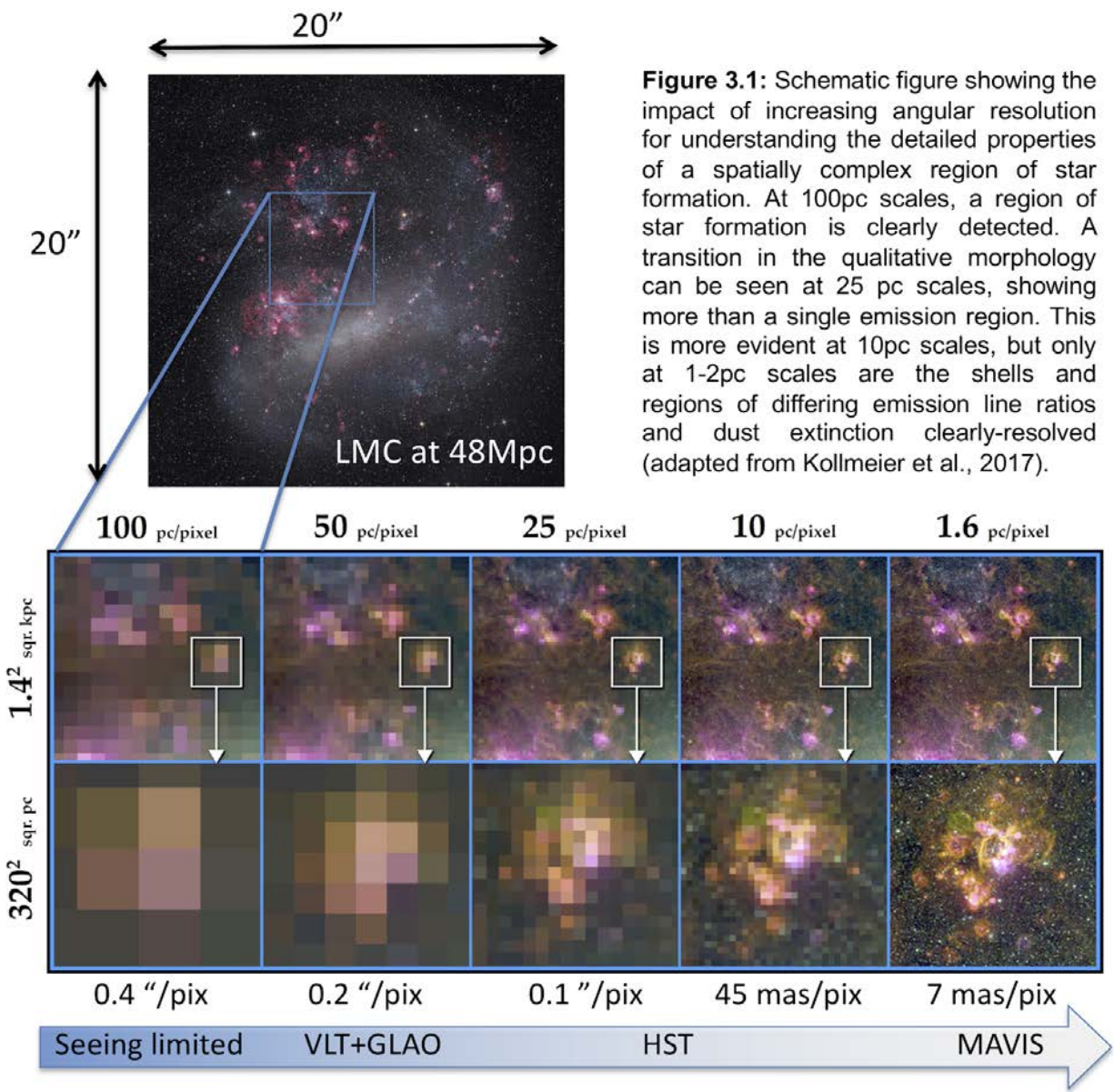

**Figure 3.1:** Schematic figure showing the impact of increasing angular resolution for understanding the detailed properties of a spatially complex region of star formation. At 100pc scales, a region of star formation is clearly detected. A transition in the qualitative morphology can be seen at 25 pc scales, showing more than a single emission region. This is more evident at 10pc scales, but only at 1-2pc scales are the shells and regions of differing emission line ratios and dust extinction clearly-resolved (adapted from Kollmeier et al., 2017).

These properties as measured in the Milky Way provide the best quantitative constraints on the most uncertain galaxy formation physics. However, to make full use of these advances, we must be able to compare these findings on physically similar sub-parsec scales to galaxies other than the Milky Way, and in environments that do not share the cosmic history of the Galactic neighbourhood. Working near the optical diffraction limit of the VLT, MAVIS will push the sub-parsec resolution regime well beyond the local group, opening new regimes of galaxy environments and morphologies to explore on these scales, and driving a new wave of quantitative tests of galaxy formation models.

## Confronting the sub-grid physics of galaxies

Identifying the processes regulating star formation in galaxies is an outstanding and highly active area of research in astrophysics. It is well-established that some form of energy feedback is required to prevent over-cooling of gas and a subsequent excess of star formation in galaxy formation models. It is also understood that this source of feedback may vary with galaxy mass, and that environmental effects, such as interactions with other galaxies in close proximity or with the hot intra-cluster medium of galaxy clusters, seem also to play a role. While the results of these effects are manifest in the global (kpc-scale) properties of galaxies, as demonstrated by the various relationships between total mass, star formation, gas surface density, angular momentum, etc., the physical processes themselves (such as stellar-wind or AGN-driven feedback) originate on far smaller (of order parsec) scales. As such, the diagnostic power of observations to constrain and distinguish between different physical models of galaxy formation must also lie in understanding galaxies on these scales.

Figure 3.1 illustrates this point, showing a schematic view of the LMC as observed at a distance of 48Mpc. The left-most middle inserts show what would be obtained with the sampling and typical image quality of MUSE on the VLT, both seeing-limited and with GLAO, and both revealing the clumpy nature of the ionized gas emission. The bottom insert shows more detail for one 'knot' of star formation. As we move down in spatial scales, below 50pc pixels we obtain a fundamentally different view of this knot, which becomes apparent as a conglomerate of separate regions. Only at the <10 parsec scales provided by MAVIS is it possible to distinguish the networks of shocks and ionization fronts, giving critical insight into the true nature of the ISM phases and star formation energetics. MAVIS will provide images with < 2pc pixel scales and spectroscopy with < 5-10pc spaxel scales out to 50Mpc, encompassing essentially every major environment and galaxy mass and morphology class at the scale of single HII regions.

## 3.2 Understanding the origin of early-type galaxy envelopes

**Science goal:** How do early-type galaxies form their outer envelopes?

**Program details:** Apply deep MAVIS imaging capabilities to derive optical colours of individual ancient stars in the outer parts of early-type galaxies out to 10Mpc

**Key observation requirements:** The ability to obtain point-source photometry for 30th-magnitude sources in optical broad-band filters, in crowded fields, and with sufficient field of view to build deep colour-magnitude diagrams.

**Uniqueness of MAVIS:** The ability to measure 30th magnitude sources in modest exposure times, and the potential to combine optical with infrared colours from ELT, make this a unique science case for MAVIS.

Around half the stellar mass in the Universe resides in so-called 'early-type' galaxies (ETGs) - objects largely devoid of any ongoing star formation, and which have smooth appearances and high stellar densities. In addition to accounting for a large portion of the current day stellar mass, these ancient stars also belong to the same population that contributed to reionizing the Universe at early times. Our understanding of this important stellar population has come almost exclusively from integrated light spectral observations. These have shown the dense central regions of ETGs to be ancient, metal rich, and in particular having a higher [α/Fe] than the solar neighbourhood. Recent observations also indicate a different stellar initial mass function (IMF) than the canonical Milky Way function (Conroy & van Dokkum 2012; Cappellari et al. 2012). The central regions of ETGs therefore clearly represent an epoch of star formation distinct from the solar neighbourhood, both in time and in physical conditions, and are thought to relate to the earliest 'compact nuggets' of star-forming galaxies at the highest observable redshifts.

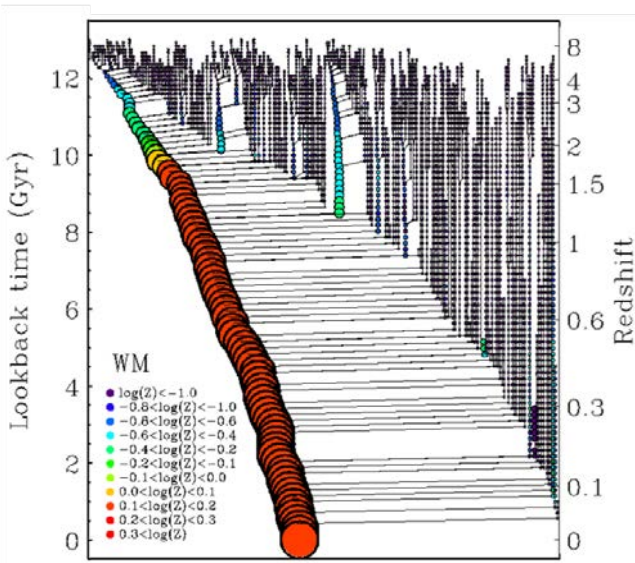

**Figure 3.2:** *Simulated 'merger tree' showing the growth in mass (shown by symbol size) of a typical ETG over cosmic time. Note that the large mass of this galaxy is the result of accreting a very large number of low-mass galaxies, with correspondingly low metallicity (symbol colour). These objects are disrupted when the main galaxy potential is comparable to the binding energy of the satellite, therefore lower-mass galaxies, with lower metallicities, are predicted to deposit their stars at larger radii. (From Hirschmann et al. 2015)*

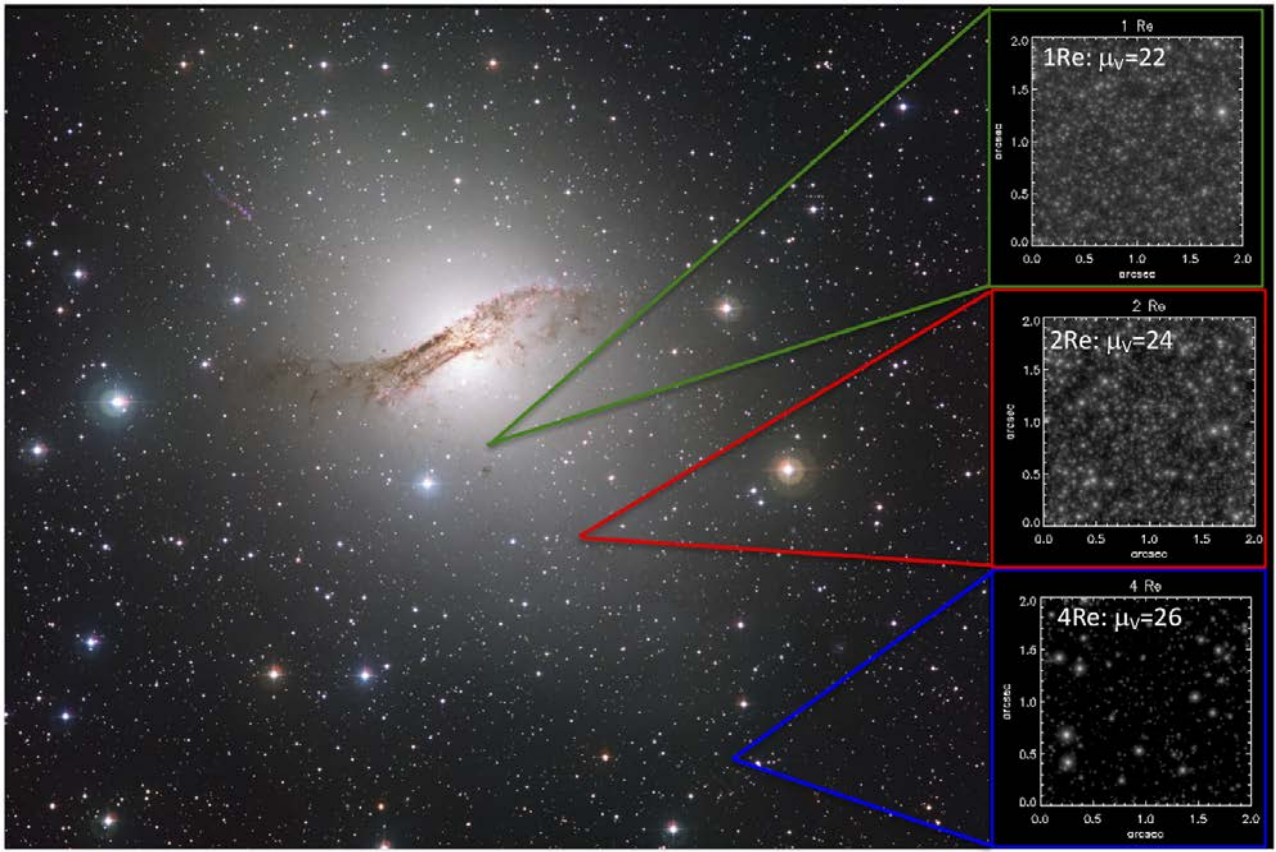

**Figure 3.3.** *Early-type galaxy Centaurus A, at a distance of 4Mpc, with simulated central 2" thumbnails of MAVIS V-band images at different radii (inserts), based on 10hr integrations. Main image credit: ESO*

The sizes and masses of current-day ETGs, however, imply a strong evolution of both properties since that earliest epoch. A foundation stone of current galaxy formation theories is that ETGs continue to build their size and mass not through star formation, as more common disk galaxies do, but through the ongoing accretion of (typically gas-poor) neighbouring galaxies, and in particular, of low-mass satellite galaxies (Fig. 3.2). Disrupting low-mass galaxies in the outer regions of ETGs can conveniently explain both the mass and size growth, and the incidence of metallicity gradients, since lower mass systems have lower metallicity, and are deposited at larger radius due to lower binding energies.

This scenario, however, is largely untested, with the ages of the outer envelopes a key unknown to distinguish accretion versus in situ formation. The outer regions of ETGs are notoriously difficult to measure in integrated light, as the surface brightnesses are too low (typically $\mu_V$>26). Even with state-of-the-art instruments like MUSE, obtaining spectroscopy with sufficient S/N to break the degeneracy of age and metallicity, and distinguish between different stellar populations older than ~10 Gyr is not possible. MAVIS allows progress in this field by instead resolving individual stars, allowing the significantly more accurate method of colour-magnitude diagrams (CMDs) to be applied.

Fig. 3.3 presents a simulated case study applying this method to the outer regions of Centaurus A, located around 4 Mpc from the Milky Way. Three fields are shown, to illustrate the impact of moving to sparser fields. This also allows us to address both the faintest limits in sparse fields, and where crowding may start to impinge on the photometric accuracy. At 2 effective radii ($R_e$), the surface brightness ($\mu \approx 22$ mag/sq arcsec) makes source confusion a potentially limiting factor, increasing photometric uncertainties. However, Fig. 3.3 demonstrates that populations of 10 and 14 Gyr can still be easily distinguished. This implies that a similar accuracy will be comfortably reached in the lower surface brightness outer regions ($\mu \geq 26$ mag/sq arcsec) where integrated light spectroscopic studies struggle. For example, at this surface brightness it would take around 70 hours of integration with MUSE (assuming the optimistic case that all 90,000 spectra in each exposure are co-added with only Poisson noise) to reach sufficient S/N to separate these old populations. By contrast, 1-sigma errors of less than 0.2 magnitudes at the distance of Centaurus A can be obtained in a few hours with MAVIS.

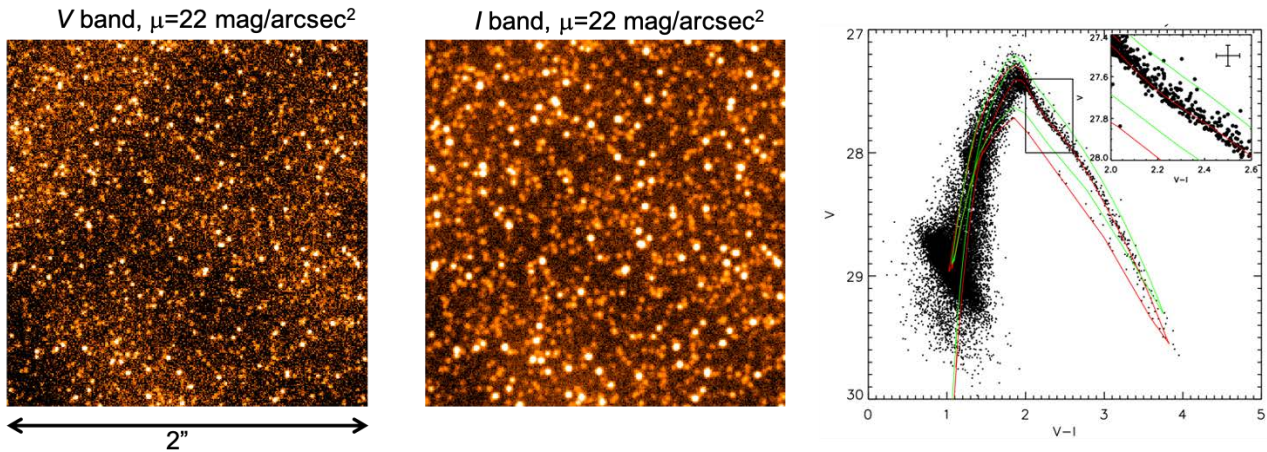

**Figure 3.4.** *Left and Centre: V- and I-band thumbnails of the more crowded 2Re pointing shown in Fig. 3.3. Right: derived CMD from the full MAVIS field, showing that the 14 Gyr (red) and 10 Gyr (green) isochrones can be clearly distinguished. Such age fidelity from integrated light spectroscopy is prohibitively expensive in the lower-surface brightness outer regions, even with MUSE.*

## 3.3 Solving the origin of the UV-excess in nearby early-type galaxies

**Science goal:** What stellar evolutionary process gives rise to the UV-luminous population in ancient galaxies?

**Program details:** Apply deep MAVIS imaging capabilities at blue wavelengths to constrain the Horizontal Branch morphology of early-type galaxies out to 5-10Mpc

**Key observation requirements:** The ability to obtain point-source photometry for 28-30th-magnitude sources in B- and V-band filters, in crowded fields, and with sufficient field of view to build deep colour-magnitude diagrams.

**Uniqueness of MAVIS:** The ability to measure 28-30th magnitude B-V colours in modest exposure times, make this a unique science case for MAVIS.

It has been known for almost 50 years that inactive and passive early-type galaxies (ETGs) emit a surprisingly significant amount of far-ultraviolet radiation (the so-called UV excess, or UVX, see O'Connell 1999), typically attributed to some population of hot, old stars. The most likely candidate stars producing the UVX are hot horizontal-branch (HB) stars and later stellar phases (P-AGB, P-EAGB, AGB-Manqué, Greggio and Renzini 1990). Individual hot HB stars have been detected with HST (Brown et al., 1998, 2000) in M32 and in the bulge of M31. However colour information of these stars is still lacking, hence their temperature is not known, nor their evolutionary phase. MAVIS, with its resolution of ~20 mas, a field of view of 30"x30", and optical coverage of the B, V, R, I bands, offers an ideal opportunity to go a step further than HST in understanding the properties of the stars responsible for the UVX in ETGs, their luminosity function, evolution, and dependence on age and metallicity.

In order to illustrate how MAVIS can be used to discriminate between the different candidate stars, Fig. 3.5 shows evolutionary tracks of the possible contributors to the UVX of ETGs. The thin black line shows the isochrone of a solar metallicity, 10 Gyr old stellar population up to the tip of the Red Giant Branch, while the colored lines show where stars from the same stellar population in more advanced evolutionary stages (i.e. HB and post-HB) could be found. Violet points show the canonical evolution (HB + post AGB); green shows the evolution of HB + post EAGB; and the other tracks are relative to hot HB stars and their AGB-manqué progeny. The different tracks pertain to different masses of HB stars, which decrease going from the red to the blue part of the CMD. Different evolutionary masses of HB stars can be realized in stellar populations with the same age and metallicity by increasing the Helium abundance and/or the mass lost on the Red Giant Branch. The thick black diagonal line indicates the approximate MAVIS detection limit for an exposure time of 10,000 sec. and a S/N=5 for an ETG with a distance modulus of 26.0. Clearly MAVIS observations would allow us to discriminate between the different candidates, and to accurately determine their evolutionary phase.

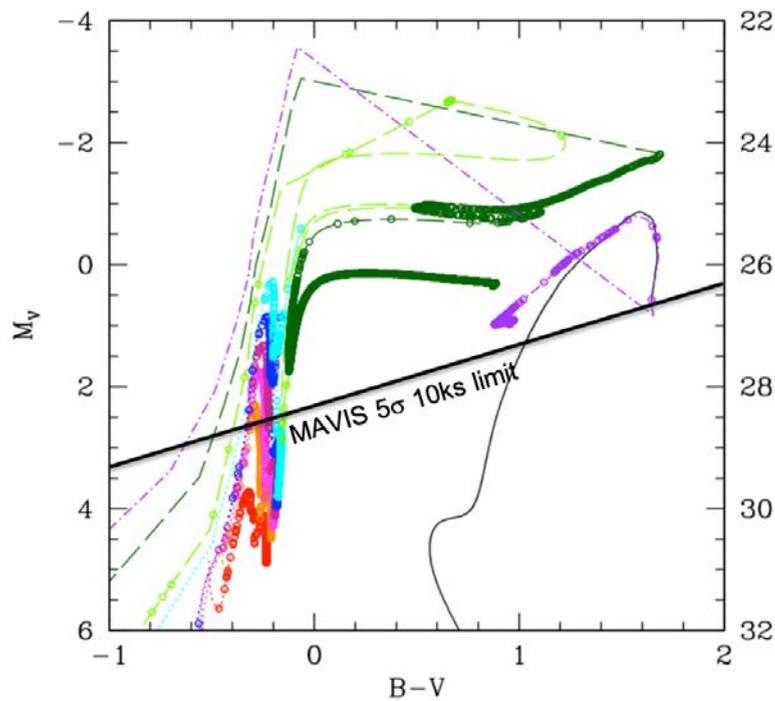

***Figure 3.5:*** *Colour-magnitude diagram showing evolutionary tracks of the possible contributors to the UVX in ETGs. See text for details.*

This case makes use of the B- and V-bands, thereby making the classical colour-magnitude diagram ($M_V$ vs. B-V). Blue coverage is optimal for this science, and indeed the overlap of the tracks in Fig. 3.5, crowding effects, and the faintness of low-mass candidates (red points) would all be further reduced with access to even bluer bands (e.g. U-band) at comparable spatial resolution and sensitivity. This may be possible with MAVIS, but we focus here on B and V, as these have a more secure performance. V and R bands could also be used, but with a reduction in sensitivity to the different UVX candidates.

## Case Study: Resolved Stellar Populations of Early-Type Galaxies

**Key Questions:** What are the stellar ages and metallicities in the envelopes of Early-Type Galaxies (ETGs)? How do ETGs build their outer envelopes?

**Why MAVIS?** MAVIS will allow the deepest and sharpest optical images ever obtained for point sources. Optical CMDs provide the sharpest tool to dissect the age and metallicity properties of ETG envelopes. The surface brightness regimes are too faint for integrated light spectroscopy.

**How will MAVIS be used:** MAVIS in imaging mode will be used to obtain deep exposures in V and I bands and multiple radii in a sample of nearby ETGs.

**Field selection:** MAVIS will routinely enable point source limiting AB magnitudes of 30.0 to be obtained in optical broad bands. This puts the 'knee' of the CMD, where age accuracy < 1-2 Gyr can be obtained, within reach out to 10Mpc. According to HyperLEDA, there are at least 10 galaxies within 10 Mpc of the Milky Way with ETG morphology and stellar mass greater than $10^9 \, M_\odot$ observable from Paranal.

**Exposure time:** To obtain a 5σ limiting mag for point sources with MAVIS of 28 (30) mag in the V and I band requires 2 and 1.5 (10 and 9) hours exposure time, respectively. Assuming three radial locations within each galaxy, measuring CMD-derived age and metallicity profiles for all 10 ETGs would take 200 hours.

## 3.4 Cepheids as probes of galaxy chemistry and the cosmological distance scale

**Science goal:** How do the chemical enrichment histories of other late-type galaxy disks compare with the Milky Way? Can metallicity effects explain the tension between near-field and CMB cosmology, or do we need an additional source of dark radiation in the early universe?

**Program details:** Apply deep MAVIS spectroscopic capabilities at blue wavelengths in crowded fields to measure the metallicity of bright Cepheid variable stars in disk galaxies out to 30Mpc.

**Key observation requirements:** Blue (V-band) spectroscopy at high spatial-resolution (due to crowded fields), high spectral resolution (to measure abundances), and high sensitivity (faint sources).

**Uniqueness of MAVIS:** Combination of sky coverage, spatial- and spectral-resolution, and peak sensitivity.

Classical Cepheids are pulsating stars of 4-10 $M_{\odot}$, obeying well-defined relations between their intrinsic luminosity and their pulsation period. Such Period-Luminosity (PL) relations make them fundamental calibrators of the cosmic distance scale. Besides this important role, Cepheids also have a number of properties that make them the best stellar proxy available for the gas-phase chemical composition of disc galaxies. In fact, they are young (< 600 Myr, Bono et al. 2005), age-datable by their period, luminous, hence visible to large distances, and their surface temperatures (~5,000 K) allow for precise abundance determinations of many different chemical elements (da Silva et al. 2016). Because they are so young, they still sit relatively close to their birth place, and hence their chemical composition reflects that of the Interstellar Medium (ISM) at the same location. This means that high-resolution spectroscopy of Cepheids in distant disc galaxies with MAVIS will enable us to map their ISM in many different chemical elements for the first time. The main goal of this proposal is to obtain such a map for 13 disc galaxies up to 30 Mpc away and for which many (10–400) Cepheids have already been identified with HST photometry. This data will allow us to tackle two fundamental open problems described in the following.

### 3.4.1 Chemical evolution of late-type galaxy disks

The present-time chemical composition of the ISM of each galaxy encodes its evolutionary history, and provides the most fundamental observable against which we can test up-to-date chemical evolution models. More specifically, the spatial distribution of the heavy elements within a galaxy holds clues to its star formation history, gas inflow and outflow, stellar yields and initial mass function (Matteucci, 2003; Chiappini et al. 2001; Feltzing et al 2013). In the past 20 years, the efforts of the scientific community have focused on the measure of the mean radial metallicity gradient, i.e. the iron distribution as a function of the distance from the galactic center.

While this is a fundamental observable to test the so-called "inside-out" formation scenario of galactic evolution models, recent works (Sanchez et al. 2014; Ho et al. 2015) have shown that the scatter around the metallicity gradients between galaxies is almost as large as the mean of the metallicity gradient, thus suggesting that a simple linear metallicity gradient does not fully describe the metallicity distribution in these galaxies. Similarly, recent studies (e.g. Hayden et al. 2015; Jacobson et al. 2016) on the Milky Way (MW) have shown that the chemical abundance pattern follow a more complex spatial structure, consequence of the secular galactic evolution processes (e.g. mixing and radial migration due to non axis-symmetric gravitational potential, heating due to mergers, waving etc.). In this context, it is fundamental to compare the MW to external spiral galaxies, for which we can measure all the structural parameters and relate the abundance pattern to their global properties, such as the total mass.

However, while it is now possible to measure abundances of many elements for millions of individual stars in the MW and its satellite thanks to large spectroscopic surveys [e.g. APOGEE (Majewski et al. 2017), Gaia-ESO (Randich et al. 2013), GALAH (De Silva et al. 2015) etc.], only global abundances can be measured for distant galaxies to date. In fact, integrated properties of galaxies at redshift z > 0.1 can be obtained via Integral Field Unit (IFU) spectroscopy (Sanchez-Menguiano et al. 2018, and references therein), while direct measurements of the metallicity are still possible for intermediate-distance galaxies by detecting emission lines of HII regions. Such regions are good probes of the gas-phase metallicity, because they are ionised by young massive OB stars, but they only allow for measurements of a few elements (Oxygen, Nitrogen), and even such abundances are affected by the assumptions made on the gas atomic parameters and the adopted model for the HII region (Bresolin 2014). Yet, a full characterization of the chemical content in many elements is fundamental for improving our understanding on chemical enrichment in different environments and its implications for chemical evolution in the context of hierarchical galaxy assembly. This is achievable with MAVIS by adopting bright stellar probes of the ISM, such as Cepheids, which can be directly compared to their counterparts in the MW and nearby galaxies.

### 3.4.2 The calibration of the distance scale

A direct, high precision measurement of the local value of the Hubble constant ($H_0$) provides independent constraints on the spatial curvature of the Universe, it's dark energy content, and neutrino physics. Yet, there is a 3.8σ disagreement between the value of $H_0$ determined from modeling the CMB anisotropies (Planck Collaboration, 2016) and the one measured via the Cepheid-based distance ladder (Riess et al. 2016, 2018). This tension can be explained either by considering a non-standard physics in the cosmological model, or by performing a full scale re-analysis of the systematics in the Cepheids-based distance scale.

In the first case, a suggested solution is the inclusion of an additional neutrino-like species which can act as an additional source of dark radiation in the early Universe (Wyman et al. 2014). In the second case, it is necessary to further improve the accuracy of Cepheids' distances to external spiral galaxies, as Cepheids are used to calibrate the peak luminosity of Supernovae type Ia (SNeIa) that are, then, used to measure $H_0$ out in the Hubble flow. In this context, the largest systematic uncertainty is the relatively unconstrained role of metallicity in modulating the Cepheids' PL relation. In fact, the metallicity adopted for Cepheids in external galaxies may be incorrect, and, even if the metallicity is correctly measured, the metallicity effects on the PL relation may be underestimated.

To estimate metallicity, current methods rely on the conversion of Oxygen abundances measured from HII regions to predicted stellar iron abundance at the Cepheids' location. However, this method can introduce errors on the Cepheids' metallicity estimate, due either to systematics in the HII emission lines analysis or to the intrinsic scatter around the mean metallicity gradients. The second issue of mitigating the metallicity effects on PL relation studies can be addressed by using only spiral galaxies with a similar metallicity, in such a way that the metallicity correction to the PL is cancelled out. But once again, this assumption relies on the metallicity derived from the HII regions. **A direct measurement of Cepheids' individual metallicity offers the clear advantage of being independent on any assumption, either on the gas conditions or on the metallicity distribution in these galaxies.**

### 3.4.3 Observational Needs

The starting point of our survey is the list of Cepheids identified in 13 spiral galaxies by the SH0ES (SNe, H0, for the Equation of State of dark energy) program visible from the VLT and shown in Fig. 3.6. In fact, by measuring the abundances of Cepheids' in these galaxies, we can also satisfy the criteria needed for the abundance project, as the target list consists of intermediate-distance spiral galaxies, mostly face-on. For each of these galaxies we plan to obtain high-resolution spectroscopy (R=5,000-10,000 - Fig. 3.7) of all the Cepheids already known in these galaxies in order to measure their detailed chemical abundances. The details about the position on the Sky, the distance in Mpc and the number of Cepheids identified in each galaxy are also given in the table of Figure 3.6. The luminosity of the Cepheids with a pulsation period of 30 days is given in the last column of the table, and it can be considered the faint end of the population of Cepheids in our target list. In fact, the sample of Cepheids identified with the HST photometry is biased towards long period Cepheids, because they are the most luminous.

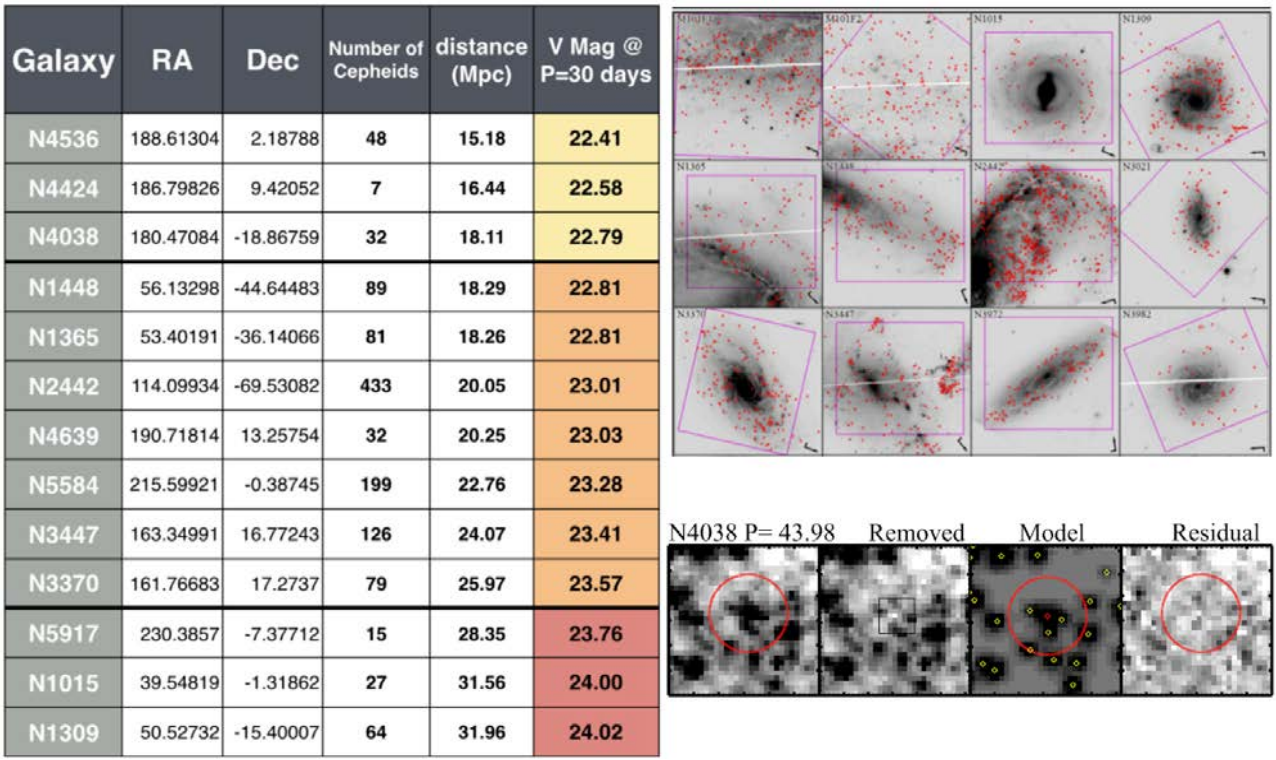

| Galaxy | RA | Dec | Number of Cepheids | distance (Mpc) | V Mag @ P=30 days |
|--------|-----|-----|--------------------|----------------|--------------------|
| N4536 | 188.61304 | 2.18788 | 48 | 15.18 | 22.41 |
| N4424 | 186.79826 | 9.42052 | 7 | 16.44 | 22.58 |
| N4038 | 180.47084 | -18.86759 | 32 | 18.11 | 22.79 |
| N1448 | 56.13298 | -44.64483 | 89 | 18.29 | 22.81 |
| N1365 | 53.40191 | -36.14066 | 81 | 18.26 | 22.81 |
| N2442 | 114.09934 | -69.53082 | 433 | 20.05 | 23.01 |
| N4639 | 190.71814 | 13.25754 | 32 | 20.25 | 23.03 |
| N5584 | 215.59921 | -0.38745 | 199 | 22.76 | 23.28 |
| N3447 | 163.34991 | 16.77243 | 126 | 24.07 | 23.41 |
| N3370 | 161.76683 | 17.2737 | 79 | 25.97 | 23.57 |
| N5917 | 230.3857 | -7.37712 | 15 | 28.35 | 23.76 |
| N1015 | 39.54819 | -1.31862 | 27 | 31.56 | 24.00 |
| N1309 | 50.52732 | -15.40007 | 64 | 31.96 | 24.02 |

*Figure 3.6: Left – List of the target galaxies, with the name, the position on the Sky and the number of Cepheids already identified in it Riess et al. (2016). In the last two columns we indicate the distance in Mpc of each galaxy and the apparent magnitudes of Cepheids in that galaxy with a period of 30 days. This is the limiting magnitude to be achieved for MAVIS spectroscopic observations. We will use the HR-Blue mode for the galaxies in yellow, for which we will be able to derive detailed chemical abundances; while for the galaxies in orange and red, we will use the LR-Blue mode to obtain accurate stellar iron abundances. Right-Top: HST images of all the galaxies used in Riess et al. (2016). Among these, only 13 are observable from the VLT. The red dots indicate the Cepheids in each galaxy, while the magenta square indicates a field of view of 2.7' × 2.7'. Right-Bottom: Example of Cepheid extraction, showing crowded 1"x1" field.*

Moreover, by taking advantage of the unprecedented spatial resolution of MAVIS, we also plan to identify new variables through multi-epoch photometry in at least two bands (V, I; limiting magnitude V=27 in ≤1hr, with S/N>10). In fact, the identification, classification and analysis of variable stars is best carried out in the optical bands, where the pulsation amplitude is significantly higher with respect to the one at longer wavelengths. The newly identified variables can then also be included for spectroscopic targeting.

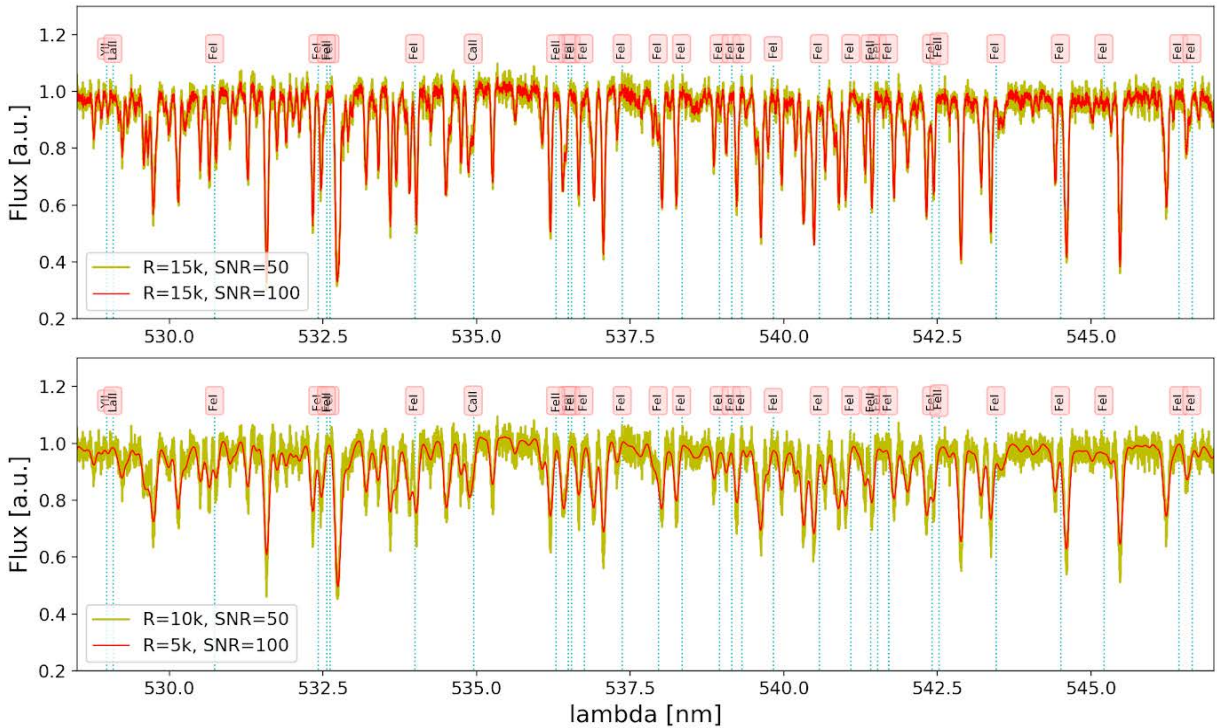

**Figure 3.7:** *Example of a typical Milky Way Cepheid spectrum ($T_{eff}$=5,370K, log(g)=1.6, $V_t$=2kms$^{-1}$) observed with different MAVIS-like resolutions (indicative of MAVIS LR-Blue and HR-Blue modes) and S/N, to illustrate the strength and quality of various spectral indicators. Spectral lines of iron and alpha-elements are labelled. Note that Cepheids in extragalactic environments are expected to be more metal poor than the example one shown here, which is based on an ESO2.2m/FEROS spectrum of a Galactic, solar-metallicity Cepheid, degraded to mimic MAVIS resolution.*

## 3.5 Resolving the physics of ram-pressure stripping

**Science goal:** How do dense environments alter the ISM conditions in galaxies, driving morphological evolution in galaxy clusters?

**Program details:** Deep MAVIS observations of galaxies interacting with their intra-cluster medium. Use imaging to identify young stellar associations, determine their age, and trace the distribution of dust lanes as well as ionised gas. Use IFU spectroscopy to trace both stellar and multi-phase gas kinematics.

**Key observation requirements:** High angular resolution multi-band photometry to identify recent star clusters (<10-20pc). Spectroscopy with sufficient velocity resolution to distinguish cold/warm ISM phases and measure the velocity dispersion of newly formed stellar associations (<20 km/s).

**Uniqueness of MAVIS:** No other instrument has sufficient angular resolution combined with spectral resolution to distinguish the physical (~10pc) and dynamical (<20 km/s) scales involved in ram-pressure stripping.

The discovery, three decades ago, of a significant population of atomic hydrogen (HI) deficient galaxies in nearby clusters provided the first clear observational evidence for removal of the interstellar medium (ISM) in high-density environments (Haynes et al. 1984).

Since then, thanks to the improvement of ground- and space-based facilities, it has become possible not only to quantify more accurately the effect of the cluster environment on the HI content of galaxies (Cortese et al. 2011), but also to show that environmental mechanisms can affect other components of the ISM, such as molecular hydrogen, dust, and metals. Indeed, both large statistical studies and detailed investigations of peculiar galaxies have shown that HI-deficient systems are also, at fixed stellar mass, H2- and dust-deficient when compared to isolated galaxies (Vollmer et al. 2005; Fumagalli et al. 2009; Cortese et al. 2016). However, the effect on dust grains and molecules appears to be less dramatic than on the HI component, supporting a scenario in which the ISM removal is more efficient in the outer, HI-dominated, star-forming disc – as expected in the case of hydro-dynamical mechanisms, such as ram pressure (Gunn & Gott 1972).

The more recent discovery of Jellyfish galaxies (Cortese et al. 2007; Fig. 3.8) has further demonstrated the importance of ram-pressure stripping in the evolution of cluster galaxies but, at the same time, has highlighted critical areas in which our understanding of the physical processes affecting the ISM of infalling cluster galaxies is still very much limited. The presence of extragalactic star-forming regions, dust streams and giant molecular clouds demonstrate that the cold ISM can survive even when removed from the disk. Perhaps even more surprisingly, the turbulence in the wake of the stripped tail appears to be able to condensate the stripped atomic gas into molecules and trigger the formation of new stars, but the details of this process are still eluding the astronomical community (Hester et al. 2010; Tonnensen & Bryan 2009). How can cold gas survive and condense in such a harsh environment? Are the physical processes regulating the efficiency of star formation in the intra-cluster medium the same as in the galaxy's disk? Can stars formed in the wake of the tail fall back onto the galaxy?

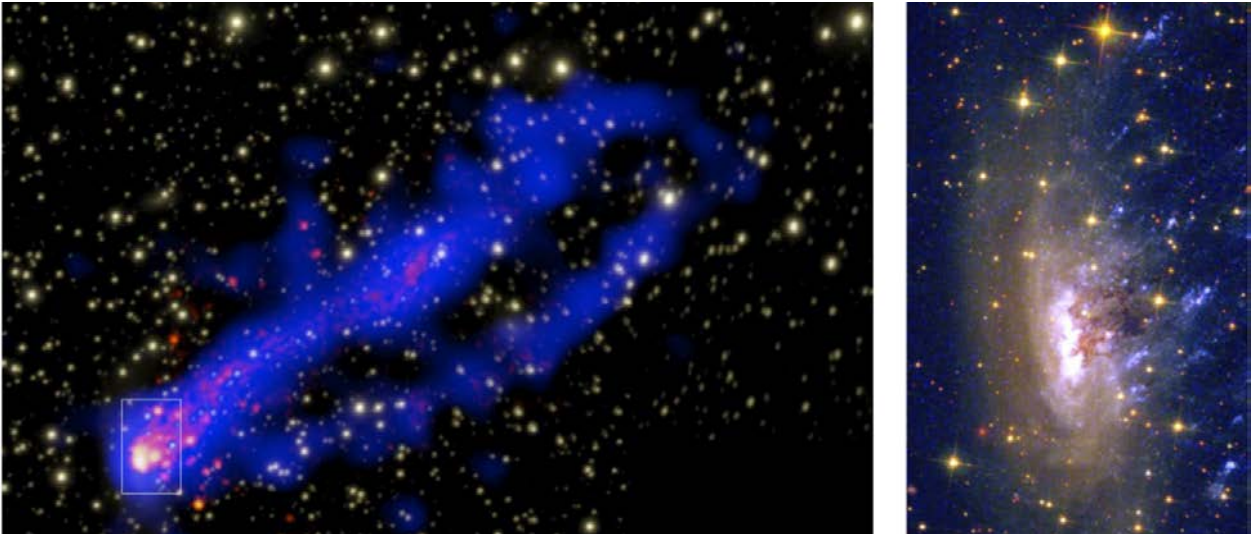

***Figure 3.8*** *Left: Composite X-ray Chandra (blue) / Hα (red) image of Jelly-Fish galaxy ESO 137-001 in the Norma cluster. Right: HST image of the disk of ESO 137-001, showing the extent of the stellar component and the detailed filamentary structure of extra-planar dust and young stars. From Jachym et al. (2014).*

In the last decade, high-resolution imaging of nearby galaxies obtained with the Hubble Space Telescope and the Atacama Large Millimetre Array have started to provide us with some clues on the physics affecting stripped material in clusters. The discovery of dust filaments departing from newly formed star clusters (see Fig. 3.9) is tantalising evidence for the presence of strong magnetic fields driving the morphological distribution of the stripped gas, and perhaps shielding it from the hot intra-cluster medium (Kenney et al. 2015). Similarly, the displacement observed between Hα-emitting ionised gas and individual stellar associations formed in the tails (Jachym et al. 2014) suggests that the stripped star-forming gas clouds keep being affected by ram pressure, depositing newly formed stars in the stripped tails.

The advent of MUSE improved dramatically our understanding of ram-pressure. Detailed studies on jellyfish galaxies in nearby clusters (Fumagalli et al. 2014; Fossati et al. 2016; Consolandi et al. 2017) and the GASP Large Programme of jellyfish galaxies at z~0.05 (Poggianti et al. 2017) provided fundamental insight on the ionized gas tails. GASP is to date the first systematic study of RPS in nearby galaxies; GASP results are however hampered by the spatial resolution of MUSE seeing-limited observations, that corresponds to ~1kpc. High resolution observations are still very challenging and are available only for a handful of cluster galaxies. Even more problematic is the lack of detailed kinematic information for the stellar and ionized gas components affected by ram pressure. Indeed, without knowing the three dimensional distribution of the various phases of the ISM it is impossible to reconstruct the past history and predict the future evolution of the stripped gas and stars.

Our knowledge of the effects of ram pressure on galaxy evolution is even more limited for galaxies outside clusters. In recent years, indirect evidence supporting ram pressure as the primary driver of star formation quenching has been presented (Brown et al. 2017), but direct incontrovertible proof is still missing. This is mainly because ram pressure is expected to be milder in groups than in clusters, and thus unable to produce the dramatic features observed in galaxies infalling into Virgo or Coma. If ram pressure is really at play in groups, it would not produce gas and dust tails several tens of kpc long, but would create streams of physical sizes similar to those observed in cluster galaxies when stripping takes place in the inner one effective radius (e.g., 100 pc, Fig. 3.9). Thus, a large statistical survey of group galaxies at HST-like spatial resolution has the potential to provide ultimate proof for the importance of ram pressure stripping in groups.

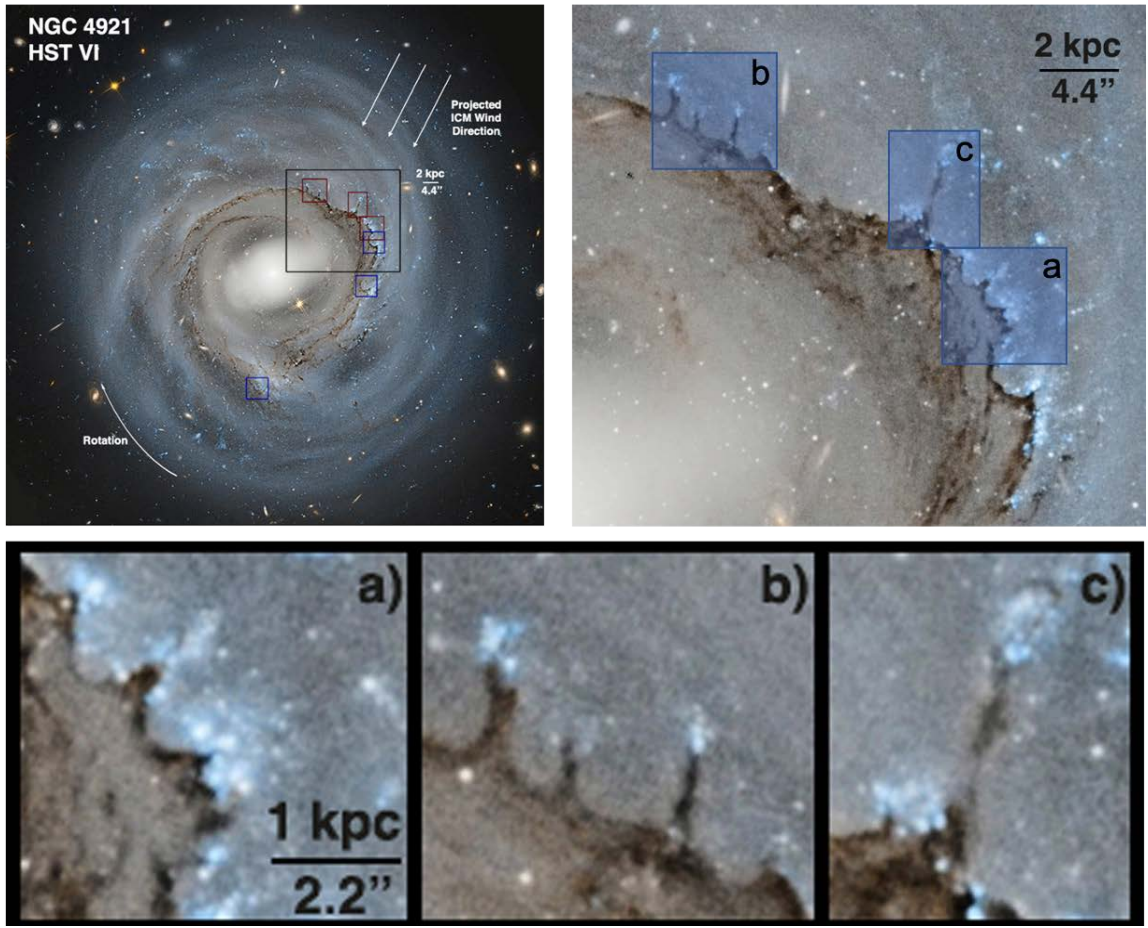

**Figure 3.9:** *Top Left - Deep HST V + I image of the Coma spiral NGC 4921, indicating the relative orientation of the galaxy's spin and systemic motion through the intra-cluster medium (ICM). The top-right and bottom panels zoom-in on regions where ram pressure is directly affecting the star-forming disk. This north-west side of the galaxy harbours a 20 kpc dust front formed through ram pressure. A series of dust filaments connecting the dust front to star clusters in the striped part of the disk are clearly visible. The continuity of these filaments indicates that the decoupling between various phases of the interstellar medium is partly inhibited, suggesting that magnetic fields may be playing an important role. The top-right image corresponds approximately to the MAVIS imaging field of view, with the inserts comparable in size to the MAVIS IFU 'Coarse' (50mas sampling ≈20pc/spaxel) field of view. Adapted from Kenney et al. (2015).*

MAVIS - with both imaging and spectroscopy capabilities – will advance this field significantly. In terms of imaging, most of the improvement with respect to current facilities will come from the significantly larger collecting area of the VLT, making it possible to target large samples in reasonable amounts of time. In terms of wavelength coverage, having a broad-band filter in the blue (ideally u-band) would maximise our ability to trace young stars, and narrow-band filters would allow us to trace ionised gas. The narrow-band filters are particularly important as they could be more efficient than spectroscopy to identify interesting features for follow-up study.

The ability to determine the kinematics of gas and stars on ≈10-20 parsec scale is where MAVIS will really shine, and provide a unique synergy with what ALMA will likely do for molecular gas in the next decade. As shown in Fig. 3.9, regions of interest are typically on ≈1-3" scales, with the scales of spatial complexity well suited to the MAVIS IFU. Spectral resolution of around 10000 is needed to be able to quantify small velocity differences (at the level of 10-20 km/s) between different gas phases and determine the typical velocity dispersion of newly formed stellar associations. This resolution is provided at Hα and [OI] by the MAVIS HR-Red mode, and for Hβ and [OIII] in HR-Blue mode, as needed.

## 3.6 Dynamical measurements of supermassive black holes and nuclear star clusters in low mass galaxies

**Science goal:** Can supermassive black holes (SMBHs) provide definitive constraints on the formation pathways of compact systems? What is the relationship between nuclear star clusters and SMBHs?

**Program details:** High resolution (spatial and spectral) IFU spectroscopy is necessary to accurately measure the central kinematics of the galaxies within the SBH sphere of influence. High resolution imaging is also needed to measure galaxy and nuclear cluster photometric properties.

**Key observation requirements:** High angular resolution multi-band photometry to identify and measure sizes of nuclear star clusters (<20pc). Spectroscopy with sufficient velocity resolution to measure dynamical signature of SMBH in low mass systems (<20 km/s).

**Uniqueness of MAVIS:** No other instrument has sufficient angular resolution combined with spectral resolution to distinguish the physical (~10pc) and dynamical (<20 km/s) scales involved in measuring nuclear star clusters and SMBH's in compact and low mass galaxies.

It is generally accepted that the evolution of galaxies is closely entwined with their nuclear properties, and in particular with the presence of a supermassive black hole (SMBH). As the SMBH grows over its lifetime through the accretion of gas, it can inject massive amounts of energy back into the host galaxy and its surroundings, both mechanically removing and radiatively heating the gas reservoir, and subsequently regulating star formation in galaxies. However, this picture is built primarily on the existence of an empirical scaling relationship between the masses of SMBHs and their hosts - a relationship that is heavily biased towards the brighter and more massive end of the galaxy population, and completely misses the crucial regime of lower-mass black holes, where the most likely (but so far unknown) progenitors of more massive SMBHs live. This bias is wholly driven by observational limitations. SMBH mass measurements require high-angular resolution dynamical measurements, preferring galaxies that are nearby (more easily resolved SMBH sphere of influence), high surface brightness (to achieve the S/N required for accurate stellar kinematics), centrally cusped (for on-axis AO guiding on the nucleus), and massive (allowing lower spectral resolution to be used). With it's large sky coverage and flexible NGS configurability, combined crucially with its high spectral resolution, MAVIS is perfectly suited to lifting these selection biases, and pushing SMBH research into important, and unexplored, regions of parameter space. This section highlights key advances in lower-mass systems, but we note that MAVIS will allow large, complete/unbiased samples of galaxies at all masses to have their nuclear regions dissected in unprecedented detail.

### 3.6.1 SMBHs in compact systems

In the distribution of mass and size for stellar systems in the Universe (Fig. 3.10), the realm of compact galaxies at the lower-mass end is populated by two distinct types of galaxies. At the lowest and most compact end, there are the ultra-compact dwarfs (UCDs; Hilker et al. 1999, Drinkwater et al. 2003). They share the stellar mass regime of dwarf spheroidals ($10^6 < M_*/M_\odot < 10^8$) but with much (factor 10-100 times) smaller sizes ($10 < R_e < 100$ pc). They are closely followed by the family of compact elliptical galaxies (cEs), which instead share the stellar mass regime with dwarf ellipticals ($10^8 < M_*/M_\odot < 10^{10}$) but with compact sizes of $100 < R_e < 1,000$ pc. They also have high stellar densities that are similar to those found in the cores of early-type galaxies or the bulges of spiral systems. In fact, both families follow the luminosity– and mass–size relations described by the bright and massive ellipticals alone (e.g. Brodie et al. 2011), branching off from the dwarf ellipticals and dwarf spheroidals paths.

Nonetheless, both families are also scarce in the local Universe, with only a couple hundred reported to date (e.g. Penny et al. 2014; Norris et al. 2014; Chilingarian & Zolotukhin 2015), posing a big question mark in the galaxy formation paradigm. They have been proposed to be the remnant product of larger and more massive galaxies (dwarf ellipticals for the UCDs and elliptical; i.e. Bekki, Couch & Drinkwater (2001), Pfeffer et al. (2016) or spiral galaxies for the cEs) that had their outer stars stripped. This seems to be supported by the fact that the majority of them are in the vicinity of larger, more massive hosts or located in high-density environments. The smoking gun for this scenario is that a number of them have been 'caught in the act', showing clear signs of interaction and disruption with their host (e.g. Paudel et al. 2013). However, such claims are also challenged by the discovery of a small number of isolated compact systems that have no plausible host in the vicinity (e.g. Huxor, Phillipps & Price 2013; Paudel & Ree 2014). In that case, cEs have been depicted as the unstripped, low–mass and faint end of elliptical galaxies (e.g. Martinovic & Micic 2017) and UCDs the unstripped, high-mass end of Globular clusters (e.g. Bekki, Couch & Drinkwater 2001; Mieske, Hilker & Misgeld 2012).

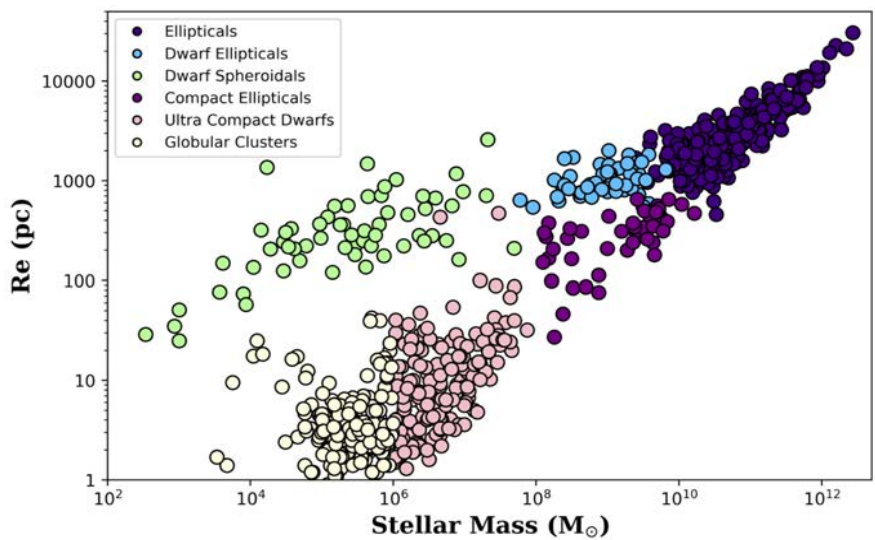

***Figure 3.10:*** *Mass–size relation diagram: the well-known trend of galaxies being larger as they are more massive is shown in this diagram. It shows the location of the different families of giant elliptical galaxies, dEs, dSphs, cEs, UCDs and globular clusters. It is clear that both cEs and UCDs deviate from the dwarf families and follow very tightly the relation defined by the ellipticals alone.*

We thus find once again the long-standing puzzle of nature (intrinsic origin) vs. nurture (stripping origin). Although the tidal stripping origin seems to be the most common mechanism for shaping the population of compact objects in the low mass end (e.g. Guérou et al. 2015), it is still unclear for which mass range and environment this dominates, and therefore what is the resulting abundance of intrinsic compact low-mass systems. To further investigate this topic, we recently analysed a sample of 25 cEs to reveal their origins (Ferré-Mateu et al. 2018), studying a set of discriminant tools that could differentiate between the possible origins. We were able to provide strong constraints on the origins, evolutionary stages and possible progenitors for those cEs. However such exercise was time–consuming and needed all the tools combined to provide robust results. Actually, one key discriminant was missing in such a study, a crucial one: the presence of super massive black holes (SMBHs).

When galaxies suffer a stripping event, this results in the removal of up to ~90% of the stellar mass and a noticeable change in size, whereas the central velocity dispersion and their SMBH remain unchanged (e.g. Ferré-Mateu et al. 2015). We would thus expect stripped galaxies to have much larger SMBHs than expected by their stellar masses, whereas intrinsic low-mass systems would follow the SMBH–galaxy scaling relations. In fact, both UCDs and cEs have been found to have elevated dynamical mass–to–light M/L) ratios (e.g. Hasegan 2007; Forbes et al. 2014), which could be justified by the presence of SMBHs.

Unfortunately, while SMBHs hold the definitive key to reveal the origin of low-mass compact objects, there are only a handful of directly measured SMBHs in such systems (e.g. Nowak et al. 2007; Mieske et al. 2013; Seth et al. 2014). This is mostly due to the limitations on the angular- and spectral-resolution of the current instruments both on ground-based telescopes and in space.

Galaxies with no signs of AGN, such as the majority of the cases above, can only have their SMBHs detected by gravitational effects on the dynamics of the stars inside the radius of the SMBH sphere of influence, $r_{soi}$. Since the $r_{soi}$ scales with the black hole mass (de Zeeuw 2010), high angular resolution spectral observations ($\theta_{BH} \approx 0.02''$ scales, see Figure 3.11) are needed to detect large samples of such SMBHs. In addition, to account for the full possible variety of orbital configurations in galaxies requires the use of fully-general three-integral Schwarzschild (e.g. van der Marel et al. 1998 and references therein). Such general models require integral-field kinematic constraints to limit the possible orbital structure (Cappellari & McDermid, 2005), and equivalently high- resolution imaging to constrain the central stellar mass distribution. Accurately accounting for stellar mass-to-light ratio variations (including the stellar Initial Mass Function) is also important, so broad spectral coverage to characterise the stellar population is important. Finally, low-mass stellar systems naturally exhibit lower stellar velocity dispersions (generally below 30 kms$^{-1}$), requiring moderately high spectral resolution (R≈5000) around the V-band spectral region.

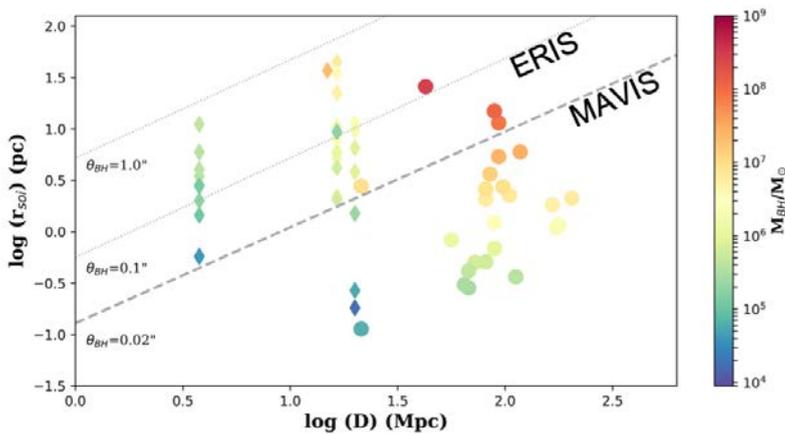

**Figure 3.11:** *Black hole radius of influence with distance to the host galaxy: the dotted lines correspond to angular sizes of 1.0" and 0.1" (e.g. typical ERIS resolution). The dashed line corresponds to an angular size of 0.02", which corresponds to the expected angular resolution of MAVIS. UCDs are shown in diamonds while cEs are shown in circles, both color–coded by their estimated SMBH (not measured).*

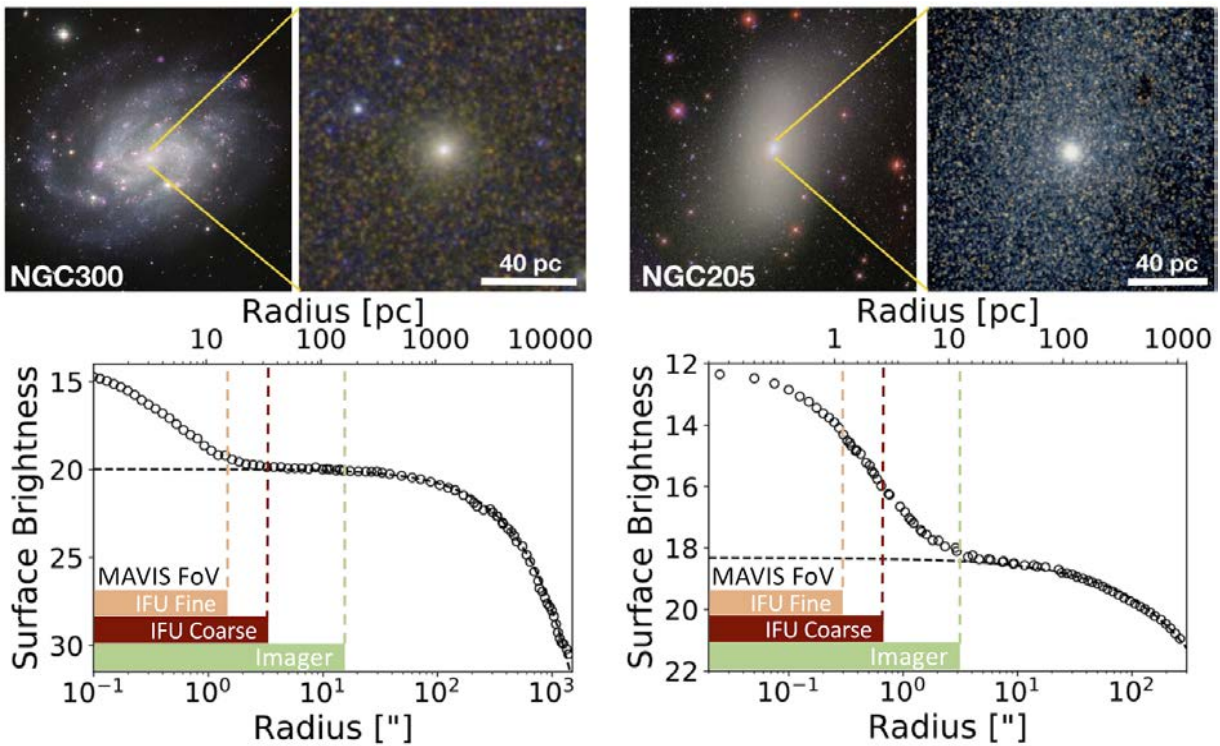

**Figure 3.12:** *The NSCs in the late-type spiral NGC300 (left) and early-type galaxy NGC205 (M110, right). The top panels show galaxy-wide images with zoom-ins into the central regions of each galaxy. The bottom panels show surface brightness profiles of the two galaxies, which in both cases indicate the presence of an NSC by a rise above the light profile of the host galaxy body. The radii covered by MAVIS IFU and imaging modes are also indicated. Adapted from Neumayer et al. 2020.*

### 3.6.2 Unveiling the connection between SMBHs and Nuclear Star Clusters

Compact stellar nuclei and central star clusters, collectively referred to here as Nuclear Star Clusters (NSCs - see Fig. 3.12), are present in ~70% of low- and intermediate-luminosity galaxies, effectively dominating the centres of galaxies in the mass range of $10^8$–$10^{10} M_\odot$. This is just below the mass range where SMBH scaling relations are well established. Indeed, both SMBHs and NSCs show tight relationships with their host galaxy mass (e.g. Ferrarese et al. 2006, Fig 3.13), leading to speculation that NSCs may evolve into SMBHs, being two different phases of the same 'Central Massive Object' (CMO). However, there are a growing number of studies showing that SMBHs and NSCs can co-exist in galaxies (Filippenko and Ho 2003; Seth et al. 2008; Nguyen et al. 2019; see Neumayer et al. 2020 for a recent review), which includes the NSC and SMBH at the centre of our own Galaxy (Genzel et al. 2010; Schödel et al. 2014). The emerging picture is one of a closely connected evolutionary path, where SMBHs and NSCs coexist and interact with one another in ways that may intimately depend on the host galaxy properties. However, demographics of the underpinning dynamical SMBH masses and the size/mass properties of the NSCs, remain poorly determined due to the observational challenge these low-mass systems present. In particular, all but a handful of NSCs are prohibitively faint, extended, or embedded in their hosts for AO systems to use as NGS sources. MAVIS, with its ability to guide on stars far off-axis, yet still provide exquisite resolution, can overcome this limitation, permitting large and unbiased samples to be studied.

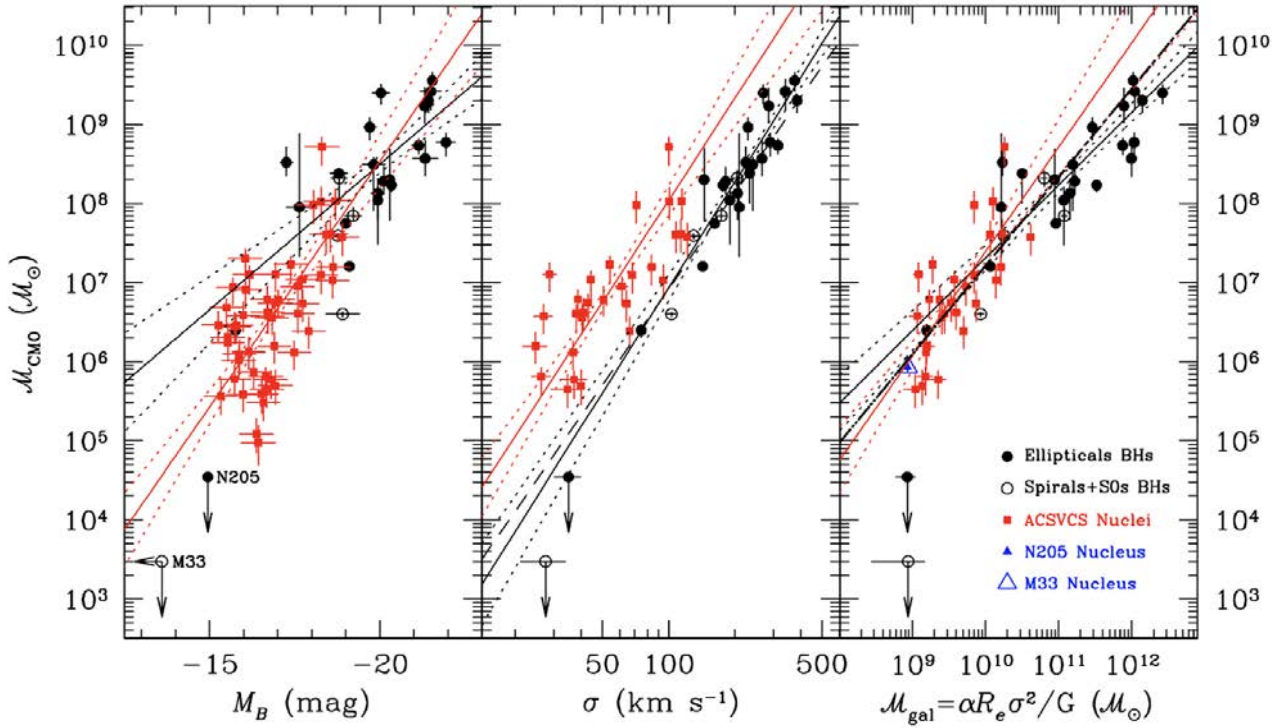

**Figure 3.13:** *Common scaling relations between SMBHs and stellar nuclei. Left: Mass of the Central Massive Object (CMO, meaning both SMBHs and stellar nuclei/NSCs) plotted against absolute blue magnitude of the host galaxy (or bulge for spiral galaxies). NSCs are shown as red squares. The supermassive black holes (SMBHs) in early-type and spiral galaxies are shown as filled and open circles respectively. Middle: CMO mass as a function of velocity dispersion of the host galaxy, measured within $R_e$. Right: CMO mass plotted against galaxy dynamical mass. Adapted from Ferrarese et al. 2006. Note that the NSC masses are all photometrically derived. Even today, 14 years after this seminal work, only a handful of galaxies in the regime of $\sigma < 100$km/s or $M_{gal} < 10^{10} M_\odot$ have dynamically-derived NSC or SMBH masses, due to lack of appropriate instrumentation.*

Most of the mass measurements for NSCs come from HST-based photometric studies, where luminosities are converted into mass using relations between the measured color and the predicted stellar mass-to-light ratio (M/L). from stellar population models (Georgiev et al. 2016; Spengler et al. 2017; Sánchez-Janssen et al. 2019). Spectral stellar population synthesis methods can also be used to estimate the (M/L). (Rossa et al. 2006; Seth et al. 2010; Kacharov et al. 2018), resulting in higher accuracy. Dynamical mass measurements are considered the most accurate way to estimate NSC masses, comparing the measured (typically integrated) velocity dispersion to a dynamical model for the stellar potential based on the luminosity profile of the NSC. Since the integrated velocity dispersions of NSCs are typically low (< 20 − 30 km/s), the spectral resolution of the observations needs to be sufficiently high (R ≈ 5000, e.g. Walcher et al. 2005; Kormendy et al. 2010). MAVIS is ideally suited to this task.

While NSCs are readily observed in images of nearby galaxies (see Fig. 3.12), measuring the presence and mass of a potential SMBH residing within NSCs is significantly more challenging, requiring accurate measurements of gas or stellar motions within the black hole radius of influence (see previous section). To separate the gravitational effect of the NSC from that of the black hole, accurate high-resolution mass models for the NSC are essential, requiring both surface photometry and spectroscopy at parsec (≈ 25mas at 8Mpc) scales, providing sensitive stellar population diagnostics and (M/L). information. Combining stellar population analysis with dynamical modelling has enabled black hole mass detections below ~ $10^5 M_\odot$ in NSCs (Nguyen et al. 2019). Again, MAVIS, at optical wavelengths so crucial for stellar population information, is perfectly suited for this task.

Recent work on a volume limited sample of five $10^9$ to $10^{10}$ M$_\odot$ early-type galaxies has shown that all have evidence for central black holes (Nguyen et al. 2017, 2018, 2019), providing strong evidence that, at this mass range where NSC occupation peaks, many co-exist with massive black holes. Our knowledge of NSCs in lower-mass *late-type* galaxies, however, is woefully lacking. The impressive results from samples of NSCs in low-mass early types clearly highlight the importance for obtaining comparable (i.e., large and homogeneously selected) samples also for lower-mass late-type galaxies. This data should also help test the environmental dependence of NSC formation. Furthermore, understanding the stellar populations of these NSCs, and determining if they more closely resemble those in low-mass early types or higher-mass late-types would also help us understand NSC formation more generally. Combined MAVIS imaging and spectroscopy can play a major role in this effort.

Future facilities promise to greatly improve our understanding of NSCs. With the impending launch of JWST, a better census of NSCs in higher mass galaxies will be possible through high resolution mid-infrared observations of dusty galaxy cores. Also, the identification of faint AGN enabled by JWST will aid in understanding the frequency of black hole / NSC co-existence in lower mass galaxies (e.g., Dumont et al. 2019). LSST will also play an important role in characterizing the population of black holes in low mass galaxies through statistics on tidal disruption events, and variability of AGN (e.g., Baldassare et al. 2018). However, high angular- and spectral-resolution capabilities will still be key to progressing in this field in a systematic way at low masses. ELT will provide unprecedented capabilities in this regard, but will lack the high-quality stellar population information available in the blue. MAVIS can therefore play a pivotal role in this area, both alone, and in synergy with e.g. HARMONI and MICADO on the ELT.

# 4 Star Clusters over Cosmic time

## 4.1 Introduction

Star clusters are key constituents in the study of many fields of astrophysics, from star formation, to stellar and galactic evolution, to cosmology (for a review, Krumholz, McKee, and Bland-Hawthorn 2019). They are ubiquitous: observed from the very nearby Universe (in our Galaxy and in Local Group galaxies) to the high-redshift galaxies. No longer studied only in isolation, modern astronomy research highlights star clusters as corner-stone objects to improve our understanding of the Universe across cosmic time.

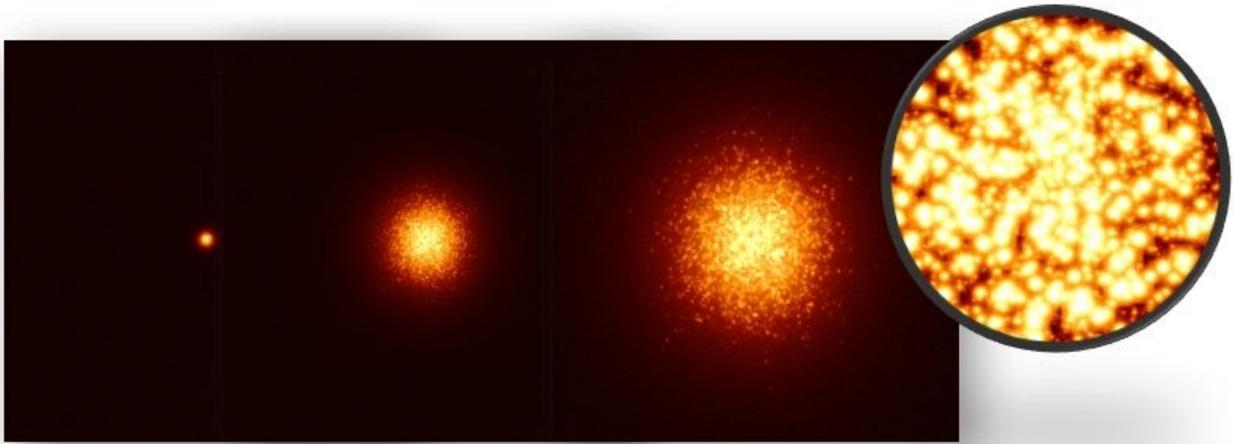

**Figure 4.1**: *Simulated image of a star cluster from 10 Mpc to 10 kpc (from the left to the right), in the FoV of MAVIS 30"x30".*

MAVIS offers exciting opportunities to study star clusters at different time and spatial scales. The combined capabilities of photometry at very high spatial resolution with low- and high-resolution spectroscopy opens possibilities to explore parameters previously not possible. In particular we highlight the key science areas of most impact outlined by these Big Questions:

- *What is the fate of high-redshift star formation clumps? Which is the nature and origin of the giant star forming clumps observed at z>1?*
- *What is the link of star forming clumps at different z with the present-time cluster population? How the first globular clusters formed at high-z?*
- *Is there a universal Cluster mass function? Does it vary with time, environment and metallicity?*
- *What can we learn about galaxy evolution from their cluster populations? How does the spatial distribution of metallicity vary with time?*
- *What regulates the emergence of multiple populations in star clusters? Do massive clusters contain intermediate mass black holes?*
- *Is the initial mass function universal? Does it depend on metallicity?*
- *What is the influence of metallicity on the evolution of massive stars observed in young massive clusters?*

We explore each key area in more detail throughout this chapter, highlighting the niche discovery space opened up by MAVIS. While we separate the key science areas for clarity, it must be stressed that they are not standalone studies. The unique capabilities of MAVIS opens up new areas of interdisciplinary science and addresses global parameters that impact much broader areas of Astrophysics, such as the universality of the initial mass function, dependence on metallicity, environment, cluster dissolution and role in the building of galaxies. Fundamentally, by allowing us to study the details of objects previously unresolvable both near and far, we envisage MAVIS will make leaps and bounds in making decisive links across Galactic and extragalactic astronomy, and open new fields of astronomy research.

## 4.2 Galactic stars clusters: Revealing the mystery of Globular Clusters

**Science goal:** Characterising cores of globular clusters: extent of abundance anomalies, multiple stellar populations and core binary fraction

**Program details:** Spectroscopy of individual stars in cores of Galactic globular clusters over a range of cluster mass and metallicity parameters. Periodic observations for radial velocity measurements.

**Key observation requirements:** High resolution spectroscopy (R =10000-15000), pre-imaging with MAVIS

**Uniqueness of MAVIS:** The high-spectral resolution and blue wavelength coverage, e.g., with respect to MUSE, allows a detailed chemical and kinematical characterization of single stars in Galactic Globular cluster cores from the main sequence to the RGB.

Galactic globular clusters (GGCs) are the oldest known objects in the Universe. They are perhaps the most complex stellar populations and remain unexplained by current models of cluster formation, evolution and stellar nucleosynthesis. After decades of dedicated studies, the community has run out of options with existing instrumentation. With high spatial and spectral resolution, MAVIS presents the first novel opportunity to explore globular cluster Cores - the missing link to resolve this decades long mystery. As per the simulated images of MAVIS (Fig. 4.2), we can target individual stars within the cores of GGCs. Specifically, MAVIS enables us to answer two key questions:

- ***What is the extent of abundance anomalies down to the main sequence in the cores of GGCs and are they consistent with the photometrical observed multiple stellar populations?***
- ***What is the binary fraction in cores of GGCs and what role do binaries play in the framework of multiple stellar populations?***

Characterising the multiple populations in GGCs by examining the relationships across (i) radial distribution (ii) chemical composition and (iii) evolutionary stage as well as (iv) binary fraction, in order to find the missing links to the formation of these clusters is the primary objective and science driver.

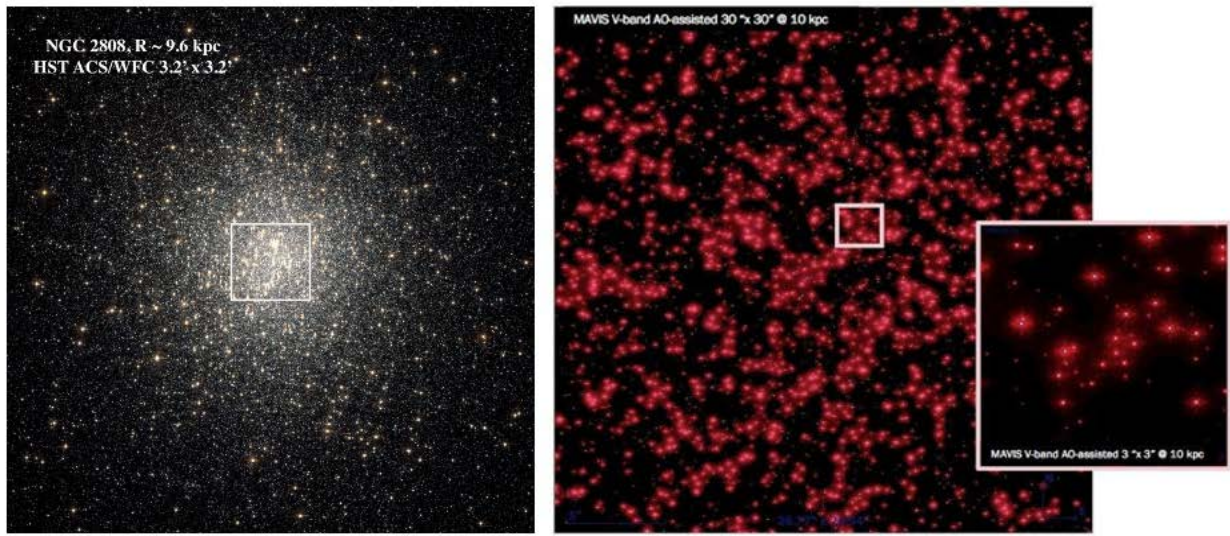

***Figure 4.2:*** *HST image of GGC NGC 2808 compared to simulated MAVIS images of the central regions of a comparable globular star cluster at the distance of 10 kpc. The angular size of the middle image is ~30"x30", while the zoomed image is ~3"x3" corresponding to the MAVIS IFU for spectroscopy.*

### 4.2.1 The origins of multiple populations and chemical peculiarities

Spectra of stars in GGCs exhibit anti-correlations in the light-element abundances, the most notorious being the Na-O abundance anti-correlation (e.g., Gratton et al. 2019 and references therein). In several peculiar cases, star-to-star metallicity and neutron-capture element variations, such as Ba abundances have also been observed (e.g., Marino et al. 2011, Yong et al. 2014, Marino et al. 2017). No other stellar population displays such peculiarities. HST based photometry shows evidence of multiple evolutionary sequences in the colour-magnitude diagrams (CMDs) of most GGCs (Milone et al. 2017, see Fig. 4.3). This includes multiple giant branches as well as multiple main sequences, which combined with the chemical peculiarities all add to the mystery. The currently held view is that GGCs host multiple stellar populations, but how such systems formed and evolved remains a mystery.

Current instrumentation limitations prevent the study of the high densities of cluster cores. All globular cluster spectroscopic studies to date are based on stellar samples from the outer regions of the clusters. The cores of GGCs remain largely unexplored or limited to the tip of the red giant branch that suffer statistical limitations (e.g., D'Orazi et al. 2010 on Ba-stars, Lucatello et al., 2015). Therefore the radial distribution of GGC stars has not been explored and is a missing link to formation scenarios and subsequent dynamical evolution (e.g. Bastian et al. 2015, D'Ercole et al. 2008). MAVIS is capable of providing a first look into the chemical content of these previously inaccessible samples.

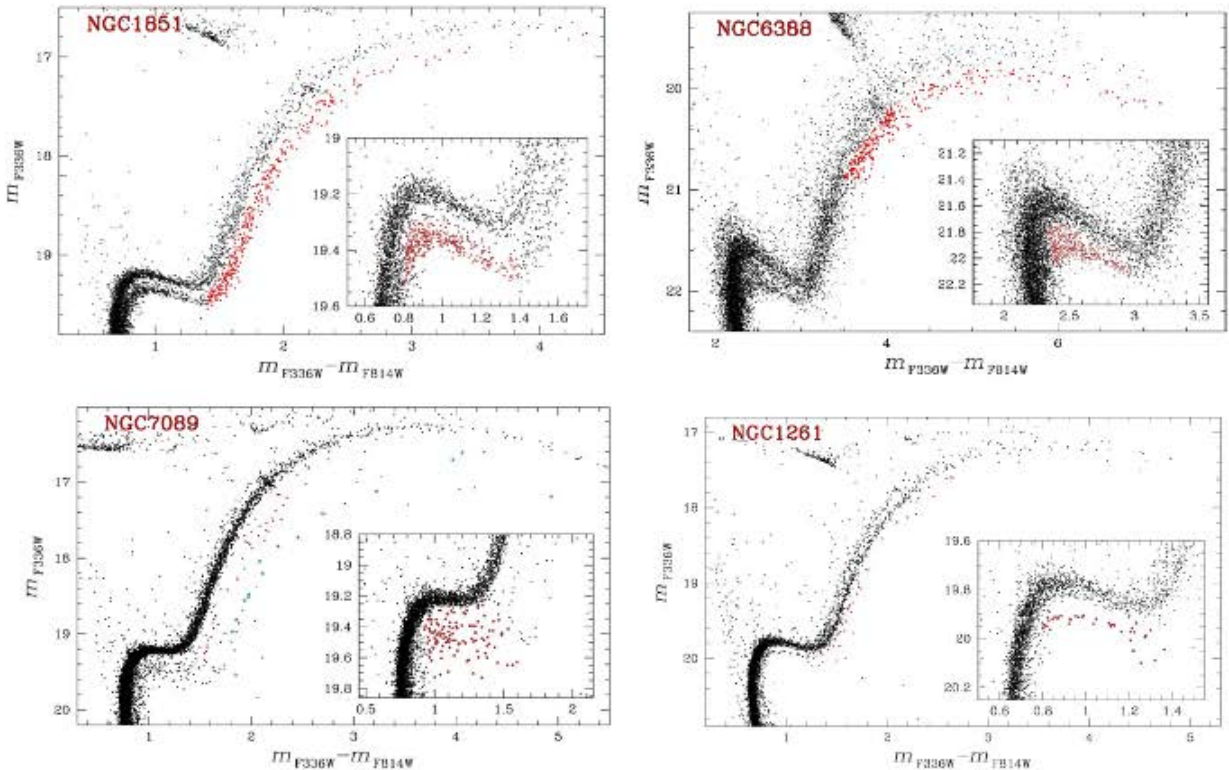

**Figure 4.3:** *CMDs from HST photometry for a representative sample of GGCs from Milone et al., 2017 highlighting multiple populations spanning the main sequence to the tip of the giant branch.*

## 4.2.2 Core Binaries in the context of cluster formation

A secondary factor to the missing link to GGC formation is the binary fraction within the cores and its relationship to the chemical peculiarities of multiple stellar populations (e.g. Heggie and Giersz, 2014 and references therein). For any given mass pair, binaries sink into the globular cluster centre compared to single stars of the same component mass; this is simply as a result of the equipartition energy. This process is faster where the cluster relaxation time (i.e., the characteristic time between star-star interaction, which is dependent on stellar density) is shorter. Therefore, in the densest parts of GCCs we expect to find most of the binaries, especially for the most massive cases.

As dynamical evolution proceeds, the three body encounters eject the lightest stars, and leave the most massive binaries concentrated in the core, which become harder and harder at each non-destructive encounter, with a period distribution expected to peak below few 10 days (e.g., Heggie et al. 2008). Given the older age of GGCs, where a sizable number of massive stars have already evolved, most remaining binary counterparts are likely non-luminous (white dwarfs, neutron stars, and black holes). These are therefore difficult to detect when non-interacting.

A radial velocity survey of GGC cores will shed light for a census on binary rate and mass-period distribution. ***The combined spatial and spectral resolution capability of MAVIS offers a unique opportunity to conduct a systematic census of these core binaries.*** Photometry via stellar models can provide the mass of the luminous component, while the binary period will provide the mass of the dark component. Binary studies in the core of GGCs will link closely with the studies of multiple populations and chemical peculiarities. Specifically the ability to characterize the mass-period distribution of binaries with massive dark component for each discrete stellar population as well explore the chemical composition of the luminous component in binaries compared with the single (non-binary) stars.

## 4.2.3 Technical requirements and Observational outline

For the first time, **MAVIS with its AO-assisted spatial resolution combined with its high spectral resolution** capabilities provides a unique opportunity to explore the cores of GGCs, from radial velocities to chemical abundances of individual stars. The key instrument requirements that make MAVIS a cutting edge facility to solve the mystery of GGCs are summarized below.

**Spectral resolution and Wavelength:** Spectroscopy at a resolution of R=15,000 or higher, allows the detection of elements in three key nucleosynthesis families; light elements such as C, N and O, alpha elements such as Na, Mg and Al, and neutron capture elements such as Ba, Y, La, Nd and Eu. A lower spectral resolution would severely hinder the science exploitation for this field of study, as weak spectral features would not be detectable. The lower resolution is the reason why e.g. MUSE cannot be utilised for the proposed science (see Fig 4.5). Wavelength coverage to a blue limit of 400nm is needed for studying the above mentioned range of chemical elements. Notable spectral features are the CH bands around 410-430nm, neutron-capture features around 400-470nm and atomic features of Li, Na, O, Mg, Al are seen around 470-680nm. Note that continuous large wavelength coverage is not needed.

**Signal to noise requirements and Efficiency:** It is important to reach the top of the main sequence which spans a magnitude: B ~18 - 24 depending on cluster (see also Fig. 4.3). For radial velocity measurements we require a minimum SNR of 10 for individual epochs. For chemical abundance analysis we require a minimum SNR over 30 at the wavelength of interest to derive abundances of around 0.1dex precision (see Table 4.1 and Fig 4.4 for synthesis on simulated spectra). This precision is needed to identify and distinguish between the multiple populations. For a timely completion of a large-scale survey of GGCs, we therefore require that the instrument delivers an end to end efficiency such that SNR = 10 at B = 22 is achieved within 1hr exposure. A lower efficiency would result in longer exposures and a larger observing program.

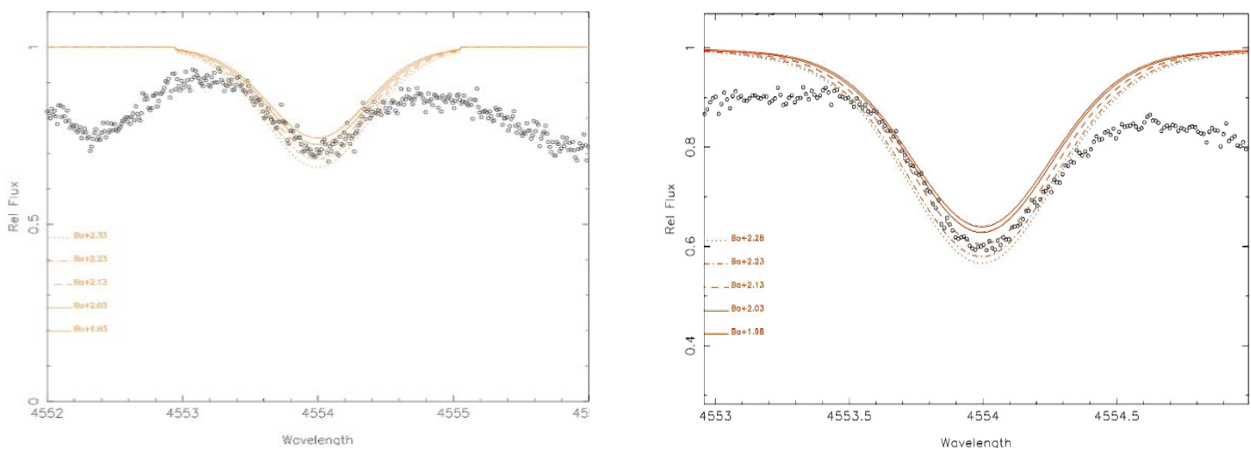

***Figure 4.4:*** *Example spectrum synthesis runs for the Ba line at 455.4nm using MAVIS simulated spectra for SNR = 30 at R=10,000 (left) and R=15,000 (right). The open circles are the simulated spectra mimicking the observed spectrum, while the various solid, dotted and dashed lines are synthesis runs at different abundances of Ba.*

**Targets and observational outline:** There are over 50 GGCs with HST photometry suitable for spectroscopy with MAVIS and more possible with MAVIS photometry. The photometric information will be used to derive global properties of the targets, including the photospheric temperatures, gravity and preliminary metallicity. The spectroscopic observations with MAVIS will enable individual element abundance measurements. A spectroscopic observing campaign would aim to collect at least 10 targets per population per cluster. This would also require multiple pointings of the 3x3 arcsec IFU to cover both the number of targets as well as spatial coverage of the cluster core. For each pointing, single one-hour exposures would provide sufficient signal for radial velocity measurements. Repeat observations over 10-20 epochs are needed to detect any binary activities as the period distribution peaks around 10 days. The stacking of these multi-epoch spectra would provide the SNR needed for abundance measurements.

***Table 4.1:*** *Typical abundance precision (in dex) for the neutron-capture element Ba with spectral synthesis using MAVIS simulated spectra for different spectral resolution and signal to noise ratios.*

|          | SNR ~ 150 | SNR ~ 100 | SNR ~ 80 | SNR ~ 30 |
|----------|-----------|-----------|----------|----------|
| R=10,000 | 0.12      | 0.15      | 0.15     | 0.23     |
| R=15,000 | 0.05      | 0.08      | 0.08     | 0.12     |
| R=20,000 | 0.05      | 0.05      | 0.08     | 0.10     |

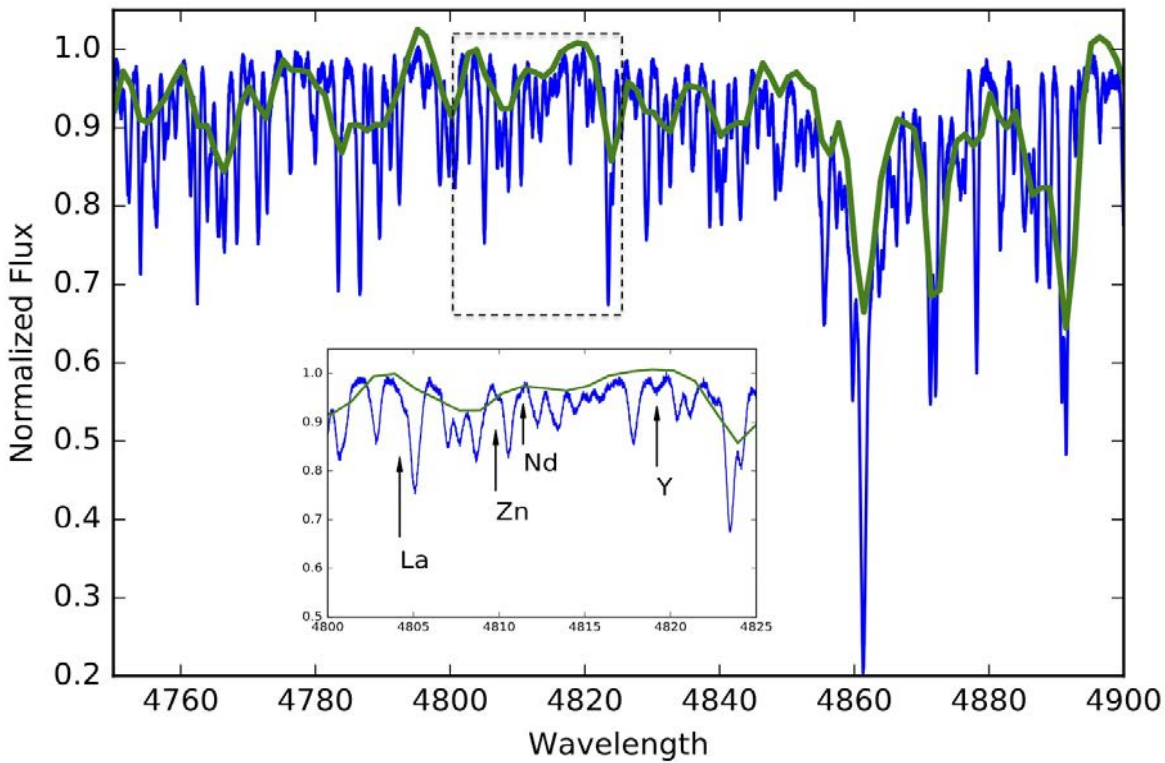

***Figure 4.5:*** *Overlay of MAVIS simulated Solar spectra (blue) compared to MUSE spectra (green; from Ivanov et al., 2019) of a similar type star with solar metallicity. Insert plot is a zoom-in of the dashed spectral region, pointing several neutron-capture features which are weak but detectable with MAVIS, but impossible at MUSE resolution. Further, note MUSE does not have spectra below 475nm to compare the light element features available in that region, which will be possible with MAVIS.*

We note that globular clusters are some of the oldest known systems in the Universe for which an age can be provided reliably. ***Therefore the push to understand these objects in greater detail with the latest instrumentation is a must for the development of astronomy broadly.*** As the methodology and spectral analysis techniques are already well developed, the study of cores of GGCs will be a highly scientifically productive use of MAVIS.

### Case Study: MAVIS view on the unexplored cores of GGCs

**Key Questions:** What is the extent of abundance anomalies down to the main sequence stars in the cores of GGCs? Are they consistent with multiple stellar populations? What is the binary fraction in cores of GGCs? What role do binaries play in the framework of multiple populations?

**Why MAVIS?** MAVIS is the only instrument to date that will have sufficient spatial resolution to resolve and collect light from individual stars in the cores of GGCs at high spectral resolution covering the blue wavelengths needed to measure detailed individual chemical abundances.

**How will MAVIS be used:** MAVIS will be used in its High Resolution spectroscopic mode to collect spectra of individual GGC stars, with the aim of collecting at least 10 targets per population per cluster. Multiple pointing of the ~3x3 arcsec IFU will be needed to cover the number of targets as well as spatial coverage of the cluster core, depending on cluster size and distance. Multi-epoch observations will be carried out to monitor radial velocity variations.

**Field selection:** There are over 50 GGCs with HST photometry suitable for spectroscopy with MAVIS and more possible with MAVIS photometry.

**Exposure time:** The main sequence turn-off magnitudes vary from approx. 18- 24 in B. Typically approx. one-hour exposure with MAVIS is sufficient to measure radial velocities at SN > 10. Repeat observations over 10-20 epochs are needed to detect binary activities as period distribution peaks around 10 days. The stacking of these multi-epoch spectra would provide the SNR needed for abundance measurements.

## 4.3 The hunt for seed black holes in globular clusters

**Science goal:** Potential detection of intermediate-mass black holes in globular cluster centres; precision astrometry and proper motion measurements in the crowded cluster centre

**Program details:** High precision multi-epoch astrometry in the central regions of Milky Way globular clusters, using MAVIS imaging mode in multiple photometric bands. Potential to combine with MAVIS spectroscopy, in addition to Gaia, historic HST, and MUSE data.

**Key observation requirements:** Near-diffraction limited imaging in cluster centres coupled with well-characterised instrumental distortion, permitting 150µas astrometry (goal 50µas).

**Uniqueness of MAVIS:** The extremely high angular resolution of MAVIS allows accurate proper motions to be measured beyond the crowding limit of HST; High sensitivity of MAVIS will allow well-populated proper motion histograms, even in small fields.

Observations in many wavelengths (including gravitational waves) have confirmed the existence of black holes (BHs) in two mass ranges: stellar-mass BHs (5-40 $M_\odot$) and supermassive BHs ($10^6$-$10^{10}\,M_\odot$). Interestingly, the possible presence of BHs populating the intermediate-mass range (IMBHs, $10^2$-$10^5\,M_\odot$) remains a highly debated issue. IMBHs could be the seed for the formation of the more massive SMBHs (Ebisuzaki et al. 2001), therefore finding them would provide key information on the assembly history of the Universe and on the formation and evolution of the first structures (Volonteri et al. 2003).

A promising location to search for IMBHs is the centre of globular clusters (GCs). Indeed, it has been suggested that a runaway collapse in the cores of these systems could generate IMBHs (Portegies Zwart et al. 2004), or that the BH mass can build up due to stellar mergers taking place in these dense environments (Giersz et al. 2015). In addition, the extrapolation of the empirical relation between the mass of SMBHs and the velocity dispersion of their host also suggests that IMBHs could be found in GCs (Ferrarese & Merritt 2000). Unfortunately, direct detection of IMBHs in GCs is extremely challenging (van der Marel 2004). GCs are almost gas-free, thus X-ray and radio emissions from gas accretion on the IMBH are expected to be faint: in fact, deep searches for such emission have only provided upper limits on the mass of IMBHs (Maccarone & Servillat 2008; Strader et al. 2012).

Another possibility would be to look for the signatures in the spatial and kinematical distribution of stars produced by the presence of an IMBH (Bahcall & Wolf 1976). However, the presence of a central shallow cusp in the surface brightness profile (Baumgardt et al. 2005; Trenti et al. 2007) could also be due to other effects like mass segregation, core collapse, or the presence of binary stars (Vesperini & Trenti 2010), and the expected rise in the velocity dispersion profile towards the centre of the cluster could be caused by the presence of radial anisotropy (van der Marel & Anderson 2010) or of a centrally concentrated population of stellar-mass BHs (Lützgendorf et al. 2013, Zocchi et al. 2019).

Fortunately, all the above explanations for the signatures related to the presence of a central IMBH have an impact on the dynamics of the cluster as a whole: by studying the global dynamics of the GCs, together with their central kinematics, it will be possible to disentangle these contributions.

The measurement of line-of-sight (LOS) velocities of individual stars is demanding: spectroscopic measurements require large telescope time, and the crowding at the centre undermines their accuracy even when using the Hubble Space Telescope (HST, van der Marel et al. 2002; Watkins et al. 2015 - see Fig. 4.6). An alternative is to measure the LOS velocity dispersion by means of integral-field units (IFUs), the downside being that the result is affected by shot noise from the brightest giant stars. These methods can sometimes provide discrepant results: an example is the controversy around the cluster NGC 6388, with the IFU measurements by Lützgendorf et al. (2015) suggesting the presence of an IMBH with mass $2.8 \times 10^4$ $M_\odot$ and Lanzoni et al. (2013) finding an upper limit of ~$2.3 \times 10^3$ $M_\odot$ with individual LOS velocities. A significant improvement in data quality is obtained when considering internal proper motion (PM) measurements (Bellini et al. 2014, 2018). PMs offer many advantages over LOS velocity studies: (1) because no spectroscopic measurement is involved, fainter stars can be studied, yielding better statistics on the kinematic quantities of interest; (2) the stars can be measured individually, contrarily to IFU measurements, thus avoiding disproportionate contributions from bright giants; (3) two components of velocity are measured (tangential and radial) at once.

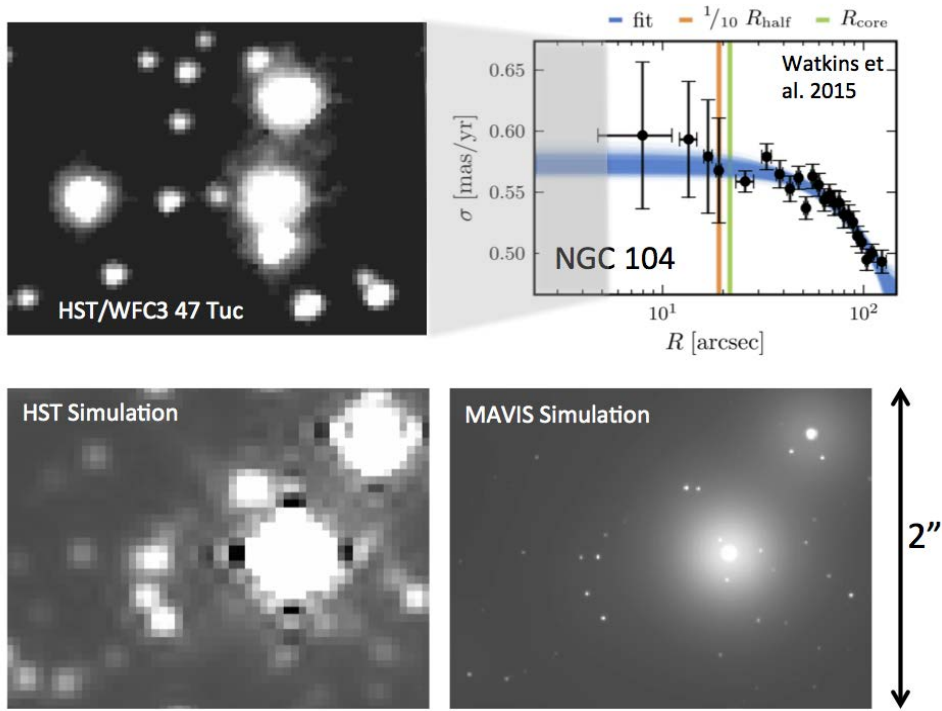

**Figure 4.6:** *Top-right: Velocity dispersion profile for globular cluster 47 Tucanae from HST proper motion data presented in Watkins et al. (2015). The presence of an IMBH would increase the central dispersion within the central few arcseconds. However, the HST data (top-left) becomes confusion limited in this regime. Bottom-left: Simulated HST (F555W, 50mas/pix) image of a similar stellar density to the centre of 47 Tuc, compared to the same field imaged with MAVIS (bottom-right, V-band, 7mas/pix). All images are approximately the same angular size.*

In particular, PMs directly constrain the velocity dispersion anisotropy of a GC, breaking the mass-anisotropy degeneracy (Binney & Mamon 1982). Without this piece of information a robust detection of an IMBH is not viable. Therefore, it is fundamental to sample the 2D kinematics not only in the very central regions, but where the anisotropy signal is expected to reach its maximum (Zocchi et al. 2017), at intermediate radial distances. For this reason, not only the superb spatial resolution of MAVIS, but also its wide field of view are crucial for these observations, outperforming the capabilities of HST for the global assessment of the presence of an IMBH. Moreover, because of the possibility to reach stars with faint magnitudes (down to ~30 mag), MAVIS will allow us to measure the radial dependence of the stellar mass function in detail, and to probe the dynamics of stars with different masses in the centre of GCs. This information is necessary to determine the degree of mass segregation of a GC, and thus its evolutionary stage, helping to distinguish the effects of a population of stellar-mass BHs from those of an IMBH (Bianchini et al. 2017).

## 4.3.1 Targets and Observational Outline

Proper motion measurements require exquisite accuracy to determine the often sub-pixel difference in the positions of stars from images taken over many years. The higher the spatial resolution and intrinsic motion of the objects under study, the shorter the timespan of the (at least) two epochs at which the objects need to be studied. For GCs, traditional (ground-based) astrometry has required epochs spanning years to tens of years , while with Gaia or HST often year long epochs are sufficient. The spatial resolution of MAVIS will allow for the measurement of proper motions to take place on timescales similar to Gaia and HST.

To perform proper motion measurements accurately, we must be able to distinguish instrumental effects from true stellar motion on the smallest (sub-pixel) scales. To achieve baselines on the order the same as Gaia or HST, we must be able to separate these two contributions to the measured motion. To address this issue, two key calibration factors need to be met: high-precision point-spread function (PSF) models must be created either in post-processing or delivered alongside the data, and high-precision geometric-distortion (GD) solutions must be determined to remove the effect of the static distortion present in the AOM. The expected high variability of the PSF in AO detectors, both spatially across the field of view (FoV) and from one image to the next, will not be a dramatic issue for this particular science case: as GCs contain thousands of stars that can be observed in each individual exposure within a FoV of 30"x30" with MAVIS. These stars can be used to obtain high-precision, spatially-varying empirical PSF models for each exposure, as is routinely done for HST (Bellini et al. 2013; Bellini et al. 2017), reaching a precision in the PM determination of a fraction of pixel. To correct for GDs, auto-calibration techniques can be employed (Libralato et al. 2015). Moreover, having at disposal thousands of stars and appropriate PSF models will allow us to minimize even the most subtle observation-dependent variations of the GD. The GD is expected to only have a marginal impact in narrow-field AO detectors, compared to the centroiding accuracy, but external catalogues can be used in a second step to further validate the GD correction, as was done for GeMS (Massari et al. 2016a, 2016b).

Only crowded stellar fields – either in stellar clusters or in the field of external galaxies – allow for a sufficient number (thousands) of calibrating stars in the 30"x30" FoV of MAVIS, especially considering that no optical instrument can rival the photometric depth allowed by MAVIS. Previous observations to guide the astrometry could be rather sparse or completely absent, however this is not the case for GCs, where external astrometric calibration objects are numerous thanks to Gaia (Gaia Collaboration et al. 2016).

To estimate the number of Milky Way GCs for which MAVIS could detect a central IMBH, we use the intrinsic velocity dispersion as a guide for what precision will be needed, and their distance to translate this dispersion into proper motion terms. In order to detect an IMBH, MAVIS must be able to measure the dispersion at a level equal to or smaller than the average cluster dispersion. In Fig. 4.7,  we show the apparent proper motion dispersions of MW GCs observable from Paranal, taken from the catalogue of Harris (2010), plotted as a function of helio-centric distance and including curves of constant intrinsic dispersion. Only GCs with measured dispersions are shown (60/158 GCs). This simple exercise shows that to build a meaningful sample of MW GCs (>20 objects) for which a central IMBH could be detected over a reasonable time baseline, an astrometric precision of 150µas is required. To significantly improve on this (i.e. double the sample size), an astrometric precision of 50µas would be needed. These respective values are therefore adopted as the MAVIS requirement and goal astrometric precisions, driven by the need for MAVIS to undertake dedicated astrometric surveys of MW GCs in the central regions inaccessible by both Gaia and HST.

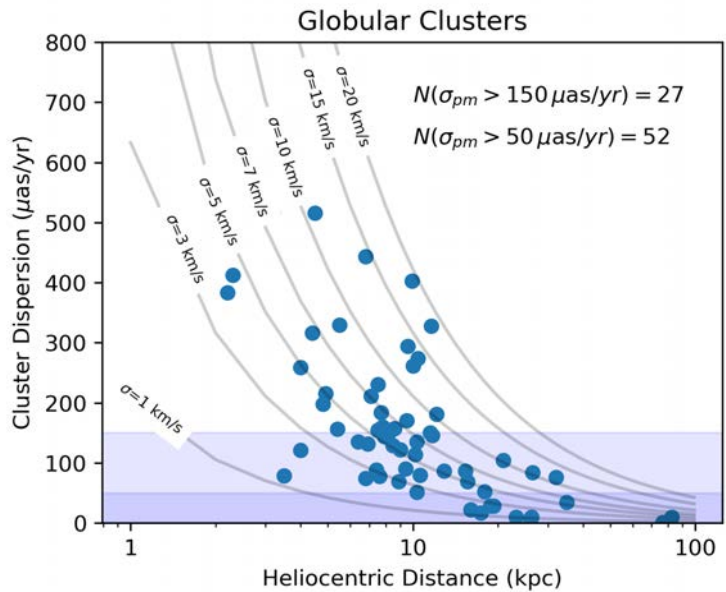

**Figure 4.7:** *Central dispersions of several MW GCs as a function of cluster distance. The shaded regions indicate the requirement and goal for MAVIS astrometric precision (150μas and 50μas respectively), giving an indication of IMBH detectability. The number of GCs above these limits are also indicated.*

### 4.3.2 Simulating the Astrometric Performance of MAVIS

Precision astrometry with MCAO instrumentation is a relatively new area, and brings a number of new challenges and uncertainties compared to, for example, the stable space environment of HST. To understand whether the level of precision outlined above can be met, we have built a tool to simulate the astrometric capabilities of the current MAVIS design. The details of the tool can be found in the Astrometric Error Budget in Section 4 of the accompanying document AD1. In the following section we use the MAVIS astrometric simulator to investigate the possibility of MAVIS detecting IMBHs in globular clusters. Included in the simulation is a field variable PSF, a small residual tip-tilt jitter from an imperfect NGS constellation and a static field distortion from the AOM.

We supplied the astrometric simulator (Monty et al. *in prep.*) with a catalogue of stellar positions, magnitudes and proper motions generated using the N-body code described in Baumgardt, 2017. A model of the globular cluster NGC 3201 was chosen as it is close enough (5 kpc) such that MAVIS could probe the cluster core. Two models were run through the astrometric simulator, one with one with a central 1500 M$_\odot$ IMBH corresponding to 1% of the cluster mass, and one without. A ten year epoch was simulated using monochromatic images in the V band, at the peak of MAVIS' predicted performance. Astrometric measurements were recovered directly from the simulated images using the FIND routine in DAOPhot (Stetson 1987, 1994). The results of this exploration are summarised in Fig. 4.8 and 4.9.

In Fig. 4.8, we show the predicted proper motion precision of MAVIS, highlighting the predicted capabilities of MAVIS given some of the more stringent astrometric requirements. Fig. 4.9 shows the recovered cluster velocity profile over a ten year epoch as a function of radial distance. The measured dispersion from stars fainter than V=22 mag are shown, which are the most challenging case for proper motions. The recovered measurements are compared against the input N-body models both with (green region) and without (red region) the central IMBH. Fig. 4.9 also includes a sample simulated image of the cluster highlighting the spatially variable PSF present throughout the simulation.

Bearing in mind the caveats associated with this simulation, namely the optimistic but not completely un-realistic treatment of tip-tilt jitter and static distortion, the results in Fig. 4.9 are encouraging for MAVIS. MAVIS will be able to measure stellar velocities in the crucial central few arcseconds as accurately from proper motions as other instruments will at much larger radius for spectroscopic radial velocities (e.g. MUSE delivers ~5km/s radial velocity errors). We can also conclude from this simulation that MAVIS has the potential to detect modestly sized IMBHs in GC centres. Finally, a handful of high-velocity stars (~30 km/s) were also recovered in the centre of the MAVIS FoV, highlighting an additional potential capability of MAVIS, direct measurement of an IMBH mass through orbital fitting, analogous to the Milky Way Galactic Centre. Ultimately, this simulation provides an exciting glimpse into the astrometric capabilities of MAVIS; both for indirect detection of IMBHs in GCs, from thousands of accurate stellar proper motions, and for direct detection of an IMBH through a handful of accurately measured high-velocity stars.

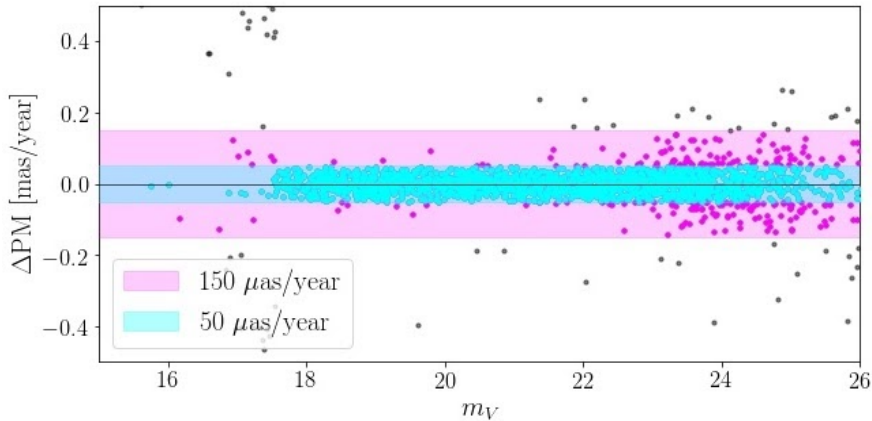

**Figure 4.8:** *Predicted proper motion precision for a nearby globular cluster (NGC 3201), in the presence of a field variable PSF, ideal tip-tilt residual error and static distortion. Proper motions were measured over a ten year epoch.*

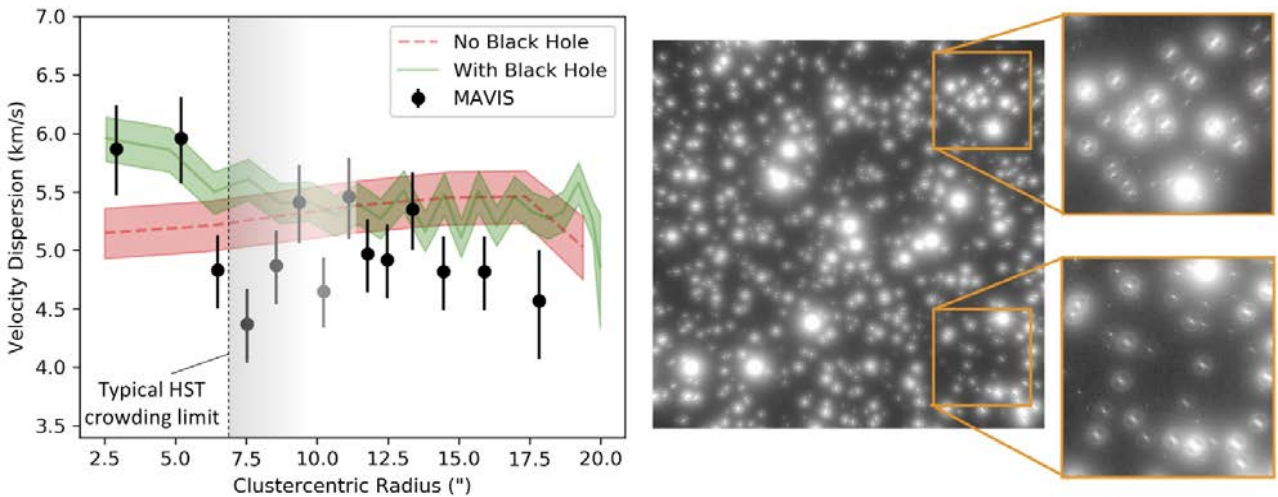

**Figure 4.9:** *(Left) Velocity dispersion profile recovered from simulated images over a ten year epoch. Black symbols represent measured stars with a V band magnitude less than 22. The green region denotes the input N-body model with the 1500 $M_\odot$ IMBH, using the same magnitude limit. The red region represents the N-body model without an IMBH. Vertical dashed line indicates the typical radial limit of HST proper motion measurements, which miss the crucial central regions where the IMBH signature is maximum. (Right) a simulated image of NGC 3201 highlighting the spatially variable PSF present in the image throughout analysis.*

## 4.4 Stars clusters in the Local Universe: Tracers of star formation histories and chemical evolution

**Science goal:** Characterization of star clusters in nearby galaxies: ages, distances, metallicity, detailed chemical abundances.

**Program details:** AO-assisted imaging with photometry and spectroscopy of individual stars in clusters located in galaxies of the Local Group, integrated photometry and spectroscopy of star clusters in the Local Universe, up to ~10 Mpc.

**Key observation requirements:** High- and low-resolution spectroscopy (R=10000-12000 and R=5000), AO-assisted imaging in FoV of ~30"x30" (or even smaller for extragalactic star clusters).

**Uniqueness of MAVIS:** The high-spectral resolution of MAVIS, e.g., with respect to MUSE, allows us to obtain a detailed chemical and kinematic characterization of single stars in the resolved stellar population (for cool stars up to 100-200 kpc, and hot stars up to several Mpc).

Understanding the formation and evolution of galaxies, and in particular of the Milky Way, is one of the major puzzles of astrophysics: a detailed physical scenario is still missing, and it requires the joint effort of observations and theories (e.g.,Pilkington et al. 2012, Thompson et al. 2018). The evolution with time of the chemical content of a galaxy is one of the topic constraints to understand the most relevant mechanisms driving galactic formation and evolution (e.g., Matteucci & Greggio 1986, Chiappini et al. 2001, Magrini et al. 2009, Prantzos et al. 2009, Grisoni et al. 2018, Maiolino & Mannucci 2019). In principle, it should be possible to trace back the past composition of a galaxy (as a function of both time and position) by measuring abundances of stars born in different epochs. Most stars do not modify in their photospheric abundances the composition of the interstellar medium from which they were formed, with the exception of giant and supergiant stars which, for some elements, can be contaminated by elements produced in the stellar interior (see, e.g., Lagarde et al. 2012). However, the ages of individual stars are very difficult to measure.

Star clusters are groups of stars born at the same time from the same molecular cloud, and in the case of open clusters, they share the same chemical composition. Unlike field stars, ages (and distances) of star clusters can be determined with good accuracy by comparing their colour-magnitude diagrams with theoretical isochrones (Randich et al. 2018). Star clusters have indeed proven to be excellent tracers of the chemical evolution of our Galaxy (e.g., Friel et al. 2002, Sestito et al. 2006, 2008, Magrini et al. 2017, 2018, Donor et al. 2020). Large spectroscopic surveys, such as the Gaia-ESO Survey (Gilmore et al. 2012, Randich & Gilmore 2013) and APOGEE and APOGEE-2 (Majewski et al. 2017, Zasowski et al. 2017) have devoted a large amount of observational time to collect spectroscopic data of Galactic star clusters bracketing a large varieties of conditions in terms, for instance, of mass, galactocentric position, age. These observations have allowed us to obtain the largest sample of homogenous measurements in stars of open clusters, including: radial velocities, stellar parameters, and abundances of elements belonging to different nucleosynthesis channels (as, for instance, α-elements, iron-peak elements, neutron-capture elements) that are allowing to constrain several aspects of the chemical evolution of our Galaxy (e.g., Duffau et al. 2017, Spina et al. 2017, Casali et al. 2019, 2020, Donor et al. 2020).

The challenge is now to move towards nearby galaxies and to investigate, with comparable precision, the star cluster populations in different environments. With a deep study of star clusters in external galaxies, we aim at answering a variety of questions related to several fields, from star formation to cluster populations and chemical evolution:

- ***Is the star formation clustered? Are there preferred environments for clustered star formation?***
- ***on which time scales do star clusters disperse? How do they evolve? Are the cluster mass and initial mass functions universal? How does the cluster formation efficiency depend on metallicity?***
- ***Is there an age-metallicity relationship in star clusters? How does the spatial distribution of metallicity in different galaxies vary with time?***

The combined MAVIS AO-assisted photometric and spectroscopic capabilities will allow us to fully address these questions. In Table 4.2, we present the list of galaxies whose star clusters population is accessible with MAVIS, in both photometric, low-resolution and high-resolution spectroscopic modes. In the Table, we show the typical magnitudes of four different spectral types, from F-G-K giant stars (typical stellar population of clusters older than 0.3 Gyr) to O-B stars, present in the youngest massive clusters (YMC), at the distance of the listed galaxies. As described in Appendix A1, the limiting magnitudes for MAVIS spectroscopy, with an exposure time of 3600s and corresponding S/N=10 and a 3x3 spatial binning are: 21.0 for LR-blue @500nm, 21.7 for LR-red @700nm, 19.7 for HR-blue @500nm, and 20.8 for HR-red @700nm. Based on these limits, Table 4.2 indicates with different colours the spectral types accessible via the different MAVIS modes.

For the AO-assisted imaging, we need a spatial resolution from 25 to 50 mas, to separate single stars, and to limit the effect of the unresolved fainter stellar population. In Fig. 4.10 we present simulated V-band images of a star cluster located at distances ~10 kpc, 50 kpc, 1 Mpc and 3 Mpc.

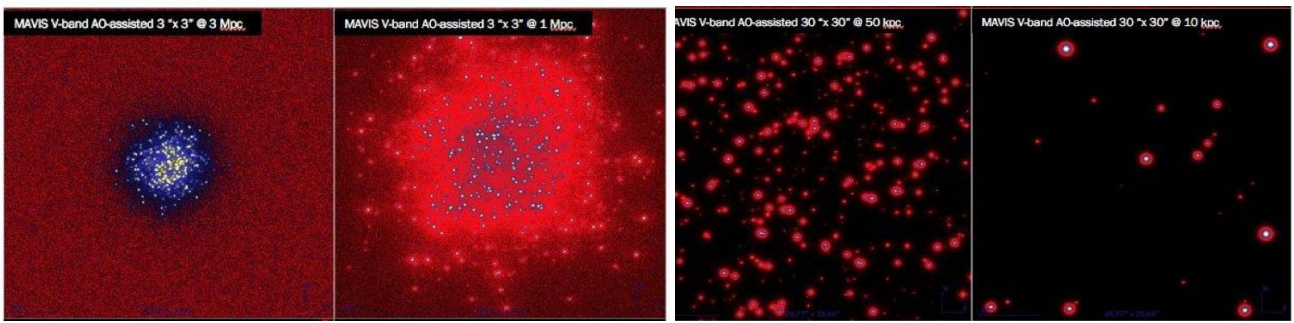

***Figure 4.10:*** *Simulated V-band images of an old star cluster (1,000 M$_\odot$, half-light radius of 5 pc, texp=1 hr) at the distance of 3 Mpc, 1 Mpc, 50 kpc, 10 kpc.*

| | Nearby Galaxies | distance (kpc) | distance modulus | G5III-K0III (clump old clusters) | O5V (MS young clusters) | B0V (MS young clusters) | B5V (MS young clusters) | angular size" |
|---|---|---|---|---|---|---|---|---|
| Local Group | Sagittarius | 30 | 17.39 | 17.89 | 11.59 | 13.29 | 16.29 | 68.76 |
| Local Group | LMC | 50 | 18.49 | 18.99 | 12.69 | 14.39 | 17.39 | 41.25 |
| Local Group | SMC | 60 | 18.89 | 19.39 | 13.09 | 14.79 | 17.79 | 34.38 |
| Local Group | Sculptor | 90 | 19.77 | 20.27 | 13.97 | 15.67 | 18.67 | 22.92 |
| Local Group | Sextans | 90 | 19.77 | 20.27 | 13.97 | 15.67 | 18.67 | 22.92 |
| Local Group | Carina | 100 | 20.00 | 20.50 | 14.20 | 15.90 | 18.90 | 20.63 |
| Local Group | Fornax | 140 | 20.73 | 21.23 | 14.93 | 16.63 | 19.63 | 14.73 |
| Local Group | Phoenix | 400 | 23.01 | 23.51 | 17.21 | 18.91 | 21.91 | 5.16 |
| Local Group | NGC6822 | 500 | 23.49 | 23.99 | 17.69 | 19.39 | 22.39 | 4.13 |
| Local Group | IC1613 | 720 | 24.29 | 24.79 | 18.49 | 20.19 | 23.19 | 2.86 |
| Local Group | Tucana | 870 | 24.70 | 25.20 | 18.90 | 20.60 | 23.60 | 2.37 |
| Local Group | WLM | 930 | 24.84 | 25.34 | 19.04 | 20.74 | 23.74 | 2.22 |
| Local Group | Aquarius | 1020 | 25.04 | 25.54 | 19.24 | 20.94 | 23.94 | 2.02 |
| Sextans group | Sextans A | 1500 | 25.88 | 26.38 | 20.08 | 21.78 | 24.78 | 1.38 |
| Sextans group | Sextans B | 1500 | 25.88 | 26.38 | 20.08 | 21.78 | 24.78 | 1.38 |
| Sextans group | NGC3109 | 1500 | 25.88 | 26.38 | 20.08 | 21.78 | 24.78 | 1.38 |
| Local Uni | NGC5253 | 3150 | 27.49 | 27.99 | 21.69 | 23.39 | 26.39 | 0.65 |
| Local Uni | NGC7793 | 3440 | 27.68 | 28.18 | 21.88 | 23.58 | 26.58 | 0.60 |
| Local Uni | NGC1313 | 4390 | 28.21 | 28.71 | 22.41 | 24.11 | 27.11 | 0.47 |
| Local Uni | NGC1705 | 5100 | 28.54 | 29.04 | 22.74 | 24.44 | 27.44 | 0.40 |
| Local Uni | IC4247 | 5110 | 28.54 | 29.04 | 22.74 | 24.44 | 27.44 | 0.40 |
| Local Uni | NGC1433 | 8300 | 29.60 | 30.10 | 23.80 | 25.50 | 28.50 | 0.25 |
| Local Uni | NGC1291 | 10400 | 30.09 | 30.59 | 24.29 | 25.99 | 28.99 | 0.20 |

| MAVIS Mode |
|---|
| Imaging |
| HR-Blue |
| HR-Red |
| LR-Blue |
| LR Red |

***Table 4.2:*** *list of galaxies accessible with MAVIS in imaging and spectroscopic modes (low and high spectral resolutions), with stars divided per spectral types.*

In the very nearby galaxies, within 100 kpc, we will be able to study the whole star cluster population, reaching the most luminous stars of both young and old clusters, with photometry and high-resolution spectroscopy.

An example is the star cluster population in the Fornax galaxy, with its six globular clusters (see, e.g., Hendricks et al. 2016), whose detailed chemistry might give important constraints on the mechanisms of star cluster survival (e.g., Wang et al. 2019). A simulated image of a cluster at the distance of the Fornax dwarf galaxy is shown in Fig. 4.11.

Among the nearby galaxies, the star cluster population in the Magellanic Clouds (MCs) will be among the most important targets for MAVIS. Their star cluster populations show important differences with respect to the Galactic one, both for the metallicity and for the age and mass ranges (in the LMC are many massive clusters, with a deficit of clusters of ages between roughly 3 and 10 Gyr, while in the SMC the age distribution is more continuous (Da Costa & Hatzidimitriou 1998; Mighell et al. 1998), with two distinct episodes of cluster formation that occurred 2 and 8 Gyr ago (Rich et al. 2000).

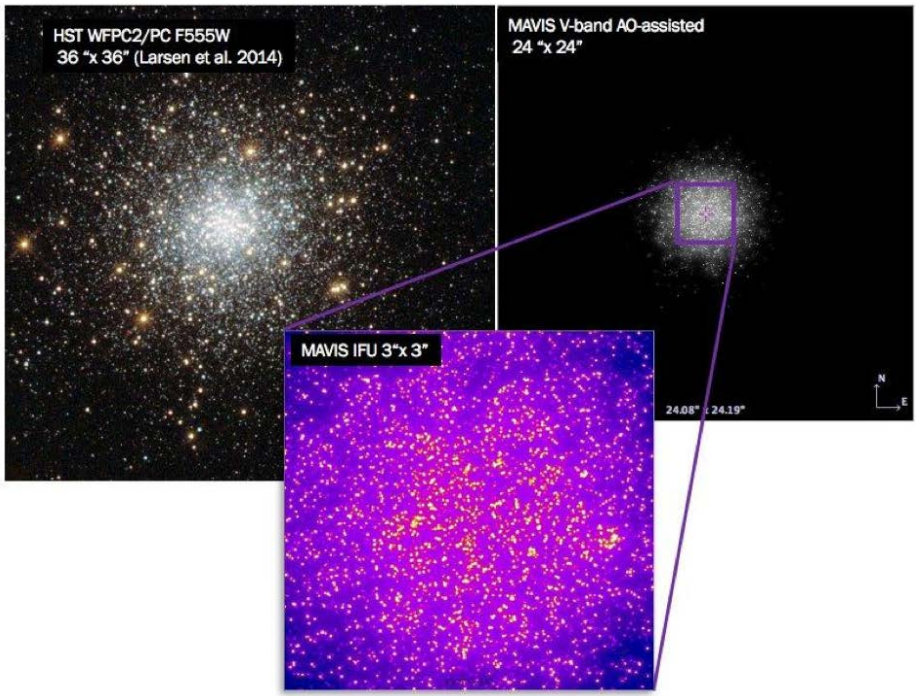

***Figure 4.11:*** *HST observations of a globular cluster in the Fornax dwarf spheroidal galaxy, compared with simulated observations of MAVIS V-band, with a zoom on the central regions (FoV 3"x3", corresponding to the MAVIS spectroscopic mode).*

***The MAVIS photometry and spectroscopy of the MC star clusters will contribute to our understanding of several aspects of their formation and evolution, and of their host galaxy.*** In Fig 4.12, we show the simulated I-B-V band images of the central regions of a YMC located at the distance of ~60 kpc. Young massive star clusters (1–5 Myr) in the LMC and SMC are optimal sites to derive the IMF: the field-of-view of the imager will be adequate to observe clusters with diameters ~7-9 pc at the distance of the MCs, and thanks to the high-spatial resolution, individual stars can be resolved along the main sequence (MS). A recent determination of the IMF in the cluster 30 Doradus, in the Large Magellanic Cloud, sampled the mass range down to ~ 1 $M_\odot$ (Andersen et al. 2009). The standard IMF exhibits a flattening for M<1$M_\odot$ and peaks somewhere between 1-0.1 $M_\odot$ (Scalo 1998, Chabrier 2003). To detect variations in the IMF it is indeed necessary to sample the mass range at least down to 0.1$M_\odot$, where metallicity effects could also be detectable. This will be possible with MAVIS: the AO-assisted imaging will allow us to reach for pre-Main sequence stars of 0.1 $M_\odot$ in 1 Myr old clusters (V~28.8 and R~27.4 at the distance of the LMC, V~29.3 and R~27.8 at the distance of the SMC) in 1 hr exposure-time with a SNR~10.

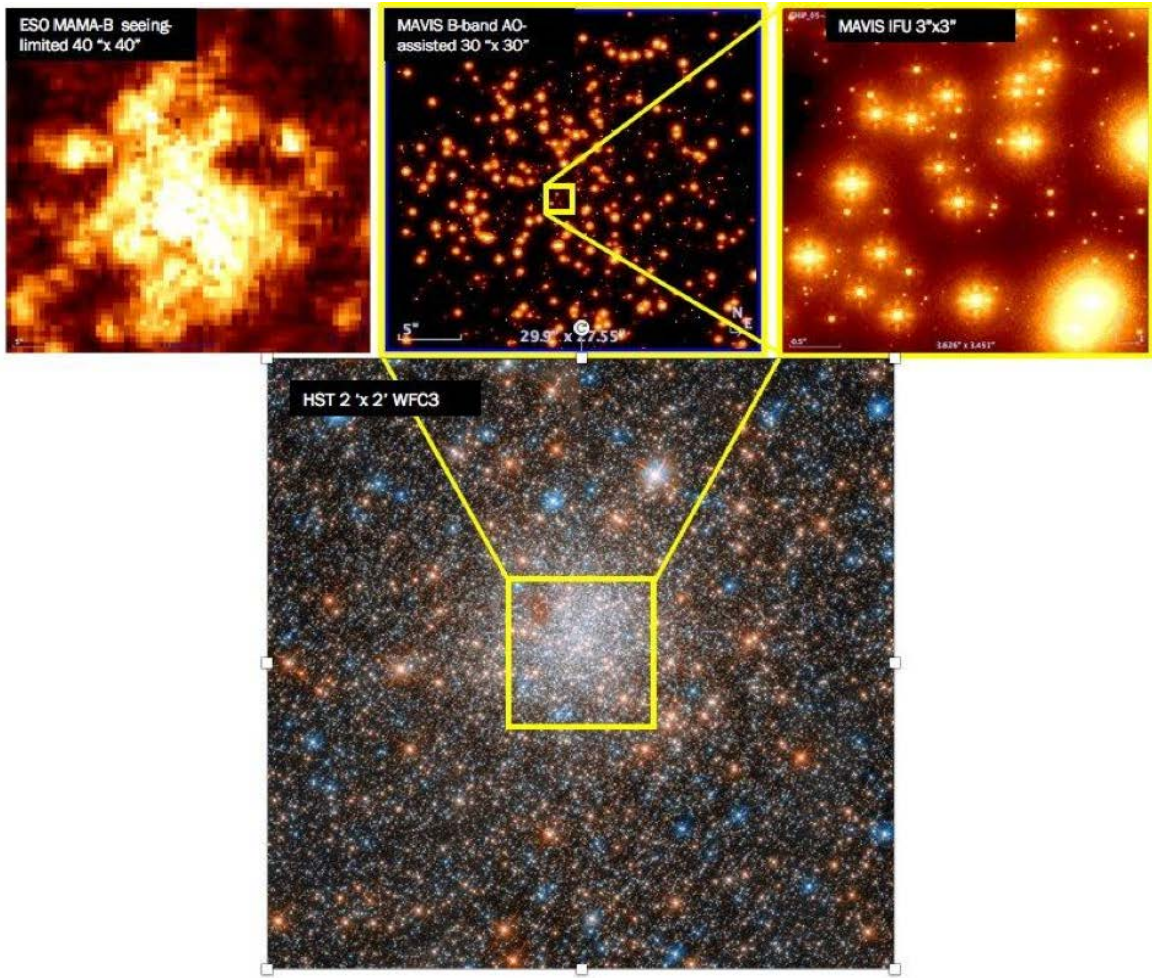

***Figure 4.12****: ESO archive image (from Aladin, MAMA-B), simulated B band MAVIS image, with t=3600 s in a FoV 30"x30", zoom on the central 3"x3" field for NGC1898, in the Large Magellanic Cloud. Bottom image from WFC3 HST, FoV 2'x2'.*

In addition, the high-resolution spectroscopic mode, which will extend in the HR-blue mode towards the blue at 425.0 nm, will allow to characterize the most massive stars (for which we can obtain SNR > 100 in 1 hr exposure time) and to derive the chemical composition from spectra of the lower mass cluster members.

The combined studies of star clusters with different ages will allow us to constrain the age-metallicity relationship (AMR). which is particularly important not only because of their proximity, but also because the MCs and the Milky Way form a system of interacting galaxies. It is therefore important to trace the influence of this interaction on the star formation and the chemical evolution (see, e.g., Livanou et al. 2013). In Fig. 4.13, we show the age-metallicity relationship derived for the field stars in the LMC and SMC, compared with the AMR from their star clusters, in which a clear enrichment in the global metallicity is present in both galaxies. ***MAVIS in its HR-spectroscopic mode will allow us to obtain spectra of the brightest stars in both young and old clusters with very high SNR (see Table 4.2 for limiting magnitudes of stars of different spectral types)****.* The high spectral resolution (R=15000) and wide spectral coverage (in the blue and in the red) will permit to measure many elements, such α and r-process elements, to probe fundamental aspects such as the efficiency and time-scale of the star formation episodes, and to fully characterize and constrain their chemical evolution (see, e.g., Colucci et al. 2012 for integrated MC clusters).

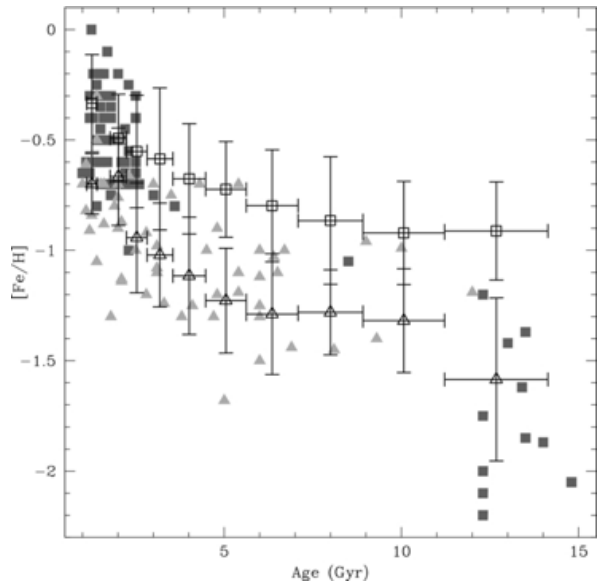

***Figure 4.13:*** *Composite field age-metallicity relationships (AMRs) of the LMC (open boxes) and SMC (open triangles). Their respective cluster AMRs are also drawn with filled boxes (LMC) and filled triangles (SMC). Figure from Piatti et al. (2012).*

Finally, the high-resolution spectroscopy of individual stars in massive clusters in the MCs might shed light on the debated origin of the multiple populations in Galactic globular clusters (see, e.g., Milone et al. 2018), as shown in Fig.4.14. With MAVIS we will reach in 1 hr exposure time a SNR~10 in the MS (mv~20), and a SNR > 100 at the MS turn-off and in the clump. The expected SNR for the high-resolution spectra will allow us to possibly detect chemical peculiarities and anti-correlations in the multiple populations of MC massive star clusters.

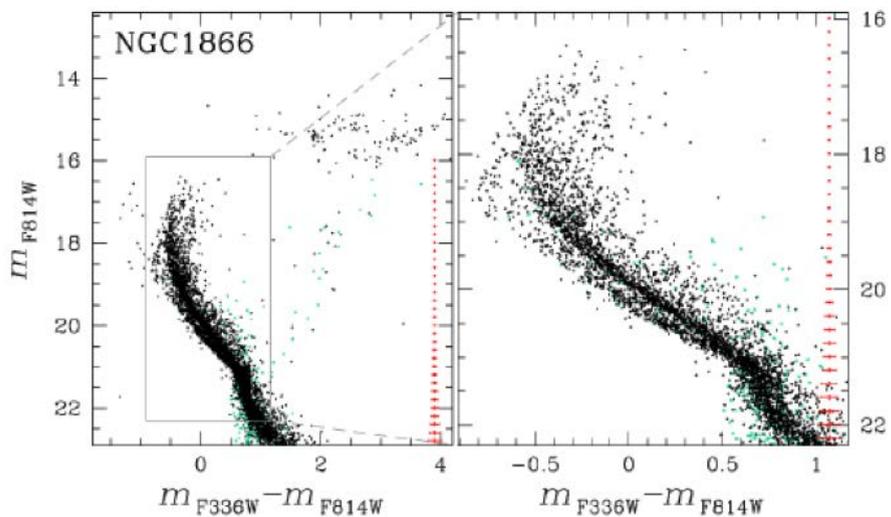

***Figure 4.14:*** *CMD of NGC1866, star cluster in the MCs, with an extended Main-sequence Turn-off, show multiple populations. Figure from Milone et al. (2018).*

In Fig. 4.15, we show two simulated spectra of a star ($T_{eff}$ =5800 K, log g=4.4) at R=15000 for two SNR ~10 and ~35. We show a limited spectral range, to highlight some absorption lines, marked with the atomic number of the corresponding elements. Even at the lower SNR, which can be obtained with just an hour of integration, many important lines can be detected.

With the combination of MAVIS AO-assisted imaging and high-resolution spectroscopic mode, we will be able to obtain complete colour-magnitude diagrams of star clusters in the Local Universe, and to derive abundances of several elements belonging to the different nucleosynthesis channels. Considering the star clusters located within ~100 kpc, we will combine the information from FGK stars, deriving abundances of at least Fe, Mg, Ca, Co, Cr, Na, Ni, Sc, Ti, V, and possibly of the faintest lines of Y and Zr, with the information from OB stars, which can be observed up to the outskirt of the Local Group.

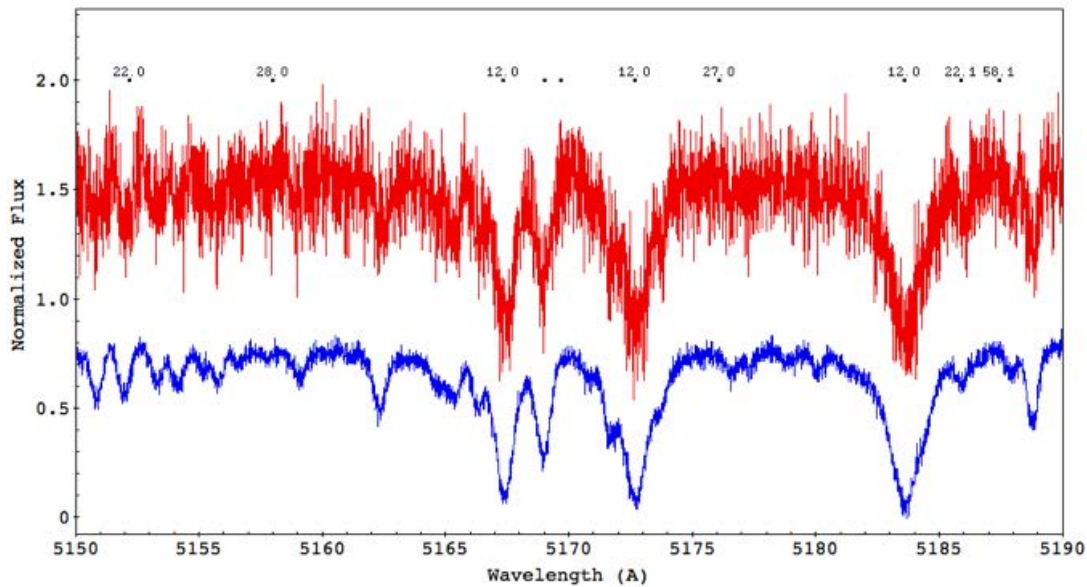

***Figure 4.15***: *Simulated spectra at R=15000 of a G-type star at SNR~10 (red spectrum) and SNR~35 (blue spectrum). Some absorption lines are marked and labelled with the atomic number of the corresponding element.*

### Case Study: Revealing the formation history of dwarf galaxies

**Key Questions:** Which is the origin of the star cluster population in the Fornax dwarf galaxy? What is the nature of Fornax 4, the nuclear star cluster? How do their composition compare with the field stellar population? which is the balance between accreted and in-situ formation?

**Why MAVIS?** MAVIS will allow us to investigate the cluster population in the Fornax dwarf galaxy, combining its AO-assisted imaging and spectroscopic capabilities. For the first time, combining the AO-assisted imaging and the 3"x3" IFU in the high-resolution spectroscopic mode, MAVIS will allow us to obtain resolved colour-magnitude diagrams to derive precise cluster age, and to measure abundances from the brightest stars of several elements belonging to the different nucleosynthesis channels, e.g. Fe, Mg, Ca, Co, Cr, Na, Ni, Sc, Ti, V (and possibly of Y and Zr).

**How will MAVIS be used:** The photometric information will be used to derive global properties of the targets which allow us to build the cluster CDM and HR diagrams. The spectroscopic observations with MAVIS will enable individual element abundance measurements.

**Field selection:** There are 6 known star clusters located in the Fornax dSph halo, which are suitable for spectroscopy and photometry with MAVIS. Fornax 4 (shown in the Figure) is probably the most interesting object, since it might be a nuclear star cluster. The MAVIS FoV is perfectly suited to study the whole clusters, while its IFU to to obtain a spectroscopic follow-up of a large number of resolved member stars in the crowded central regions to confirm the metallicity spread recently found by Martocchia et al. (2020).

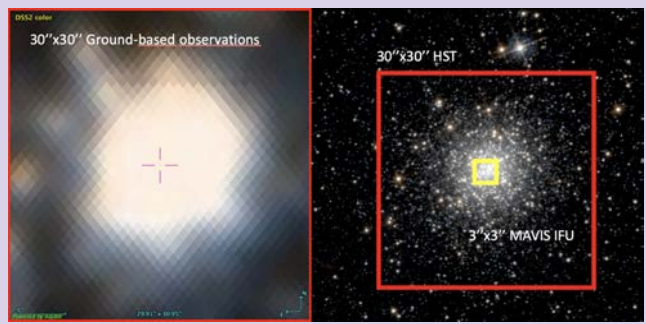

**Exposure time:** Single one-hour exposures would provide SNR~10 for the brightest stars in the clusters in the spectroscopic mode (m~21), and would allow us to map the whole cluster sequence in imaging mode.

# 4.5 Massive stars in metal poor environments

**Science goal:** The physical properties of massive stars at low metallicities, akin to those in the early Universe

**Program details:** Spectroscopy of individual massive stars in galaxies of the Local Group within a distance of 2-3 Mpc.

**Key observation requirements:** Multi-object spectroscopy with MAVIS from 425 to 680 nm. R ~ 15,000 for abundance studies; R ≥ 5000 for wind studies; S/N ≥ 35 per resolution element at 450 nm.

**Uniqueness of MAVIS:** the high-spectral resolution, especially in the blue, e.g., with respect to MUSE, allows a detailed chemical and kinematical characterization of single stars in resolved stellar populations. For hot massive stars, MAVIS will reach stars in galaxies up to several Mpc.

With stellar masses in the range of ten to several hundreds of solar masses, massive stars are among the most important cosmic engines, driving the evolution of galaxies throughout the history of the Universe. They begin their lives as OB stars and evolve in a few million years, ending up as core-collapse or pair-instability supernovae. As such, they are the progenitors of neutron stars, pulsars, magnetars and black holes, and are believed to be the source of long, soft gamma-ray bursts. On cosmological scales, they have been suggested to be the main cause of the last major phase change of the Universe, reionization, when the intergalactic medium hydrogen went from being fully neutral to fully ionized (Jarosik et al., 2011; Robertson et al., 2010). Despite their importance, the formation and evolution of massive stars remain poorly understood and many crucial questions lack satisfactory answers.

Moreover, in order to understand and quantify the role of massive stars in a Universe ever growing in metallicity, **it is necessary to describe the variation of their physical properties as a function of the chemical composition**. The metal-poor regime is indeed subject to a particularly growing interest in order to understand the conditions of earlier cosmic epochs, and to ultimately extrapolate the prescriptions for physical properties to interpret the intermediate- and high-redshift Universe, and eventually connect to the pristine First Stars.

The most pressing questions can be summarized as:

- *What is the upper mass limit on the IMF? Does it depend on metallicity?*
- *What kind of outflows do metal-poor massive stars experience?*
- *How do massive stars evolve in metal-poor environments?*

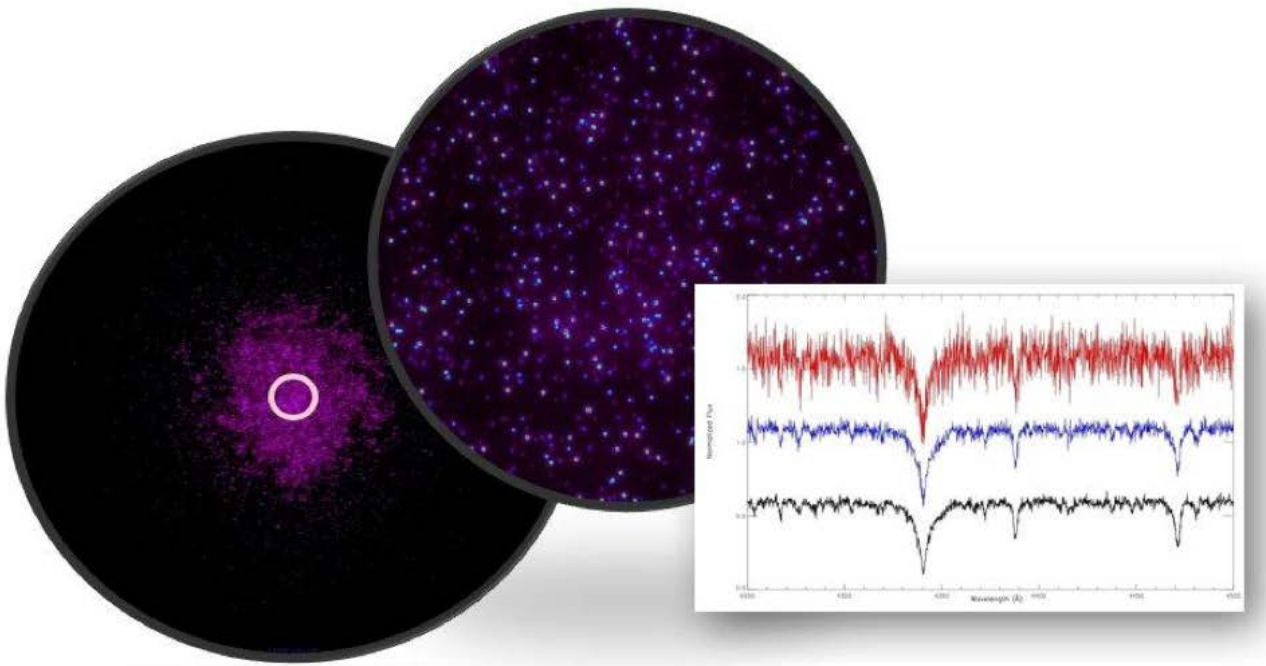

*Figure 4.16: Simulated images of a massive cluster (10000 stars) at the distance of 50 kpc. The left circle contains the FoV with radius ~30", while the central circle ~3". The right panel, there are simulated spectra of hot stars for the high-resolution mode of MAVIS.*

For all these questions, massive stars in the Small Magellanic Cloud (SMC, 1/5 $Z_\odot$) currently serve as templates for the low-metallicity objects in the early Universe (e.g. Ramachandran et al. 2020). In other words, our experimental knowledge of the properties of metal-poor massive stars is drawn from one single point in metallicity.

However, the 1/5 $Z_\odot$ metallicity of the SMC is not representative of the Universe past redshift z=1 (Madau & Dickinson 2014). Moreover, the theoretical framework for lower metallicities predicts substantial differences in the evolutionary pathways (Meynet & Maeder 2002, Eldridge et al. 2008, Szécsi al. 2015, Groh et al. 2019) with impact in life-overall feedback and end-products.

As of today, the SMC marks not only a metallicity but also a distance frontier, and a sizable leap down in metallicity requires reaching distances in the range 1-3 Mpc, that is the outer Local Group and surroundings. To name a few, very promising Local Group and nearby dwarf irregular galaxies (dIrr) with 1/10 $Z_\odot$ (Sextans A, 1.3 Mpc away, Camacho et al. 2016), 1/20 $Z_\odot$ (SagDIG, 1.1 Mpc, Garcia 2018) and 1/30 $Z_\odot$ (Leo P, 1.6 Mpc, Evans et al. 2019) shall be priority targets for MAVIS observing programs on massive stars.

***Our goal is to use the multiplex capability of MAVIS to identify massive star populations in these metal-poor galaxies of the local Universe, down to completeness and to perform quantitative studies of the individual massive stars.*** These goals require the outstanding spatial resolution and sensitivity of MAVIS. High-quality spectroscopy over 425–550 nm of individual luminous stars with mid to high resolution (R=5000 to R=15000) will then constrain their physical properties (temperatures, luminosities), surface abundances and mass-loss of their winds. Only with such information can we fully characterize the nature of these stars, and test theoretical predictions for their properties, and provide a firm basis for extrapolation to massive star populations of the early Universe.

### 4.5.1 Formation of massive stars

Our understanding of massive stars formation suffers significant gaps ranging from the formation of individual stars, to how the upper initial mass function (IMF) builds, and whether/how the environment plays a role in it. For instance, the most massive stars known as of today (~ 150$M_\odot$) have been found at the core of the Tarantula Nebula in the LMC (Crowther et al. 2010, see Fig. 4.17, Bestenlehner et al. 2020) and have 0.4 $Z_\odot$ metallicity. The integrated light of unresolved, metal-poor starburst also shows evidence of stars over 100 $M_\odot$ (e.g. Wofford et al. 2014).

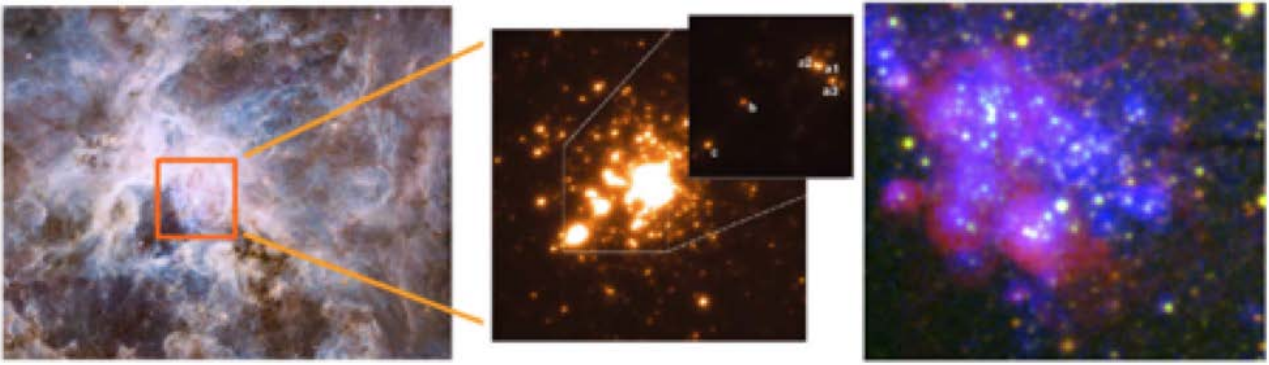

***Figure 4.17:*** *Left and middle panels: The R136a cluster at the heart of the Tarantula nebula, in the LMC, hosts the most massive stars known in the local Universe (Crowther et al. 2010). Right: The Local Group 1/10 $Z_\odot$ galaxy Sextans A hosts HII shells equivalent in size, but no star more massive than 60 $M_\odot$ has been detected yet (from Garcia et al. 2019, with permission).*

**For any galaxy out to ~3 Mpc, surveys of typically 30 - 50 hours built on MAVIS very high sensitivity, will deliver unprecedented inventory of the full zoo of O-B-A supergiants, planetary nebulae, rare LBV-WN/Ofpe-B[e]-WN-WC stars, and H II regions.** These surveys will further provide simultaneously complete spectroscopic (R ~ 5000) samples for the quantitative study of these object classes (see next sections). This will provide unprecedented constraints on the IMF of both dense, and starburst regions, including comprehensive library of stellar spectra for stars brighter than B2 V on the main-sequence, which is the low-mass end of the massive star regime, in metal poor environments.

30 Doradus-like concentrations of massive stars will be disentangled in galaxies out to 1-2 Mpc, allowing a deep census of a variety of massive star populations. Completeness will be reachable down to magnitudes corresponding to the B2-B5 spectral type on the main-sequence. Six hours spectroscopic programs with AO assisted integral-field spectrograph of MAVIS, in its high and low resolution mode (R~15000 and R~5000, respectively), will deliver spectra, perfectly suited to perform detailed spectral modeling, from which we will obtain wind and stellar parameters, including stellar masses for the whole stellar population. This analysis will reveal how massive are the most massive stars in the clusters, and whether their cores shelter the most massive stars, like in the Tarantula Nebula.

### 4.5.2 Stellar winds at low metallicity

It is well-known that mass loss, rotation, and binarity are major drivers of every phase of massive-star evolution (e.g., Maeder & Meynet 2000; Puls et al. 2008; Langer 2012). These processes are strongly coupled, making the prediction of massive star evolution a very difficult task (Groh et al. 2013). Large uncertainties affect mass loss processes and rates beyond the main sequence, when the stars go through a RSG, LBV or WR phase, depending on their initial masses. Despite many theoretical and observational studies, no global consensus has been achieved on the details of the evolution and properties of massive stars.

Mass loss through radiatively-driven winds (RDW) strips the hydrogen envelope and angular momentum from the most massive stars, drastically altering their evolution and subsequent end-points (e.g. Vink 2012), with direct implications for feedback and chemical evolution. This is because the wind inherits a strong dependence on metal content that two massive stars born with the same initial mass but different metallicity can follow very distinct evolutionary pathways (Chiosi & Maeder 1986).

Theory predicts that RDWs should become weaker as their metallicity decreases (Vink, et al. 2001, Björklund et al. 2020). The metallicity dependence has been confirmed on the observational side down to SMC metallicity (0.2 $Z_\odot$). The winds of most metal-poor hot stars require a special formalism (Kudritzki 2002) but the latter lacks observational constraints for sub-SMC metallicities. Optical observations on 8-10 m telescopes of a handful of stars in star-forming Local Group galaxies indicated that winds were stronger than predicted by theory at the galaxy metallicities, which was soon contradicted by HST UV observations of the same objects. This is a major setback before implementing theoretical recipes in evolution models.

Because of these contradictions, the nature and properties of stellar winds beyond the SMC metallicity regime remain essentially uncharted territory. **AO assisted 3D spectroscopy with MAVIS of massive stars in Local Group and nearby dIrr southern galaxies, is the required next step to make progress.** The medium resolution (R = 5000) mode will be crucial to study the Hα line in the spectrum of massive stars (in particular when Hα is in emission or as a P Cygni profile), to confirm the presence of winds, and to constrain mass loss rates and velocity fields, as shown in Fig. 4.18.

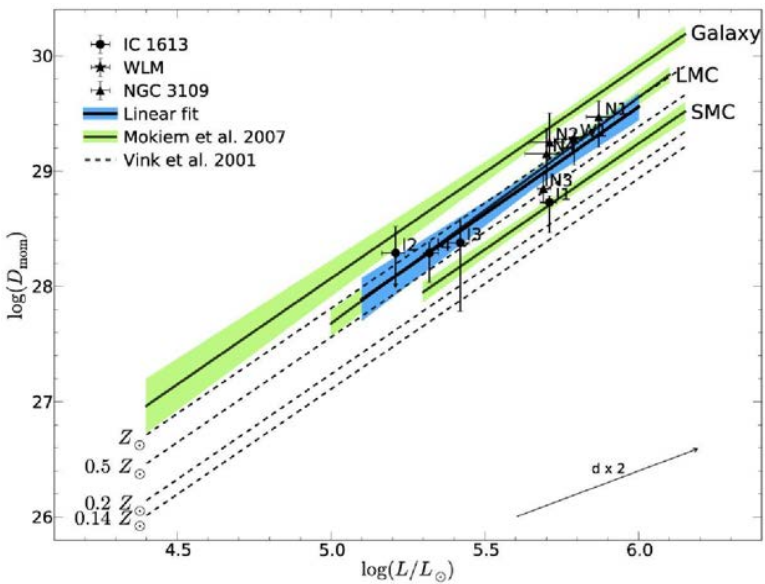

*Figure 4.18:* The momentum carried by the wind depends on stellar luminosity and metallicity. Solid lines indicate linear fit to empirical results from Mokiem et al. (2007) for Galactic, LMC and SMC stars. Theoretical predictions from Vink et al. (2001) are also indicated (dashed lines). The optical studies of stars with Z = 1/7 $Z_\odot$ suggested that their winds are as strong as LMC analogs. From Tramper et al. (2014).

### 4.5.3 Evolution of massive stars

The various phases of the evolution of massive stars, their interrelations and the relative numbers of different classes of massive stars should be accounted for by evolutionary models. The latter should also draw paths that depend on metallicity, mass loss, rotational velocity or mass exchange in binary systems (Langer 2012).

Evolutionary tracks that properly deal with rotation and the complicated physics of massive stars have been extensively calculated for the Milky Way, LMC and SMC (Ekström et al. 2012), Population III stars (Marigo et al. 2003, Ekström et al. 2008, Yoon et al. 2012), and for intermediate 1/50 $Z_\odot$ metallicities (Szécsi et al. 2015). Significant changes are expected in the evolution of metal-poor massive stars, some of them with tremendous impact on ionizing fluxes. For metal-poor massive stars with large amounts of core angular momentum (i.e. rotating sufficiently fast), vigorous mixing occurs in fast timescales from the core to the surface, to the point that the star is mixed to a homogeneous composition (Maeder 1987; Yoon & Langer, 2005). A star undergoing chemical homogeneous evolution (CHE) will either evolve into an envelope-inflated RSG, or stay compact in the regime of high effective temperatures of the Hertzsprung-Russell diagram (HRD), (e.g. Brott et al. 2011; Szécsi et al. 2015). This effect is expected to be magnified at sub-SMC metallicities. Testing this critical prediction about the evolution of massive stars at low metallicity become in range with MAVIS. This boils down to settling the case of how these stars acquire/maintain their rotation rate and how this links with internal mixing processes, or the properties of stellar winds.

Again, this is of crucial importance as there is overwhelming evidence that evolved massive stars with large amounts of angular momentum and mixing are natural progenitors of LGRBs (Langer 2012). Current scenarios, such as the collapsar model (Woosley 1993) indeed require a massive star following CHE. There is also a known preference of LGRBs and super luminous supernovae for metal-poor galaxies (Lunnan et al. 2014, Chen et al. 2017) which is yet another clue on the specific evolution of metal-poor massive stars.

Such fast-rotating, evolved massive stars have been identified in the MCs (Walborn et al. 2010). Observational biases aside (discovery of these stars is more complete in MCs because of lower extinction than in the Galaxy), this observed preference for rapid rotation in low-metallicity environments may be related to the significantly reduced removal of mass and angular momentum during the stars evolution, caused by the strong metallicity dependence of mass-loss via line-driven stellar winds (Vink et al. 2001).

To make progress, we need to undertake systematic searches for such fast-rotating stars beyond the SMC, i.e. in more extreme metallicity regimes. MAVIS spectroscopic capabilities allow us to explore the origin of fast rotating massive stars, and determine whether they are (or have been) members of binary systems or instead have evolved as single objects. The spectral resolution of MAVIS is indeed well suited for radial velocity searches in such massive objects. In addition, the proposed mechanisms for producing fast rotating O-type supergiants, single vs. binary, may leave disparate abundance signatures (de Mink et al. 2009). The differences expected for N/C and N/O abundance ratios from single vs. binary scenarios are larger than the typical measurement uncertainties, and the location of stars in a N/C–N/O diagram (Langer 2012) can be used for disentangling which mechanism is responsible for the observed surface abundances. To discriminate between the two scenarios requires to obtain spectra with good signal-to-noise ratio (SNR > 35), at R ~ 15,000 to measure weak CNO lines accurately and disentangle blends (see Fig. 4.19), well within reach of MAVIS for massive stars within 1-2 Mpc.

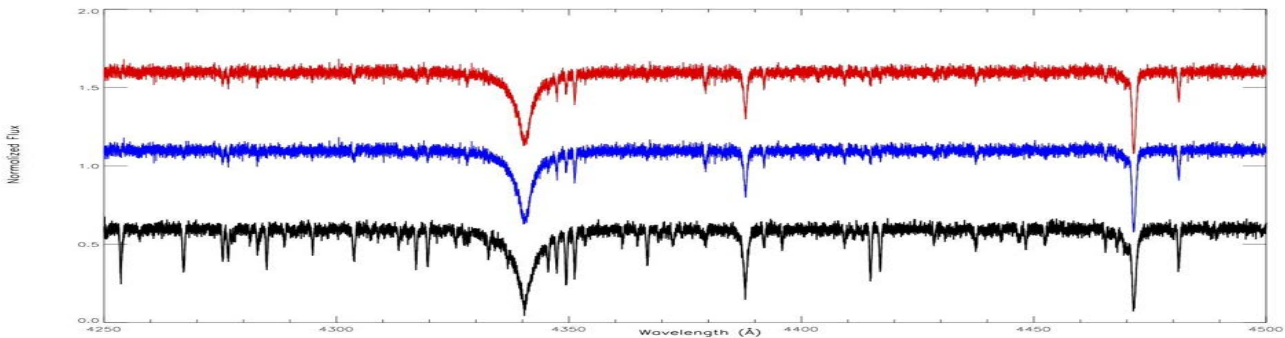

**Figure 4.19:** *Simulated spectra at R=15000 of O-type dwarfs with $T_{eff}$=30kK (black), 35kK (blue) and 40kK (red). SNR=50 is used. The strongest absorption lines are those of H , HeI at 438.7nm 447.1nm. Several lines of ions such as CIII, OIII,, NIII, MgII are also visible.*

## 4.6 Resolving star forming clumps at intermediate z

**Science goal:** What is the size, luminosity, mass, SFR, age of star forming clumps at intermediate redshift? What is the fate of the high-redshift stellar clumps? Are they contributing to the bulge formation? Are they the progenitors of today's globular clusters?

**Program details:** Deep optical photometry and IFU spectroscopy of a large sample of clumpy galaxies at redshift 1-2

**Key observation requirements:** Multi-band imaging, up to 900-1000 nm, will allow to detect rest-frame near-UV (up to z=2.5) and V-band (up to z~1) emission, probing young and intermediate age stellar populations. R~5000 IFU spectroscopy to detect emission lines and spectral features (e.g. [OII]3727, Balmer lines, [OIII]5007, break at 4000 Å) that are needed to characterize star-forming regions and to estimate stellar ages.

**Uniqueness of MAVIS:** The unique large sky-coverage and high spatial resolution of MAVIS will allow us to probe -for the first time- clumps at sub-kpc scale in large, with statistically significant samples of clumpy galaxies.

### 4.6.1 Resolving star forming clumps within galaxy discs

Over 2/3 of the stars in our Universe formed in galaxies at redshift z = 1−3 (Hopkins & Beacom 2006). During this epoch galaxies appear highly turbulent, thick, gas-rich and exhibit irregular morphologies, dominated by clumpy star forming regions. Star formation in the super-high pressure environment of clumps might therefore be the dominant mode of star formation in the Universe.

Clumps may also play a role in the formation of bulges, and thus the morphological transformation of galaxies (reviewed in Bournaud et al 2016). In this scenario, the bright, high surface density star-forming region generates a massive, gravitationally bound structure of stars. These stellar clumps then sink to the galaxy center via gravitational friction over a period of a few hundred million years. If clumps are sufficiently massive, and if there are enough of them in a galaxy, the aggregate mass will then be sufficient to constitute a bulge. It has also been suggested that these high-redshift stellar clumps may be the sites where today's globular clusters have formed. If this scenario is confirmed, observations of massive clumps at intermediate redshift could probe under which gas pressure and density conditions globular clusters have formed. Thus, any further constraint on the clump populations and ISM conditions in galaxies at the peak of the cosmic star formation history (which coincide with the peak of the globular cluster formation history), would also represent a paramount step forward in solving the globular cluster puzzle.

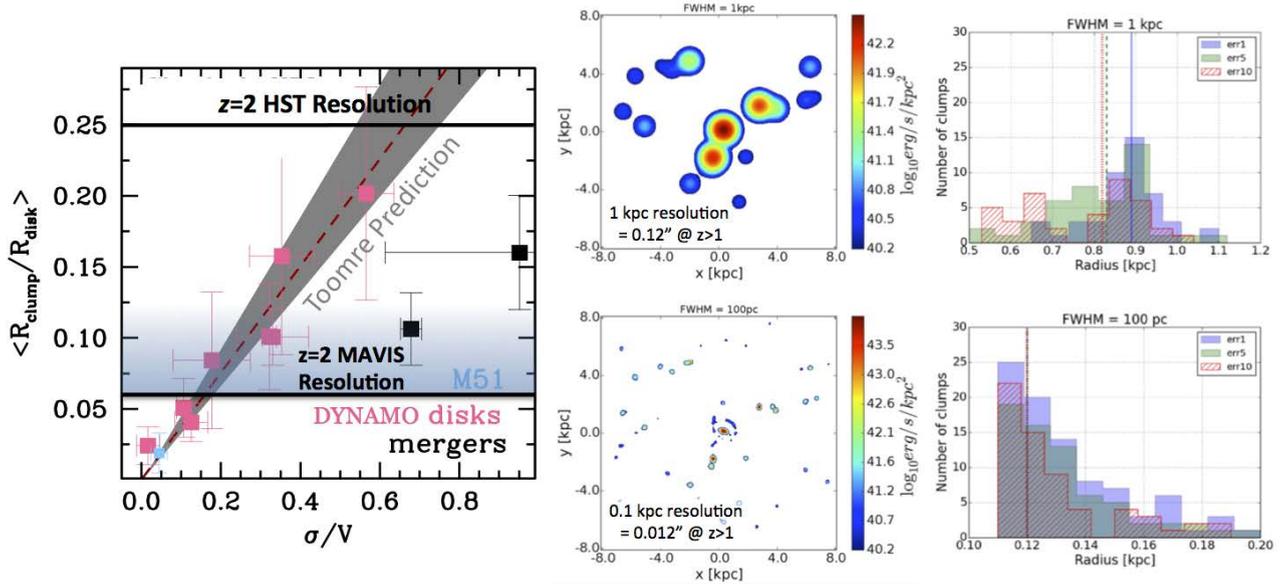

**Figure 4.20:** *Left: The average sizes of star forming regions are compared to galaxy kinematics for the DYNAMO survey (Fisher et al. 2017). The grey diagonal wedge region represents the prediction from Toomre-style instability arguments. The black, horizontal lines indicate where the most extreme z > 1 galaxies would have resolved Toomre scales with HST (upper) and MAVIS (lower). Right two columns: Simulated observations of a star-forming disk galaxy by Tamburello et al. (2017). The top panels show the observed Hα emission with 1 kpc resolution (left) and the inferred size distribution (right, colours indicate different error levels). The bottom panels are with 100pc resolution. The number of clumps and their size distribution is drastically overestimated without sub-kiloparsec resolution.*

There is still limited information about these star forming "clumps" and the main difficulty is imposed by the spatial resolution of the available instrumentation. State-of-the art observing facilities allow for ~0.1 arcsec resolution, which corresponds to just under 1 kpc at z=1-2. Recent work by several research groups showed that clump masses, sizes and SFRs might be highly overestimated by observations at ~1 kpc resolution. To date, the only possible way to probe sub-kpc scales is by using lensed galaxies and several studies (e.g. Jones et al. 2010, Livermore et al. 2012, 2015, Cava et al. 2018) find that typical clump sizes range 0.1-0.5 kpc, as much as an order-of-magnitude smaller than un-lensed galaxies at the same redshift.

Both theory and simulation also predict clump sizes smaller than what is observed at z ~ 2. Instability theory predicts that for clumpy galaxies $R_{clump}/R_{disk}$ = a/3(σ/V), where $R_{disk}$ is the size of the disk, a is a geometric constant ranging from 1 to 2, and σ/V is the ratio of the velocity dispersion to the rotation velocity (see Fisher et al. 2017a for more detail). In Fig. 4.20 we show this prediction in comparison to clumps in the DYNAMO survey of low-redshift galaxies. DYNAMO disks have a strong, linear correlation between clump sizes and kinematics that is completely consistent with the prediction. In this diagram we also show a line representing 1 kpc resolution in a typical disk galaxy at z ~ 2 with Rdisk ≈ 4 kpc. Main-sequence galaxies at z ~ 1 − 3 have V/σ of 0.2-0.4 (Wisnioski et al 2015). This is completely unresolved by the best current facilities. It is important to realize that this prediction is based on Toomre-style arguments. The Toomre scale predicts the most probable spatial scale for collapsing structures. According to Toomre theory the maximum size scale that will collapse is twice the scale shown here. Therefore for the vast majority of galaxies at z < 2, present facilities like HST or traditional AO systems cannot even probe the largest size scales expected to collapse.

The ages of high-redshift stellar clumps are so far uncertain. Even the detailed analysis of the Cosmic Snake clumpy galaxy (Cava et al. 2018) has allowed only very rough age estimates, because of the well-known age-extinction degeneracy encountered in SED fitting. The detected clumps show ages of ~ 90 Myr for the most distant clumps from the galactic center, and much older clumps (~ 300 Myr) closer to the galactic center. These age determinations are apparently in contrast with the Hα emission detections from the HST narrow-band imaging obtained by Livermore et al. (2015) and which point to clump ages shorter than 10 Myr. However, the direct link between the star-forming H II regions detected in Hα and the stellar clumps detected in rest-frame UV broad-band images remains to be done.

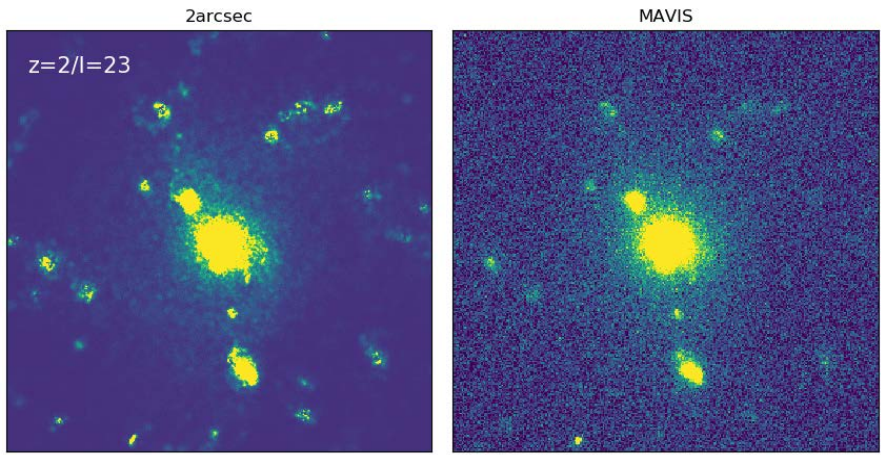

**Figure 4.21:** *Left: Source model from Tamburello et al. (2017) of a clumpy galaxy at z=2. Right: simulation of a 5h observation of the model galaxy with MAVIS in the I-band assuming an integrated AB magnitude of 23. The size of the image is 2", corresponding to ~17kpc at z=2.*

At redshift z=1-2 1 arcsecond corresponds to ~8 kpc and therefore MAVIS imaging would allow for measurement of clump properties in the UV with 150 pc resolution. This is similar to what is currently achieved with lensed galaxies and with HST imaging of local clumps (e.g. DYNAMO). As an example, Fig. 4.21 shows a simulated I-band observation with MAVIS of a clumpy galaxy at z=2, based on the model galaxies from Tamburello et al. (2017). The spectroscopy of MAVIS, available at the same spatial resolution as the broad-band imaging, will enable us to obtain spectra of individual stellar clumps, selected in MAVIS rest-frame UV broad-band images. We expect that ~ 70% of the clumps at z ~ 1 will have i- and z-band AB magnitudes < 29 (Dessauges-Zavadsky et al. 2017), and will thus be bright enough to be observed spectroscopically. The detection of the spectral continuum will help to derive the clump ages from the 4000 Å break index, D(4000). The break is expected to be small for young stellar populations and large for old, metal-rich stellar clumps. Together with the Hδ absorption line expected to arise in systems that experienced a burst of star formation, we can use the $Hδ_A$ index to even more tightly constrain the stellar clump ages (Kauffmann et al. 2003). The two spectral indices will be covered by the MAVIS spectral wavelength range up to z ~ 1.

These measurements will allow us to determine whether the ages of clumps are as short as a few Myr, or whether the clumps are longer lived with ages of a few tens to hundreds Myr, or whether clumps span a range of ages (see Adamo et al. 2013). We will also more reliably probe the possible clump age segregation with the galactocentric distance. If confirmed, this will provide important clues for the clump migration toward the galactic center. Moreover, the clump lifetime is a key property for testing several scenarios that attempt to explain how globular clusters can form multiple chemically enriched stellar populations (Bastian & Lardo 2017). Stellar clusters forming within long-lived clumps will possibly be able to retain and pollute the immediate ISM and form new chemical enriched stars that will stay gravitationally bound within the clusters.

## 4.6.2 Star formation outside galaxy discs

Galaxy disks are the usual cradle for star formation but recent observations (see Poggiant et al. 2019 and refs therein) have shown that massive star forming regions are common in the tails of ram-pressure stripped (RPS) galaxies. RPS is the removal of interstellar gas due to the hydrodynamical interaction with the hot intergalactic medium and it is believed to have a strong impact on galaxy populations in dense environments such as galaxy groups and, especially, clusters. SF clumps in RPS galaxies offer a unique opportunity to study the star formation process under extreme conditions, in the absence of an underlying disk and embedded within the hot intracluster medium. Most of the results obtained so far on these systems is based on a few local galaxies observed with MUSE. High-resolution observations with HST were carried out so far only for a very minor number of galaxies. Yet, a comprehensive, high spatial resolution study of these systems is missing.

The first systematic study of SF regions in the tails of a large sample of RPS galaxies was presented by Poggianti et al. (2019), based on the GASP survey of local (z~0.05) RPS galaxies with MUSE. The unique sensitivity, spatial resolution and sky coverage of MAVIS will allow us to probe for the first time SF in the tails of RPS galaxies at intermediate redshift, at the epoch of formation of galaxy clusters (z~1). This would be fundamental as the infalling rate of galaxies in clusters increases with redshift and therefore the number of expected RPS galaxies is much higher at intermediate redshift with respect to what observed at low-redhist.

MAVIS imaging and IFU spectroscopy at intermediate resolution (R~5000) are needed to study the sizes, stellar masses and ages of the clumps and their clustering hierarchy, investigating whether the hierarchical structure in the tails follows the one in disks. In this way it will be possible to study the clump scaling relations, explore the universality of the star formation process and verify whether a disk is irrelevant for star formation, as hinted by RPS galaxy results so far. The size of RPS tails are up to 100 kpc and therefore a few MAVIS IFU pointings would cover the whole extension of the extraplanar gas tails. MUSE allows to detect diffuse gas emission in the tails of RPS at intermediate redshift with an exposure time of a few hours. With similar exposure times, the highest spatial resolution of MAVIS would therefore allow to resolve individual SF complexes and size and study their physical properties.

# 4.7 The origin of the first star clusters: star forming clumps at high-z

**Science goal:** Identify globular cluster precursors up to the reionization epoch and quantify their ionizing photon production efficiency.

**Program details:** "Unpack" high-z galaxies, identify star-forming clumps (~100 pc size) hosting gravitationally-bound YMCs (<20 pc scale) at cosmological distance up to z=7 by measuring the dynamical age of the involved systems.

**Key observation requirements:** AO-assisted imaging with FoV of ~30"x30" and wavelength coverage up to 0.9-1.0 um is desirable to access UV continuum up to z=7. Low-resolution IFU spectroscopy (R~5000) to perform two dimensional maps of high ionization emission lines of the most magnified gravitationally lensed arcs.

**Uniqueness of MAVIS:** The significantly enhanced sky coverage of MAVIS will allow us to place it at the focus of the best cosmic lenses (not feasible with MUSE). R~5000 is necessary to unveil the often complex Ly-alpha profile and detect ionized channels optically thin to Lyman continuum radiation, e.g. narrow Lya emission at systemic velocity (not feasible with current R<3500 of MUSE).

The observational investigation of star-formation at high redshift (z>2, within the first 3 Gyrs) at very small physical scales (1-100 pc), including star-forming complexes and YMCs, is a new challenge in observational cosmology (e.g., Vanzella et al. 2017b). Thanks to strong gravitational lensing, the possibility to catch and study globular clusters precursors (GCP) is becoming a real fact (e.g., Pozzetti et al. 2019, Vanzella et al. 2017b,c, 2019a, 2020; see also Bouwens et al. 2017).

The luminosity function of GCPs has also been addressed for the first time and their possible contribution to the ionising background is still under debate (e.g., Boylan-Kolchin 2018). While still at the beginning, the open issues of (1) GC formation (e.g., Bastian & Lardo 2018; Renzini et al. 2015; Calura et al. 2019) and (2) of what sources caused reionization (e.g., Dayal & Ferrara 2018) can be addressed with the same observational approach, at least from the high-z perspective. This might be the consequence of the fact that some of the extremely faint sources possibly dominating the ionising background seem to match the ultraviolet appearance of YMCs. In turn, these are likely to also include GCPs, whose specific properties (stellar mass, luminosity, densities) might shed light on the different formation scenarios (Bastian & Lardo 2018; Renzini et al. 2015; Zick et al. 2018). Star-formation is observed to be hierarchically organized in the Local Universe (e.g., Adamo et al. 2015, 2017, 2020) and hydrodynamical simulations suggest that star-formation continues to follow such behavior along cosmic time. In particular, the star-formation occurring in gravitationally-bound young star clusters is expected to increase with redshift (>10(30)% at z>3(6)), as well as the characteristic cluster mass at the knee of the of the initial cluster mass function (dubbed truncation mass, Pfeffer et al. 2018, Bastian & Lardo 2018). Under such a scenario, it is therefore reasonable to expect that any facility/tool able to increase the angular resolution will eventually unveil such tiny SF regions in which the substantial occurrence of star clusters would prominently emerge at high redshift.

### 4.7.1 MAVIS on non-lensed fields: star forming complexes at 100-200 pc scale

***MAVIS will be a key instrument allowing us to easily recognize the SF clumps at ~200 pc scale in non-lensed fields.*** In particular, the 20 mas FHWM at lambda~500 (800) nm would correspond to 160 (110) pc at z=2 (6) with the rest-fame spectral range of 130-300 nm at z=2 (400-900 nm observed frame) and <130 nm at z=6 (< 900nm observed frame). Barely resolved (or not resolved) SF knots at high-z - after a suitable PSF deconvolution - will allow us to set upper limits on their ultraviolet size down to a single pixel (e.g., Vanzella et al. 2019, 2020). In this case, the single 7.5 mas/pixel expected for MAVIS would probe 40 (60) pc at rest-frame λ ~ 130 (300) nm and at z=6 (2), therefore approaching single star-forming complexes. ***MAVIS will routinely "unpack" high-z galaxies in the ultraviolet, addressing the details of their constituents.*** The interaction between the single giant HII regions made of OB association and star clusters can be addressed with IFU spectroscopy at the level of 100-200 pc scale, by performing two-dimensional maps of the ionization parameter, monitoring the opacity of the interstellar medium, and address the location of the most massive and hot stars, typically segregated in star clusters and their central core.

### 4.7.2 MAVIS on lensed fields: gravitational bounded star clusters at 1-20 pc scale.

A definitive quantum leap will be performed when combining MAVIS with strong lensing (SL) magnification. Currently, HST imaging (30 mas/pix) coupled with SL allows us to identify very nucleated tiny sources at cosmological distance (e.g., Vanzella et al. 2017; Bouwens et al. 2017), probing a few tens pc scale. Most probably, the best examples of young massive star clusters (or even GCP) at such distances have been reported by Vanzella et al. (2019, 2020) at z=2.4 and z=6.14 by placing firm upper limits on the sizes of 20 and 13 pc, respectively. Current studies, however, are at the very beginning and limited by HST angular resolution. MAVIS at 20 mas (7.5 mas/pix) will relax by a factor ~ 4 the required lensing magnification to achieve the same results reported above (allowing to significantly increase the accessible volume, Fig. 4.22). If applied to the above lensed (and super-lensed) targets, MAVIS will probe 1-10 pc scale. The relatively wide field of view of MAVIS will also allow us to capture "clusters of clusters" in a single shot, unveiling the expected clustering signal during formation of such objects (Pozzetti et al. 2019).

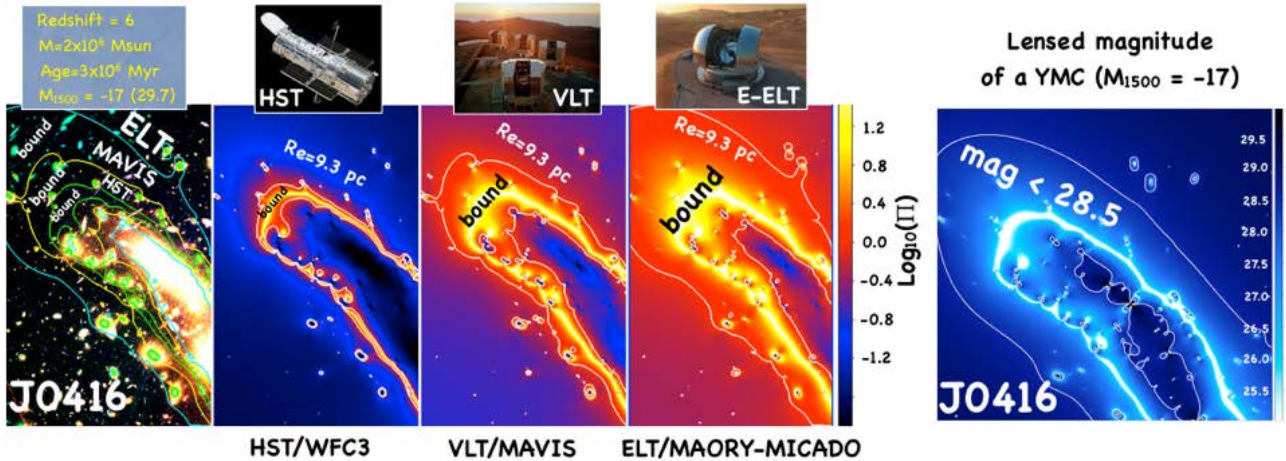

**Figure 4.22:** *Left: Lensed regions marked with contours within which different instruments can probe a 3 Myr old gravitationally-bound star cluster at z=6 with M~2x10⁶ Msun (having radius ~ 10 pc). Right: the color coded lensed (observed) magnitude of the aforementioned star cluster (Muv = -17). The white contour outlines the area with magnitude brighter than 28.5.*

*For the first time, with MAVIS, SF clumps will be unpacked to unprecedented small spatial scales approaching the single star clusters (effective radius, $R_e$<20 pc), eventually recognized as gravitationally-bound systems.* Indeed, following Gieles (2011) for the calculation of the dynamical age, MAVIS can probe 10⁶ Msun star clusters up to z=6, under suitable conditions (see Figs. 4.22 and 4.23). This will be unprecedented. Similar results have been currently obtained with HST in the very rare super-lensed systems, for which the magnification can exceed x 30-50.

MAVIS can target point-like sources with I=28 at SNR~10 in a reasonable amount of time ($t_{exp}$ < 2 h), allowing for the detection of the faint (lensed) YMC population. In addition, MAVIS Low-resolution IFU spectroscopy (R~5000, for sky-sub) AO-assisted is key to perform two dimensional maps of high ionization emission lines of the most magnified gravitationally lensed arcs, looking for the location of the most massive and hot stars: wavelength coverage of IFU as close as possible to 1um, to follow, e.g., CIV 1550 (CIII 1909) up to z=5(4), and Ly-alpha up to z=7. Typical line fluxes down to 5x10⁻¹⁸ cgs can be reached with MAVIS with SNR>5.

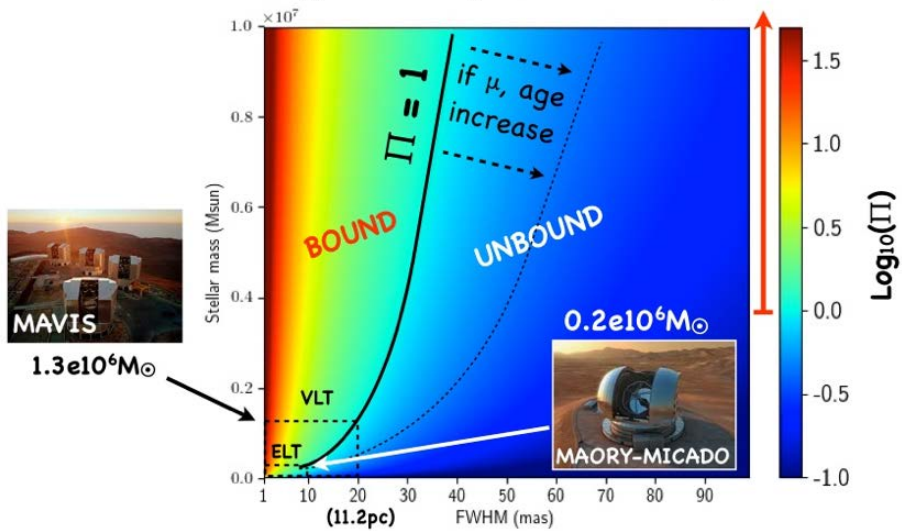

**Figure 4.23:** *Region in which gravitationally-bound (dynamical age, ⊓ > 1, color-coded) YMCs can be probed as a function of their stellar mass and AO capabilities (FWHM, mas). Bound and unbound regions are marked for a 5 Myr old star cluster at z=6 and magnification = 10.*

An example (at the moment the only one) of a gravitationally bound star cluster identified at cosmological distance has been presented by Vanzella et al. (2020) studying the Sunburst arc (Rivera-Thorsen et al. 2019), shown in Fig. 4.24. The tangential stretch provided by the strong lensing amplification allowed to constrain the size of a young (3 Myr old, and a few $10^6$ $M_\odot$) SF-knot of magnitude R~22 down to 20 pc size. Interestingly, such knot also shows a leakage of ionizing radiation (Lyman continuum, LyC) into the IGM, that is a relevant feature when studying the analog populations of reionizing sources at much higher-z (where the LyC cannot be observed directly). The detection of LyC photons is fully consistent with the observed spectral features associated with the presence of hot and massive stars (including WR stars) dominating the ionizing radiation field. Targeting such super-lensed objects (dozens of them are already identified and future wide field survey will increase dramatically the statistics) with MAVIS will allow us to reach the pc scale, as it will be the case for the Sunburst complex: 2.5 (6.6) pc with 7.5 (20) mas (see Fig. 4.24). For the first time this will also allow us to make a census of the foming star clusters present in such a galaxy, and therefore compute the cluster formation efficiency, a quantity currently computed only for SF galaxies in the local Universe. MAVIS, complemented by ELT (e.g., MAORY-MICADO), can, for the first time, explore such a quantity along the cosmic epochs. Concluding, MAVIS imaging will unpack routinely high-z galaxies down to 100-200 pc scale in non-lensed fields, approaching the single most massive star clusters. In lensed fields the single star clusters will be easily identified as gravitationally-bound objects (down to $10^5$ $M_\odot$) in the range 2<z<7. In super-lensed fields the 2-4 pc scale will be probed. The interplay between the most prominent engines in terms of SFR-density, ionizing photon producers, and feedback (e.g. Bik et al. 2015), affecting the opacity/geometry of the ISM of their hosting galaxies will be the key argument for galaxy evolution and for exploring (by analogy) if such systems played a dominant role during reionization.

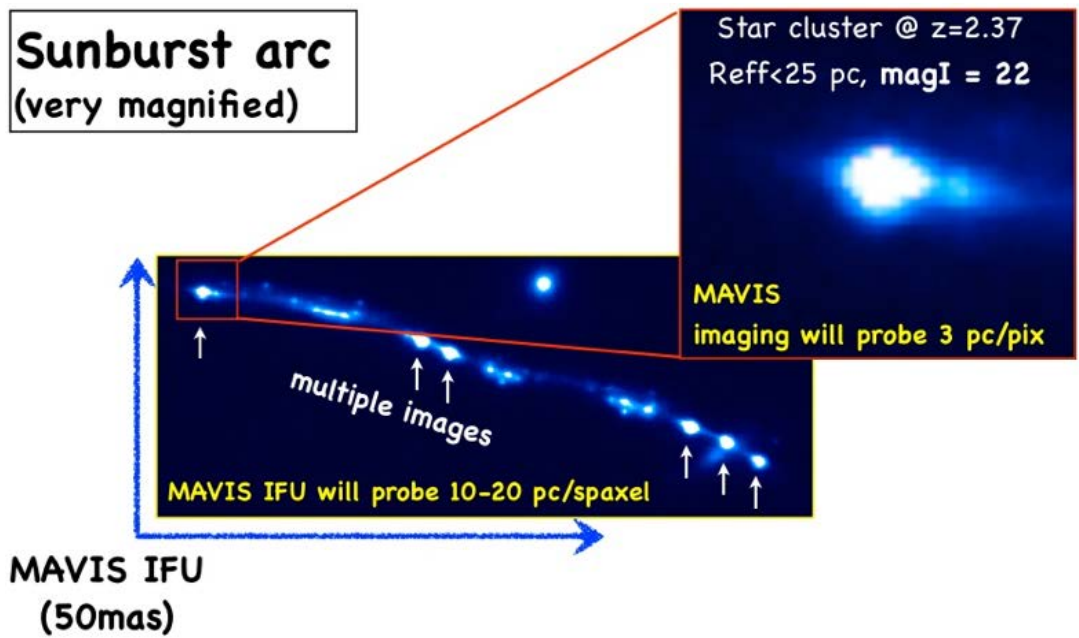

**Figure 4.24:** *An example of a superlensed system (Rivera-Thorsen et al. 2019; Vanzella et al. 2020) with magnitude I=22 representing currently the first cosmological YMC, for which MAVIS will probe 3 pc/pix with imaging and 10-20 pc/spaxel with IFU spectroscopy. Only MAVIS with its enhanced sky coverage can target such superlensed systems with a fully operational AO. One hour exposure on the I=22 star cluster (now barely resolved with HST) will allow us to sample its light profile (~3 pc) and perform ionization maps of the arc down to 10-20 pc.*

# 5 The Birth, Life, and Death of Stars and Their Planets

## 5.1 Introduction

Star formation, stellar evolution, and the formation and evolution of planets, are important building blocks in the framework of the Universe. In the following sections, we present a few of the many cases where MAVIS will shed light on the various aspects of the life and death of stars that are not clarified yet.

## 5.2 Young Stars

**Science goal:** Investigating the evolution of proto-planetary disks in young stars and how this depends on star parameters such as metallicity, on the presence of jets, and on (substellar) companions.

**Program details:** Spectroscopy and/or multi-band imaging (for fainter targets) of young stellar objects in star-forming regions in the solar neighborhood (close-by T Tauri and Herbig AeBe objects will be targeted for jets), and of young clusters with low metallicity in the outer parts of our Galaxy and in the Magellanic Clouds.

**Key observation requirements:** Medium/high resolution spectroscopy to better characterize objects, measure accretion, and optimize stellar continuum subtraction. Large field of view to improve observation efficiency.

**Uniqueness of MAVIS:** Sky coverage (laser guide stars allow us to observe many targets too faint for SCAO instruments) and large field of view. Unique combination of integral field spectroscopy, sensitivity, medium/high spectral resolution, and high angular resolution.

During their pre-Main Sequence evolution all low-mass stars show evidence of mass accretion from a disk, which in a few million years dissipates forming a planetary system (Hartmann et al. 1998, Meyer et al. 2007). Disk properties show huge and still not well understood variations both for stars of similar mass and over the mass spectrum. These results indicate how we are still missing in our picture crucial ingredients that regulate the disk physics and evolution and its possible connection to the properties of the environment. This is particularly important for understanding the formation of planetary (and more in general sub-stellar) companions, which occurs within the disks. MAVIS will offer not only a unique combination of high-angular resolution, spectroscopic capabilities, and high sensitivity, but will also provide a huge improvement in sky coverage, with much less stringent constraints on the brightness of natural guide stars for the employ of AO correction. This is a strong advantage for AO observations of young stellar objects, which are typically located in regions with high values of extinction. This means that MAVIS will provide for the first time AO-assisted observations on large areas of the star-forming regions that are still unexplored up to now in terms of high-angular resolution investigations. These include the accretion process and disk dispersal in low-metallicity environments, the ejection of jets and winds and their role in removing angular momentum from the star-disk system, and the formation channels of sub-stellar companions.

### 5.2.1 Star and planet formation in low-metallicity environments

During the last decades, extensive and detailed multi-band surveys of star-forming regions located in the solar neighborhood (distance < 1 kpc) (see Reipurth et al. 2008ab for a review of the literature on Galactic star forming regions) have been carried out with the aim of characterizing the stellar population in young clusters to put constraints on the accretion process and on the mechanisms driving disk dispersal and planet formation.

However, all the nearby young clusters have metallicities close to the solar one (Spina et al. 2017), so these observations cannot provide information on the star formation process in low-metallicity environments. Moreover, observations of nearby galaxies show that many stars form in very massive and dense star clusters ($M \sim 10^4$-$10^5 M_\odot$, stellar density $> 10^3$ stars pc$^{-3}$), which are not present in the solar neighborhood. The properties of disks and their evolution through accretion may strongly depend on the cluster mass and density, since the ultra-violet radiation produced by massive OB stars may induce disk photoevaporation and two-body dynamical interactions can truncate disks (Johnstone et al. 1998, Rosotti et al. 2014).

Observations of distant young clusters in the outer Galaxy have provided the first evidence that most stars forming in a low-metallicity environment experience disk dispersal at an earlier stage (in less than about 1 Myr, see Fig. 5.1 left) than those forming in solar metallicity environments, where disk dispersal may last up to 5-6 Myr (e.g. Fedele et al. 2010). This would indicate that metallicity plays a pivotal role on the time-scale available for planet formation and on the (debated) dependence of exoplanet frequency around metal-rich stars. Further support to this scenario has been provided by studies conducted in low-metallicity star-forming clouds of the Magellanic Clouds with HST (e.g. De Marchi et al. 2010, 2013, 2017, and references therein), which suggest that metal-poor stars accrete at higher rates than solar-metallicity stars in nearby Galactic star-forming regions (Fig. 5.1 right). These results require however further validation as they are are based on photometric data only, which may be subject to larger systematic errors than spectroscopic data; for instance, photometric ages derived via color-magnitude diagrams may be affected by the emission produced by the accretion shock or the disk and by differential extinction and mass accretion rates derived from narrow-band images (e.g. Hα flux) may be contaminated by unresolved emission from the interstellar gas or the chromosphere.

On this basis, MAVIS can provide a major step forward in the study of distant and massive young clusters in low-metallicity environments by combining high-angular-resolution, which is required to limit the problem of source crowding, and sensitivity. The most suitable targets for this project are young clusters in the outer Galaxy (at d~5 kpc), which show typical metallicities down to about 0.2 $Z_\odot$ (such targets can be selected from the list of HII regions in the outer Galaxy (Sharpless 1959), for instance Sh 2-284 in Dolidze 25) and the Magellanic Clouds, which display even lower metallicities (Z~0.009 the Large Magellanic Cloud, Z~0.0005 the Small Magellanic Cloud) and are populated by several clusters younger than 100 Myr with mass larger than $10^4$ $M_\odot$ (MacKey & Gilmore 2003ab, Portegies-Zwart et al. 2010).

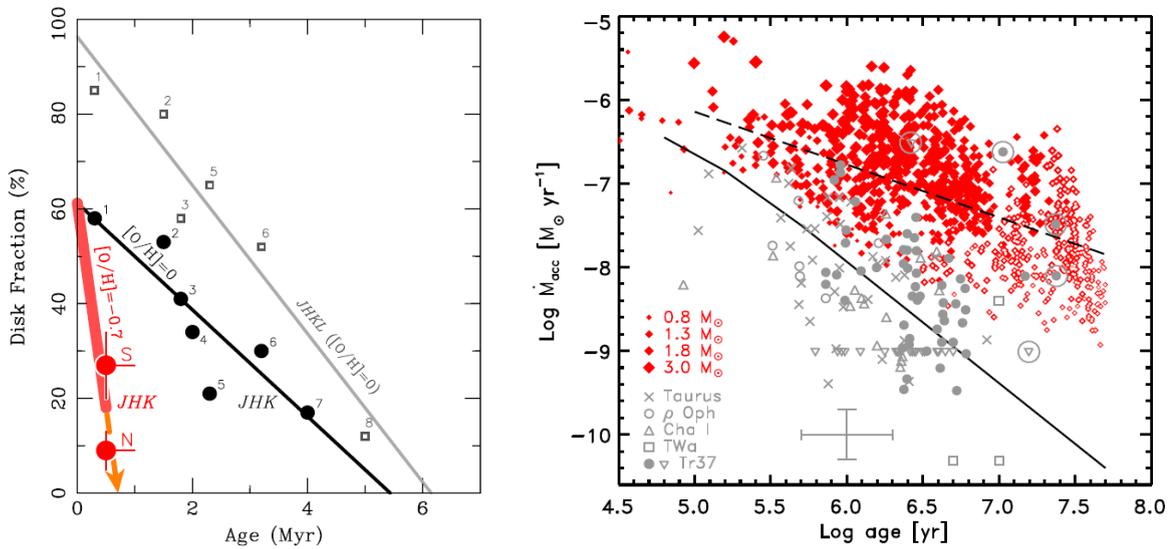

***Figure 5.1:*** *(Left) disk fraction measured through JHK images in the low-metallicity Digel Cloud 2 clusters (red) is plotted as a function of cluster age and compared to JHK and JHKL disk fraction measured in other young embedded clusters with solar metallicity (black and grey). The black and gray lines (least-squares fits) indicate the disk fraction evolution under the solar metallicity, while the red line shows the proposed disk fraction evolution in the low-metallicity environment. Such a rapid disk evolution suggests that metal-poor stars accrete at higher rates than solar-metallicity stars. Adapted from Yasui et al. (2009). (Right) Mass accretion rates ($\dot{M}_{acc}$) estimated for metal-poor pre-main sequence stars of 30 Dor in the Large Magellanic Cloud (red points), as a function of stellar age. Grey symbols refer to various solar-metallicity Galactic T-Tauri stars taken from the literature. The solid line shows the relationship between $\dot{M}_{acc}$ and time predicted by models of the evolution of a viscous disk (Hartmann et al. 1998), whereas the dashed line represents the best fit to the 30 Dor data. Adapted from De Marchi et al. (2017).*

A MAVIS survey of these clusters can be carried out in imaging mode, with observations in wide-band filters (e.g. B, V, R, I), which will be used to infer the source colors and characterize the single objects. Images taken with narrow-band filters such as Hα and CaII can be used to measure the mass accretion rate (Alcalá et al. 2017), even for stars too faint to be observed with the spectrographic mode. However, the availability of a spectroscopic mode with medium resolution (R~5000-10000), as the one proposed for the MAVIS IFU, would ensure a more reliable characterization of the sources (even those actively accreting), with the possibility to derive temperatures and luminosities by fitting the spectrum with a model that takes into account the photospheric emission, the extinction, and the emission of a slab of hydrogen heated by the accretion shock (e.g. Alcalá et al. 2014). This method will also allow us to derive accretion luminosity and rates. Furthermore, the spectroscopic data would enable us to derive a simultaneous estimate of the metallicity directly from the data and would provide a more efficient continuum removal for the use of line tracers. The fluxes of Hα and CaII lines, plus additional emission tracers such as other Balmer and HeI lines, can be used to derive a further estimate of the accretion luminosity via the empirical relationships calibrated on accreting T Tauri stars (e.g. Alcalá et al. 2017).

Finally, the increased sensitivity of MAVIS with respect to previous observations will allow us to probe stars in these clusters with masses well below the current limit of ~1M$_\odot$, to investigate the possible correlation between metallicity and the Initial Mass Function at low masses, where turbulent fragmentation, which depends on the cooling rate and thus on metal abundance, dominates (Kroupa 2001).

## 5.2.2 Formation of jets from young stars and angular momentum problem

A fundamental requirement for the formation of a new star is that angular momentum is removed from the circumstellar disk. Since collisional particle viscosity alone is too low to efficiently transport angular momentum from the inner to the outer disk (Turner et al. 2014), and turbulence due to magneto-rotational instabilities appears to be suppressed when realistic disk conditions are considered (Bai & Stone 2013), jets may represent a valid solution to the angular momentum problem in young stellar objects, via its vertical transport with the jet material. This is understood through the action of a large-scale magnetic field, anchored in the rotating star-disk system, along which the jets are launched (e.g. Blandford & Payne 1982, Ferreira 2013). The currently favoured model is the "disk-wind", accelerated from the inner few astronomical units of the disks (Pudritz et al. 2007; Frank et al. 2014). If this mechanism is confirmed, it would imply that the long-standing angular momentum problem can be solved, and also that disks are strongly magnetized, with important implications for planetary formation (Turner et al. 2014, Frank et al. 2014, Baruteau et al. 2014).

In this context, observing the very base of the jets is therefore a crucial aspect for assessing their magneto-centrifugal formation mechanism and determining the region where they originate and thus their feedback on the disk. Previous high-angular resolution observations with HST and AO instruments have shown that jets reach quasi cylindrical collimation with a narrow width (<4 au) on scales <30 au from the source (see e.g. compilation in Ray et al. 2007), but being limited to typical angular resolutions of 0".1-0".2 have not yet been able to resolve the collimation scales, nor the acceleration scales of the jets. To investigate the morphology and kinematics of the jet launch zone we need to probe the jet at distances lower than 5-10 au from the driving source, which correspond to angular separations smaller than 35-70 mas for YSOs in close-by star-forming clouds (e.g. Taurus at ~140 pc). With its combination of high spatial resolution and spectroscopic capabilities MAVIS can provide unprecedented observations of the base of the jets to directly unveil their onset mechanism.

Stellar jets (Fig. 5.2) emit mostly in atomic permitted and forbidden lines, with almost no continuum (Maurri et al. 2014). Relevant optical lines that can be used to map the jets are Hα, [OII]λλ3726,3729, [OI]λ6300, [OI]λ5577, [OIII]λ5007, [SII]λλ6713,6731, [SII]λ4076, [SII]λ4069, [NI]λ5198, [NII]λ5755, [NII]λ6583. The proposed MAVIS IFU module will allow us not only to observe all these tracers, but also to obtain an optimal subtraction of the star continuum light, which is the main problem for imaging the base of the jet because of the high contrast between the jet emission lines and the stellar continuum, which can be as high as $10^3$-$10^4$. The low resolution mode of MAVIS around R=5000 is sufficient to separate the different emission lines, including the lines in the [OII] and/or [SII] doublets, which is necessary to derive the electron density. The IFU at this resolving power will also produce images of the jet in separate velocity channels for each line, disentangling the different velocity components, e.g. imaging 4 to 6 velocity intervals within the same line, as in jets the line profiles are broad (200-400 km/s). This will allow us to obtain valuable information concerning the kinematic of the outflow and provide an important test of the magneto-centrifugal model, according to which the flow has an onion-like kinematic structure, with progressively slower components launched from regions in the disk at larger distance from the star. This would give in turn clues to the structure of the disk.

Finally, the combination of the observed lines offers the possibility to derive the physical conditions of the gas (ionization, electron density, temperature) from the maps of the line ratios, using dedicated and well tested spectral diagnostic techniques (Maurri et al. 2014, Bacciotti, F. & Eislöffel 1998, Giannini et al. 2019). The main goal here is to derive the total density in the flow, which combined with the determined jet poloidal velocity gives the mass outflow rate, the fundamental quantity for the jet dynamics.

A direct measurement of the rotation of the jet, which in combination with the mass outflow rate would provide the angular momentum transported by the flow, can be obtained with MAVIS provided that high enough spectral resolution is available. The expected difference in velocity between the two sides of the rotating jet is however below 15 km/s, so a spectral resolution of at least R=15000 would be in principle required to investigate this crucial parameter. Hints of the rotation of the jets have been obtained using HST/STIS (e.g. Bacciotti et al. 2002, Coffey et al. 2004, Coffey et al. 2012; Fig. 5.3). Even though the limited spectral resolution of STIS (R~6000 in the optical) does not allow one to determine directly the velocity shifts expected for rotation, a proper Gaussian fit or cross-correlation of the line profiles at opposed positions can reveal a velocity gradient.

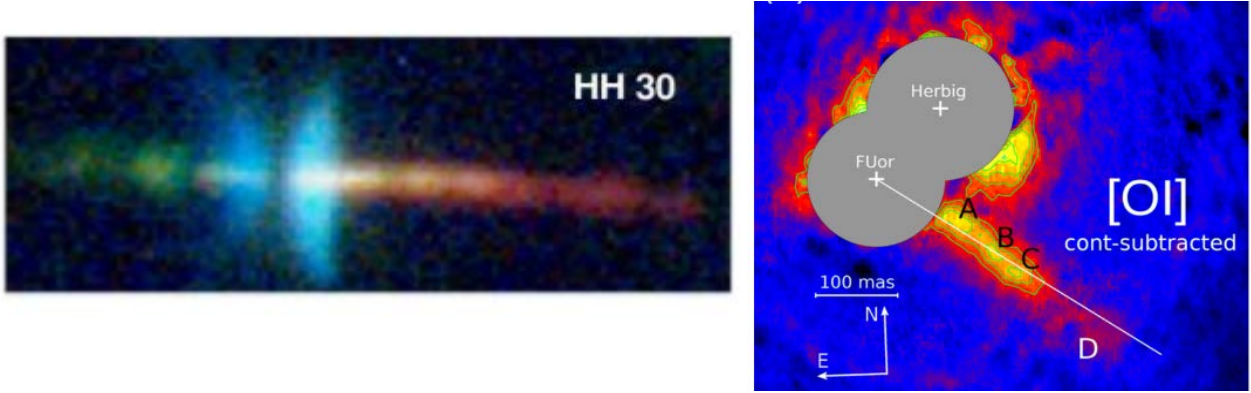

**Figure 5.2:** *Left: optical image of the HH30 disk and stellar jet taken with HST/WFPC2 (adapted from Ray et al. 1996). The two cusp-shaped blue nebulae are caused by scattered light from the star, while the dark lane between them is the disk. Right: VLT/SPHERE [OI]λ6300 image of the jet from the FUor component of ZCMa with ZIMPOL (Antoniucci et al. 2016), at 30 mas angular resolution.*

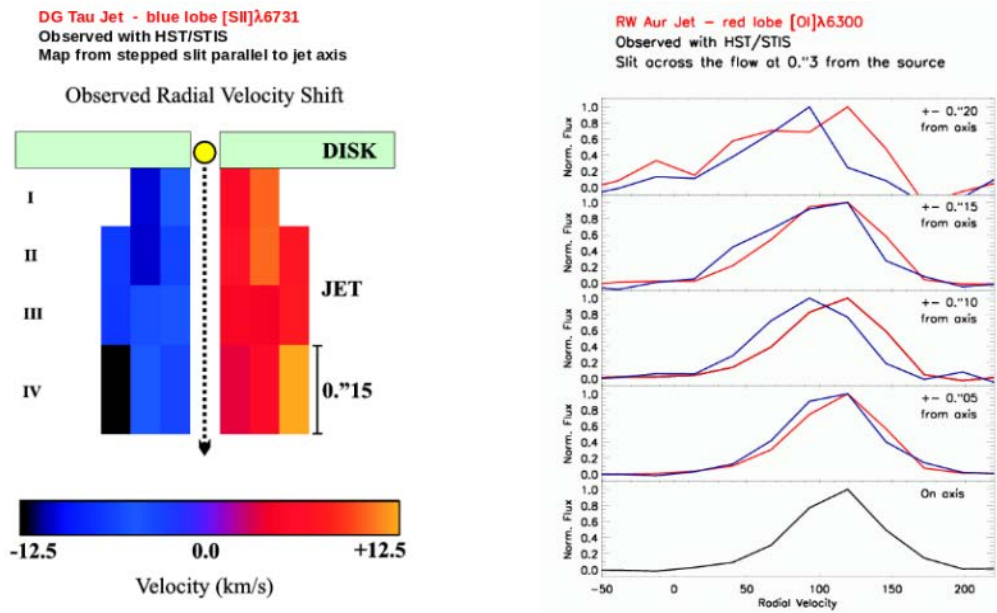

**Figure 5.3:** *HST/STIS detection of rotation signatures. Left: map of the observed radial velocity shift in the DG Tau jet, indicating rotation around the axis [34]. Right: cuts of the spectral line map in the RW Aur jet at positions opposed to the central axis, for larger distances from the axis, going from bottom to top (Coffey et al. 2004).*

The low resolution mode of MAVIS, with a similar spectral resolution as HST/STIS (but combined with higher spatial resolution) can enable the same kind of analysis on line tracers detected with high SNR. However, a more substantial improvement with respect to HST would definitely be obtained with the higher resolution mode proposed for the blue arm (R~15000), at least for jets that show emission from tracers covered by this setting (e.g. [OIII]λ5007, [SII]λ4076, [SII]λ4069, or [NI]λ5198).

The best targets for the proposed investigation of jets are YSOs in near-by star-forming regions (Taurus, Lupus), which are relatively bright (V>10, H=8-14) and show typical flux of line emission tracers of the order $10^{-14}$ -$10^{-15}$ erg/s/cm$^2$.

## 5.2.3 Formation of sub-stellar companions

Sub-stellar companions (SSCs) orbiting solar-mass young stars are interesting objects not only due to the issue concerning their formation mechanism, but also because they can effectively influence the lifetime and properties of protoplanetary disks. Despite the increasing number of SSCs detected in the last few years, especially at large separations (order of 100 au from their hosts, e.g. Deacon et al. 2016, Lafrenière et al. 2008, Naud et al. 2014, Schmidt et al. 2008, Zhou et al. 2014), the formation channel of these objects remains not well established. They could form within the disk of their primary, through mechanisms such as core-accretion (e.g. Pollack et al. 1996) or disk-instabilities (e.g. Boss 2011). Alternatively, SSCs might form via cloud fragmentation and collapse (as expected for isolated brown dwarfs) and might just represent the low-mass end of sources in multiple systems. It is plausible that these mechanisms might be at work simultaneously, with different efficiency at different separations from the primary.

Models generally predict different time-scales and rates for the formation of SSCs (e.g. Bowler et al. 2011, Stamatellos & Herczeg 2015), so that measuring their mass accretion rate can be used to discriminate between the proposed formation channels and investigate their relative efficiency.

With its high spatial resolution and sensitivity MAVIS is the perfect instrument to perform a multiplicity survey in near-by (0.2-1 kpc) young clusters to characterize accreting sub-stellar companions of solar-mass stars. MAVIS will: 1) make a statistical analysis of the frequency of sub-stellar companions, covering a range of separations that have been poorly investigated so far, i.e. from a few au to about 40 au; 2) derive the mass accretion rates ($M_{acc}$) of the identified companions using accretion tracers like Hα, to determine whether their formation is consistent with direct fragmentation from collapsing molecular cores, as in isolated brown dwarfs, or it is more compatible with the formation channels proposed for giant planets (i.e. core-accretion or disk-instability models); 3) combine the results of this analysis with highly complementary JWST infrared data, which will provide information on the disks around the primary stars, so as to statistically analyze the impact of SSCs on the disk occurrence and lifetime (see Cheetham et al. 2015) and analyze the relationship between their accretion rate and disk properties.

Based on the sensitivity estimates for MAVIS, we expect to be able to observe Hα fluxes as low as $3\times10^{-17}$ erg/s/cm$^2$ at S/N~5 in 1h of observation. For targets at 500 pc, these fluxes would correspond to accretion luminosities values ($L_{acc}$) down to $2\times10^{-6}$ L$_\odot$, if we adopt the relationship that connects Hα luminosity and $L_{acc}$ in T Tauri stars (Alcalá et al. 2017). These values of the accretion luminosity on brown dwarf companions would indicate accretion rates of the order of $10^{-13}$-$10^{-14}$ M$_\odot$/yr ($M_{acc}$~$L_{acc}$R$_*$/GM$_*$). Suitable clusters for the proposed analysis would be for instance the Orion Nebula Cluster (d=440 pc), the σ Orionis cloud (d=360 pc), the ρ Ophiuchi star-forming region (d=130 pc), NGC 2264 (d=750 pc), and the clusters in Vela-D (d=700 pc).

As for the frequency of SSCs, in a recent work about multiplicity in the ρ Oph cloud, Cheetham et al. (2015) observed a value of $7_{-5}^{+8}$ % for brown dwarf companions (13-80 M$_J$) in a range of separations between 1.3 and 780 au. Albeit still largely incomplete, these studies suggest that typical values between 5% and 10% for the frequency of sub-stellar companions in young star-forming clouds can be expected. A survey with MAVIS will definitely allow us to assess these numbers, but more importantly will reveal what fraction of the detected SSCs are actively accreting, which is fundamental for understanding their formation mechanism. The survey will also provide accurate measurements of the global binary fraction in the investigated star-forming clouds, i.e. not only limited to the sub-stellar mass regime. This information would be used to study the universality of the stellar multiplicity (see, e.g., Duchêne et al. 2018 and references therein).

While the survey can be carried out in imaging-only mode (using wide-band filters to characterize the objects and Hα observations to derive the accretion rate of the targets), with the employ of the IFU module, even at low spectral resolutions (R~5000), we will be able to obtain a much more reliable characterization of the identified companions through comparison with template models. Moreover, the IFU would allow us to use multiple emission tracers, such as Balmer, HeI, and CaII lines, to derive the accretion rates (adopting the empirical relationships based on accreting T Tauri stars, e.g. Alcalá et al. 2017), while ensuring also a more efficient continuum removal to improve the contrast and obtain a better flux estimate.

## 5.3 Evolved Stars

**Science goal:** Characterisation of the nature, evolution, and circumstellar environment of the precursors of white dwarf merger systems.

**Program details:** AO-assisted imaging with photometry and spectroscopy of evolved binaries with a second-generation proto-planetary disk, and white dwarf binaries.

**Key observation requirements:** AO-assisted imaging with a PSF of ~20-25 mas in V and high spatial resolution. High- and low-resolution spectroscopy (R=10000-12000, and R=5000)

**Uniqueness of MAVIS:** The high contrast imaging capabilities that allow to resolve second-generation proto-planetary disks such as those commonly seen around post-AGB binaries. The deep imaging limit, e.g., with respect to SPHERE, will allow for the study of a large sample of these faint objects. The high-spectral resolution,  e.g., with respect to MUSE,  allows a detailed chemical characterisation of the evolved binary systems.

The late stages of evolution for low- to intermediate-mass single stars is a rapid transition from the Asymptotic Giant Branch (AGB) phase through the transient post-AGB phase and towards the Planetary Nebula Phase (PNe) before the stellar remnant cools down as a White Dwarf (WD). Although this evolution scheme seems straight forward, there is little understanding from first principles of the different important physical processes that govern these evolutionary phases. More importantly, the evolution of any star, especially during and after the giant phases (Red Giant Branch [RGB] phase and AGB phase), can be strongly affected if the star is in a binary system. Binary interaction alters the intrinsic properties (such as: chemical composition, pulsation, mass-loss, dust-formation, circumstellar envelope morphology etc.) of the star and plays a dominant role in determining its ultimate fate. A plethora of peculiar objects ranging from the spectacular thermonuclear novae, supernovae type Ia, sub-luminous supernovae, gravitational wave sources, etc. to less energetic systems such as subdwarf B stars, barium stars, bipolar planetary nebulae, etc., result from mass transfer in binary stars. Therefore, an in-depth study on poorly understood processes that govern the late stages of stellar evolution, including binary interaction processes, is essential to constrain stellar evolution as well as the chemical evolution of the Universe. In the following sections, two compelling science cases that can be addressed with MAVIS are illustrated.

### 5.3.1 Protoplanetary disks around evolved stars

At the end of their asymptotic giant branch (AGB) phases stars lose their envelope in the form of a shell. Interestingly, post- Asymptotic Giant Branch (post-AGB) binaries display disk-like infrared excess in their spectral energy distribution (SED; see Fig 5.1, e.g. De Ruyter et al. 2006). This indicates that, because of poorly understood dynamical interactions with the inner binary, part of the expelled matter from the evolved star forms a circumbinary disk of dust and gas. Keplerian circumstellar disks are commonly detected around post-AGB binaries with binary orbital time scales of the order of 1 yr to several years (Van Winckel. 2003, De Ruyter et al. 2006, Van Winckel et al. 2009, Bujarrabal et al. 2017, Kamath et al. 2014, Kamath et al. 2015). A long life time of the disks is supported by the high degree of dust processing (grain growth, crystallization; Gielen et al. 2011, Hillen et al. 2015). The evolution of the (binary) star and the formation and evolution of the disk are closely coupled. The primary's AGB phase was abruptly terminated due to mass loss induced by a poorly understood binary interaction process. Part of the ejected material was forced into a circumbinary disk. As the post-AGB star evolves into a hot stellar remnant, the disk disperses and the ejected gas may become ionized and shine as a planetary nebula (PN). Interactions between the circumbinary disk and the binary can be strong and are not well understood. For example, the observed binary periods in these systems do not match the theoretically predicted values (Nie et al. 2012, Manick et al. 2017). While the presence of stable dusty circumbinary disks in these systems are now well established, their formation, structure and evolution are not understood. The disks, however, do appear to be a key ingredient in our understanding of the late evolution of a very significant binary population and they may play even a crucial role in the (orbital) evolution of the central stars.

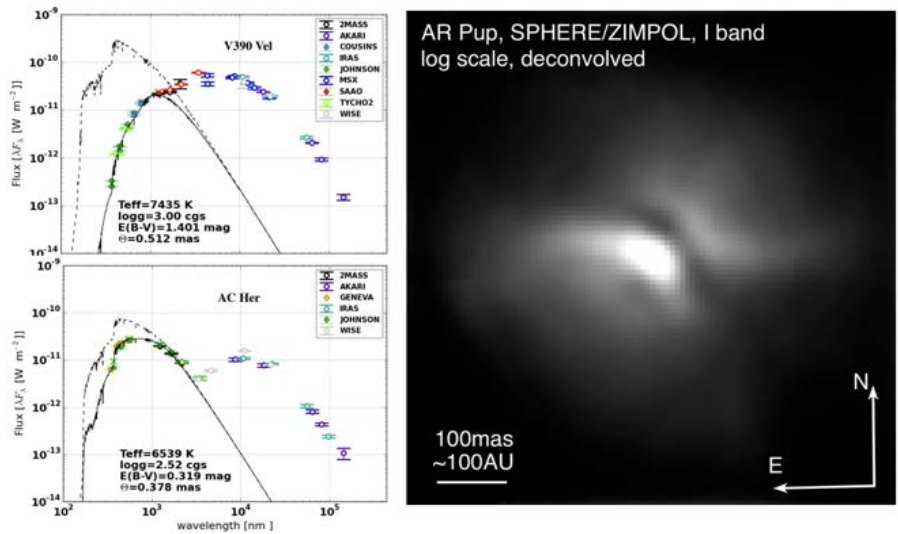

**Figure 5.4:** *(Left panel) Well sampled SEDs of two objects in our target list. Top-left: V390 Vel, showing the strongly obscured star (solid line vs. dashed line for the expected stellar spectrum) and the infrared excess. Bottom-left: AC Her, where the inner rim is at significantly larger distance from the central source than the sublimation radius. (Right Panel) VLTI/SPHERE image of AR Pup in the I band which shows a clearly resolved disk (Ertel et al., 2019). This is the only image of a pAGB disk other than the Red Rectangle so far.*

Using interferometry in combination with radiative transfer (RT) models, we found strong, extended scattered light components in our best studied, face-on systems 89Her and V390Vel; (e.g., Hillen et al. 2015). This cannot be explained by scattering on the disk surface alone. It may be interpreted as scattering on dust that is likely present in the narrow-angle Hα outflows that are detected with phase-resolved spectroscopy in several systems (e.g., Gorlova et al. 2015). Outflows are observed originating from the secondary star (Bollen et al. 2017) and from the disk (Bujarrabal et al. 2017). The outflow can originate either from stellar mass loss or from disk dispersal, However, the origin of the outflows, and the launching mechanism remains unknown. Thus, locating its origin and determining its nature constrains the evolutionary time scale of the system. In face-on systems, an outflow that is vertical to the disk midplane is seen face-on and blends with the disk. Thus, extracting parameters of either component is strongly degenerate and possibly biased. It requires an edge-on system to unambiguously separate the outflow from the disk and to study both components independently. An additional advantage of an edge-on system is that the bright stellar light is blocked by the disk, so that high contrast is not needed to directly image its close environment (but extreme angular resolution such as that offered by MAVIS is required).

Binarity of central stars of PNe seems the most promising scenario to break the spherical symmetry of AGB envelopes (De Marco, O. 2009). Common envelope evolution can only explain a small fraction of these systems (~15%; Miszalski et al. 2009). Only recently, binaries with periods of the order of years have been discovered in PNe (Van Winckel et al. 2014). This makes systems like AR Pup prime suspects. Their study is thus critical for the understanding of PN formation.

The Red Rectangle is the first post-AGB disk to be spatially resolved due to its small distance (710 pc) and peculiar bright extended red emission (Bujarrabal et al. 2003). The more nearby disk around the AGB star L2 Pup, imaged with SPHERE (Kervella et al. 2015) shows that L2 Pup possibly has a sub-stellar (~1 Jupiter mass) companion, therefore different to the post-AGB systems with disks. The inner rim of the disk around IRAS08544-4431, a pAGB binary system, was resolved in the H-band using interferometric imaging using PIONIER (Hillen et al. 2017). Attempts to image other post-AGB disks, in the optical, with Hubble remained futile due to the limited angular resolution. In our recent study, using SPHERE/VLT, we were able to resolve the edge-on disk, in the optical, AR Pup, another post-AGB binary system (see Fig. 5.4; Ertel et al., 2019). These studies show a large diversity of systems in terms of disk size and evolutionary state of the primary star (post-Red Giant Branch, AGB or post-AGB). Imaging several post-AGB disk systems is a crucial step to a more general picture of their properties and evolution. Additionally, disks around post-AGB binaries are considered to be similar to protoplanetary disks around PMS stars. For instance, the infrared excess and Keplerian disk rotation were first observed around protoplanetary disks around PMS stars. However, these are not the only properties disks around post-AGB are sharing with protoplanetary disks. As in protoplanetary disks, post-AGB disks show signs of dust grain growth and crystallisation (Gielen et al. 2011). However, the extent of the similarities (and/or differences) of these disks is yet to be studied in detail. This will provide a gateway to studying the similarities and differences in processes such as the launching of jets and outflows, observed both in evolved binaries and young stellar objects.

MAVIS will allow for the first comprehensive – yet inexpensive – high angular resolution multi-wavelength imaging survey of post-AGB wide binaries with circumstellar disks. The inner parts of the disk where most scattered light is expected, will be measured from the high-resolution images in the U, B, V, R, and I bands. Potential outflow will be identified as an extended structure perpendicular to the disk mid-plane in case of edge-on disks. So far, no studies have been made with high-resolution images in the U, B bands. MAVIS will allow for investigating the flux contribution in the U, B bands from the disk- central star system. Spectra covering the entire optical and near-IR wavelength regime will be useful to characterise the stellar parameters of the evolved binaries; required for modeling the MAVIS observations with state-of-the-art circumbinary disk radiative transfer model MCMax (2D or 3D if needed; Min et al. 2009). A more comprehensive picture of the disk will be drawn by detailed radiative transfer modeling of the system's SED. For that, the spectroscopically derived stellar parameters, and the radial extent and dust properties will give critical input to break degeneracies (e.g., dust mass and optical depth vs. grain size and radial extent).

With this study we will resolve an ideal, representative subsample of disks in order to study the inner structures and their relation to the different SED types. Thanks to the multi-wavelength imaging and spectroscopic data, we will constrain the dust grain properties. The grain size in an outflow will determine its origin. Very small grains (<1 micron) form in a stellar wind. Larger grains (1 micron to cm-sized) cannot form in the immediate vicinity of the star and must originate from the disk. We will be able to resolve several structures: the stable Keplerian disc, the potential disc wind and eventually the bipolar outflow.

Our ultimate aim is to make the first inventory of the inner structures of these second generation of stable discs to understand the structure and evolution of these discs and their role in the evolution of the binary system. MAVIS will be the only high-order AO system, on an 8-m telescope, in the Southern hemisphere that can reach our required angular resolution (~25 mas in V) at visible wavelengths (covering the UBVRI bands). While VLT/SPHERE provides (PSF ~20mas in V) with ZIMPOL, the sky coverage limits us to very bright targets (V mag < ~12). This only allows less than ~5% of the entire post-AGB population (in our Galaxy, LMC, and SMC) to be studied. The majority of the evolved binary systems targeted in this study are fainter than V mag ~12. By allowing targets with V mag >12 to be resolved, MAVIS provides the unique opportunity to study a statistically large and comprehensive sample of evolved systems. This is currently not possible using any other instrument/telescope. This is crucial as there exists a large diversity of systems in terms of disk size and evolutionary state of the primary star (post-Red Giant Branch, AGB or post-AGB). Imaging several post-AGB disk systems, as proposed in this study, will produce an inventory of the structural components of pAGB binaries with disks. This is a crucial step to arrive at a more systematic picture of their properties and evolution.

## 5.3.2 White dwarf binaries

White dwarfs (WDs), the end state of 95% of all stars, and the current state of most of the stars ever formed with initial mass M > 1.2 Msun, are important and useful in a multitude of applications: as tracers of Galactic structure and star-formation history, as sources of gravitational waves, as progenitors of type Ia supernovae (SNeIa), and as fossil probes of the planetary systems around stars. However, only a tiny and incomplete fraction of the local (< 1 kpc) population of WDs is known at present. Most of the ~30,000 known WDs have been discovered by the Sloan Survey, as an accidental by-product of observations targeted at color-selected quasar candidates.

This situation is changing, as we speak, thanks to Gaia. Gaia parallaxes (i.e. distances) permit reliable identification of WDs by means of their location on the HR diagram. From the recent (April 2018) Gaia Data Release 2, Gentile Fusillo et al. (2018) have identified, out of 1.7 billion DR2 sources, nearly half a million WD candidates in the 8-21 mag range in the Gaia G band (~4000-9000 Å) out to a distance of ~1 kpc , ~260,000 of them high- confidence candidates, and ~17,000 WDs within 100 pc, where the census is essentially volume-limited (i.e. it includes all WDs, including the oldest and coolest ones). By mission end (~2020), these numbers could grow by ~50%, as Gaia scans the sky uniformly and statistical parallax errors decrease. Gaia is thus a superb WD discovery machine. However, the low-resolution (R = 100) Gaia spectra will not be sufficient in most cases to determine the WD atmospheric composition (H or He) by means of the gravity/pressure- sensitive Balmer and He lines, preventing also accurate determinations of WD mass, radius, and cooling age, parameters essential, e.g., for deriving the WD luminosity function and the local star-formation history.

Low-resolution (R=2,000-5,000) spectroscopy of the Gaia WD sample is therefore planned by several upcoming large spectroscopic surveys (WEAVE, DESI, 4MOST), but most notably SDSS-V (2020-2025, the fifth Sloan incarnation, Kollmeier et al. 2017), which will obtain spectra for a significant fraction of all Gaia WD candidates. Furthermore, SDSS-V will observe each WD on multiple epochs, permitting the detection, via radial-velocity variations of the photo-primary WD, of short-orbit WD binaries with component separations of up to a few $R_\odot$ (periods up to about 40 hr, RV changes above ~250 km/s, Badenes & Maoz 2012).

Indeed, studies of current samples of WDs suggest that about 10% of WDs are in double-WD (DWD) systems, with separations in the range ~1-1,000 Rsun, roughly evenly distributed per decade in separation (Maoz et al. 2018). The gravitational-wave-loss driven mergers of the close-binary members of this population are a favoured (but unproved) explanation for the phenomenon of Type Ia supernovae (SNeIa), whose progenitors' identity is a major unsolved problem (Maoz, Mannucci & Nelemans 2014). Furthermore, such close double WDs will be both the dominant foreground sources of gravitational waves for the space-based LISA (setting its noise floor) and also LISA's individual calibration sources. The thousands of Gaia WDs identified as DWDs by SDSS-V by means of their RV variations will permit estimating their merger rate (which can be compared to the SN Ia rate), as well as laying the groundwork for LISA.

Furthermore, Gaia will also directly detect or order 10,000 wide-orbit (0.01-3 AU, i.e., above 20 Rsun) DWDs by means of the astrometric wobble that the often-unseen companion WD induces in the position of the photo-primary WD. The census of DWDs over all binary separations (close-orbit from RV wobble, wide-orbit from astrometric wobble) will thus permit characterizing the DWD distribution over the full range in separation and in component masses. In addition to the applications already noted (SNeIa, LISA) this distribution is key to understanding binary formation and evolution (particularly the enigmatic common-envelope phase), and the formation of systems of WDs in accreting and interacting binaries. However, the full identification of the components of Gaia's astrometric DWD candidate systems will require high-spatial- resolution optical spectroscopy. This is where MAVIS comes in, as detailed further below.

Gaia will also identify very-wide binaries consisting of a WD and an evolved star (a sub- giant or a giant) for which ages can be determined from the photometry and the spectra, if they can be spatially resolved. These systems can help populate the important but still poorly known initial-final-mass relation for WDs. Similarly, in WD-MS wide binaries (at least 10% of all binaries), the WD provides a clock and the MS star a measurement of metallicity, for a reconstruction of Galactic metal-production history (e.g. Fouesneau et al. 2018). Again, MAVIS can uniquely get spectra for the WDs in these binary systems.

Among the ~25,000 WDs in the eventual Gaia 100-pc sample, where typically WDs are near the Gaia 20-mag completeness limit, Gaia's astrometric precision is ~1mas (improving to 0.2 mas at 17 mag, where there are still significant numbers of WDs in the 100 pc sample). Most WDs, including those in DWD systems, have masses in a narrow range of ~0.6+/-0.2 $M_\odot$. Despite the similar masses of the two WDs in a DWD, most often the luminosities of the two WDs are very different, because the ages are different and the surface areas are different (lower-mass WDs are bigger), and therefore most often only one WD is directly detected (not necessarily the more- or the less-massive one). The Gaia detected astrometric wobble in most DWD cases will therefore be comparable to the binary separation. Gaia will find in the 100 pc sample, by means of astrometric wobble, about 500 DWDs with separations of ~0.3-3 AU and periods of ~2 month to 5 years (the Gaia mission lifetime). Most of these systems will have, at a known epoch during their orbits, separations greater than the ~20 mas. MAVIS can therefore resolve out and directly detect the faint WD companion (even if it is extremely faint, owing to the deep MAVIS 29.5-mag imaging limit). The companion's luminosity (at the Gaia-based distance) and color, both based on the MAVIS photometry, already largely sets the companion WD's mass (over-constraining the kinematic mass measurement based on the Gaia astrometric orbit) and the WD's cooling age. MAVIS can then easily also get separated spectra of both WDs. The R~5000 MAVIS resolution is sufficient to determine what type of WD each one is (DA: H atmosphere, DB: He atmosphere, etc.) and to measure the profiles of e.g. the Balmer lines, which give an independent estimate of the effective temperature Teff, gravity log g, and again the mass and radius. In summary, MAVIS has the potential to produce a complete and fully physically characterized sample of several hundred DWD systems with ~1-AU separations, a sample that will be uniquely useful for answering a host of questions related to binary evolution.

While performing the observations outlined above, MAVIS would automatically also survey the primary WD's surrounding for additional faint companions out to 15 arcsec (considering the 30 arcsec diameter MAVIS field of view) with spectroscopy of these companions possible out to 1.5 arcsec (considering the 3 arcsec diameter IFU field). This will reveal tertiary companions (WDs, M stars, etc.) that were missed by Gaia's limits in depth (20-21 mag) and angular resolution (> 200 mas), at physical separations of 1-1000 AU from the primary WD. Another worthwhile direction would be to image, with MAVIS, a sample of WDs from the Gaia 100 pc sample without astrometric wobble indications of a ~1-AU companion. This data would probe for binary companions, again in the 1- 1,000 AU range where Gaia is insensitive.

Binaries consisting of a WD plus a main-sequence star or a WD plus an evolved star (sub-giant or giant) are also common and potentially useful. While very wide (>> 200 mas) binary systems of this kind will be identified by Gaia via their common proper motions, MAVIS will be able to get the spatially resolved spectra of the star and the WD. Furthermore, MAVIS can survey samples of main-sequence and evolved stars to search for WD companions in the 20-200 mas separation range that is invisible to Gaia and, again, get their spectra. A sample of such systems spanning binary separations of 1-10 AU would be particularly illuminating for understanding the outcome of the common-envelope (CE) phase in terms of component masses and separation, in such systems that have so-far encountered only one such phase, since the border between CE- and no-CE evolution is likely in this range.

## 5.4 Solar System

**Science goals:** Characterization of the 3D structure and dynamics of planetary atmospheres at high spatial resolutions and different time scales. Investigation of surface and atmospheric composition of solar system objects (including comets, rings, planets) at large heliocentric distance. Monitoring of newly discovered very faint objects.

**Program details:** Spectroscopy and high contrast imaging in different bands of planetary atmospheres to derive composition and infer short and long term variability. Spectroscopy of cometary activity of the distant coma. Spectroscopy and photometry of distant and faint objects of the solar system, i.e. KBOs and rings.

**Key observation requirements:** Low resolution spectroscopy (R=5000) and spatially resolved imaging. $10^{-2}$ to 5 arcsec FoV; higher resolution spectroscopy (R=15000) to partially resolve ro-vibrational bands.

**Uniqueness of MAVIS:** High contrast imaging capabilities and flexible sky coverage to observe faint and/or distant objects of the solar system.

Our Solar System offers the opportunity to investigate different objects, including planets, comets, minor bodies as well as other objects in the outer solar system, that can provide information on planetary formation. The continual discovery of new objects within our Solar System, and those coming from outside recently discovered, poses the question about how much material has been released during the primordial phases of the Solar System formation, and how it is distributed with solar distance. By investigating the composition of different objects in the Solar System, through spectroscopy in the visible and near infrared, it is possible to infer their origin, and the link to planets.

In the visible spectral range, peculiar features, related to composition or dynamics, are observed on the above mentioned bodies of the Solar System, and hence this spectral band is fundamental to understand composition and physical properties of objects in the Solar System.

In the following, some of the compelling science cases that can be addressed with MAVIS are illustrated.

### 5.4.1 Comets

Comets are the more pristine objects in our Solar System that are thought to represent the planetary building blocks. Their investigation represents an insight on formation and evolution of our Solar System. They are classified in Short and Long Period Comets (SPC and LPC), according to their dynamical properties. Quite recently, one so-called interstellar comet, 2I/Borisov, has been discovered (MPEC 2019-R106; 2019 September 11). Although it seems to be completely similar to comets belonging to our solar system, according to composition and cometary activity, its orbit is hyperbolic, with e>3, marking its nature as an extrasolar comet. This discovery increases the interest in comets, in order to better understand their composition, and shed light on their role within the solar system.

At present only a few comets have been largely studied, mainly through dedicated space missions (see for example the Rosetta mission). Ground-based surveys at several telescopes have significantly increased our knowledge on these objects, focusing on activity monitoring of the distant coma when comets approach the inner solar system. Sublimation of water ice is the dominant process that controls the cometary activity while approaching the inner solar system. At visible and near infrared wavelengths active comets' emission is dominated by sunlight scattered by gas and dust in the coma. Continuum is dominated by the light scattered from dust grains while gaseous species fluorescence is observed in emission. The flux scattered by comet's dust is determined by means of the $A(\vartheta)f\rho$ parameter (A'Hearn et al., 1984) where $A(\vartheta)$ is the albedo measured at phase $\vartheta$, f is the dust filling factor and $\rho$ is the radius of the circular aperture. This quantity is proportional to the dust production rate (Fink and Rubin 2012).

Gaseous species in the coma are identifiable by means of the wavelength of their fluorescence emissions. The most common species observed from ground in the visible range are summarized in Table 1. The strongest ones include CN (0-0) and (1-0) at 388 nm and 421 nm respectively, $C_3$ at 405 nm and atomic Oxygen (green line at 578 nm, red doublet of forbidden emission at 630-636 nm). Some organic materials can be distinguished by means of emissions in the 350-370 nm range (see Fig. 5.5 as example). By using fluorescence g-factor (photons/s/mol at 1 AU) it is possible to model the emission band flux as $F=(Qhcg\rho)/(8\lambda v\Delta^2 r_h^2)$ where Q is the production rate of the molecule (mol/s), $\rho$ is the radius of the aperture projected to the distance of the comet, h is the Planck constant, c is the speed of light, $\lambda$ is the band central wavelength, v is the velocity of expansion of the gas, $\Delta$ is the observer-comet distance, and $r_h$ is the heliocentric distance of the comet (AU). In general both the dust production rate and the gas production rate depend from heliocentric distance and from local activity (bursts, jets) happening on the nucleus.

The variations reflect compositional differences as well as different origin in the early solar system and evolution, heterogeneities in individual nuclei, and evolutionary processing driven from sun-insolation. High resolution spectra (R = $\lambda/\Delta\lambda$ = 60,000 see Fig. 5.6) show a rich spread of features that will be partially resolved by MAVIS HR-Blue (R=15000), allowing us to observe high J (>100) roto-vibrational bands from which derive the rotational temperature of the different molecules. These measurements are valuable to infer the photochemical transition between parent and daughter species which dissociate due to the solar UV radiation at different altitudes above the nucleus.

Some activity has also been observed on comets as far as 7 AU from the sun (Mazzotta Epifani et al. 2009) and also beyond. At these distances, activity is mainly due to highly volatile species, like CO and $CO_2$ that generate a dust coma around the cometary nucleus. The monitoring of this kind of activity would reveal further information on the comet's composition and age. The low resolution spectroscopic mode foreseen for MAVIS (about R~5000) will likely allow accessing the coma activity of comets that are approaching the inner Solar System, and also when they are still distant.

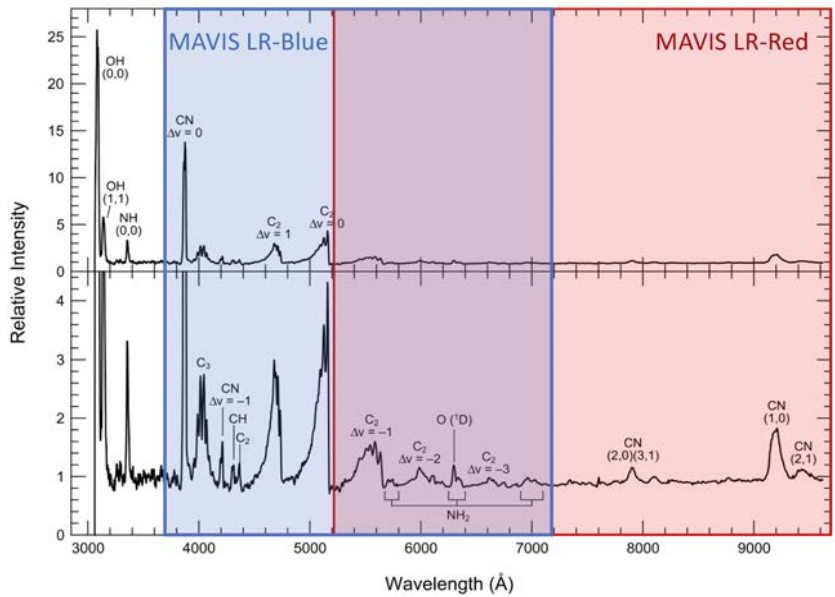

**Figure 5.5:** *Typical VIS-NIR emissions of gaseous species for comet 109P/Swift-Tuttle (from Feldman et al., 2004). In particular, the blue wavelength range of MAVIS LR-Blue provides unique coverage of the CN, CH, $C_2$ and $C_3$ features, unobservable with any existing high-angular resolution IFU.*

| Species* | Transition | System Name | Wavelength (Å) |
|---|---|---|---|
| OH | $A^2\Sigma^+$–$X^2\Pi_i$ (0,0) | | 3085 |
| CN | $B^2\Sigma^+$–$X^2\Sigma^+$ (0,0) | Violet | 3883 |
| | $A^2\Pi$–$X^2\Sigma^+$ (2,0) | Red | 7873 |
| $C_2$ | $d^3\Pi_g$–$a^3\Pi_u$ (0,0) | Swan | 5165 |
| | $A^1\Pi_u$–$X^1\Sigma_g^+$ (3,0) | Phillips | 7715 |
| | $D^1\Sigma_u^+$–$X^1\Sigma_g^+$ (0,0) | Mulliken | 2313 |
| $C_3$ | $\tilde{A}^1\Pi_u$–$\tilde{X}^1\Sigma_g^+$ | Comet Head Group | 3440–4100 |
| CH | $A^2\Delta$–$X^2\Pi$ (0,0) | | 4314 |
| | $B^2\Sigma^-$–$X^2\Pi$ (0,0) | | 3871, 3889 |
| CS | $A^1\Pi$–$X^1\Sigma^+$ (0,0) | | 2576 |
| NH | $A^3\Pi_i$–$X^3\Sigma^-$ (0,0) | | 3360 |
| $NH_2$ | $\tilde{A}^2A_1$–$\tilde{X}^2B_1$ | | 4500–7350 |
| O I $^1D$ | $^1D$–$^3P$ | | 6300, 6364 |
| O I $^1S$ | $^1S$–$^1D$ | | 5577 |
| C I $^1D$ | $^1D$–$^3P$ | | 9823, 9849 |
| $CO^+$ | $B^2\Sigma^+$–$X^2\Sigma^+$ (0,0) | First Negative | 2190 |
| | $A^2\Pi$–$X^2\Sigma^+$ (2,0) | Comet Tail | 4273 |
| $CO_2^+$ | $\tilde{B}^2\Sigma_u^+$–$\tilde{X}^2\Pi_g$ | | 2883, 2896 |
| | $\tilde{A}^2\Pi_u$–$\tilde{X}^2\Pi_g$ | Fox-Duffendack-Barker | 2800–5000 |
| $CH^+$ | $A^1\Pi$–$X^1\Sigma^+$ (0,0) | Douglas-Herzberg | 4225, 4237 |
| $OH^+$ | $A^3\Pi$–$X^3\Sigma^-$ (0,0) | | 3565 |
| $H_2O^+$ | $\tilde{A}^2A_1$–$\tilde{X}^2B_1$ | | 4270–7540 |
| $N_2^+$ | $B^2\Sigma^+$–$X^2\Sigma^+$ (0,0) | First Negative | 3914 |

**Table 5.1:** *List of major gaseous fluorescence emissions in the VIS-NIR spectral range (from Feldman et al., 2004)*

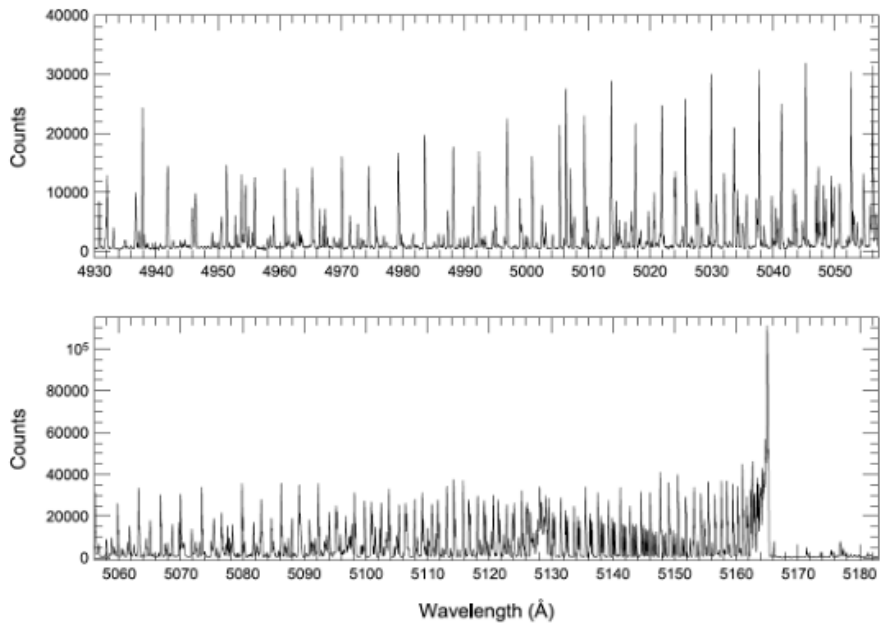

**Figure 5.6:** *High spectral resolution spectra of Comet 122P/deVico showing two-thirds of the $C_2$ $\Delta v = 0$ band observed at a resolving power $\lambda/\Delta\lambda = 60{,}000$ (from Feldman et al., 2004).*

## 5.4.2 Outer Solar System objects: Kuiper Belt Objects and rings

Planetary rings are common features in the outer Solar System where all giant planets (Jupiter, Saturn, Uranus, Neptune) are circled by similar objects (Tiscareno and Murray, 2018). While our knowledge of Saturn's rings is quite extensive (Cuzzi et al., 2018) thanks to the wealth of data returned by the 13 years-long Cassini mission, our knowledge of the remaining ring systems are limited by the sparse observations returned by Voyagers probes during fast flybys of the outer planets. In these cases only telescopic observations are available. Apart from inferring structure, composition and particles' physical properties, the studies of ring systems are valuable to shed light on the formation and evolution processes of giant planets and their complex satellite systems. Among the others, composition and total mass of ring systems are key parameters to understand ring ages. Two hypotheses about formation processes are currently discussed: "primordial rings" in which they are remnants of the protoplanetary disk from which the planet accreted or "young rings" in which they formed from the disruption of one or more satellites migrated within the planet's Roche limit. There are two arguments used to quantify the ring's age: the amount of "pollution" and the ring spreading due to viscosity. Being located beyond the frost line, rings are in general dominated by water ice resulting in high albedo. Nevertheless, at visible wavelengths rings appear red-colored (Filacchione et al., 2012) in contrast with water ice which is neutral to blue in reflected light. This is a strong indication of the presence of contaminants (silicates, organic material, carbon) mixed with water ice particles (Ciarniello et al., 2019) carried by meteoritic bombardment (Cuzzi and Estrada, 1998). Once meteoritic flux is inferred, ring age can be derived from the degree of redness. MAVIS has the capabilities to measure ring visible color at different radial positions allowing to trace compositional variability through it. Ring mass, related to optical depth, plays a major role in the dynamical stability of the entire system through viscosity (Salmon et al., 2010). If rings are "massive", the transfer of angular momentum due to the ring's self gravity causes high viscosity. On the contrary, if rings are "light", collisional processes among finite particle aggregates dominate resulting in a low viscosity. The dynamical stability of the disk is assessed by means of Toomre (1964) theory. MAVIS has the capability to measure ring optical depth, a quantity correlated to ring mass, by means of stellar occultations.

Within our solar system a ring system has been discovered also around Centaurus asteroid 10199 Chariklo, orbiting beyond Saturn and grazing Uranus' orbit. With an elongated shape of 296 × 264 × 204 km, Chariklo is the largest Centaurus object. By using stellar occultation, Braga-Ribas et al. (2014) were able to detect two rings with radii 396 and 405 km and widths of about 7 km and 3.5 km, respectively (Fig. 5.7). This discovery opens the search for further similar ring systems in the realms of the outer solar system objects.

Moving beyond the Centaur region, we can find Kuiper Belt Objects (KBOs), i.e. bodies located beyond Neptune's orbit, that count more than 100,000 members, with diameters > 100 km (Jewitt et al., 1998; Choi et al., 2003). Despite the expected large number of KBOs, to date their discovery is still limited to the biggest ones, because these objects are quite faint and difficult to be observed with the ground-based facilities available. At present, KBOs as faint as 24[th] magnitude are accessible from ground (Elliot et al., 2005). However, it is estimated that a large fraction of smaller and fainter objects (with diameters smaller than 100 km and magnitudes lesser than 24[th]) is still uncovered. Dedicated surveys with instruments able to investigate the sky at very deep magnitudes would certainly increase the number of objects beyond Neptune's orbit. This is fundamental to gain a better picture of the material distribution in this Solar System region, and finally to increase our knowledge in the Solar System formation.

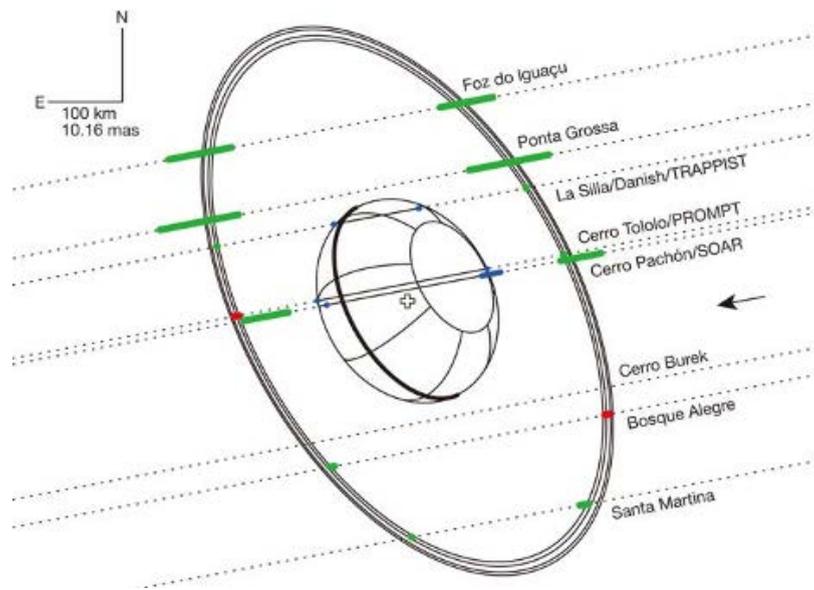

***Figure 5.7:*** *Chariklo ring system as detected from ground-based stellar occultations, from Braga-Ribas et al. (2014).*

For some of these objects, photometric and spectroscopic observations are available in the visible and near infrared, from which it can be inferred the surface composition (Luu and Jewitt, 1996; Green et al., 1997; Jewitt and Luu, 1998; Barucci et al., 1999; Davies et al., 2000; Barucci et al., 2000; Jewitt and Luu, 2001; Barkume et al., 2008) and physical properties. A large diversification of KBOs composition has been suggested, from the large variability in the spectral properties and photometric colors obtained in the visible and near infrared (Jewitt et al., 2001). The reported variation in the KBO population is quite difficult to explain, considering that these objects seem to originate from the same region. Hence, collisional resurfacing (Luu and Jewitt, 1996) or cometary activity (Hainaut et al., 2000) should be considered.

This scientific topic would largely benefit from new imaging and spectroscopic observations of large KBOs, with the aim of surface characterization in the visible range. The imaging capability of MAVIS, coupled with the high contrast with respect to background, will allow the follow up of very faint objects, newly discovered, to guarantee their orbit definition.

### 5.4.3 Outer planets: Uranus and Neptune

Uranus and Neptune are the two members of the Icy Giants planet class. In comparison against Gas Giants (Jupiter and Saturn), they are characterized by smaller masses (14.6 and 17.2 $M_\oplus$ instead of 317.8 and 95.2) and a much higher content of heavy elements (enrichment with respect to solar composition presumably about 30 instead of 3-8). These two facts combined have implications on the deep structure of these bodies and, in turn, on the energy transport mechanisms from their deep interiors to the observable atmosphere.

Temperatures in the upper troposphere of Icy Giants are so cold (60K) that they remain largely depleted in all minor components, excluding noble gases. Observable features are indeed represented by aerosol layers produced by the condensation of different molecules. Outermost layers (approx. 0.1 – 1 bars) are composed of $CH_4$, while deeper levels (1-3 bars) are presumably formed by $H_2S$ (Irwin et al., 2018) while in the Neptune case also $NH_3$ remains a candidate (Irwin et al., 2009). These aerosol layers are used as tracers to study the atmosphere dynamics.

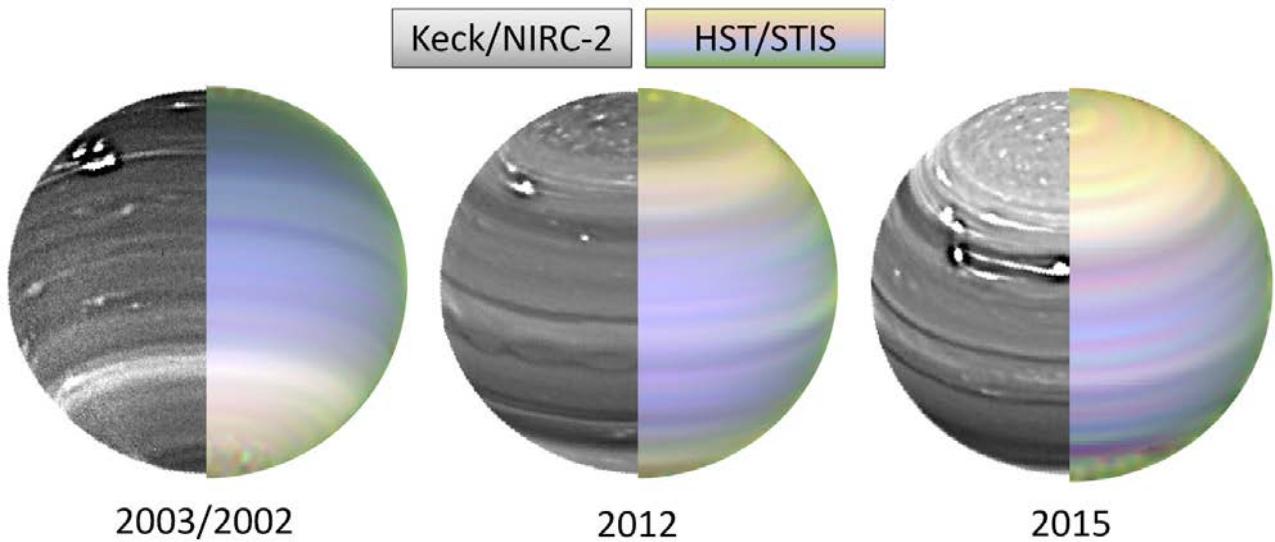

**Figure 5.8:** *Multiwavelength monitoring of Uranus using NIR AO-corrected imaging (Keck/NIRC-2 in H-band, left hemisphere) and optical spectroscopy using a scanned long-slit (HST/STIS in I band windows, right hemisphere) at comparable spatial resolution (~0.1"). MAVIS can provide 4-5 times higher spatial resolution, matching what would be delivered by ELT/MICADO. Adapted from Sromovsky et al. 2018.*

Both planets present a clear zonal pattern of clouds. On Uranus there is a clear disruption of zonal pattern beyond about 60°S, given the patchy appearance of the entire polar region (Sromovsky et al., 2015). The retrieved zonal wind fields present two, hemisphere symmetric, prograde jets at about 60° latitude and retrograde jet on the equator (Kaspi et al., 2013). Jets are strong: on Neptune they exceed 300 m/s, the highest values reported to date in the entire solar System. Available information suggests that wind patterns and thermal structure of the upper troposphere of Uranus were subject only to minor changes from the Voyager era to at least 2011 (Orton et al., 2015), despite the strong seasonal variations caused by the extreme axial tilt of the planet. The structure of wind fields is consistent with the available Uranus air temperatures latitudinal cross section: at 0.1 bar, two hemispheric symmetric minima are found at 40°N and 40°S, with maxima of comparable amplitude at the equator and over both poles. These data are overall consistent with upwelling at intermediate latitudes and subsidence at high latitudes and at the equator.

Icy giants' atmospheres are characterized by a rich phenomenology. Albeit Uranus shows very little details during the 1986 Voyager encounter (occurred in the vicinity of the summer solstice), atmospheric phenomena (mostly in the form of distinctive bright high altitude clouds, presumably formed of methane ice) has been observed more and more frequently moving toward the Northern spring equinox in 2008 and in subsequent years (De Pater et al., 2014). A bright polar cap on the Southern hemisphere has progressively disappeared while moving toward the equinox. Another notable, transient feature of Uranus was the Uranus Dark Spot (UDS), a darker area observed at intermediate latitudes in 2006 with a size of about 1500 km (Hammel et al., 2009). An hypothesis on its nature sees it as the result of an anticyclonic system, creating an area of depleted cloud coverage at its center.

Neptune displayed a richer appearance already during the Voyager encounter (Smith et al., 1989). The abrupt appearance of very bright clouds has been documented continuously after initial observations from the Hubble Space Telescope. These clouds systems appear often much larger, brighter and longitudinally extended than their counterparts in Uranus (Hueso et al., 2017). A Great Dark Spot was observed in detail by Voyager, about five times larger than Uranus Dark Spot described above. It was surrounded by brighter high altitude clouds and completely disappeared before Hubble Neptune observations in 1994. Another Neptune dark spot has been observed since 2015 in the Southern hemisphere, again accompanied by brighter high clouds. As in the Uranus case, these features are interpreted as regions of cloud clearance that expose deeper atmospheric layers.

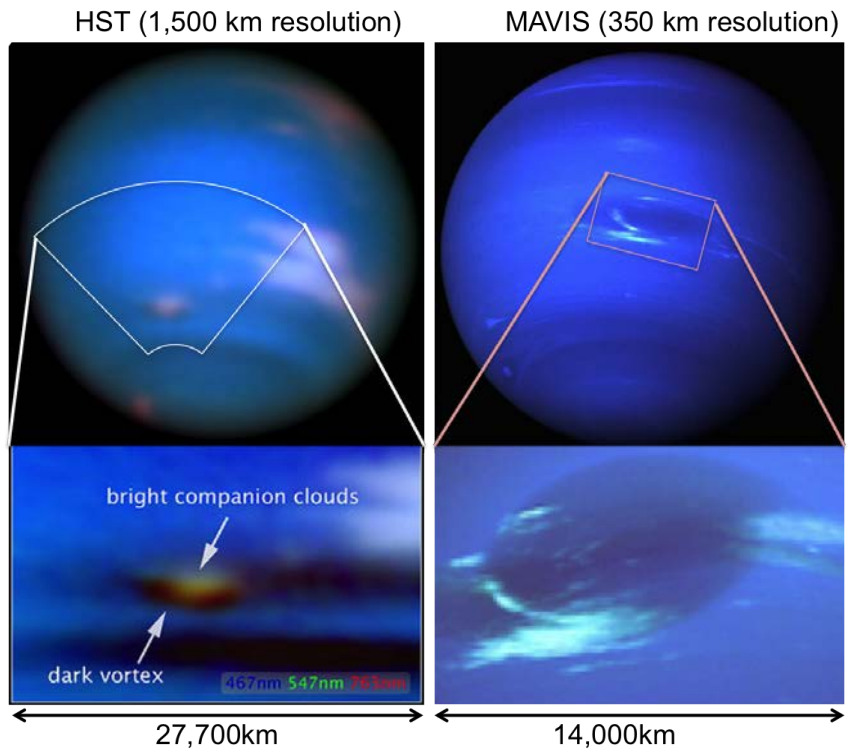

HST (1,500 km resolution)   MAVIS (350 km resolution)

27,700km   14,000km

*Figure 5.9:* *MAVIS will provide a few hundred km resolution on the icy giants' atmospheres, surpassing HST resolution. Comparable observations were recently demonstrated during MUSE-NFM commissioning. MAVIS will improve this even further, with superior AO correction, and Nyquist-sampled imaging.*

Sromovsky et al. (2009) demonstrated for Uranus that joint effects of Rayleigh scattering and methane absorption (in lower troposphere) create a fairly good distribution of weighting functions peaks between 0.5 and 1 μm, covering the range between 0.7 and 7 bar and encompassing therefore the vertical region where most external aerosol decks are expected to occur. Spectra in this range are therefore capable to provide, upon inversion, important constraints on vertical distribution of clouds. This information, once extended over large areas and coupled with horizontal dynamic patterns, represent a fundamental benchmark for validating the circulation models for gas giants and, in a larger perspective, for exoplanets.

Experience gained with Juno/JIRAM data demonstrates that in giant planets' atmospheres, turbulent phenomena create very non-uniform cloud conditions down to the scale of a few tens of kilometers. More realistically, we can use as a reference the spectral study of features such as the UDS at about one tenth of its size, to get at least an idea of its internal variability. This translates into a required spatial resolution of ~300 km at Uranus, i.e. 0.02 arcsec. In this field, it would be important to achieve for Uranus a S/N ratio at least equal to 10 around 870 nm (if needed, with spectral binning) where Uranus has a I/F[3] of about 0.01.

---

[3] Spectral intensity I/F is the ratio, as a function of wavelength, of the observed radiance to that of a normally-illuminated 'Lambertian' (ideal, diffusing) reflector at the same distance from the sun as the target.

# 6 Science Requirements

This section provides motivation for the key MAVIS science requirements, as derived from the science cases presented here. This includes explicit reference to the most relevant sub-chapters, where appropriate.

## 6.1 Sky Coverage

MAVIS is a facility instrument at the VLT, implying a general-purpose instrument that should be generally applicable to a wide range of science topics, and therefore capable of targeting a broad range of astrophysical targets, found in a variety of environments, and with the additional important capability of building statistically significant samples of objects. This automatically implies that MAVIS shall be able to deliver its image quality requirements over a significant fraction of the sky.

Moreover, such a general purpose facility instrument on a large aperture telescope should expect to regularly target faint sources in key extragalactic fields, which are typically chosen to be free of bright stars and foreground contamination from the Milky Way. This implies that MAVIS should be able to deliver its specified image quality even for fields typically found towards the Galactic Pole. The Galactic Pole is therefore taken as the limiting environment for defining sky coverage, with a requirement that at least 50% of such fields should be observable with nominal performance. This of course implies a lower limit on the coverage of MAVIS over the full sky.

AO correction strongly depends on the brightness of the available natural guide stars. Through detailed simulations (see AD2), we have shown that, even with an optimised design, the distribution of available stars in the night sky (in terms of location on the sky and brightness) physically excludes the possibility to provide diffraction-limited image quality with 50% probability at the Galactic Pole. However, accepting that formal Strehl ratio is generally not the defining metric for scientific utility, and in fact that image quality defined in terms of FWHM and Ensquared Energy is more practical, we find that performance consistent with the requirements of our science cases can be met for at least 50% of Galactic Pole fields.

This results in the following requirement, which applies to imaging and IFU modes:

| **R-MAVIS-SCI-1 Sky Coverage** | | |
|---|---|---|
| *MAVIS, in at least 50% of random pointings within 10 degrees of the Galactic pole, shall provide AO correction leading to an absolute ensquared energy, in V band, of at least 15% within a 50 mas square aperture.* | | |
| **Driving Science Case Chapters** | **Requirement Rationale** | **Other Relevant Chapters** |
| 2.2 Kinematics of the first disk galaxies | Parent samples are drawn from Deep Fields, selected to have few bright stars | 2.2, 2.3, 2.4, 2.6, 3.4, 4.3, 4.4, 4.6, 5.2, 5.3 |
| 3.6 Supermassive black holes in low mass galaxies | Targets have nuclei too faint/extended for guiding, and have extended (>10") envelopes | |
| 4.7 Origin of first star clusters at high-z | Limited number of quality targets; multiple lensed images of same sources need observed | |
| 5.4 Solar System | Time-sensitive observations of sources too faint for on-axis guiding | |

## 6.2 Strehl Ratio

As described in the Introduction to this document, there is a growing and urgent need to provide a comprehensive, high-angular resolution optical capability on a large-aperture telescope. Scientifically, this is driven by a wealth of discovery potential in unexplored regimes of crowded fields, complex spatial structures, and sensitivity. Strategically, such a facility replaces and extends the capabilities provided by the deteriorating Hubble Space Telescope (still one of the most over-subscribed telescopes in existence), and provides essential optical coverage at comparable angular resolutions and sensitivities as ELTs will provide in the infrared. Technically, the VLT Adaptive Optics Facility has demonstrated its huge potential for providing AO corrections at optical wavelengths; yet so far it is lacking a general purpose, high sky coverage, optical instrument capable of addressing the scientific needs.

A driving purpose for MAVIS is to provide high angular resolution imaging and spectroscopy at optical wavelengths, with a high sky coverage fraction allowing general-purpose applications, and a large corrected field. In general, the most demanding applications, in terms of angular resolution, will make use of MAVIS in imaging mode, which then provides the limiting cases. Image quality then impacts on three key aspects: 1) Point-source sensitivity, 2) Confusion limit, and 3) Astrometric precision.

| R-MAVIS-SCI-2 Strehl Ratio | | |
|---|---|---|
| *MAVIS, for a flat spectrum source in V-band, over a 15 minute observation, shall provide an average Strehl ratio ≥10% (goal 15%) over the science field. Performance at longer wavelengths is based on the assumption of constant wave front error.* | | |
| **Driving Science Case Chapters** | **Requirement Rationale** | **Other Relevant Chapters** |
| 2.5 Morphology of the first galaxies | Strehl ratio and small FWHM are a key driver of photometric depth | 4.4, 4.5, 4.6, 4.7, 5.4 |
| 3.2/3.3 Resolved stellar populations in ETGs | Distant old stellar populations are faint, benefitting from sharp PSF; stars with most interesting abundances live in crowded inner regions | |
| 4.2 Abundance ratios in globular cluster centres | IFU spectroscopy in highly crowded regions | |
| 4.3 The hunt for seed black holes in globular clusters | Needs precision photocenters in dense stellar field, and down to faint crowding limit to populate velocity histograms | |
| 5.2/5.3 Circumstellar environments | Requires best possible inner-working angle to probe immediate vicinity of stars | |

## 6.3 Strehl Ratio Variation

As discussed above, wide-field imaging capabilities are fundamental for many MAVIS science cases, given the large corrected field provided by MCAO. However, to fulfill the science requirements the PSF variations across the field of view have to be minimized. This is particularly relevant for accurate photometry and astrometry.

| R-MAVIS-SCI-3 Strehl Ratio Variation | | |
|---|---|---|
| *MAVIS, at V-band, shall have an RMS variation of Strehl ratio over the science field less than 10% of the average. Performance at longer wavelengths shall scale on the assumption of constant wave front error.* | | |
| **Driving Science Case Chapters** | **Requirement Rationale** | **Other Relevant Chapters** |
| 2.5 Morphology of the first galaxies | Need to maximise volume probed, this maintain PSF across full field; maintain accurate morphological information regardless of field position | 2.5, 3.2, 3.3, 4.3, 4.4, 4.6, 5.2, 5.3, 5.4 |
| 3.2 Resolved stellar populations in ETG envelopes | Maintain high photometric depth across sparse fields to populate CMDs | |
| 4.3 The hunt for seed black holes in globular clusters | Maintain astrometric accuracy (PSF regularity) to edges of field for complete proper-motion profiles | |

## 6.4 Wavelength Range

The science cases described here, and put forward in the MAVIS White Papers, require a broad range of wavelengths, from the far blue to the far red of what is considered the 'optical' wavelength range. Achieving a significant AO correction at the bluest wavelengths will not be an easy task, but the significantly-improved performance with respect to seeing-limited observations will still be of significant value in enabling novel science. At the red end, AO correction is easier, and we consider more the complementarity with existing facilities (e.g. continuous wavelength coverage into the infrared), understanding that JWST and ELT will dominate in terms of sensitivity and angular resolution redwards of 1μm (see Fig. 1.4).

### Imaging

MAVIS will enable very significant scientific progress for point-source and compact objects, by virtue of the significant concentration of light enabled by the MCAO system. Such objects comprise both stars and compact emission regions such as star-forming clumps. In both cases, there is significant value in accessing the bluest wavelengths possible, as these are high value for e.g. temperature (and therefore age) diagnostics in stellar populations (e.g .Chapters 3.2, 4.4) and hot evolved stars (Chapter 3.3); and for emission lines in narrow bands, in particular the [OII] (λλ 3727,3729 Å) doublet - a strong tracer of star-formation.

In terms of delivered performance, the uncertain blue AO correction, coupled with the challenge of optimising the full system for a broad range of wavelength, make delivering full performance bluer than V-band a goal, rather than a requirement. The science cases presented here will benefit from U-band coverage (including narrow-band [OII] capability) and strongly benefit from the more achievable B-band region, but nonetheless are feasible as described with photometry limited to as blue as V-band.

## Spectroscopy

One of the strong points of MAVIS spectroscopy is the extensive wavelength coverage, including blue and red ranges that are not accessible by other optical facilities, e.g. MUSE. The blue end will be particularly important for the study of hot stars, and will make accessible several key features as the CN/CH bands at ~4100-4300Å (used as chemical clocks in globular clusters - Chapter 4.2.3); the neutron-capture elements features between 4000-4700Å (useful to trace the star formation history time scale in stellar populations - Chapters 4.2, 4.4); the CaII H and K (3969Å, 3934Å) lines (useful as a tracer of chromospheric activity and mass accretion rates - Chapter 5.2.1); and the [OII]λλ3727,3729Å doublet as a density diagnostic gas in stellar jets (Chapter 5.2.2).

Extragalactic astronomy will also benefit from this extended blue wavelength range, giving access to important emission and absorption line features not observable with MUSE at low redshift, such as the important ionised gas diagnostics [OII]λλ3727,3729Å doublet and the auroral [OIII]λ4363Å line. Bluer coverage also provides lower-redshift access to redshifted UV lines, for example making it possible to observe Lyα emission from as recently as z~2.

On the red side, the extended wavelength range will be fundamental to study the first galaxies in the reionization epoch at around z~7, covering a gap between the current MUSE capabilities and the forthcoming ERIS NIR spectroscopy at the VLT. In particular, each Angstrom gained at λ>9300Å corresponds to an additional 2-3x$10^5$ years of cosmic lookback time, eventually reaching 720 Myr after the Big Bang at z=7.22 (Lyα at 1 μm, see Chapter 4.7). In addition to probing the reionization of the Universe, the red end of MAVIS will also extend to z~0.5 the ISM and kinematical studies in intermediate redshift galaxies, granting access to Hα emission at those cosmic epochs (Chapters 2.2, 2.3).

Fig. 6.1 illustrates the key spectral features covered by the MAVIS wavelength range as a function of their redshift. This figure also indicates which spectral resolution modes will cover those lines - see section 6.9. Note that the spectral window from Hβ to [SII] (important for determining ionised gas physical properties) is simultaneously accessible in a single LR-Blue or LR-Red configuration at all possible redshifts.

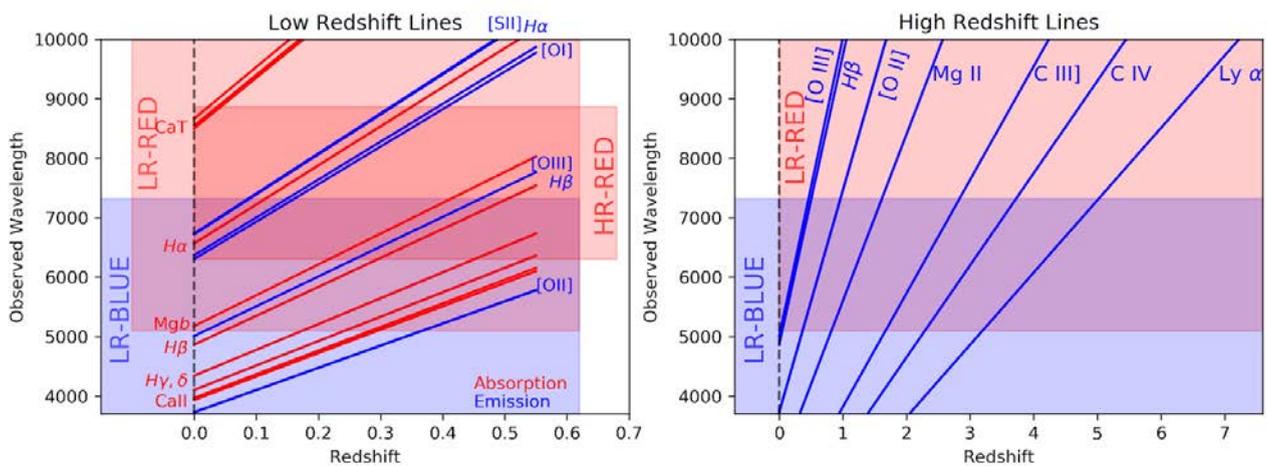

**Fig 6.1**: *Diagonal lines indicate major absorption (red) and emission (blue) features plotted as a function of redshift, with the overlapping wavelength ranges of MAVIS modes shown as shaded regions. The left panel shows low redshift coverage; the right panel shows a larger range of redshift, where rest-frame UV lines enter the optical range. HR-Blue mode is not shown, as it is mostly designed for local, non-redshifted features. Likewise, only the LR modes are shown for the highest redshift range (right plot).*

## R-MAVIS-SCI-4 Wavelength Range

*MAVIS shall provide imaging and spectroscopic capabilities from 370-950 nm (goal: 370-1000 nm).*

| Driving Science Case Chapters | Requirement Rationale | Other Relevant Chapters |
|---|---|---|
| 2.5 Morphology of the first galaxies | Detecting photometric drop-outs at z>6 to determine low-mass UV luminosity function | 2.2, 2.3, 2.4, 3.2, 4.6, 4.7, 5.3 |
| 2.6 Spectroscopy of the first galaxies in the epoch of reionization | Detecting and characterising Lyα emission systems at z≥7 with imaging and spectroscopy | |
| 3.3 Resolved stellar populations: UV-upturn in ETGs | Targets blue stars; B band gives better contrast and diagnostics of hot, old stars | |
| 3.5 Resolving the physics of ram-pressure stripping | [OII] λ370nm at z≈0 as SF/ISM/electron density diagnostic (NB imaging, spectroscopy) | |
| 4.2 Galactic stars clusters: Revealing the mystery of Globular Clusters | Access to heavy element abundances around 400-470nm | |
| 5.2 Circumstellar environment of young stars | CaII H&K lines (396.9nm, 393.4nm) to derive mass accretion rates | |

## 6.5 Imaging Field of View

As discussed already, the wide field capabilities of MAVIS are fundamental to significantly increase the observing efficiency of the instrument, maximizing the number of objects per field. For example, the foreseen 30"x30" square arcsec field of view will, for example, enable the simultaneous observation of ~600 galaxies in a deep extragalactic field, or to maximize the number of cluster members once studying stars in open and globular clusters in our Galaxy and in the Magellanic Clouds, and to fully cover individual star clusters in nearby galaxies. Moreover, a large continuous field of view is fundamental to study spatially extended and complex objects, like nearby star clusters or galaxies.

| R-MAVIS-SCI-5 Imaging FoV | | |
|---|---|---|
| MAVIS shall provide an unvignetted imaging field of view of at least 29x29 arcseconds. | | |
| **Driving Science Case Chapters** | **Requirement Rationale** | **Other Relevant Chapters** |
| 2.4 Resolving the physical properties of classical bulges and pseudo bulges | Bulges are ≤1.5kpc scale radius = 30" diameter at Fornax cluster | 3.3, 3.4, 3.5, 4.2, 4.4, 4.7, 5.4 |
| 2.5 Morphology of the first galaxies | Need to maximise volume probed, which goes as field diameter$^2$ at fixed depth | |
| 3.2 Resolved stellar populations in ETG envelopes | 30" gives few thousand RGB stars from old population at surface brightness 26mag/sq.arcsec, enough for CMD | |
| 4.3 The hunt for seed black holes in globular clusters | 30" gives sufficient radial coverage to give contrast in radial proper motion profile between stellar cluster core and IMBH | |

## 6.6 Imaging Sensitivity

The limiting magnitudes in the imaging mode (around m~29-30, for more details see Appendix A) will allow us to obtain unprecedented observations.  In fact, the point-source imaging sensitivity and high tolerance to crowding are clearly among the most important achievements of MAVIS, since it will exceed that of the Hubble Space Telescope, giving an order of magnitude higher depth, and with higher angular resolution. Moreover, this capability is fundamental to the complementarity with the ELT in the infrared using MICADO, with MAVIS having the fundamental benefit of the limited sky-background at the optical wavelengths, permitting comparable sensitivity at optical wavelengths.

In particular, the unprecedented optical depth of MAVIS imaging, combined with its superior spatial resolution,  will allow access to new frontiers in stellar science, enabling the detection of resolved , cool giant stars up to 10 Mpc and O-B stars even at much larger distances. This is fundamentally important, as it extends the range of morphologies and environments of galaxies for which resolved stellar population analysis is possible. The ability to work in extremely crowded environments allows this sensitivity to be exploited in dense regions like star clusters. For example, MAVIS will be able to map the whole stellar sequence of the densest core regions of Galactic and Magellanic Cloud stars clusters, as well as to resolve their primeval progenitors, i.e. the compact sources in lensed galaxies at high redshift.

In extragalactic astronomy, MAVIS will be able to image the faintest galaxies ever detected, e.g. being able to detect in 10 hrs integration compact high-z galaxies up to I=30.4, about 1 mag deeper than the Hubble Ultra Deep Field.

Finally, the imaging depth of MAVIS will allow the follow up of very faint, newly discovered objects in our Solar System, such as Comets, KBOs, TNOs etc, to guarantee their orbit definition, surface and environment characterization in the visible range.

| R-MAVIS-SCI-6 Imaging Sensitivity | | |
|---|---|---|
| *MAVIS, in a one hour total exposure time, shall be able to image an isolated point source of mV=29 with a SNR≥5.* | | |
| **Driving Science Case Chapters** | **Requirement Rationale** | **Other Relevant Chapters** |
| 2.5 Morphology of the first galaxies | Need to maximise volume probed, which is proportional to photometric depth; push to low mass systems beyond Hubble Ultra Deep Field (<29.5mag) | 2.4, 3.3, 3.4, 4.2, 4.3,  5.2, 5.3, 5.4 |
| 3.2 Resolved stellar populations in ETG envelopes | Detect RGB of ancient stellar populations within 10Mpc requires limiting AB magnitude of V=29-30 | |
| 4.4 Stars clusters in the Local Universe | Detecting down to 0.1Msun stars in the L/SMC for IMF studies requires V>29 ABmag | |
| 4.7 The origin of the first star clusters at high-z | Detecting lensed young massive clusters forming at z>7 requires limiting magnitudes of V>29 | |

## 6.7 Astrometric Performance

A principal case for high precision astrometry is to probe regions inaccessible to existing instruments, such as HST and GAIA. One such compelling case is the search for intermediate mass black holes (IMBHs) in low mass stellar systems, and globular clusters in particular. In Chapter 4.3, it was shown that in order to establish a statistically meaningful constraint on the general presence or absence of IMBHs in Milky Way globular clusters from stellar dynamics, a minimum precision of 150µas would be necessary, given the distance distribution of such clusters. Bringing this precision to 50µas would double that sample size, and is a well-motivated goal.

This astrometric precision, with a combination of internal precision and external accuracy based on Gaia and HST data, will allow MAVIS to perform 2D kinematics not only in the very central regions of globular clusters, but also at intermediate radial distances, revealing the contrasting signal of the presence of an IMBH. The superb spatial resolution of MAVIS and its wide field of view are crucial for these observations, outperforming the capabilities of HST for the global assessment of the presence of an IMBH. In addition, reaching the faintest stars (down to ~30 mag), MAVIS will allow us to measure the radial dependence of the stellar mass function in detail, and to probe the dynamics of stars with different masses in the centre of GCs.

| R-MAVIS-SCI-7 Astrometric Performance | | |
|---|---|---|
| *MAVIS, assuming on-image reference sources are present to enable the plate-scale and rotation calibration to at least 0.01%, shall enable astrometric measurements such that the relative position of two unresolved, isolated sources, detected with an SNR>200, separated by up to 1 arcsecond, can be measured to within 150 micro arcseconds (goal 50 micro arcseconds).* | | |
| **Driving Science Case Chapters** | **Requirement Rationale** | **Other Relevant Chapters** |
| 4.3 The hunt for seed black holes in globular clusters | 150µas/yr corresponds to a characteristic globular cluster velocity dispersion of 5-10km/s at 10kpc from the sun. This precision allows a sample of order 20 objects. A goal of 50µas/yr permits more than doubling the potential targets. | |

## 6.8 Spectroscopy Field of View

In the local Universe, the study of the densest regions of nearby star clusters (in the Galaxy - see Fig. 4.2 and in the Magellanic Clouds - see Fig. 4.17) require an approximately 3" diameter IFU FoV to cover the region of interest, which implicitly makes it suitable for such clusters at larger distances. Most regions of interest for detailed spectroscopy in nearby (<100Mpc) galaxies (HII regions, nuclear star clusters, coherent ISM filaments, etc. are generally on <few hundred parsec scales) are also well suited to an IFU of this field size. A similar field of view seems appropriate also for solar system science: as an example, Neptune diameter is ~2.3".

At cosmologically important distances, typical disk galaxies have half-light radii ranging from ~3-15 kpc, at z = 0.5, which corresponds to ~0.5-2.5" (see e.g. Fig. 2.5). For the same mass, early-type galaxies have correspondingly smaller sizes at any given epoch, so are also encompassed by this field of view. Moreover, the apparent (projected) shape of galaxies on the sky, given by their minor-to-major axis ratio, is b/a≈0.6-0.7 (Padilla & Strauss 2008), indicating there may be advantages to a non-square field of view.

9 square arcsec represents the minimum required for many of the science cases, from stellar to extragalactic applications. The baseline design for MAVIS (using the 'Fine' spaxel size of 25mas) has a rectangular field with a √2 aspect ratio (3.6" x 2.5"), comparable to the general apparent flattening of galaxies. By employing the 'Coarse' spaxel size of 50mas, the IFU FoV becomes proportionally larger, covering an area of 36 square arcsec (7.2" x 5"), which may be applicable for lower surface brightness applications, and when the level of PSF correction can tolerate coarser sampling. The shape of the IFU will be investigated further in phase B, also considering potential PA restrictions from having the IFU offset from the field centre.

| R-MAVIS-SCI-8 Spectroscopy FoV | | |
|---|---|---|
| *MAVIS shall provide spectroscopic capabilities over a field of view area of at least 9 square arcseconds.* | | |
| **Driving Science Case Chapters** | **Requirement Rationale** | **Other Relevant Chapters** |
| 2.2 Kinematics of the first disk galaxies in the epoch of Cosmic Change | Typical disk galaxies have half-light radii of ~3-15 kpc at z = 0.5, corresponding to diameters of 1-5 arcsec | 2.3, 2.4, 2.6, 4.2, 4.3, 4.5, 4.6, 4.7, 5.4 |
| 3.5 Resolving the physics of ram-pressure stripping | Coherent fine structure features (filaments and star formation) on ram pressure fronts within galaxies are few x 100pc to 1kpc in size, corresponding to <3" where MAVIS has ~10pc physical resolution (≤100Mpc). | |
| 3.6 Supermassive black holes and stellar nuclei in low mass galaxies | Need resolved dynamics of nuclear star clusters (on 10-100pc scales) together with immediately surrounding regions of the host galaxy. For feasible distances of 3-30Mpc, requires a minimum field of 3". | |
| 4.4 Stars clusters in the Local Universe | Star cluster half-light radii of 5pc beyond Local Group at distances feasible for MAVIS spectroscopy (1-10Mpc) projects to 2-0.2" diameter. | |
| 5.2/5.3 Circumstellar environments | Circumstellar jets and disks exist on 10s to few 100s of AU scales, corresponding to 0.05-2" at the distance of the Taurus star forming complex. | |

# 6.9 Spectroscopy Spectral Resolution

As discussed previously, MAVIS will be a general-purpose instrument that should be applicable to a broad range of science topics, as motivated by the science cases presented here, and by the larger collection of MAVIS White Papers. In general terms, this includes both bright compact sources, for which detailed kinematics (both in terms of velocity shifts of individual stars, and dispersions for integrated light measurements) and faint absorption/emission line detection is required; and fainter, diffuse objects where lower resolution is adequate to resolve emission line shapes. Combined with the broad wavelength coverage requirements, and considering the need for simultaneous wavelength coverage for certain science cases, four different spectroscopic setups are foreseen for the instrument, being 'high' and 'low' spectral resolution modes with 'red' and 'blue' coverage for each. Motivation for these modes is provided in more detail below.

A low resolution setup will thus cover the full wavelength range at R~5000 with two gratings (LR-Blue and LR-Red). This resolution is the minimum required for optimal sky line subtraction in the red and accurate stellar kinematic estimates in galaxies, and resolving non-circular motion through line spectral shape analysis (see e.g. Fig 7.3). A spectral resolution R~5000 is also key to probe signatures of emission lines of down a few dozens km/s FWHM in the highest redshift galaxies. In particular Lyα multi-peaked profiles (e.g. Vanzella et al. 2020) require R~5000 in order to allow proper radiation transfer modeling and recognize narrow features as signatures of transparent ionized channels for the escaping of ionizing photons. Moreover, it is ideal for fainter, diffuse targets, like external galaxies or their sub-components.

However, this resolution is not high enough for other science cases, such as the chemical abundances in resolved stellar populations. In Table 6.1, we show the number of detectable absorption lines per element (in an ideal condition of high SNR and fixed spectral coverage). There is a larger difference between R=5000 and R=10000 than between R=10000 and the following higher spectral resolutions. Thus the high-resolution MAVIS mode with R>12k in the blue and R>9k in the red will allow a complete chemical characterization of resolved stellar populations.

| Spectral resolution | *Fe* | *Al* | *Ba* | *Ca* | *Ce* | *Co* | *Cr* | *Cu* | *Eu* | *La* | *Mg* | *Na* | *Ni* | *Sc* | *Si* | *Sr* | *Ti* | *V* | *Y* | *Zn* | *Zr* |
|---|---|---|---|---|---|---|---|---|---|---|---|---|---|---|---|---|---|---|---|---|---|
| *5000* | 5 | 0 | 2 | 4 | 1 | 3 | 2 | 1 | 1 | 1 | 1 | 1 | 5 | 5 | 2 | 0 | 5 | 3 | 2 | 1 | 0 |
| *10000* | 23 | 2 | 7 | 20 | 4 | 23 | 21 | 13 | 11 | 11 | 7 | 18 | 22 | 22 | 20 | 13 | 23 | 19 | 18 | 5 | 0 |
| *15000* | 23 | 3 | 7 | 22 | 7 | 23 | 23 | 15 | 15 | 14 | 4 | 21 | 23 | 23 | 23 | 12 | 23 | 21 | 23 | 2 | 1 |
| *20000* | 23 | 4 | 8 | 22 | 11 | 22 | 23 | 13 | 14 | 21 | 16 | 20 | 23 | 23 | 22 | 10 | 23 | 23 | 23 | 8 | 4 |
| *50000* | 23 | 11 | 12 | 22 | 17 | 21 | 22 | 15 | 10 | 21 | 18 | 22 | 22 | 21 | 9 | 22 | 21 | 21 | 21 | 14 | 11 |

**Table 6.1**: *Detectable absorption lines at different spectral resolutions for a F-G-K star. Elements are colour-coded by their nucleosynthesis channels. In light blue the iron peak elements, in orange the neutron-capture elements (both from s- and r-process), in light red the alpha-elements, produced (maily) by the core collapse SNe, in light pink, the elements with more complex origin.*

In Fig. 6.2 we compare the effective wavelength coverage of the instrument at R=10k, 15k, and 20k. Assuming a typical SNR~40, we observe a reduction of the number of observable elements and number of lines per element for decreasing resolution. However, the foreseen MAVIS high resolution mode in the blue at R=12k is, indeed, a perfect compromise between wavelength coverage (reaching also the blue part of the spectra - ideal also for metal poor stars and hot O-B stars) and spectral resolving power (allowing the detection of many unblended lines of key elements up to several kpc - see Fig. 6.3). In addition, the combination of the HR-mode in the blue, with the HR-mode in red, at R~9k, will allow a complete spectral characterization of different stellar populations, including variable stars (see Figure 6.4).

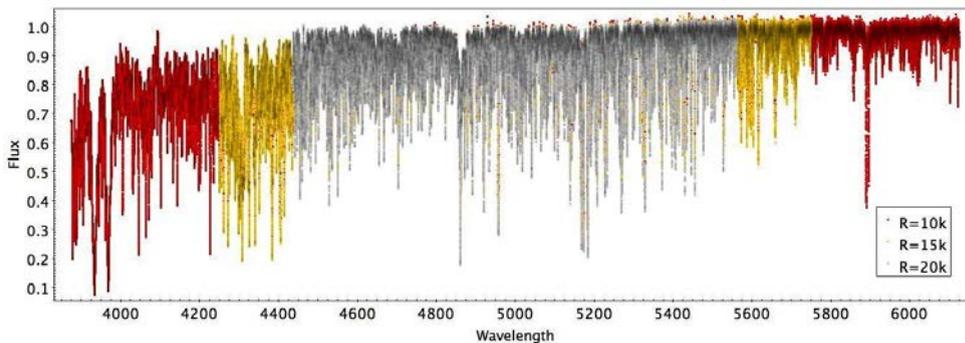

**Figure 6.2**: *The effective wavelength coverages at different spectral resolutions. The selected MAVIS HR-mode in the blue is the best compromise between wavelength range and resolving power.*

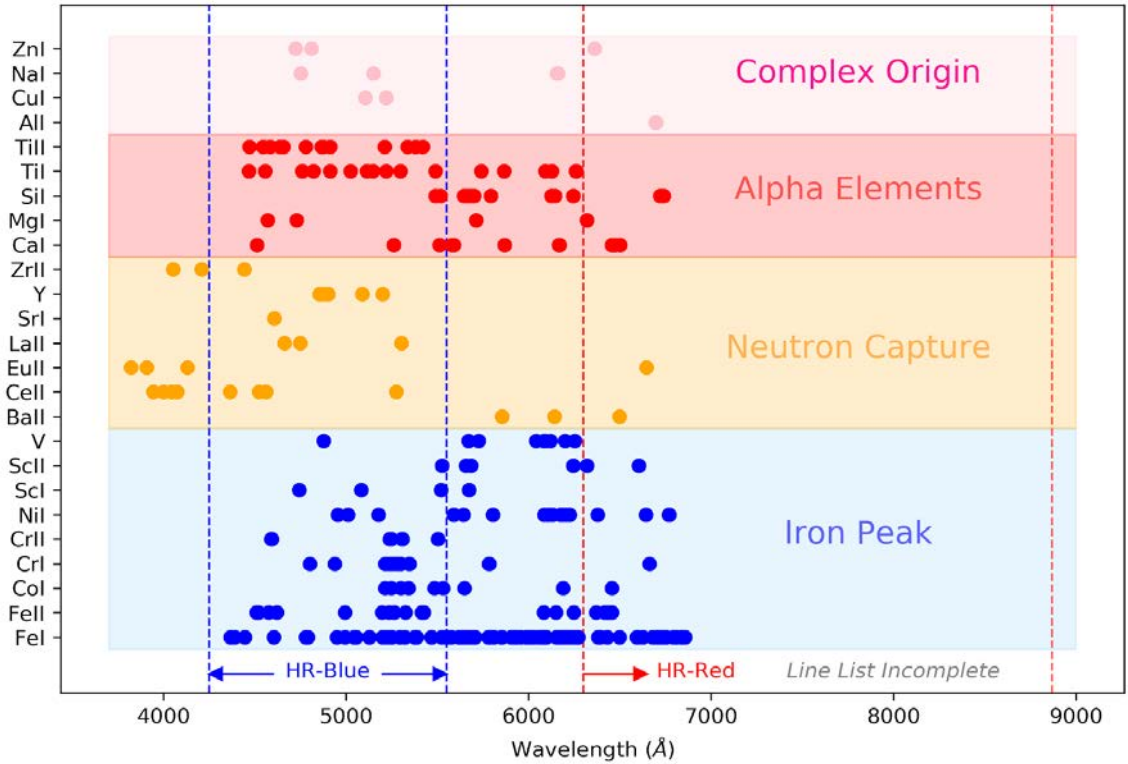

**Figure 6.3:** *Spectral features measurable by MAVIS high spectral resolution modes, indicating the same elements groups as Table 6.1: blue - Iron Peak; orange - r- and s-process neutron capture elements; red - alpha elements from core-collapse SNe; and pink - elements with more complex origin. The HR-Blue range is an excellent compromise across the nucleosynthetic channels, HR-Red is indicated, as it will be suitable for specific studies, but the line list used is not complete in that range, and is truncated at 7000Å.*

Studies of ionized gas in the local Universe ($z \approx 0$) find that disk galaxies can have velocity dispersion as low as $\sigma_{gas} \approx 10$ km s$^{-1}$, which is a fundamental lower limit set by the thermal broadening of ionized gas in HII regions. This can be measured with a spectral resolution of R~9000 or higher, with the principal spectral line of interest being Hα 6563Å. Because there are multiple bright ionized gas emission lines available (Hβ 4868Å, [OIII] 5007Å), dynamical studies of intermediate redshift galaxies can also be carried on at that resolution with spectral coverage of 6300-8700Å, with the HR-Red setup.

The resulting spectral modes of MAVIS are summarised in Fig. 6.5.

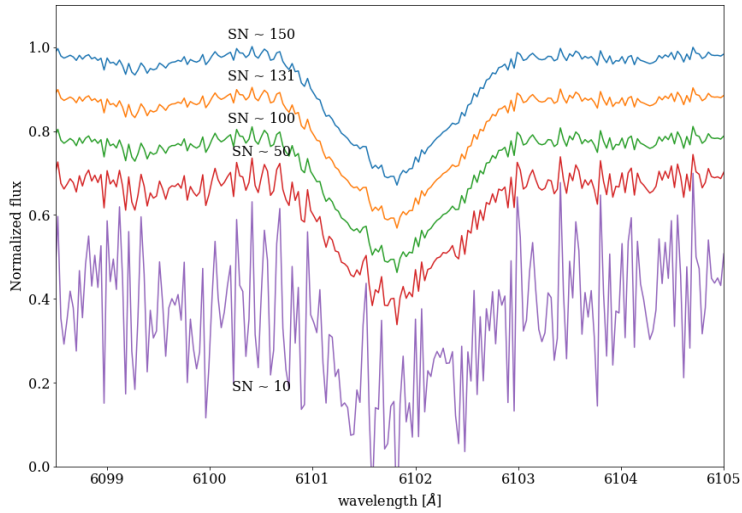

**Figure 6.4**: *Spectra of a Cepheid star at the MAVIS resolution (HR-Red mode) at different SNR.*

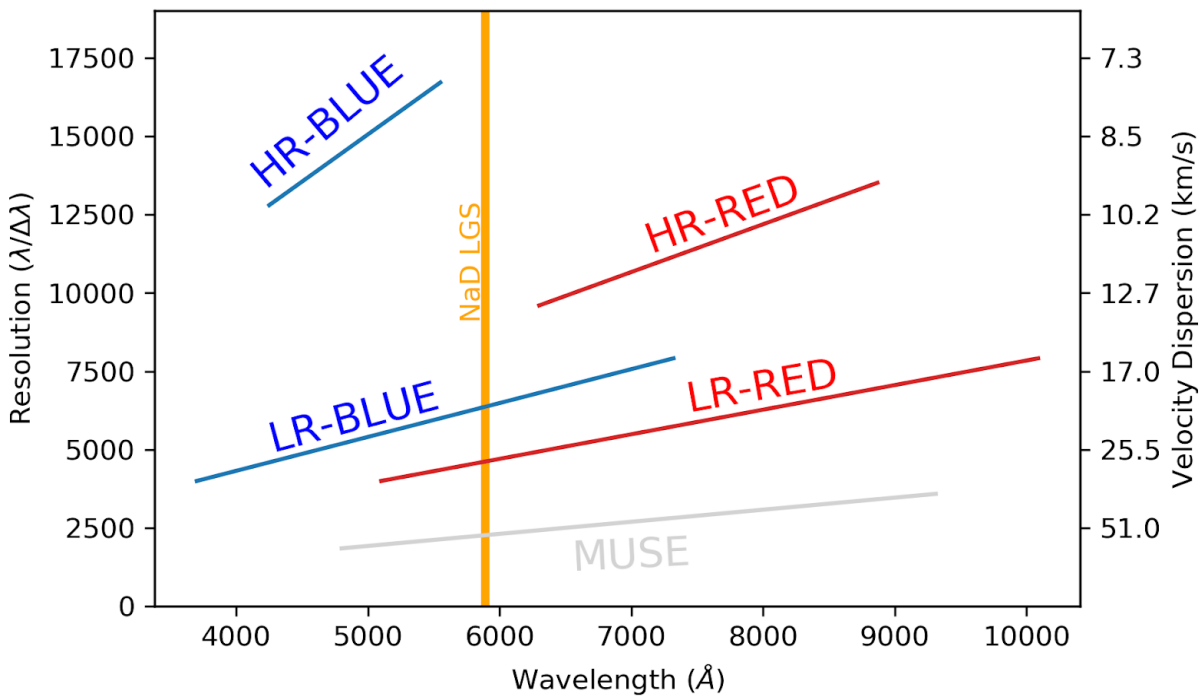

**Figure 6.5**: *Summary of MAVIS spectral modes, showing wavelength coverage and spectral resolution. Left axis indicates resolution, computed as wavelength/FWHM. Right axis shows these values in terms of velocity dispersion, indicating the minimum line widths that can be securely measured. MUSE is overplotted for comparison.*

## R-MAVIS-SCI-9 Spectroscopy Spectral Resolution

*MAVIS shall provide four spectroscopic spectral resolution configurations, as follows:*

| **R-MAVIS-SCI-9.1 LR-Blue** | *An average spectroscopic spectral resolution of R>5000 across the wavelength range 370-720 nm* |
|---|---|
| **R-MAVIS-SCI-9.2 HR-Blue** | *A minimum spectroscopic spectral resolution of R=12000 across the wavelength range 425-550 nm* |
| **R-MAVIS-SCI-9.3 LR-Red** | *An average spectroscopic spectral resolution of R>5000 across the wavelength range 510-950 nm (goal 510-1000 nm)* |
| **R-MAVIS-SCI-9.4 HR-Red** | *A minimum spectroscopic spectral resolution of R=9000 across the wavelength range 630-880 nm* |

| **Driving Science Case Chapters** | **Requirement Rationale** | **Other Relevant Chapters** |
|---|---|---|
| **R-MAVIS-SCI-9.1 LR-Blue** | | |
| 2.3 Resolving ISM variations across cosmic times | R=5000 is the minimum required for optimal sky line subtraction, and to resolve line doublets and non-circular motion through line spectral shape analysis. | 2.6, 4.4, 4.5, 5.2, 5.3, 5.4 |
| 3.5 Resolving the physics of ram-pressure stripping | Splitting the [OII] $\lambda 3727,3729\text{ÅÅ}$ doublet at z≈0 requires minimum resolution of ≈4000 | |
| 3.6 Supermassive black holes and stellar nuclei in low mass galaxies | Mass scale for black holes and nuclear star clusters is ≈$10^5 M_\odot$, corresponding to velocity dispersion $\sigma$≈20-40kms$^{-1}$ | |
| **R-MAVIS-SCI-9.2 HR-Blue** | | |
| 4.2 Galactic Globular Clusters; | The high-spectral resolution, e.g., with respect to MUSE, allows a detailed chemical and kinematical characterization of single stars in resolved stellar populations. | 3.6, 4.5, 5.3 |
| 4.4 Star clusters in the local Universe; | | |
| 5.2 Disks and jets around young stars | Rotation gradients across jets and disks are ≤15km-s, requiring R≈15000 in forbidden emission lines (e.g. [OIII]) | |
| **R-MAVIS-SCI-9.3 LR-Red** | | |
| 2.3 Resolving ISM variations across cosmic times | OH sky line subtraction >700nm; resolve emission line doublets; and detecting non-circular motion through line spectral shape analysis all require R>5000 | 2.2, 2.4, 2.6, 4.4, 4.6, 5.2, 5.3, 5.4 |
| 4.7 Origin of first star clusters at high-z | Complex Ly$\alpha$ profiles have structure on <25kms$^{-1}$ scales | |
| **R-MAVIS-SCI-9.4 HR-Red** | | |
| 2.2 Kinematics of the first disk galaxies in the epoch of Cosmic Change | A spectral resolution of R~9000 is required to measure velocity dispersions down to the thermal broadening of HII regions, $\sigma_{gas} \approx 10$ km s$^{-1}$ | 3.5, 3.6, 4.4, 5.2, 5.3 |

# 6.10 Spectroscopy Spatial Resolution

As shown by the simulations in Chapter 4, a spatial resolution of 20-25 mas allows stars to be resolved in clusters up to several Mpc. The enhanced point-source sensitivity and faint confusion limit of MAVIS will finally allow us to study galaxies star-by-star, out to the nearest galaxy groups and clusters. Thanks to the combination of MAVIS imaging capabilities with IFU spectroscopy, we will remove, through the analysis of colour magnitude diagrams, the degeneracy related to the integrated light studies, and to feed the spectroscopic analysis with direct input from the photometry.

On the other hand, fainter diffuse/extended targets, such as high-z galaxies, will benefit from a coarser spatial scale, to be able to detect lower surface brightness extended structures (since S/N is proportional to square spaxel side-length). This will allow the ISM and kinematical studies of intermediate redshift galaxies (see e.g Sect. 2.2 and 2.3), as well the detection and spectral characterization of the first galaxies in the reionization epoch (see Sect. 2.6), while maintaining relevant spatial details. A coarser scale is also well suited to faint NGS configurations, where the PSF requirement is to concentrate 15% of the energy within a 50mas aperture.

To accommodate these two regimes - finer sampling for point-source fields with bright NGS and good corrections; and coarse sampling for faint extended sources, and where the PSF correction is near the limiting faint-NGS case - we define two spaxel scales: a 'Fine' sampling mode employing 25mas spaxels (giving the 3.6"x2.5" field of view); and a 'Coarse' sampling mode with 50mas spaxels (giving a 7.2"x5.0" field of view). We note that the spectral resolution and spatial sampling are decoupled, such that all four spectral modes can operate with both spaxel sizes at the same spectral resolution.

We further note that both spatial scales have the potential to somewhat undersample the spatial PSF. This is not uncommon in IFUs, which generally aim to balance preserving both spatial information and spectral S/N (on top of the additional cost of smaller spaxels in detector space for a given field of view). As noted above, for extended sources, S/N is proportional to the square pixel side length, and astrophysically relevant surface brightnesses are sufficiently low that Nyquist sampling the diffraction limit (of any telescope size) generally results in prohibitive exposure times. Even for point sources, Nyquist sampling has a very significant impact on sensitivity. For example, in Fig. 6.6 we illustrate that Nyquist sampling the typical MAVIS PSF captures only half the ensquared energy of a 25mas spaxel for point sources, requiring >40% more exposure time for the same S/N, and with limited benefits. Such sampling gives more than an order of magnitude less flux for extended sources, pushing exposures times up by a factor 3-4.

We therefore aim to find a balance between preserving the corrected PSF, and permitting a useful sensitivity for different science applications. We also note the success of 3D extraction methods, such as 'PampelMuse' (Kamann 2019), which make use of a high-resolution, well-sampled image to centroid and extract sources in 3D from lower resolution/sampling datacubes. The combination of imaging and spectroscopy obtained with similar observing conditions, PSF, and wavelengths would make this a powerful technique to extract the most from crowded fields with MAVIS, despite non-Nyquist IFU spaxels.

The precise spaxel sizes will be further explored in Phase B as the system performance becomes clearer, and accurate numerical models of the instrument and data are developed. But the overall concept of having both a 'fine' and 'coarse' plate scale is well-motivated by the relevant science cases.

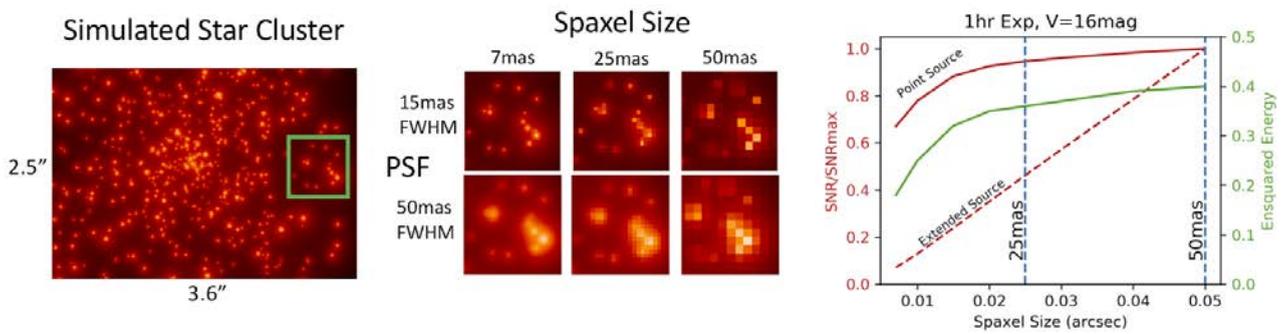

**Figure 6.6:** *Left - Simulated star cluster in the MAVIS IFU field of view. Green box shows the field used in the middle thumbnails illustrating the Fine and Coarse IFU spaxel sizes for two different PSF extremes. Right - Growth curves of S/N and ensquared energy with spaxel size for a typical MAVIS source brightness.*

| R-MAVIS-SCI-10 Spectroscopy Spatial Resolution | | |
|---|---|---|
| *MAVIS shall support two spectroscopic spatial configurations:* | | |
| **R-MAVIS-SCI-10.1 Spectroscopy Fine Sampling** | | *20-25 mas square spaxels* |
| **R-MAVIS-SCI-10.2 Spectroscopy Coarse Sampling** | | *50 mas square spaxels* |
| **Driving Science Case Chapters** | **Requirement Rationale** | **Other Relevant Chapters** |
| **R-MAVIS-SCI-10.1 Spectroscopy Fine Sampling** | | |
| 4.2 Galactic star clusters: revealing the mystery of Globular Clusters<br><br>4.4 Star clusters in the local Universe: tracers of star formation history and chemical evolution | MAVIS is the only instrument to date that will have sufficient spatial resolution to resolve and collect light from individual stars in the cores of GGCs | 3.5, 3.6, 4.4, 4.5, 4.6, 5.2, 5.3, 5.4 |
| **R-MAVIS-SCI-10.2 Spectroscopy Coarse Sampling** | | |
| 2.2 Kinematics of the first disk galaxies in the epoch of Cosmic Change<br><br>2.6 Spectroscopy of the first galaxies in the epoch of reionization | 50 mas sampling is more appropriate for diffuse, low surface brightness objects | 2.3, 2.4, 4.7, 5.2 |

## 6.11 Spectroscopy Sensitivity

A key advance of IFU spectroscopy at high angular and spectral resolutions is to measure elemental abundances in resolved stellar populations (especially in crowded regions and in star clusters) in a large variety of objects and over a wide range of distances. Some examples are shown in Table 4.2, and in the science cases illustrated in Chapter 4. These highlight the need for R=15000 spectroscopy at blue wavelengths with S/N≥10-20 using modest (~1hr) exposure times on sources including the MS of Galactic Globular clusters (MSTO at m~18) and (with even higher S/N) their HB and RGB stars (m~15 and ~12, respectively); extragalactic clusters in Magellanic clouds (typical MSTO m~20, RC stars m~19), RC and RGB stars in the star clusters of nearby galaxy up to ~150 kpc (see, e.g. the science case on the star cluster population in the dSph Fornax), and finally O and B stars up to ~2 Mpc and ~1 Mpc, respectively.

Measuring accurate global metallicity and [alpha/Fe] in the crowded star clusters of local Universe galaxies (< few Mpc, see Table 4.2) requires comparable quality spectroscopy with R≈5000 on individual main sequence OB stars in young clusters, having m~21-22.

Limiting magnitudes for MAVIS spectroscopy at various wavelengths are provided in Appendix A. Considering the 25mas 'Fine' spaxel size with a 3x3 spatial binning on point sources, S/N=10 can be achieved in one hour with the HR-Blue (R=15000) mode on m~19.6 sources, well matched to the Milky Way and Local Group sources. For a similar observation, LR-Blue and LR-Red modes provide R=5000 spectroscopy at S/N=10 for sources of m~22.6 (@500nm) and m~23.3 (@700nm), respectively.

In terms of extended sources, for a typical main-sequence galaxy observed at z=0.4, a 3 hour on-source exposure with MAVIS will be sufficient to detect Hα emission at SN > 3 per spaxel out to the disk scale radius using the low-resolution spectral mode, allowing spatial resolved studies of kinematics and ISM variation properties in galaxies in the main epoch of cosmic change. Moreover, the foreseen sensitivities will allow to detect Lyα line emission from the galaxies in the epoch of reionization, with typical line fluxes down to 5x10$^{-18}$ erg s$^{-1}$cm$^{-2}$.

| R-MAVIS-SCI-11 Spectroscopy Sensitivity | | |
|---|---|---|
| *MAVIS, in a one hour exposure time, shall be able to detect an isolated continuum point source having an effective magnitude at 550 nm of mAB=21.0 (goal: mAB=21.5), with a SNR≥10 per resolution element.* | | |
| **Driving Science Case Chapters** | **Requirement Rationale** | **Other Relevant Chapters** |
| 2.3 Resolving ISM variations across cosmic times | Spatially resolved spectroscopy of typical z~0.4 galaxy, with SNR>3 per spaxel in Hα out to the effective radius | 2.2, 2.4, 2.6, 3.4, 3.5, 3.6, 4.2, 4.5, 4.6, 4.7, 5.2, 5.3, 5.4 |
| 4.4 Star clusters in the local Universe: tracers of star formation history and chemical evolution | Detect and resolve single stars in extragalactic clusters in the Local Group at R=15000, and out to ~3 Mpc with R=5000 | |

## 6.12 Summary of baseline design compliance with science cases

As described in Section 1, prior to Phase A the MAVIS Consortium ran a 'White Paper' process to gather potential science cases for an instrument with high angular resolution optical imaging and spectroscopy capabilities, guided by (but not limited to) the ESO Top Level Requirements (TLRs) available at that time. This process was not prescriptive about instrument requirements, and allowed the community to explore the instrument capabilities they would need to best accomplish their science goals. Over the course of Phase A, this broad set of capabilities has been distilled to create the MAVIS baseline requirements. This has been done in consultation with the Consortium science team and broader MAVIS community, in particular via weekly science team meetings, and interactive community workshops in Australia and Italy.

Tab. 6.2 presents a summary of the requested instrument capabilities resulting from the whole White Paper process, where science cases have been thematically grouped for presentation purposes. The relevant TLRs from ESO are also indicated, as are the MAVIS baseline design specifications. The Table shows how the MAVIS current baseline is compliant not only to the science cases presented in this document, but also with the vast majority of the science presented in the white papers in the pre-phase A stage.

This is highlighted in Tab. 6.3, where a more detailed view of how compliant the MAVIS baseline design is with each of the white papers. Overall, the baseline design has a high level of compliance across these reference science cases. Imaging satisfies essentially all cases in full, with two cases preferring a larger field of view, but are still feasible with additional pointings. For the spectrograph, there is also a very high level of compliance (>75%), considering the diversity of science cases proposed. Where compliance is considered 'partial' or lower for the spectrograph spatial mode, White Papers indicate a benefit from spatial multiplexing, either as aperture MOS or Multi-IFU. For spectral modes, reduced compliance was generally because a lower spectral resolution was preferred. However, in most of those cases the science remains feasible at lower observing efficiency: with additional pointings, since a larger spatial multiplex was requested, and/or integration time, when lower spectral resolution was preferred. 'Total' compliance scores for imaging and spectroscopy modes are also provided, which represents a combination of the different characteristics, considering also the severity of the impact on the actual science. Some cases have a high dependence on multiplexing, for example, whereas for others, the emphasis is on spectral resolution. This further confirms the exceptional general-purpose nature of MAVIS, and its potentiality in fulfilling exciting science in different fields of astronomy.

Tab. 6.3 also highlights the fact that most (32/57=56%) MAVIS white papers require **both** imaging *and* spectroscopy capabilities of MAVIS. Indeed, *less than a quarter of the White Papers can be addressed with only imaging*. We stress this point for obvious reasons. A key scientific value of MAVIS is in fact the ability to combine imaging and spectroscopy on the same target fields, with the same angular resolution, and with extensive wavelength and spectral resolution coverage.

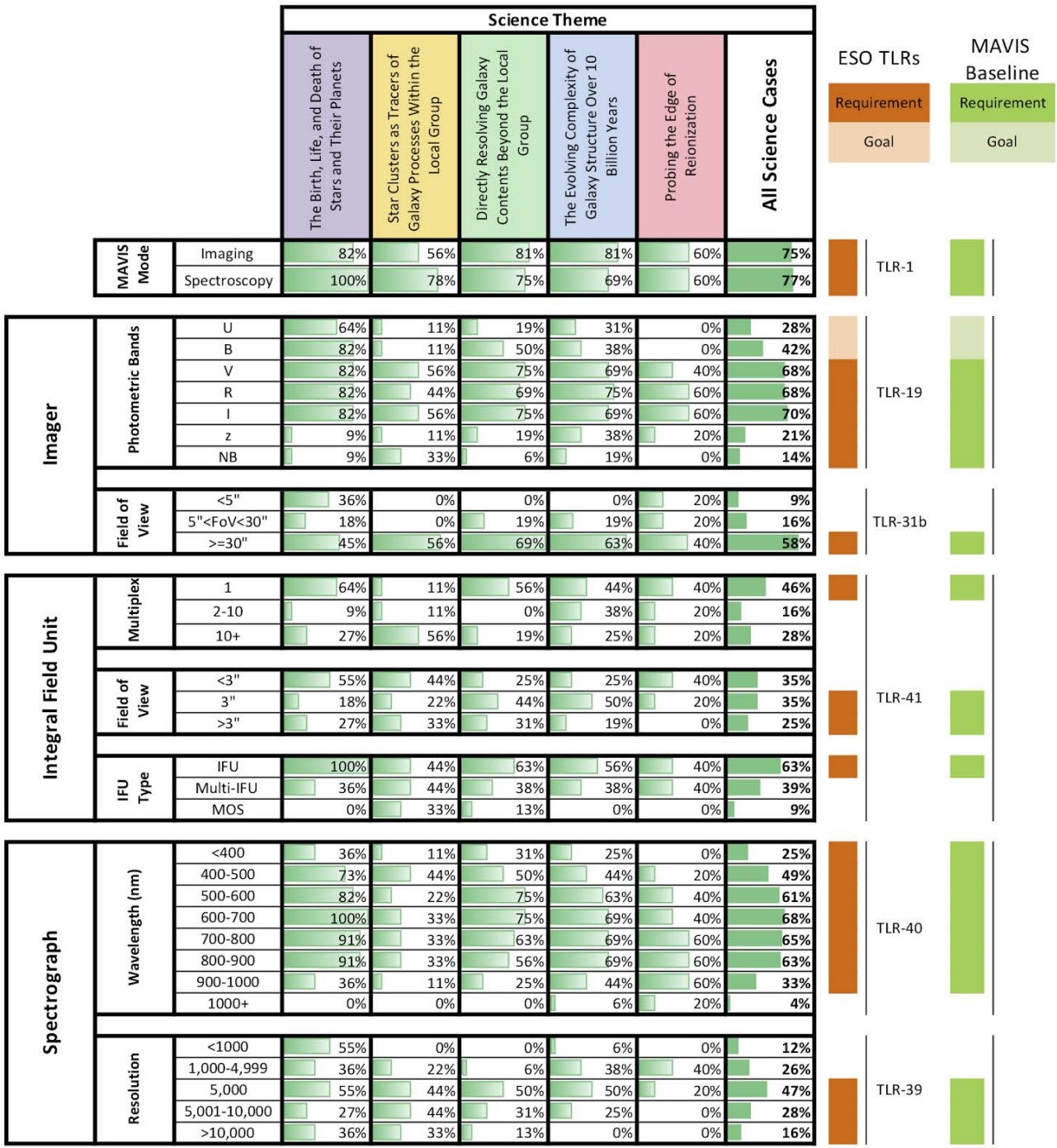

| | | Science Theme | | | | | | ESO TLRs | MAVIS Baseline |
|---|---|---|---|---|---|---|---|---|---|
| | | The Birth, Life, and Death of Stars and Their Planets | Star Clusters as Tracers of Galaxy Processes Within the Local Group | Directly Resolving Galaxy Contents Beyond the Local Group | The Evolving Complexity of Galaxy Structure Over 10 Billion Years | Probing the Edge of Reionization | All Science Cases | | |
| MAVIS Mode | Imaging | 82% | 56% | 81% | 81% | 60% | 75% | TLR-1 | |
| | Spectroscopy | 100% | 78% | 75% | 69% | 60% | 77% | | |
| Imager — Photometric Bands | U | 64% | 11% | 19% | 31% | 0% | 28% | | |
| | B | 82% | 11% | 50% | 38% | 0% | 42% | | |
| | V | 82% | 56% | 75% | 69% | 40% | 68% | | |
| | R | 82% | 44% | 69% | 75% | 60% | 68% | TLR-19 | |
| | I | 82% | 56% | 75% | 69% | 60% | 70% | | |
| | z | 9% | 11% | 19% | 38% | 20% | 21% | | |
| | NB | 9% | 33% | 6% | 19% | 0% | 14% | | |
| Imager — Field of View | <5" | 36% | 0% | 0% | 0% | 20% | 9% | | |
| | 5"<FoV<30" | 18% | 0% | 19% | 19% | 20% | 16% | TLR-31b | |
| | >=30" | 45% | 56% | 69% | 63% | 40% | 58% | | |
| Integral Field Unit — Multiplex | 1 | 64% | 11% | 56% | 44% | 40% | 46% | | |
| | 2-10 | 9% | 11% | 0% | 38% | 20% | 16% | | |
| | 10+ | 27% | 56% | 19% | 25% | 20% | 28% | | |
| Integral Field Unit — Field of View | <3" | 55% | 44% | 25% | 25% | 40% | 35% | | |
| | 3" | 18% | 22% | 44% | 50% | 20% | 35% | TLR-41 | |
| | >3" | 27% | 33% | 31% | 19% | 0% | 25% | | |
| Integral Field Unit — IFU Type | IFU | 100% | 44% | 63% | 56% | 40% | 63% | | |
| | Multi-IFU | 36% | 44% | 38% | 38% | 40% | 39% | | |
| | MOS | 0% | 33% | 13% | 0% | 0% | 9% | | |
| Spectrograph — Wavelength (nm) | <400 | 36% | 11% | 31% | 25% | 0% | 25% | | |
| | 400-500 | 73% | 44% | 50% | 44% | 20% | 49% | | |
| | 500-600 | 82% | 22% | 75% | 63% | 40% | 61% | | |
| | 600-700 | 100% | 33% | 75% | 69% | 40% | 68% | TLR-40 | |
| | 700-800 | 91% | 33% | 63% | 69% | 60% | 65% | | |
| | 800-900 | 91% | 33% | 56% | 69% | 60% | 63% | | |
| | 900-1000 | 36% | 11% | 25% | 44% | 60% | 33% | | |
| | 1000+ | 0% | 0% | 0% | 6% | 20% | 4% | | |
| Spectrograph — Resolution | <1000 | 55% | 0% | 0% | 6% | 0% | 12% | | |
| | 1,000-4,999 | 36% | 22% | 6% | 38% | 40% | 26% | | |
| | 5,000 | 55% | 44% | 50% | 50% | 20% | 47% | TLR-39 | |
| | 5,001-10,000 | 27% | 44% | 31% | 25% | 0% | 28% | | |
| | >10,000 | 36% | 33% | 13% | 0% | 0% | 16% | | |

**Table 6.2:** *Summary of White Paper instrument requirements, showing frequency of occurrence of each capability across the different science categories. On the right are bands indicating the ESO TLRs (orange), and MAVIS Phase A baseline design (green).*

| Group | White Paper Title | Ima | Spec | Filters | FoV | Total | Fine | Coarse | LR-Blue | LR-Red | HR-Blue | HR-Red | Total |
|---|---|---|---|---|---|---|---|---|---|---|---|---|---|
| The Birth, Life, and Death of Stars and Their Planets | Star and planet formation at low metallicities | | | 5 | 5 | 5 | 4 | | 5 | 5 | | | 4 |
| | Jets from young stars at high angular resolution: the launching mechanism as a solution to the angular momentum problem. | | | 5 | 5 | 5 | 5 | | | | 5 | 5 | 5 |
| | The outer Solar System: Icy Giants | | | | | | | 5 | 3 | 3 | | | 4 |
| | High Dispersion Coronagraphy for Exoplanet Characterization | | | | | | 5 | | | | | 3 | 3 |
| | Revealing the structure and evolution of the second generation protoplanetary disks around evolved stars with MAVIS | | | 5 | 5 | 5 | | | 5 | 5 | | | 5 |
| | Comets science with MAVIS | | | 5 | 5 | 5 | | 5 | 3 | 3 | | | 4 |
| | White Dwarf Binaries with MAVIS | | | 5 | 5 | 5 | 4 | 4 | | | | | 4 |
| | Observations of planetary rings with MAVIS | | | 5 | 5 | 5 | | | 3 | 3 | | | 3 |
| | The outer Solar System: KBO and Comets | | | 5 | 5 | 5 | 4 | | 5 | 5 | 5 | 5 | 5 |
| | Mass accretion in low-metallicity environments of the outer Galaxy | | | 5 | 5 | 4 | 4 | | 5 | 5 | | | 5 |
| | Frequency of accreting sub-stellar companions of low-mass stars | | | 5 | 5 | 5 | 4 | | 5 | 5 | | | 5 |
| Star Clusters as Tracers of Galaxy Processes Within the Local Group | Dynamics of resolved stellar clusters in the local group, using astrometry "calibrated" by Gaia and HST | | | 5 | 5 | 5 | | | | | | | |
| | Age, multiple populations, and stellar activity in resolved stellar clusters and stellar populations in the Galaxy with narrow-band filters between 370-450 nm | | | 5 | 5 | 5 | | | | | | | |
| | CNO abundances of Extreme Horizontal Branch stars in globular clusters | | | | | | 4 | | 5 | | 5 | | 5 |
| | Searching for Intermediate Mass BHs in Local dwarf spheroidal galaxies | | | 5 | 5 | 5 | 4 | | | 5 | | 5 | 5 |
| | Cluster white dwarfs: where the eagles dare | | | | | | 4 | | 5 | | 5 | | 5 |
| | Probing the cores of globular clusters: Abundance anomalies on the unevolved main sequence stars and impact of binarity | | | | | | 3 | | | | 5 | | 4 |
| | Testing the Universality of the Initial Mass Function with Photometry of Open Clusters in the Magellanic Clouds | | | 5 | 5 | 5 | 3 | | 5 | 5 | 5 | 5 | 5 |
| | Archaeology of the oldest stars in nearby dwarf galaxies | | | | | | 2 | | 5 | 5 | 5 | 5 | 3 |
| | Young Massive Clusters in Metal-Poor Starburst Dwarf Galaxies | | | 5 | 5 | 5 | 4 | | | 5 | | 5 | 5 |
| Directly Resolving Galaxy Contents Beyond the Local Group | Direct detection of UV-upturn hot, old stars in nearby ETGs and bulges. | | | 5 | 5 | 5 | | | | | | | |
| | Dark Matter in the smallest Dwarf Galaxies | | | 5 | 5 | 5 | 4 | 4 | 5 | 5 | | | 5 |
| | Core-collapse supernovae with MAVIS | | | 5 | 5 | 5 | | | | | | | |
| | Tracing the chemical evolution of nearby galaxies with star clusters | | | 5 | 5 | 5 | 4 | 4 | 5 | 5 | 5 | 5 | 5 |
| | The MAVIS perspective of nuclear stellar disks | | | 5 | 5 | 5 | 5 | 5 | 5 | 5 | | | 5 |
| | Resolving Super-Massive Black Holes in Compact, Low-Mass Galaxies | | | 5 | 5 | 5 | 5 | 5 | 5 | 5 | | 5 | 5 |
| | Resolved stellar populations in distant galaxies | | | 5 | 5 | 5 | | | | | | | |
| | Studying a new mode of star formation: positive feedback in galactic outflows | | | | | | 5 | 5 | 5 | 5 | | 5 | 5 |
| | The physical properties and the origin of classical bulges and pseudo-bulges in the local Universe | | | 5 | 5 | 5 | 5 | 5 | 5 | | | | 5 |
| | Dynamical measurements of supermassive black holes and nuclear star clusters with MAVIS | | | 5 | 5 | 5 | 5 | 5 | 5 | 5 | | | 5 |
| | The origin of the Hubble Sequence | | | 5 | 5 | 5 | 5 | 5 | 5 | | | | 5 |
| | Resolving the physics of ram-pressure stripping | | | 5 | 5 | 5 | 4 | 5 | 5 | | | | 5 |
| | Spectroscopy of distant Cepheids with MAVIS: A detailed chemical mapping of spiral galaxies in the Local Universe | | | 5 | 5 | 5 | 3 | 3 | | 5 | | 5 | 4 |
| | Resolved Angular Momentum with MAVIS: towards a more physical morphological classification | | | 5 | 5 | 5 | 5 | 5 | | | | | 5 |
| | Evolution of the Interstellar Medium with MAVIS | | | 5 | 5 | 5 | 4 | 5 | 5 | 5 | | | 5 |
| | Resolved stellar populations and star formation histories in the Local Universe: star-forming dwarf galaxies in the field and in clusters | | | 5 | 5 | 5 | | | | | | | |
| The Evolving Complexity of Galaxy Structure Over ~10 Billion Years | Resolving turbulence and the emergence of thin disks | | | 5 | 5 | 5 | 5 | 5 | 4 | 4 | | 5 | 5 |
| | Morphological evolution of galaxies in clusters. Bulge-disk decomposition | | | 5 | 5 | 5 | 4 | 4 | 5 | 5 | | 4 | 4 |
| | Star-forming clumps | | | 5 | 5 | 5 | | | | | | | |
| | Mass growth and size evolution of early type galaxies | | | 5 | 5 | 5 | | | | | | | |
| | Ram-pressure stripping in intermediate-z clusters | | | | | | 5 | 5 | | 5 | | | 5 |
| | Dissecting the morphological evolution through cosmic time and environment | | | 5 | 5 | 5 | | | | | | | |
| | Extreme Star Forming Regions in Galaxies at Cosmic Noon and the Nearby Universe | | | 5 | 5 | 5 | | | | | | | |
| | Probing the evolution of the star formation in galaxies up to z = 0.5 – 1.5 | | | 5 | 5 | 5 | | | | | | | |
| | Hunting absorption galaxies in emission | | | 5 | 5 | 5 | 5 | 5 | 5 | | 5 | | 5 |
| | Resolving the most compact massive galaxies across the cosmic time | | | 5 | 5 | 5 | 5 | 5 | 5 | | | | 5 |
| | Testing Cold Dark Matter with Gravitational Lensing | | | 5 | 5 | 5 | 5 | 5 | 5 | 5 | | | 5 |
| | Exploiting strong lensing clusters with MAVIS | | | 5 | 3 | 5 | 4 | 4 | 5 | 5 | | | 5 |
| | Probing stellar clumps in distant star-forming galaxies around z ~ 1 | | | 5 | 5 | 5 | 4 | 4 | 5 | | | 5 | 4 |
| | Investigating the nature of Dark Matter through galaxy dynamics at z ~ 1 – 2 with MAVIS | | | | | | 4 | 4 | | 5 | | | 4 |
| | Versatile spectroscopy on large field-of-view with MCAO | | | 5 | 3 | 4 | 4 | 4 | 5 | 5 | 5 | 5 | 5 |
| | Spatially Resolved Metal Gas Clouds | | | | | | 5 | 5 | | 5 | | | 5 |
| Probing the Edge of Reionization | A MAVIS deep field | | | 5 | 5 | 5 | | | | | | | |
| | Discovering the Nature of Superluminous and Super-Early Supernovae with MAVIS | | | | | | 5 | 5 | | 5 | | | 5 |
| | UV morphology of galaxies up to z≈°6: clumps and morphological parameters | | | | | | 4 | 4 | | 5 | | | 4 |
| | Hosts of the brightest explosions as probes of galaxy formation and evolution | | | 5 | 5 | 5 | 5 | 5 | 5 | 5 | | | 5 |
| | UV-imaging survey to resolve the morphology of z ~ 1 – 5 galaxies | | | 5 | 5 | 5 | | | | | | | |

| | Ima | Spec |
|---|---|---|
| Total fraction of cases using this mode | 79% | 77% |
| Fraction of cases using **only one mode** | 23% | 21% |
| Fraction of cases using **both modes combined** | | 56% |

**Legend**

| | |
|---|---|
| 5 | **Fully Compliant:** Science is feasible, observations are optimal |
| 4 | **Partially Compliant:** Science is feasible, but observing is less compliant |
| 3 | **Moderately Compliant:** Science is feasible, but observing is inefficient |
| 2 | **Poorly Compliant:** Science is partially feasible, but observing is inefficient |
| | **Non-Compliant:** Science is not feasible |
| | **Not Applicable:** Specific instrument mode is not needed |

**Table 6.3:** *Evaluation of how the MAVIS baseline design aligns with the full set of MAVIS White Papers, considering the main imaging and spectroscopy characteristics. Imaging mode is fully compliant with all but 2 White Papers (due to field of view size). Spectroscopy mode is fully compliant in >75% of cases. For the less compliant cases, the science remains feasible with additional pointings (since the requested a larger spatial multiplex) and/or integration time (preferring lower spectral resolution).*

# 7 Competitiveness and Complementarity

## 7.1 Sky coverage

MAVIS sits in a unique part of parameter space for general-purpose facility capabilities in terms of combining both the light gathering power AND the full angular resolution of an 8m telescope at optical wavelengths. Key to this competitive edge will be the ability for MAVIS to achieve this performance across a significant portion of the sky. As highlighted above, achieving the highest sky coverage is a key priority for MAVIS, and will set it apart in all respects from other AO instruments working in this wavelength regime. This is particularly impactful for the IFU spectroscopy, as compared with the current capabilities of MUSE Narrow Field Mode, for example, where the relatively bright (H<14) AO natural guide star must be within a radius of 3.25" of the science target, drastically reducing sky coverage.

An upgrade of IRLOS, the Infrared Shack-Hartmann MUSE wavefront sensor, is planned in the near future. This should significantly improve the limiting magnitude of the natural guide star, allowing the use of reference stars up to H<17 inside a patrol field of radius 3.35" from the science target. The upgraded MUSE wavefront sensor could possibly operate in deep mode as well, reaching H<18.5 when using unresolved targets. Although the performance in this mode is still to be evaluated, we are including this MUSE-NFM upgrade-faint in our comparisons for reference. Moreover, the MUSE-NFM patrol field can be possibly extended by pushing the science target into one of the instrument field of view corners. In this way, a separation between the target and the guide star of r<11.2" can be achieved for point sources.

To properly compare the target accessible to MAVIS and MUSE-NFM, we explore the number of galaxies in the GOODS-S, COSMOS, and UDS fields at intermediate redshift (0.1 < z < 0.5) with suitable NGS stars within the MAVIS and MUSE patrol field. In this case, we assumed an allowed guide star-galaxy separation for MUSE-NFM r < 9.08", to provide the same spatial coverage as MAVIS with the smaller pixel scale (i.e. ~3"x3" box). As shown in Fig. 7.1, out of a parent sample of ~1500 galaxies, MAVIS can observe 653 of them (~43%), while no galaxy is observable by the current MUSE-NFM. This number should increase to 11 targets with the IRLOS upgrade (0.7%), and to 44 galaxies with the best possible performances (3%), still very far from the potential new window opened by MAVIS on the high spatial resolution observations of galaxies at optical wavelengths.

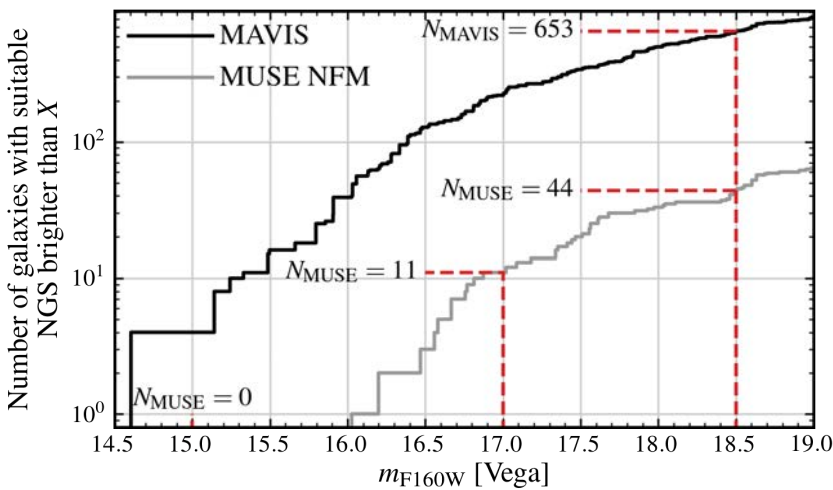

*Figure 7.1:* Comparison between the number of observable intermediate redshift galaxies in GOODS-S, COSMOS and UDS with MAVIS (black line) and MUSE (gray line) as a function of the NGS guide star magnitude. For MAVIS at least 3 stars within 60" of the field centre are required. The plotted magnitude is that of the third brightest star.

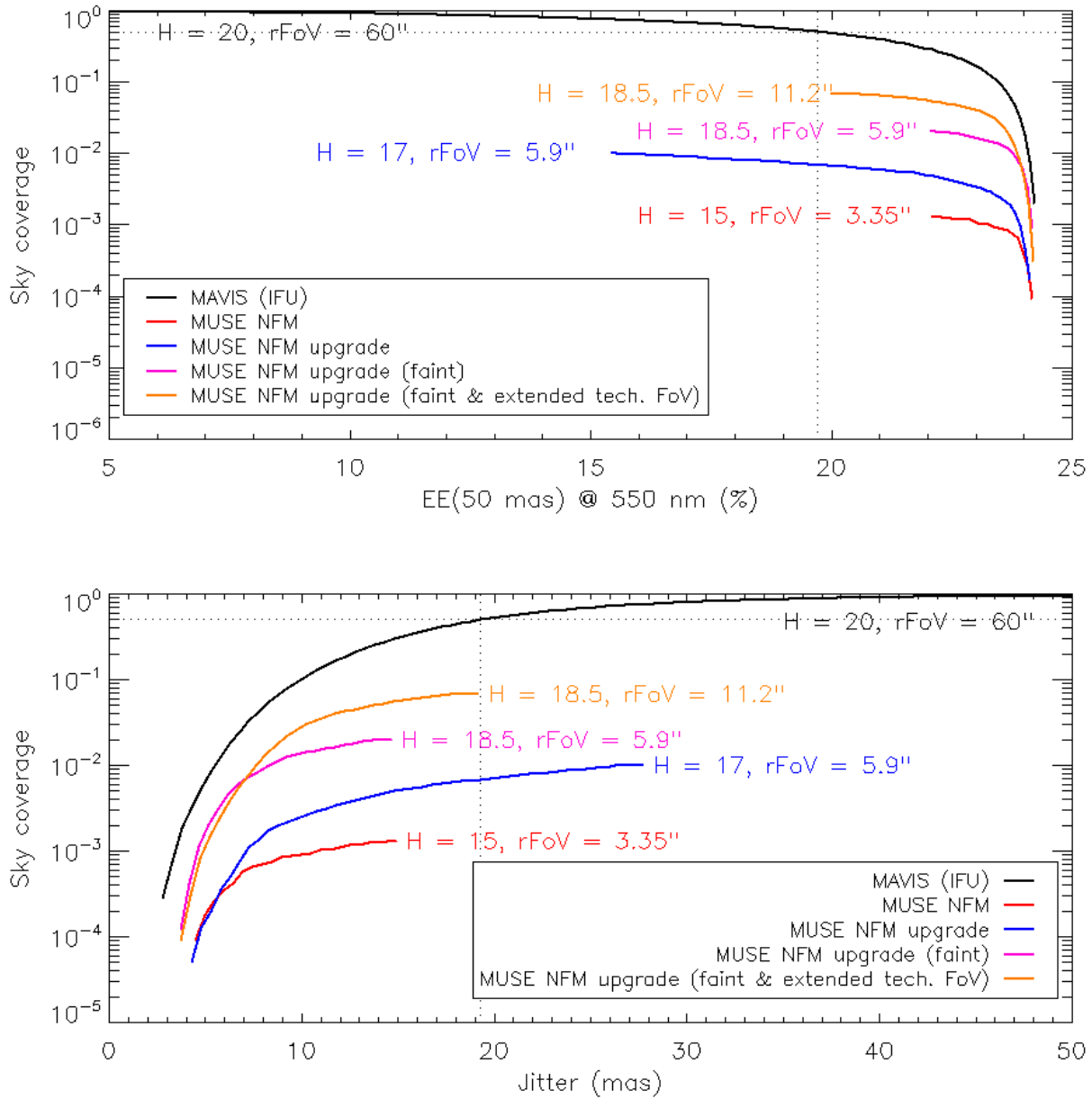

**Figure 7.2**: *The sky coverage for MAVIS (black line) and for MUSE-NFM (red line) in a 3°x3° deg field at the Galactic pole as a function of the obtained encircled energy in a 50 mas spaxel (upper panel), and as a function of the residual jitter (lower panel). The huge improvement of MAVIS sky coverage is evident. A similar comparison is also shown for the expected performances of the IRLOS MUSE planned upgrade (blue line), and for the goal performances of the MUSE upgrade with the nominal patrol field (rFoV=3.35", pink line) and extending the patrol field pushing in a corner of the field of view the (point-like) target (rFoV=11.2"; orange line).*

A limitation of this simulation is that the impact on the system performances of the shape and distance of the guide star asterisms for MAVIS, as well as of the distance and magnitude of the NGS for MUSE-NFM is not taken into account. A more realistic comparison is shown in Fig. 7.2, where the sky coverage in a 3°x3° deg field at the Galactic pole is shown as a function of Encircled Energy (EE) in a 50mas spaxel at 5500Å (top panel) and as a function of residual jitter, for both MAVIS and MUSE-NFM. Even in this case it can be seen how ~50% of sky coverage is reached for MAVIS at >15% EE, while the upgraded MUSE-NFM will be roughly two orders of magnitude worse at the same performance and science field. This clearly demonstrates that MAVIS will be an unique instrument, capable of targeting roughly half of the possible targets even in the well known extragalactic deep fields, carried out in sky regions devoid of bright stars. This is a fundamental capability to allow the characterization of the full population of astrophysical objects of interest on statistically significant samples. It also justifies why MAVIS is considered a general purpose facility instrument, with both imaging and comprehensive spectroscopic capabilities to address the wide range of science opportunities enabled by this sky coverage.

## 7.2 Imaging Capability

Another major advantage offered by MAVIS over existing optical AO-fed instruments is the capability of MCAO to provide a large field of view, with high uniformity and quality of correction. In Table 2 we compare MAVIS imaging mode with currently available instruments offering AO-fed optical imaging capabilities. For simplicity, we define the number of independent spatial elements as the field of view area divided by the area of a square patch of side length the diffraction limit of the telescope at 500nm. We note that this metric makes no accounting for the strong off-axis degradation of the PSF in single conjugate AO systems, and for those instruments represents an upper limit.

| Telescope | Visible Performance (500nm) | | | | | IR Performance (2.2um) | | | |
|---|---|---|---|---|---|---|---|---|---|
| | HST | Magellan | LBT | VLT/UT3 | VLT/AoF | ELT | TMT | GMT | JWST |
| Instrument | WFC3 / UVIS | MAG-AO / VisAO | ForeRunner / SHARK-VIS | SPHERE / ZIMPOL | MAVIS Imaging | MICADO | IRIS | GMTIFS Imaging | NIRCam |
| Field of View | 162"x162" | 8"x8" | 10"x10" | 3.5"x3.5" | 30"x30" | 50"x50" | 34"x34" | 20.4"x20.4" | 127"x127" |
| Sampling | 40mas | 8mas | 6.5mas | 7mas | 7mas | 4mas | 4mas | 5mas | 31mas |
| Resolution Element | 80mas | 20mas | 15mas | 15mas | 15mas | 14mas | 19mas | 22mas | 85mas |
| #Independent Spat. Elem. | 4.00E+06 | 1.60E+05 | 4.44E+05 | 5.44E+04 | 4.00E+06 | 1.28E+07 | 3.20E+06 | 8.60E+05 | 2.23E+05 |

***Table 7.1:*** *Evaluation of existing high-angular resolution optical (blue region) and infrared facilities (red region) in comparison to MAVIS (bold column). For simplicity, we assume that each instrument's Resolution Element is given by the diffraction limit of its respective telescope at 500nm (with the exception of HST/WFC3, which is undersampled, so we assume 2 pixels). Dividing the Field of View area by the Resolution Element gives the Number of Independent Spatial Elements as a key figure of merit. MAVIS far exceeds any existing ground-based facility by this measure, and compares well with the expected performance of ELT imagers in the infrared.*

Table 7.1 also includes equivalent information for future infrared instruments on ELTs and JWST. This shows that MAVIS, for optical wavelengths, sits in the same regime of number of independent spatial elements as these major facilities do in the infrared, further emphasising the synergy MAVIS will have there in terms of complementary information content.

We can explore the importance of complementarity further by considering the simulated performance of the key facilities in Table 7.1. Figure 7.3 presents simulated observations of an early-type galaxy at roughly one effective radius, and at the distance of Centaurus A (approximately 4Mpc). These were generated using the online Advance Exposure Time Calculator (AETC, Falomo et al. 2011, http://aetc.oapd.inaf.it) with default settings for each instrument, assuming 10 hours total integration for all cases. The first four panels show different facilities, each observing in the I-band. HST suffers from its smaller aperture, and JWST from its relatively coarse sampling and non-diffraction limited performance shortward of 1 micron. MICADO also exhibits strong PSF halos at these wavelengths, limiting sensitivity to the faintest stars. MAVIS provides a significantly sharper and well-sampled PSF by comparison, and subsequently greater point-source depth. The quality of the MAVIS image is very comparable to the MICADO performance at K-band, however, highlighting the real power of complementarity that MAVIS can bring.

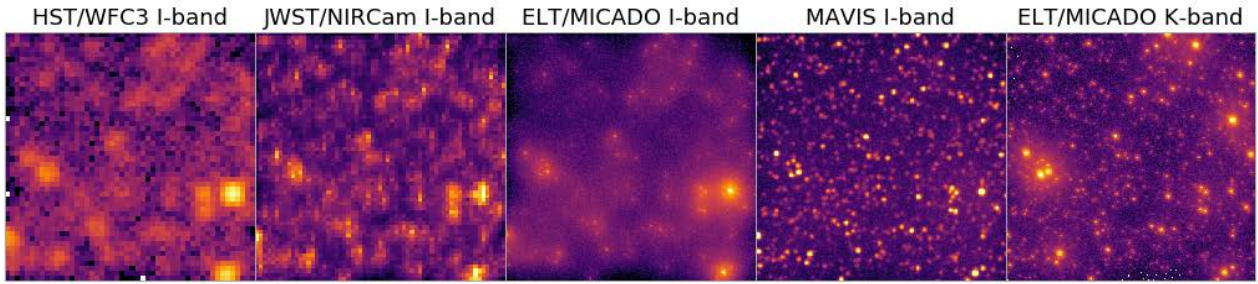

**Figure 7.3:** *Simulated 10-hour integration of a region of an early type galaxy at a distance of 4Mpc, corresponding to a surface brightness of 22 mag/sq.arcsec, taken with different facilities as indicated. The four leftmost images are for an I-band filter, chosen as it overlaps between all four facilities. The rightmost image is at K band. MAVIS provides a significant improvement over the others at I band, and delivers image quality and depth that complements very well that of ELT/MICADO in the infrared.*

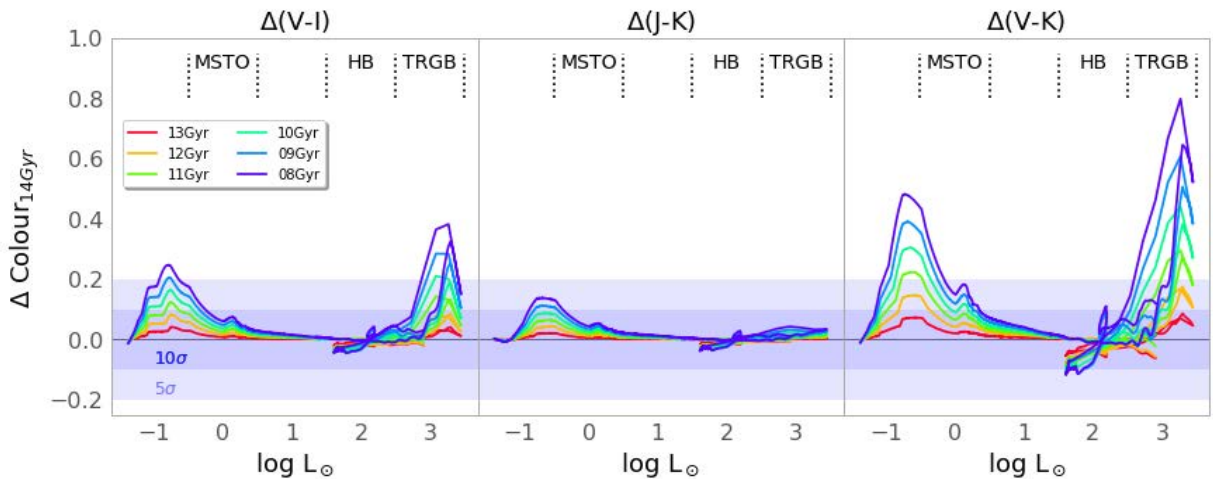

**Figure 7.4:** *Coloured lines show the difference in broad-band colours expected between a 14Gyr stellar population and younger populations as shown (all at solar metallicity). Larger deviations indicate stronger colour changes that will be easier to measure. Left panel: V-I colours measurable with MAVIS, Middle panel: J-K colours from e.g. ELT/MICADO. Right panel: Combination of optical and infrared colours. Shaded regions indicate the corresponding 10σ and 5σ photometric error level. TRGB, HB, and MSTO correspond to key characteristic regions of the Hertzsprung-Russell Colour-Magnitude Diagram, namely the Tip of the Red Giant Branch, the Horizontal Branch, and the Main Sequence Turn Off, respectively. For the expected sensitivity of MAVIS, these regions are detectable at distances of approx. 15Mpc, 5Mpc, and 1Mpc, respectively, at the 5σ level or better.*

This is shown further in Figure 7.4, which plots the change in colour between (solar metallicity) stellar population models differing in age from a 14 Gyr model in increments of 1 Gyr, as a function of stellar luminosity. This corresponds to the required relative precision in colour-magnitude diagram to distinguish these models. At the bright end (TRGB = Tip of the Red Giant Branch), optical colours accessible with MAVIS (left panel) can distinguish a 4 Gyr age difference with a S/N of 5 sigma. Comparable precision from infrared colours alone is challenging at these old ages (middle panel). But the combination of infrared and optical colours (right panel) dramatically increases the dynamic range of colour variations with age. This simple example demonstrates the power of combining optical data from MAVIS on the VLT with infrared imaging at comparable resolution and depth from the ELT.

## 7.3 Spectroscopy Capability

The recently commissioned MUSE Narrow Field Mode (NFM) has already demonstrated the utility of near-diffraction limited image quality delivered to a sensitive and powerful integral field spectrograph. In addition to the significantly more capable AO system and sky coverage than MUSE (even after the IRLOS upgrade), MAVIS offers distinct new capabilities over MUSE in several ways:

**Spectral resolution:** Probing small angular scales in most cases pushes to lower mass regimes, which in turn requires higher spectral resolution to probe the relevant kinematic scales. ESO TLRs for MAVIS set a minimum of $\lambda/\Delta\lambda$=5,000 – roughly twice the resolution of MUSE, translating directly to increased sensitivity to lower mass systems through higher precision radial velocity measurements, as well as improved capabilities for emission and absorption line strength and shape measurements. The MAVIS White Papers reinforce the utility of this spectral resolution regime, though optional lower resolution modes would also satisfy a number of science cases.

As an example, high spectral resolution is needed to adequately identify thin disks with low velocity dispersion ($\sigma_{gas} \approx 10-25$ km s$^{-1}$) at intermediate redshift, and to study their line profiles to properly disentangle the different kinematics components (see e.g. Fig. 7.5). Studies of ionized gas in the local Universe ($z \approx 0$) find that disk galaxies can have velocity dispersion as low as $\sigma_{gas} \approx 10$ km s$^{-1}$, which is a fundamental lower limit set by the thermal broadening of ionized gas in HII regions. This requires a spectral resolution of R~10000: JWST and MUSE may have sufficient sensitivity, but they will not have sufficient spectral resolution.

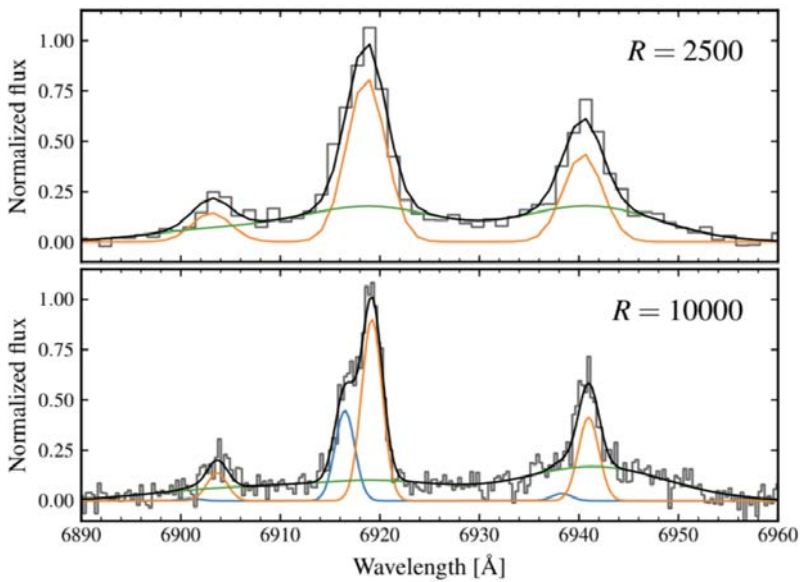

**Figure 7.5:** *Comparison of the details of the line profile in the Hα region as revealed at R=2500 using a spectrograph like MUSE and at R=10000 with a MAVIS like resolution. Multiple kinematical components are revealed only at the higher spectral resolution.*

Moreover, the high-resolution mode of MAVIS paves the road to a detailed study of the chemical composition of resolved stellar populations in the Milky Way densest regions (such as the core of globular clusters) and in extragalactic star clusters and resolved stellar populations, so far not accessible with MUSE. This is shown in Fig. 7.6, where we compare the solar spectrum at a MAVIS-like resolution and at a MUSE-like one. This is further demonstrated in Fig 7.7, which plots the flux ratio of a metal-poor star of [Fe/H] approx -1.0, divided by a solar metallicity star of similar type, for both MUSE spectra (Ivanov et al., 2019) and simulated MAVIS spectra at R = 15,000. The differences in the spectral features are much more prominent in the high resolution MAVIS simulated spectra, and are much less detectable with MUSE even for a large change in metallicity at 1.0 dex.

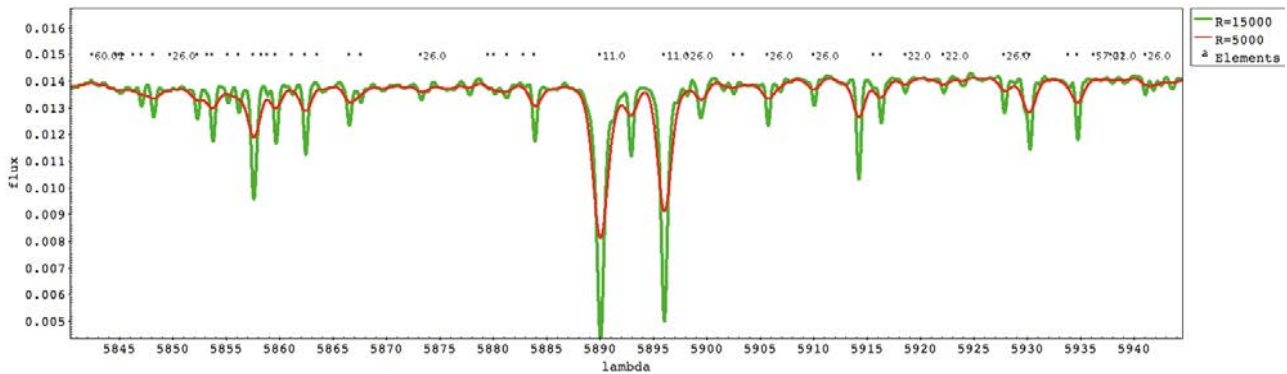

**Figure 7.6**: *Solar spectrum at MAVIS-like resolution (green) and MUSE-like resolution (red), with the atomic number of some elements.*

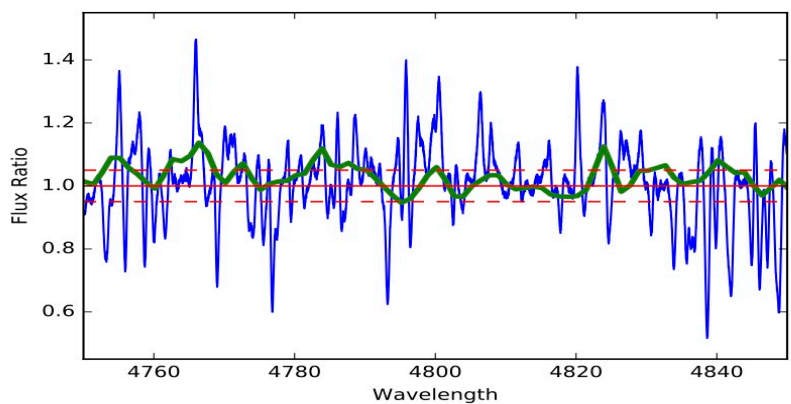

**Figure 7.7**: *The flux ratio in the Blue spectral regions for a metal-poor star of [Fe/H] ~ -1.0 divided by a solar metallicity spectrum of similar spectral type. The ratio for simulated MAVIS spectra at R = 15,000 is in Blue and ratio for MUSE data in Green. The red lines mark the +/- 5% change, corresponding to the 1σ confidence level for S/N=20. MAVIS is clearly much more sensitive to the change in metallicity, with most features changing by many factors of σ. MUSE would require a significantly higher S/N to provide the same accuracy, and most features are simply undetectable at such low resolution.*

**Blue wavelength coverage:** ESO TLRs specify significantly bluer wavelength coverage than MUSE, down to 370nm (TLR-40). Achieving a meaningful MCAO correction at such short wavelengths will be challenging. However, the scientific capability of combined high spatial resolution at blue wavelengths will be unique, accessing lower-redshift (and therefore higher surface brightness) sources of Lyα from a redshift of 2, both for emission and IGM absorption studies, and giving access to the multitude of gas and stellar diagnostic bluer than Hβ in local galaxies that are not obtainable with MUSE. For these reasons, thorough exploration of the technical and scientific trade-offs for maximising the blue coverage is a priority for MAVIS.

For stellar spectroscopy, the access to a significantly bluer wavelength coverage than MUSE will allow us detailed spectroscopic studies of the hotter O and B stars, which, thanks to their higher absolute luminosity, will be observable up to distances of several Mpc. In addition, notable spectral features, as the CN / CH bands around 420-430nm (fundamental to detected first and second population stars in globular stars and to be used as chemical clock in giant stars) and the neutron-capture features around 400-470nm (Sr: 407.7, Ba: 455.4, Eu: 412.9, all important to trace the time-scale of the star formation history in stellar populations) are available on the in the bluest part ot the spectra.

**Red wavelength coverage:** The ESO TLRs specify 950 Å as the red limit for MAVIS spectroscopy, with the goal of reaching 1 μm. Indeed, the increased red coverage offers the possibility to perfectly complement the ERIS wavelength range (covering J band to K band in IFU mode), and providing a continuous sampling in AO assisted IFU spectroscopy on the VLT from 370 Å to 2.5 μm.

The access to Lyα emission in z~7 sources is a key feature for MAVIS, with ERIS and ELT accessing even higher redshifts. This implies that extending the red coverage of MAVIS to 1 micron is important. In fact, each Angstrom gained at λ>9300 Å (MUSE red limit) corresponds to (2-3)x$10^5$ years of cosmic time, eventually reaching 720 Myr after the Big Bang at z=7.22 (Lyα at 1 μm). The wavelength range 9300-10000Å will be therefore crucial to probe the tail-end of the reionization epoch (z ~ 6.5 - 7.2). In addition, the extended wavelength range will allow observation of Hα emission up to z~0.5, boosting the diagnostic power of rest frame optical emission and absorption lines in intermediate redshift galaxies.

And finally, in the same way that MAVIS imaging will strongly leverage the infrared capabilities of MICADO on the ELT, the MAVIS integral field spectrographs will be strongly complementary to HARMONI, which will probe similar spatial scales at longer wavelengths. For low redshift studies, MAVIS will therefore allow access to the well-calibrated optical diagnostic indicators of gas and stellar emission on scales that can be directly compared to infrared data from ELTs (not to mention ALMA and JWST at even longer wavelengths).

# 7.4 Summary of key MAVIS properties

In the following, we briefly summarize the key properties discussed above that make MAVIS an unique facility among those available at ESO and at other ground based or space based observatories:

- **High sky coverage.** This is a crucial aspect of the MAVIS concept, and is really a game-changer in comparison to existing facilities. While the IRLOS upgrade for MUSE-NFM will certainly improve sky coverage, and bring the limiting NGS magnitude close to that of MAVIS, the patrol radius is very limited (~10 arcsec), giving still modest statistical sky coverage about an order of magnitude lower than MAVIS. For most science applications, the targets of interest are too faint, confused, or extended to be suitable NGS sources. The ability to utilize far off-axis NGS sources while maintaining a good correction, and to access a relatively large corrected field for both imaging and spectroscopy, provides MAVIS with a level of science utility unmatched by any previous or existing instrument.

- **Combined imaging and spectroscopy.** As has been mentioned elsewhere, the combination of imaging and spectroscopy is a crucial ingredient for most science cases - for example combining radial velocities *and* proper motions to find intermediate-mass black holes (Section 4.3), deriving stellar surface brightness profiles *and* kinematics of nuclear star clusters (Section 3.6), and measuring sizes and dynamical masses of star-forming clumps (Section 4.7). This underlines the scientific usefulness of MAVIS as a combined imaging and spectroscopic facility instrument. It is also the case that providing imaging and spectroscopy with the same instrument at similar angular resolutions and wavelengths removes a common source of systematic uncertainty when matching up different data sets, yielding increased accuracy of results.

- **Wavelength coverage and spectral resolution.** MUSE is the only available spectrograph for which a meaningful comparison can be made with MAVIS, is clearly a powerful spectral instrument with a strong track record of scientific output, and high demand from the ESO community. The strong limitations of poor sky coverage notwithstanding, MUSE-NFM draws obvious comparisons to the MAVIS spectroscopic capability. The MAVIS White Paper process, and the resulting distilled cases presented here, highlight, however, that there is a strong need for higher spectral resolution capabilities and bluer wavelength coverage, especially when moving to high angular resolution, where dynamical scales naturally become smaller (e.g. lower mass systems, higher complexity gas structure, etc.). In this respect, there is strong *complementarity* between MUSE and MAVIS, without redundancy or duplication.

- **Sensitivity.** MAVIS takes full advantage of the VLT AOF, which is itself unique in the world today. This means not only making full use of the outstanding capabilities of the 4-LGS and telescope-integrated AO, but also maximising the light-collecting power of the VLT. Design choices in MAVIS have been made to not only maximise the PSF correction, but also to maximise throughput in the science instruments; for example, imaging the output of the AO module directly without re-imaging optics to maximise transmission, and the choice of an image-slicer for the IFU to maximise throughput and image quality. In this way, MAVIS will exceed the most ambitious imaging depths obtained with HST achievable in a fraction of the time, and maintain the VLT's complementarity with the ELT working at infrared wavelengths.

# Appendix 1: MAVIS Exposure Time Calculator

## A1.1 Model components

This section describes the various components that enter into the MAVIS exposure time calculator, including the properties of the MAVIS detector, adaptive optics module (AOM), imager, and spectrograph. It also includes an outline of the various models used to describe astrophysical sources and atmospheric emission/absorption.

### A1.1.1 Detector properties

- Readout noise = RON = 3 e⁻/pixel
- Dark current   = DN = 3 e⁻/pixel/hour

### A1.1.2 Sky model

Models for atmospheric absorption and emission were generated using the Advanced Cerro Paranal Sky Model (e.g. Noll et al. 2012; Jones et al. 2016), accessed using the `skycalc_cli` interface. Emission was estimated assuming a monthly averaged solar radio flux of 1.3 MJy. Airmass, precipitable water vapour (PWV), and fractional lunar illumination (FLI) are all free parameters of the model.

### A1.1.3 Source models

Emission from astrophysical sources is included using a variety of methods, including empirical templates, theoretical stellar population synthesis models, and simple emission line models. We briefly outline these different components below.

- **Kinney et al. 1996 templates -** Kinney et al. (1996) provide ultraviolet to near-infrared spectral energy distributions (SEDs) for a variety of prototypical sources including elliptical, bulge, S0, Sa, Sb, and Sc galaxies, as well as starburst spectra with a range of dust attenuation. The resolution of these templates ranges between 6 and 10 Å FWHM
- **Flexible Stellar Population Synthesis -** We include additional templates computed using the Flexible Stellar Population Synthesis (FSPS) code of Conroy & Gunn (2009; 2010).  These allow for a broader range of analytic star-formation histories, as well as extending to higher spectral resolutions (2.54 Å FWHM) than the Kinney et al. (1996) templates described above.
- **Emission line models -** In the case of faint emission line sources, we estimated sensitivities based on a simple Gaussian emission line model, with free parameters controlling the (rest-frame) line wavelength, integrated flux, and intrinsic width.

### A1.1.4 MAVIS Throughput

Figures A1 and A2 show adopted throughput for the spectrograph and imaging subsystems, respectively. These include the adaptive optics module (AOM), detector quantum efficiency (QE), and either the spectrograph (VPH + slicer optics) or imaging filter efficiencies.

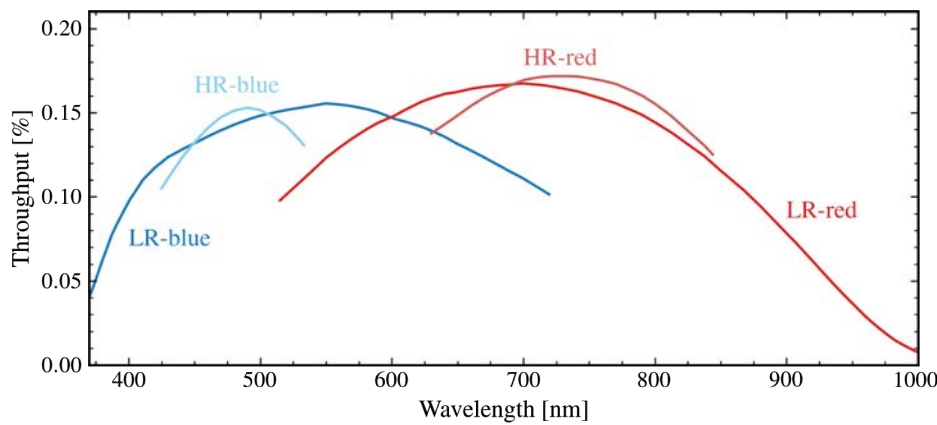

**Figure A1** - *MAVIS spectrograph throughput in each of the four spectral modes.  Throughput includes telescope, AOM, spectrograph optics, and detector quantum efficiency.*

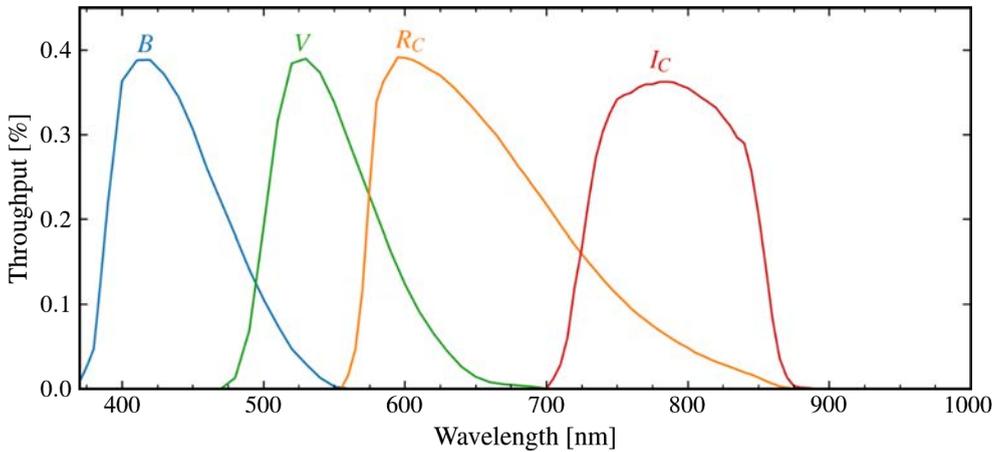

***Figure A2*** *- MAVIS imager throughput in the VBR$_c$I$_c$ filters. Throughput includes telescope, AOM, filter transmission, and detector quantum efficiency.*

## A1.1.5 Ensquared energy

We adopt estimates of the ensquared energy based on simulated MAVIS performance during 'standard conditions' as described in the Tech Spec (i.e. a seeing of 0.87 arcsec at 500 nm along the line of sight at airmass=1.16, median Cn² profile, and favourable tip-tilt/NGS asterism). The resulting ensquared energies are shown in Figure A3 for different wavelengths and spatial binnings.

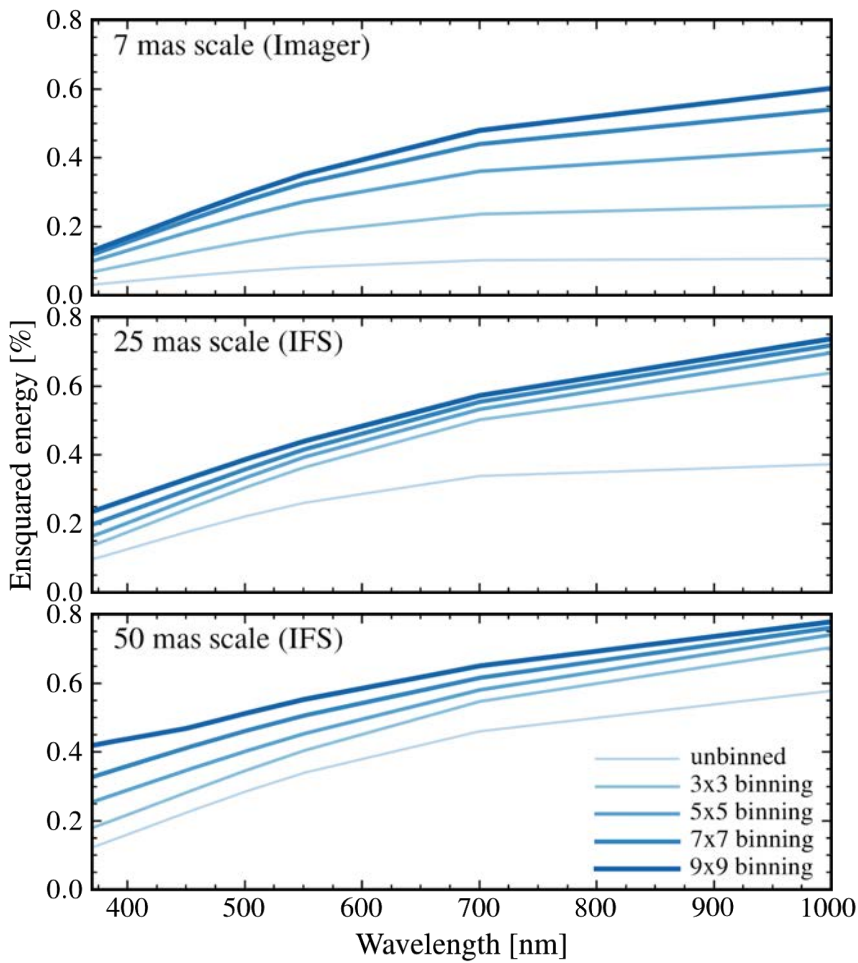

***Figure A3*** *- Ensquared energy as a function of wavelength for different spatial sampling (corresponding to the imager and spectrograph pixel scales) and pixel binning.*

## A1.2 Signal-to-noise estimation

The main formula for computing signal-to-noise per pixel, $(S/N)_{pix}$, as a function of exposure time and source flux is given by

$$(S/N)_{pix} = \frac{\sqrt{N_{dit}}\, F_O K_O t}{\sqrt{F_O K_O t + F_S K_S t + RON^2 N_{pix} + DNN_{pix} t}}\,,$$

where $F_O$ is the object flux in photon s$^{-1}$ m$^{-2}$ µm$^{-1}$, $F_S$ is the sky surface brightness in photon s$^{-1}$ m$^{-2}$ µm$^{-1}$ arcsec$^{-2}$, $t$ is the integration time per exposure, $N_{dit}$ is the number of exposures, $N_{pix}$ is the number of detector pixels and the coefficients $K_O$ and $K_S$ correct for PSF and binning effects such that

$$K_O = T_{MAVIS}(\lambda)\, Area_{UT4}\, EE(\lambda,r)\, \Delta_\lambda\,,$$

and

$$K_S = T_{MAVIS}(\lambda)\, Area_{UT4}\, A_{pix}\, N_{pix}\, \Delta_\lambda\,.$$

In the above, $T_{MAVIS}(\ )$ is the wavelength-dependent throughput shown in Figures A2 and A3, $Area_{UT4}$ is the collecting area of the UT4 mirror in square metres, $EE(\ ,r)$ is the ensquared energy as a function of wavelength and bin size (i.e. Figure A4), $A_{pix}$ is the area of an individual MAVIS spatial pixel, and is the size of a spectral bin in microns.

## A1.3 MAVIS Performance

In this section we provide tabulated performance estimates for MAVIS based on the ETC components described above. For this purpose we consider the "limiting magnitude" as the magnitude at which a $S/N$ of 5 (imager) or 10 (spectrograph) per resolution element is reached in 3600s of observation. In all cases, calculations were performed assuming observations at zenith with a new moon and PWV = 10mm. All magnitudes are given on the AB magnitude system.

### A1.3.1 Imager

Point source sensitivity, 3x3 pixel binning, 5mas TT residual.

|  | $B$ | $V$ | $R_c$ | $I_c$ |
|---|---|---|---|---|
| **Limiting magnitude** | 29.69 | 29.76 | 29.95 | 29.37 |

Extended source sensitivity, unbinned, 5mas TT residual.

|  | $B$ | $V$ | $R_c$ | $I_c$ |
|---|---|---|---|---|
| **Limiting surface brightness** | 21.87 | 21.58 | 21.62 | 20.98 |

## A1.3.2 Spectrograph

### A1.3.2.1 Limiting magnitude curves

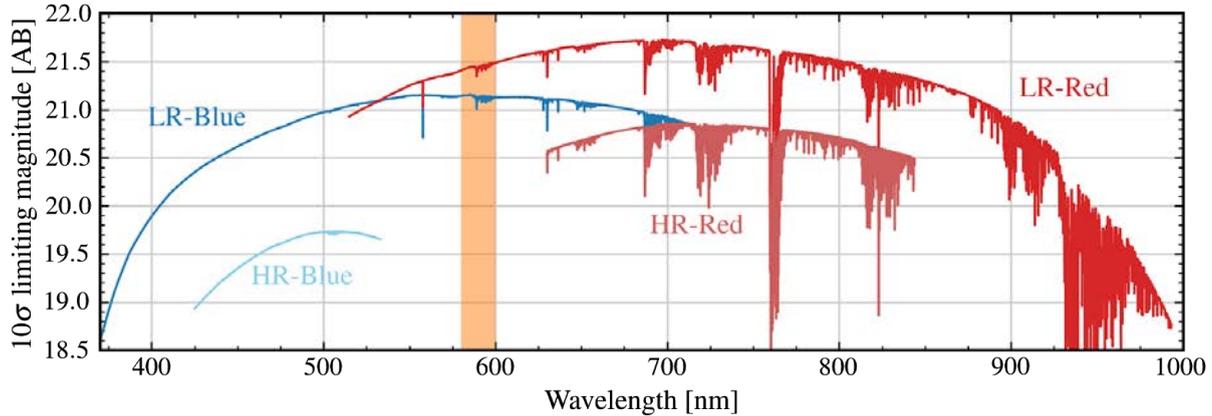

***Figure A4** - Limiting point-source magnitude as a function of wavelength for different spectrograph modes as indicated. The orange shaded region shows the approximate location of the notch filter. Solid lines show limiting magnitudes corresponding to the baseline spec (mAB=21.0@550nm).*

### A1.3.2.2 Tabulated limiting magnitudes

Point source sensitivity, 3x3 spatial binning, 5mas TT residual.

|  | spatial sampling | 370nm | 400nm | 500nm | 550nm | 600nm | 700nm | 800nm | 900nm | 1000nm |
|---|---|---|---|---|---|---|---|---|---|---|
| **LR-blue** | 25 mas | 18.51 | 19.80 | 20.88 | 21.05 | 21.02 | 20.74 |  |  |  |
|  | 50 mas | 18.90 | 20.12 | 21.09 | 21.24 | 21.21 | 20.94 |  |  |  |
| **LR-red** | 25 mas |  |  |  | 21.15 | 21.38 | 21.57 | 21.36 | 20.77 | 18.69 |
|  | 50 mas |  |  |  | 21.35 | 21.56 | 21.75 | 21.55 | 20.98 | 18.92 |
| **HR-blue** | 25 mas |  |  | 19.63 |  |  |  |  |  |  |
|  | 50 mas |  |  | 19.86 |  |  |  |  |  |  |
| **HR-red** | 25 mas |  |  |  |  |  | 20.68 | 20.56 |  |  |
|  | 50 mas |  |  |  |  |  | 20.88 | 20.76 |  |  |

Extended source sensitivity, unbinned, 5mas TT residual.

|  | spatial sampling | 370nm | 400nm | 500nm | 550nm | 600nm | 700nm | 800nm | 900nm | 1000nm |
|---|---|---|---|---|---|---|---|---|---|---|
| **LR-blue** | 25 mas | 13.37 | 14.39 | 14.87 | 14.86 | 14.71 | 14.22 | | | |
| | 50 mas | 14.87 | 15.89 | 16.37 | 16.35 | 16.20 | 15.72 | | | |
| **LR-red** | 25 mas | | | | 14.96 | 15.07 | 15.05 | 14.75 | 14.08 | 11.92 |
| | 50 mas | | | | 16.46 | 16.56 | 16.54 | 16.24 | 15.57 | 13.42 |
| **HR-blue** | 25 mas | | | 13.63 | | | | | | |
| | 50 mas | | | 15.13 | | | | | | |
| **HR-red** | 25 mas | | | | | | 14.16 | 13.95 | | |
| | 50 mas | | | | | | 15.66 | 15.45 | | |

**End of Document**